\documentclass[a4paper,hidelinks, 12pt]{article}
\usepackage[utf8]{inputenc}
\usepackage{a4wide}
\usepackage{graphics}
\usepackage{amsmath}
\usepackage{amsfonts}\usepackage{amssymb}
\usepackage{subfigure}
\usepackage{cite}
\usepackage{xcolor}
\usepackage{soul}
\usepackage[toc,page]{appendix}
\usepackage{mathtools}
\usepackage{hyperref}
\usepackage{enumitem}
\usepackage{threeparttable}
\usepackage{etoolbox}
\usepackage{longtable}
\usepackage{multirow}
\usepackage{threeparttablex}
\usepackage{booktabs}

\allowdisplaybreaks
\numberwithin{equation}{section}

\NewCommandCopy{\originalsubsection}{\subsection}
\RenewDocumentCommand{\subsection}{sd()O{#4}m}{%
  \IfBooleanTF{#1}
    {\originalsubsection*{#4}}%
    {%
     \IfNoValueTF{#2}
       {
        \renewcommand{\thesubsection}{\thesection.\arabic{subsection}}%
       }
       {
        \addtocounter{subsection}{-1}%
        \renewcommand{\thesubsection}{\thesection#2}%
       }
     \originalsubsection[#3]{#4}%
    }%
}

\makeatletter
\let\TPT@hookin\@gobble
\let\TPT@hookarg\@gobble
\newcommand{\labeltext}[2]{%
  \@bsphack
  \csname phantomsection\endcsname 
  \def\@currentlabel{#1}{\label{#2}}%
  \@esphack
}
\makeatother

\begin{document}

\begin{titlepage}
\begin{center}

{\large \bf {\boldmath$U(1)$-charged Dark Matter in three-Higgs-doublet models}}

\vskip 1cm

A. Kun\v cinas,$^{a,}$\footnote{E-mail: Anton.Kuncinas@tecnico.ulisboa.pt}
P. Osland$^{b,}$\footnote{E-mail: Per.Osland@uib.no} and 
M. N. Rebelo$^{a,}$\footnote{E-mail: rebelo@tecnico.ulisboa.pt}

\vspace{1.0cm}

$^{a}$Centro de F\'isica Te\'orica de Part\'iculas, CFTP, Departamento de F\'\i sica,\\ Instituto Superior T\'ecnico, Universidade de Lisboa,\\
Avenida Rovisco Pais nr. 1, 1049-001 Lisboa, Portugal,\\
$^{b}$Department of Physics and Technology, University of Bergen, \\
Postboks 7803, N-5020  Bergen, Norway\\
\end{center}

\vskip 2cm

\begin{abstract}

\noindent

We explore three-Higgs-doublet models that may accommodate scalar Dark Matter where the stability is based on an unbroken $U(1)$-based symmetry, rather than the familiar $\mathbb{Z}_2$ symmetry. Our aim is to classify all possible ways of embedding a $U(1)$ symmetry in a three-Higgs-doublet model. The different possibilities are presented and compared. All these models contain mass-degenerate pairs of Dark Matter candidates due to a $U(1)$ symmetry unbroken (conserved) by the vacuum. Most of these models preserve CP. In the CP-conserving case the pairs can be seen as one being even and the other being odd under CP or as having opposite charges under $U(1)$. Not all symmetries presented here were identified before in the literature, which points to the fact that there are still many open questions in three-Higgs-doublet models. We also perform a numerical exploration of the $U(1) \times U(1)$-symmetric 3HDM, this is the most general phase-invariant (real) three-Higgs-doublet model. The model contains a multi-component Dark Matter sector, with two independent mass scales. After imposing relevant experimental constraints we find that there are possible solutions throughout a broad Dark Matter mass range, 45--2000 GeV, the latter being a scan cutoff.

\end{abstract}

\end{titlepage}

\tableofcontents

\clearpage

\section{Introduction}

Different cosmological observations, based on the Standard Cosmological Model, indicate that most likely a significant part of the Universe is not made of ordinary baryonic matter. It is quite commonly assumed that the missing matter component is a hypothetical cold Dark Matter (DM). The discovery of the Higgs boson~\cite{ATLAS:2012yve,CMS:2012qbp} might be a crucial element in solving the origin of DM since the Higgs portal~\cite{Eboli:2000ze,Godbole:2003it,Patt:2006fw,Kanemura:2010sh,Chu:2011be,Djouadi:2011aa,Djouadi:2012zc,Belanger:2013kya,Curtin:2013fra,Arcadi:2017kky} models could provide an answer to the DM puzzle. In a variety of models, if kinematically allowed, Higgs boson decays into invisible channels can occur, such processes could result from a portal between the visible Standard Model (SM) sector and DM. Therefore, extending the electroweak sector may provide an answer to the DM puzzle.

One of the best studied extended-scalar-sector models with a DM candidate is the Inert Doublet Model (IDM)~\cite{Barbieri:2006dq,LopezHonorez:2006gr,Cao:2007rm}, where the DM candidate is stabilised by a $\mathbb{Z}_2$ symmetry. In the IDM the scalar sector of the SM is extended by a single $SU(2)$ Higgs doublet~\cite{Gunion:1989we,Branco:2011iw}. In the $\mathbb{Z}_2$-symmetric 2HDM, with $\mathbb{Z}_2$ also preserved by the vacuum, there are two neutral states that decouple from the gauge bosons (and from the charged scalars) in the following sense. If we denote the three neutral scalars $\{S_i,\, S_j,\, S_k\}$, and the vector bosons $Z$ and $W$ by $V$, then two couplings $S_iVV$ and $S_jVV$ vanish, as do the $S_iS_kV$ and $S_jS_kV$ couplings, and also the $S_i h^+h^-$ and $S_jh^+h^-$ couplings. But quartic couplings remain non-zero.

Although the IDM is an elegant solution, it is not capable of addressing other SM shortcomings. Bearing in mind some of the shortcomings and taking into consideration that there are no experimental hints pointing towards the IDM, one may invoke multi-Higgs-doublet models ($n$HDM) with more than two doublets. The most natural way to control the number of free parameters of $n$HDMs is to impose symmetries~\cite{Ferreira:2008zy,Ivanov:2011ae,Ivanov:2012ry,Ivanov:2014doa,Pilaftsis:2016erj,deMedeirosVarzielas:2019rrp,Darvishi:2019dbh,Doring:2024kdg}, otherwise the number of parameters would grow rapidly~\cite{Olaussen:2010aq}, and predictability would be lost. The possible discrete symmetries realisable in three-Higgs-doublet models (3HDM) were classified in Ref.~\cite{Ivanov:2012fp} and later all possible Abelian and discrete non-Abelian symmetries were classified in Ref.~\cite{Keus:2013hya}. Apart from controlling the number of free parameters, symmetries can also stabilise DM. 

In the 2HDM, the $\mathbb{Z}_2$-symmetric scalar potential can be described in terms of seven parameters. In a familiar parameterisation this amounts to taking $m_{12}^2=0$, furthermore $\lambda_5$ can be chosen real and $\lambda_6=\lambda_7=0$. A $U(1)$-symmetric potential can be represented in terms of six parameters, since it requires $\lambda_5=0$. In this context, in terms of physical observables, removing a parameter means that there is a degeneracy, or some parameter (mass or coupling) vanishes \cite{Ferreira:2020ana}. In this particular case, as compared to the $\mathbb{Z}_2$-symmetric model, the $U(1)$-symmetric model has a mass degeneracy~\cite{Pilaftsis:2016erj,Haber:2018iwr}, two mass-degenerate states with opposite CP parities. In other words, two real scalars are promoted to a complex scalar, associated with the conserved $U(1)$ charge.

There is also the possibility of breaking $\mathbb{Z}_2$ spontaneously, with both vacuum expectation values (vevs) non-zero, in which case only one neutral state decouples from the gauge bosons (and from the charged scalars). For a $U(1)$-symmetric 2HDM, if the vacuum breaks the $U(1)$, then a massless inert state (the Peccei--Quinn axion \cite{Peccei:1977hh}) emerges. In the 2HDM, softly breaking $\mathbb{Z}_2$ or $U(1)$ leads to other models~\cite{Branco:1985aq,Ferreira:2022gjh}.

Given the qualitative difference between the $\mathbb{Z}_2$-based IDM and the $U(1)$-based models within the 2HDM, it is of interest to explore the corresponding 3HDMs. Indeed, we shall show that in the neutral sector there are always mass degeneracies associated with unbroken $U(1)$ symmetries, like in the 2HDM. Depending on which symmetry remains unbroken there will be one or even two pairs of mass-degenerate neutral states. Whenever CP is conserved, these degenerate pairs have opposite CP parities: one is even and the other is odd.

Multi-Higgs models with continuous symmetries have received limited attention. In the present work we discuss these symmetries and try to answer whether or not they could accommodate  DM candidates. Actually, in the discussed models there will always be two mass-degenerate candidates stabilised by an underlying $U(1)$. Multi-component DM models~\cite{Boehm:2003ha,Ma:2006uv,Hur:2007ur,Cao:2007fy,Zurek:2008qg,Profumo:2009tb,Batell:2010bp,Liu:2011aa,Belanger:2011ww,Belanger:2012vp, Medvedev:2013vsa,Esch:2014jpa,Biswas:2015sva,Cai:2015zza,Arcadi:2016kmk,Ahmed:2017dbb,Bernal:2018aon,Poulin:2018kap,YaserAyazi:2018lrv,Bhattacharya:2018cgx,Elahi:2019jeo,Borah:2019aeq,Nanda:2019nqy,Hall:2019rld,Betancur:2020fdl,DuttaBanik:2020jrj,Chakrabarty:2021kmr,Choi:2021yps,DiazSaez:2021pmg,Hall:2021zsk,Mohamadnejad:2021tke,Yaguna:2021rds,Ho:2022erb,Das:2022oyx,BasiBeneito:2022qxd} have been scrutinised in different physical scenarios. In this work we concentrate on models with three scalar doublets. However, models with additional $SU(2)$ scalar singlets~\cite{Silveira:1985rk,McDonald:1993ex,Burgess:2000yq,Barger:2007im,Andreas:2008xy,GAMBIT:2017gge} are also very promising for solving the DM puzzle and bear many similarities with our framework. Such models could have two- and three-component scalar singlets DM~\cite{Barger:2008jx,Drozd:2011aa,Modak:2013jya,Belanger:2014bga,Bhattacharya:2016ysw,Bhattacharya:2017fid,Pandey:2017quk,Belanger:2020hyh,Yaguna:2021vhb,Belanger:2021lwd,Belanger:2022esk}.

\begin{figure}[htb]
\begin{center}
\includegraphics[scale=1.25]{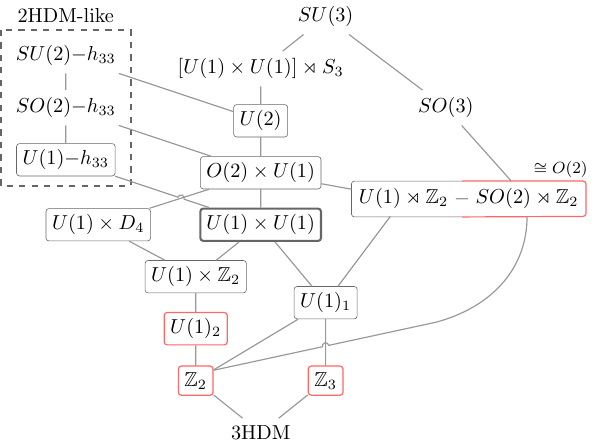}
\end{center}
\vspace*{-3mm}
\caption{Considered 3HDM cases with symmetry-breaking patterns. Symmetries increase from bottom to top, by demanding specific relations among the scalar potential coefficients, with ``3HDM" representing the most general unconstrained complex scalar potential. Lines indicate cases connected by symmetries, where the scalar potential of the upper level is included within the scalar potential of the lower level. The connection is only realised for an appropriate choice of charges for each group in each step. In the case of $O(2)$, $\{U(1) \rtimes \mathbb{Z}_2,\, SO(2) \rtimes \mathbb{Z}_2\}$, the horizontal line within the box indicates a change of basis. Cases enclosed by rectangles indicate the existence of models with a vacuum preserving the full underlying symmetry. Red rectangles indicate cases where CP violation is possible (since we are interested in classifying DM, we assume that at least one of the vevs vanishes). The upper-left panel refers to speculated 2HDM-like cases. It was pointed out that $U(1) \times D_4$ due to a non-trivial intersection of the groups might better be denoted in terms of a quotient group~\cite{Ivanonv_pr}.}
\label{Fig:Symmetry_breaking}
\end{figure}

The symmetry-breaking patterns for the discussed $U(1)$-3HDMs are provided in Figure~\ref{Fig:Symmetry_breaking} in terms of maximal symmetries. All $U(1)$-based symmetric potentials can be constructed by imposing further constraints on a $\mathbb{Z}_2$- or a $\mathbb{Z}_3$-symmetric theory. They can both lead to a $U(1)_1$-symmetric potential, whereas only $\mathbb{Z}_2$ (not $\mathbb{Z}_3$) can also lead to a $U(1)_2$-symmetric potential, with this being a maximal symmetry. Imposing further constraints, the latter can become symmetric under $U(1) \times \mathbb{Z}_2$. Both branches can lead to a $U(1)\times U(1)$-symmetric potential, which is the most general \textit{real} 3HDM. It should also be noted that additional (discrete) symmetries are possible, however only the relevant branches (containing continuous symmetries) are presented in Figure~\ref{Fig:Symmetry_breaking}. For example, both $\mathbb{Z}_2 \times \mathbb{Z}_2$ and $\mathbb{Z}_4$ can be increased to $U(1) \times \mathbb{Z}_2$.

One could also envisage the following construction: a 2HDM sector with a $U(1)$, $SO(2)$ or $SU(2)$-symmetric potential, together with a separate SM-like potential; technically, this would not be a 3HDM. The two sectors could interact via some messenger field, for example the gauge bosons or possibly via fermions. The underlying symmetry must not allow bilinear terms of one sector to couple to those of the other sector, and quartic terms combining both sectors must be forbidden. However, we are not aware of any symmetry or other mechanism that would lead to such a construction and will not discuss it further.

In Section~\ref{Sec:3HDM-review} we review the literature on 3HDMs with DM candidates, most of which is based on $\mathbb{Z}_n$ symmetries. Then, in Section~\ref{Sec:Philosophy} we outline the philosophy of our study, and in Section~\ref{Sec:Pot_U1U1} we explore features of the ${U(1)\times U(1)}$-based model of DM. Sections~\ref{Sec:Pot_U11}--\ref{Sec:Pot_U12} are devoted to less symmetric (with more free parameters) potentials, defined by $U(1)_1$, $U(1)\times\mathbb{Z}_2$ and $U(1)_2$. In Section~\ref{Sec:Pot_U1_additonal} we explore the previously covered $U(1)$-based symmetries by further on restricting the parameter space when additional permutation symmetries are applied. These constructions allow us to identify three new symmetries, which have not been previously discussed in the 3HDM literature. These symmetries along with the remaining continuous symmetries are analysed in Sections~\ref{Sec:Pot_O2_U1}--\ref{Sec:Pot_SU3}.

In Section~\ref{Sec:Different_DM_cases} we summarise DM candidates by discussing models where the imposed symmetry is unbroken by the vacuum. In Section~\ref{Sec:Pot_Zn} we cover possible implementations within the $\mathbb{Z}_2$- and $\mathbb{Z}_3$-symmetric 3HDMs since all discussed models can be presented in terms of these two with additional constraints, \textit{i.e.}, relating couplings. In section~\ref{Sec:U1U1_analysis} we review various properties of the DM candidates in the $U(1)\times U(1)$-based model. In this model there are {\it two} pairs of mass-degenerate neutral states, \textit{i.e.}, four DM candidates. After restricting the parameter space by experimental constraints coming from Particle Physics, Astrophysics and Cosmology, we found that the two mass scales (of the neutral scalars) contribute to a broad range of possible DM masses. 

\vspace{3pt}

\section{Studied 3HDMs with Dark Matter candidates}\label{Sec:3HDM-review}

In the construction of scalar DM models it is common to invoke discrete symmetries, specifically $\mathbb{Z}_n$ symmetries~\cite{Ivanov:2012hc}. We discuss $\mathbb{Z}_2$-symmetric and $\mathbb{Z}_3$-symmetric 3HDMs in Section~\ref{Sec:Pot_Zn} and further on present specific implementations in Appendices~\ref{App:Pot_Z2} and \ref{App:Pot_Z3}. Several 3HDMs that can accommodate DM are listed below:
\begin{itemize}
\item \textbf{IDM2 model with \cite{Grzadkowski:2009bt,Grzadkowski:2010au,Osland:2013sla} and without~\cite{Merchand:2019bod} CP violation}\\
The IDM2 model has two active doublets. In general, this leads to Flavour-Changing Neutral Currents (FCNC). There are several ways to control FCNC. One can assume a $\mathbb{Z}_2^\prime$ symmetry between the active doublets, with one of the doublets charged under the $\mathbb{Z}_2^\prime$ symmetry with a charge identical to that of the right-handed up-quarks. The $\mathbb{Z}_2^\prime$ is softly broken, leading to CP violation in the 2HDM. Then, an inert doublet is added, stabilised by a $\mathbb{Z}_2$ symmetry. The number of free parameters is reduced by adopting the ``dark democracy" assumption---the dark doublet couples identically to both active doublets. In light of the recent constraints from the direct DM detection (XENONnT~\cite{XENON:2023cxc} and LUX-ZEPLIN~\cite{LZ:2022lsv}), see Figure~16 of Ref.~\cite{Grzadkowski:2010au}, it looks like the allowed DM mass region is comparable to that of the IDM. The model was re-evaluated \cite{Merchand:2019bod}, assuming real couplings. The allowed DM mass regions were found to be 57--73 GeV and 500--1000 GeV.

\item \textbf{Truncated \boldmath$\mathbb{Z}_2$-3HDM with~\cite{Keus:2014jha,Keus:2014isa,Keus:2015xya,Cordero:2017owj,Dey:2023exa} and without~\cite{Cordero-Cid:2016krd,Cordero-Cid:2018man,Cordero-Cid:2020yba} CP violation} \\
The main difference of these models from the IDM2 model is that there is a single active doublet and two inert doublets. Since there is a single active doublet, one does not need to worry about FCNC (one requires an SM-like Yukawa Lagrangian). In the $\mathbb{Z}_2$-symmetric 3HDM there is a single bilinear and eight quartic phase-dependent couplings (see appendix~\ref{App:Pot_Z2}). It is argued that the phenomenology of the model does not change if only the bilinear term and the $\mathbb{Z}_2 \times \mathbb{Z}_2$-respecting quartic couplings (three in total) are considered. In order to simplify the analysis, due to the many degrees of freedom of the parameter space, couplings are further on related and numerical simplifications are assumed. Therefore, we refer to these models as ``truncated" $\mathbb{Z}_2$-3HDM. An extensive collider-oriented analysis was conducted in the references mentioned above. Several relevant cosmological cuts were also imposed. After applying the collider and cosmological cuts two allowed regions were identified. The lighter DM mass is compliant with the region around 53--75 GeV. Due to the additional inert doublet the heavy DM mass region can be as light as around 360 GeV due to the co-annihilation channels with additional inert scalars~\cite{Keus:2015xya}. As in the case of the IDM model, when considering heavy DM states (heavier than 500 GeV), the mass-splitting between the scalars coming from the inert sector should be tuned.

\item \textbf{\boldmath$\mathbb{Z}_2 \times \mathbb{Z}_2$-3HDM \cite{Hernandez-Sanchez:2020aop,Hernandez-Sanchez:2022dnn,Boto:2024tzp}}\\
The $\mathbb{Z}_2 \times \mathbb{Z}_2$-3HDM with real coefficients was studied. The symmetry is unbroken if the vacuum contains only a single non-vanishing vev. The doublet accommodating the non-zero vev is a singlet under both $\mathbb{Z}_2$ symmetries. Due to the two different underlying $\mathbb{Z}_2$ symmetries the lightest particles of the $\mathbb{Z}_2$-odd doublets are both associated with DM candidates. It should be noted that it is not possible to completely prevent the conversion processes between the two DM candidates. As a result, due to the conversion processes the dominant contribution to the relic density is made up by the lightest DM state, while the contribution of the heavier one accounts for several per cent. A limited analysis of the parameter space was performed in Refs.~\cite{Hernandez-Sanchez:2020aop,Hernandez-Sanchez:2022dnn}, with the lightest DM candidate in the mass range of 65--80~GeV, and the heavier state at the $\mathcal{O}(100)$ GeV scale. In light of the XENONnT~\cite{XENON:2023cxc} and LUX-ZEPLIN~\cite{LZ:2022lsv} direct DM detection constraints, the mass of the lightest surviving DM candidate is around 71--73 GeV. Furthermore, the upcoming fifteen-year studies of the dwarf spheroidal galaxies by Fermi-LAT~\cite{Hess:2021cdp} may completely rule out the covered parameter space. Recently, the model was re-evaluated in Ref.~\cite{Boto:2024tzp} with a broader parameter space, allowing the scalars to be as heavy as 1 TeV. The allowed DM mass regions of each sector are both similar to those of the IDM.

\item \textbf{\boldmath$\mathbb{Z}_3$-3HDM \cite{Aranda:2019vda,Hernandez-Otero:2022dxd}}\\
The only vacuum which respects the $\mathbb{Z}_3$ symmetry has two vanishing vevs. In the $\mathbb{Z}_3$-3HDM~\cite{Aranda:2019vda} there are two mass-degenerate DM candidates coming from the same inert doublet, and hence being assigned opposite CP numbers. In general, contributions of the two DM candidates to the relic density are of the same order. Later, in Ref.~\cite{Hernandez-Otero:2022dxd} the collider dynamics was studied, pointing to the fact that the final-state spectra have distinctive shapes. This model was coined Hermaphrodite DM. Since two of the DM candidates are associated with the same doublet the trilinear gauge vertices (with the $Z$ boson) have to be controlled (which coincides with maximal mixing between the inert doublets) to comply with the direct detection experiments. In Ref.~\cite{Hernandez-Otero:2022dxd} the $\mathbb{Z}_3$ symmetry was assumed to be softly broken. Then, the interaction strength of the trilinear scalar-gauge vertex becomes proportional to the soft-term coupling. Several benchmark scenarios were considered in Refs.~\cite{Aranda:2019vda,Hernandez-Otero:2022dxd}. The DM mass regions compatible with the collider and cosmological constraints were found to be 53--77 GeV or above some 420 GeV.

\item \textbf{\boldmath$S_3$-3HDM \cite{Khater:2021wcx,Kuncinas:2022whn,Kuncinas:2023hpf}}\\
In the $S_3$-symmetric 3HDM with real couplings there are three possible vacuum configurations which could result in a viable DM candidate. There are up to eight additional models if soft symmetry breaking is allowed. The DM candidate is stabilised by the $\mathbb{Z}_2$ symmetry which survives breaking of the $S_3$ symmetry. The model with real vacuum, R-II-1a (we adopt the nomenclature of Ref.~\cite{Emmanuel-Costa:2016vej}), was covered in Ref.~\cite{Khater:2021wcx}. The allowed DM mass region was found to be 53--89 GeV. Heavy DM states are not allowed as the portal coupling grows with the DM mass, while it should be close to zero for the relic density parameter to be satisfied. Another case, C-III-a, allows for spontaneous CP violation~\cite{Kuncinas:2022whn}. In this case the allowed DM mass region is 7--45 GeV. Assuming certain DM halo distribution profiles the experimental bounds from the indirect DM searches could completely rule out the C-III-a model. Adopting the most generous profile the C-III-a DM mass range is reduced to around 29--45 GeV~\cite{Kuncinas:2023hpf}. The third model is R-I-1, which is different from the other two by having only a single non-vanishing vev. In the case of complex couplings there are two additional implementations~\cite{Kuncinas:2023ycz} with vevs resembling those of the R-I-1 and C-III-a cases that could accommodate DM. 

\item \textbf{\boldmath$S_3 \rtimes \mathbb{Z}_2$-3HDM \cite{Machado:2012gxi,Fortes:2014dca}}\\
A case related to the previous one was studied in Refs.~\cite{Machado:2012gxi,Fortes:2014dca} by increasing the symmetry to ${S_3 \times \mathbb{Z}_2}$ (which should better be denoted as ${S_3 \rtimes \mathbb{Z}_2}$) and further softly breaking it to lift the mass degeneracy. As pointed out in Ref.~\cite{Das:2014fea}, see also Refs.~\cite{Emmanuel-Costa:2016vej, Kuncinas:2020wrn}, such potential, in fact, corresponds to a continuous $O(2)$ symmetry. They centered their study on a model with two vanishing vevs. The model was not systematically studied numerically, but benchmark points in the region of 40--150 GeV were provided~\cite{Fortes:2014dca}.

\item \textbf{CP4-3HDM \cite{Ivanov:2018srm}}\\
The DM candidate can also be stabilised by a CP symmetry combined with an internal symmetry. One possibility involves utilising the CP4 symmetry\footnote{ There is a misleading naming of the CP$n$ symmetries, with $n$ (an integer) indicating either the order of the CP symmetry or a running index. In Ref.~\cite{Bree:2024edl} it was suggested to use CP$x$, where $x$ is a letter.}~\cite{Ivanov:2015mwl}, see also Ref.~\cite{Haber:2018iwr}, which is assumed to be unbroken by the vacuum. The model contains two neutral mass-degenerate DM candidates protected by the underlying CP4 symmetry. In Ref.~\cite{Ivanov:2018srm} it was assumed that the thermal evolution of the two DM candidates takes place in the asymmetric regime \cite{Petraki:2013wwa,Zurek:2013wia}; however, it should be noted that the two mass-degenerate DM candidates of the CP4-3HDM are not a particle--antiparticle pair. Comparing the CP4-3HDM to the conventional asymmetric DM models, there is an additional conversion process between the DM candidates in the CP4-3HDM, which results in slightly more freedom. The CP4 asymmetric DM model is valid only at the electroweak scale since additional physics at a higher mass scale is required to explain the initial asymmetry between the DM states. A detailed scan of the parameter space was not performed.
\end{itemize}

\begin{figure}[htb]
\begin{center}
\includegraphics[scale=0.64]{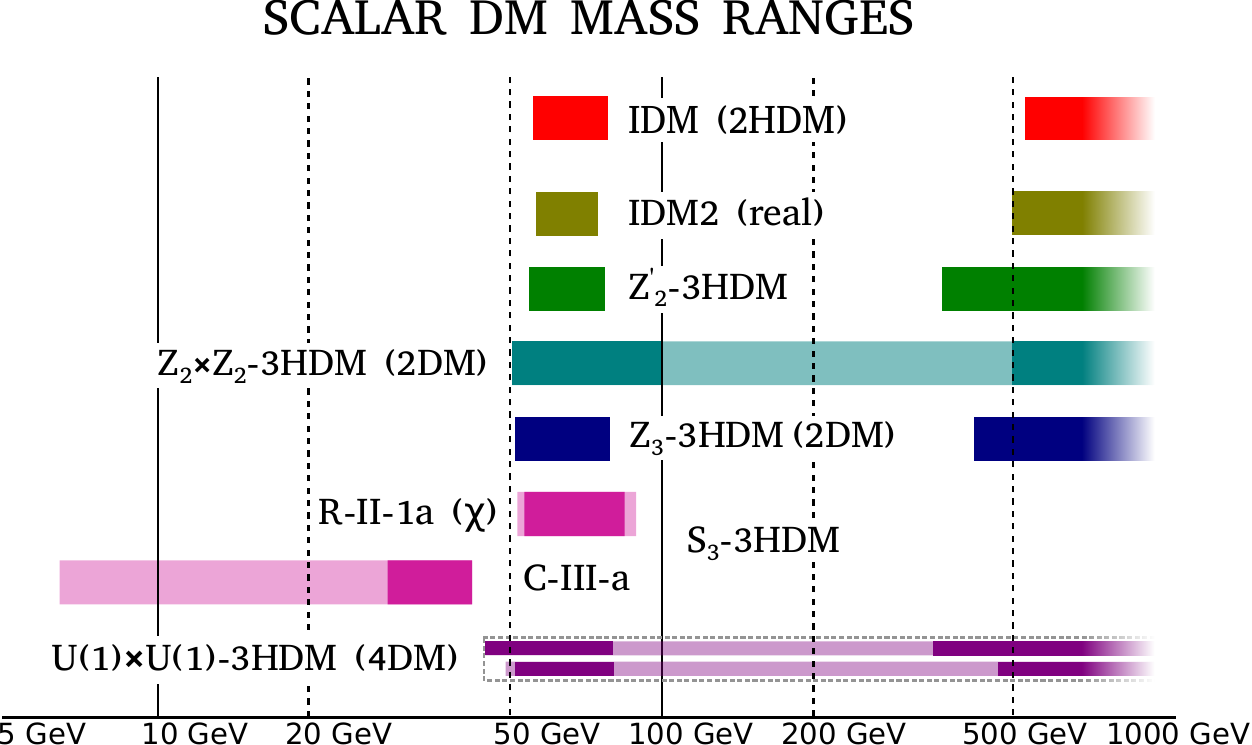}
\end{center}
\vspace*{-3mm}
\caption{Sketch of allowed DM mass ranges in the IDM and various 3HDMs up to 1 TeV. Red: IDM according to Refs.~\cite{Belyaev:2016lok,Kalinowski:2018ylg}. Olive: IDM2 with real coefficients~\cite{Merchand:2019bod}. Green: truncated $\mathbb{Z}_2$-3HDM without~\cite{Keus:2014jha,Keus:2014isa,Keus:2015xya,Cordero:2017owj,Dey:2023exa} and with~\cite{Cordero-Cid:2016krd,Cordero-Cid:2018man,Cordero-Cid:2020yba} CP violation. Teal: $\mathbb{Z}_2 \times \mathbb{Z}_2$ with two DM candidates~\cite{Boto:2024tzp}. The darker regions are consistent with solutions where a single doublet predominantly saturates the relic density. Navy: $\mathbb{Z}_3$ with two mass-degenerate DM candidates~\cite{Aranda:2019vda,Hernandez-Otero:2022dxd}. Pink: $S_3$-3HDM without CP violation, R-II-1a~\cite{Khater:2021wcx} and with CP violation, C-III-a~\cite{Kuncinas:2022whn}. The darker region is consistent with refined constraints of Ref.~\cite{Kuncinas:2023hpf}. Purple: the $U(1) \times U(1)$-symmetric model considered here. The two bands represent independent mass scales coming from different inert doublets. The darker regions are consistent with solutions when a single doublet predominantly saturates the relic density. Combinations of mass scales in the intermediate mass region, of the lighter DM candidate 100--300 GeV and the heavier one in 100--450 GeV, are excluded.}
\label{Fig:DM_mass_ranges_different_models}
\end{figure}

The DM mass ranges of these models are schematically represented in Figure~\ref{Fig:DM_mass_ranges_different_models}.

\vspace{6pt}

\section{Philosophy}\label{Sec:Philosophy}

We are interested in identifying possible DM candidates, the goal is to cover different vacua with at least one vanishing entry. We want to identify all possible cases and therefore allow for spontaneous symmetry breaking (SSB); however, we shall predominantly focus on vacua that do not spontaneously break the underlying symmetry.

In terms of Figure~\ref{Fig:Symmetry_breaking} we start in the centre with the most general phase-independent scalar potential, which is $U(1)\times U(1)$. This case is explored in Section~\ref{Sec:Pot_U1U1}. Then we relax several constraints, covering different cases presented in Figure~\ref{Fig:Symmetry_breaking}, leading to less symmetric potentials (with more allowed couplings) in Sections~\ref{Sec:Pot_U11}--\ref{Sec:Pot_U1_additonal}.

Following Ref.~\cite{Ivanov:2011ae}, there are several ways to assign $U(1)$ charges to three doublets:
\begin{subequations}\label{Eq:U11_U12_charges}
\begin{align}
\begin{split}
U(1)_1 (\alpha) \equiv {}& \mathrm{diag} \left(e^{-i \alpha},\, e^{i \alpha},\, 1 \right) \\
 ={}& e^{i \alpha} \mathrm{diag} \left(e^{-2i \alpha},\,1,\, e^{-i\alpha} \right) = e^{-i \alpha} \mathrm{diag} \left(1,\, e^{2i \alpha},\, e^{i\alpha} \right),\label{Eq:U11_U12_charges_U1a}
\end{split}\\
\begin{split}
U(1)_2 (\beta) \equiv {}& \mathrm{diag} \left(e^{i \beta/3},\, e^{i \beta/3},\, e^{-2 i \beta/3} \right)\\
={}& e^{i \beta/3}\mathrm{diag} \left(1,\, 1,\, e^{-i \beta} \right) = e^{- 2i \beta/3}\mathrm{diag} \left(e^{i \beta},\, e^{i \beta},\, 1 \right),\label{Eq:U11_U12_charges_U1b}
\end{split}
\end{align}
\end{subequations}
where the overall phase factors can be absorbed by a hypercharge rotation. An alternative way would be to choose any pair of the following assignments,
\begin{subequations}\label{Eq:U1i_charges}
\begin{align}
U(1)_{h_1} ={}&  \mathrm{diag} \left(e^{i \alpha},\,1,\,1\right),\\
U(1)_{h_2} ={}&  \mathrm{diag} \left(1,\,e^{i \alpha},\,1\right),\\
U(1)_{h_3} ={}&  \mathrm{diag} \left(1,\,1,\,e^{i \alpha}\right).
\end{align}
\end{subequations}

These $U(1)$ transformations can be expressed in terms of those given by eqs.~\eqref{Eq:U11_U12_charges}
making use of the overall $U(1)$ symmetry as:
\begin{subequations}
\begin{align}
U(1)_{h_1} ={}& e^{i \alpha/3} ~ U(1)_1 ( -{\alpha}/{2}) ~  U(1)_2 ({\alpha}/{2}), \\
U(1)_{h_2} ={}&  e^{i \alpha/3} ~ U(1)_1 ( {\alpha}/{2}) ~  U(1)_2 ({\alpha}/{2}), \\
U(1)_{h_3} ={}&  e^{i \frac{1}{3} (2 \pi +\alpha)} ~ U(1)_1 ( \pi ) ~  U(1)_2 (\pi - \alpha).
\end{align}
\end{subequations}

Subsequently in our work we take again the  $U(1) \times U(1)$ symmetry as a starting point, and consider more symmetric cases (with fewer allowed couplings), following Figure~\ref{Fig:Symmetry_breaking}. These are discussed in Sections~\ref{Sec:Pot_O2_U1}, \ref{Sec:Pot_U1_U1_S3}, \ref{Sec:Pot_U2}, \ref{Sec:Pot_SU3}. Symmetries presented in Sections~\ref{Sec:Pot_O2}, \ref{Sec:Pot_U1_Z2_S2}, \ref{Sec:Pot_SO3} do not share direct similarities (being able to relate couplings) with $U(1) \times U(1)$.
 
In Figure~\ref{Fig:Symmetry_breaking} some symmetries require combining the $U(1)$ transformations with permutations, \textit{e.g.}, applying a permutation symmetry, $h_1 \leftrightarrow h_2$, to the $U(1)_1$-symmetric 3HDM leads to an $O(2)$-symmetric potential. In the re-phasing basis of $O(2)$, the transformation
\begin{equation}
\mathcal{R} = \begin{pmatrix}
0 & e^{i \delta} & 0 \\
e^{-i \delta} & 0 &  0\\
0 & 0 & 1
\end{pmatrix},
\end{equation}
acting on the $\{h_1,\, h_2,\, h_3\}$ fields leaves the potential invariant. This illustrates how the $O(2)$ symmetry can be viewed as a $U(1)$-based symmetry. In fact, applying additional symmetries on top of the $U(1)$ symmetries, the underlying symmetries are enlarged.

\begin{figure}[htb]
\begin{center}
\includegraphics[scale=0.975]{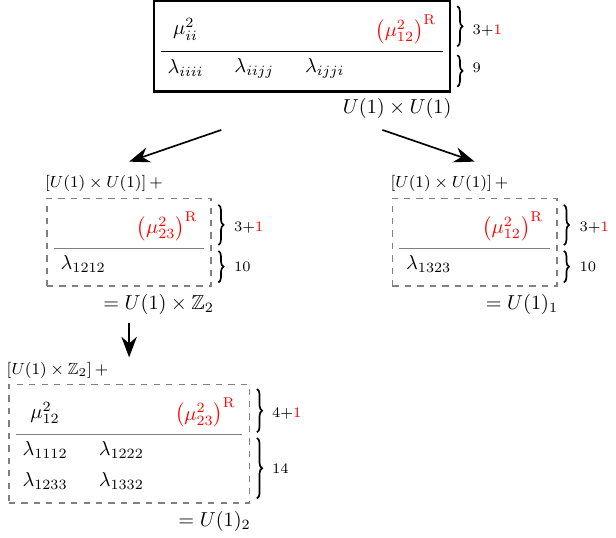}
\end{center}
\vspace*{-3mm}
\caption{Allowed couplings for some $U(1)$-based potentials. Red terms represent possible soft symmetry-breaking terms. The underlying symmetry is specified below each block. To the right of the blocks, the total numbers of bilinear and quartic terms are indicated.}
\label{Fig:Symmetries_Couplings}
\end{figure}

The scalar doublets are decomposed as
\begin{equation}\label{Eq:Extract_phase}
h_i = e^{i \sigma_i} \begin{pmatrix}
h_i^+ \\
\frac{1}{\sqrt{2}}\left( \hat{v}_i  + \eta_i + i \chi_i \right)
\end{pmatrix},
\end{equation}
where we shall assume that the vacuum, $v_i$, is complex with a phase $\sigma_i$ and the hatted variables, $\hat{v}_i$, denote absolute values. One of the $\sigma_i$ phases can always be rotated away due to the overall $U(1)_Y$ symmetry. Since we are interested in DM, we shall assume that at least one of the vevs vanishes; we do not consider cases with fine-tuning of the DM stability, \textit{e.g.}, requiring the lifetime of the DM candidate to be compatible with the age of the Universe, which could be consistent with all vevs being non-zero.  Since there is at least a single vanishing vev, one of the $\sigma_i$ phases can be absorbed, and thus we shall consider $(\hat v_1 e^{i \sigma},\, \hat v_2, 0)$ and permutations. Furthermore, we only consider cases where the neutral states of the would-be DM candidates do not mix with those of the visible sector.

There are different possible ways to check the CP properties of the model~\cite{Mendez:1991gp,Lavoura:1994fv,Botella:1994cs,Branco:2005em,Davidson:2005cw,Gunion:2005ja,Haber:2006ue,Nishi:2006tg,deMedeirosVarzielas:2016rii,Ogreid:2017alh,Trautner:2018ipq,Ivanov:2019kyh}. Allowing both complex vevs together with complex couplings, which we expand as,
\begin{equation}
\mu_{ij}^2 \equiv (\mu_{ij}^2)^\mathrm{R} + i (\mu_{ij}^2)^\mathrm{I} \text{ and } \lambda_{ijkl} \equiv \lambda_{ijkl}^\mathrm{R} + i \lambda_{ijkl}^\mathrm{I},
\end{equation}
might result in at least one unphysical phase. In the cases of spontaneous CP violation all phases can be transferred to the vevs. In all cases a change of basis can make the vevs real. Such re-phasings of the doublets do not correspond to symmetries, they are simply a choice of basis and have no physical implications. However, such change of basis would alter the form of the potential in many of the cases we studied. For example, the underlying symmetry might allow for phase-sensitive couplings, which are forced by the symmetry to be real, in some particular basis, then some couplings might split into several terms due to a phase coming from the vacuum. Whenever this happens, we prefer to keep the vacuum complex---allowing for complex coefficients together with complex vevs. This will be the case for the following symmetries: $O(2) \times U(1)$, $O(2)$, $U(1) \times D_4$, $SO(3)$. 

As stated above, we are interested in cases with at least one vanishing vev, which could be stabilised by a $U(1)$ symmetry. We only identified cases starting from the basis where the symmetry under consideration is apparent and unbroken by the vacuum. Potentially good DM candidates could still be found in scenarios with SSB. Such study is beyond the scope of this paper. Notice that there is always a change of basis which transforms $(v_1,\, v_2,\, v_3)$ into $(v_1^\prime,\, v_2^\prime,\, 0)$. In the new basis the symmetry may not be apparent.

In Figure~\ref{Fig:Symmetries_Couplings} we present a schematic representation of several considered $U(1)$-based models, with the number of bilinear and quartic couplings that respect the underlying symmetry before solving for the minimisation conditions, also including soft symmetry-breaking terms. As will be discussed in the sequel, it is sufficient to assume that the soft breaking terms are purely real for these models, explicitly denoted $\left( \mu_{ij}^2 \right)^\mathrm{R}$. The diagram should be read from top to bottom, with each subsequent block representing a lower symmetry. This indicates that each of the connected blocks includes the previous terms, \textit{e.g.}, the full list of terms for the $U(1)_2$-symmetric potential contains both the $U(1) \times U(1)$ and the $U(1) \times \mathbb{Z}_2$ terms. A full picture of the allowed terms is presented in Table~\ref{Table:Diff_Cases}.

\section{\texorpdfstring{\boldmath$U(1)\times U(1)$}{U(1) x U(1)}-symmetric 3HDM}\label{Sec:Pot_U1U1}

We start by considering the most general 3HDM with a phase-independent (real) scalar potential, which is generated by two distinct $U(1)$ symmetries. For example, we can apply simultaneously the $U(1)_1$ and $U(1)_2$ symmetries, so that the $SU(2)$ scalar doublets acquire charges
\begin{equation*}
\begin{aligned}
U(1)_1 \times U(1)_2 &: {} \mathrm{diag} \left( e^{-i \alpha},\, e^{i \alpha},\, e^{-i \beta} \right) = e^{-i \alpha} \mathrm{diag} \left( 1,\, e^{2 i \alpha},\, e^{i (\alpha-\beta)} \right)\\
& = {} e^{i \alpha} \mathrm{diag} \left( e^{-2 i \alpha},\, 1,\, e^{-i (\alpha+\beta)} \right) = e^{-i \beta} \mathrm{diag} \left( e^{-i (\alpha-\beta)},\, e^{i (\alpha+\beta)},\, 1 \right).
\end{aligned}
\end{equation*}
In this case there will be a non-trivial intersection for $\alpha = \pi$ and $\beta/3 = \pi$. One should be cautious with a direct product of the two $U(1)$ groups. Strictly speaking, it can only be defined when the only common element of the two groups is the identity element. The two phases of the two $U(1)$ groups are independent.

We shall present the scalar potential in terms of the $SU(2)$ singlets,
\begin{equation}
h_i^\dagger h_j \equiv h_{ij}.
\end{equation}
The most general phase-independent (real) scalar potential of the 3HDM is given by
\begin{equation}\label{V_U1xU1}
V_0 = \sum_i \mu_{ii}^2 h_{ii} + \sum_i \lambda_{iiii} h_{ii}^2 + \sum_{i<j} \lambda_{iijj} h_{ii} h_{jj} + \sum_{i<j} \lambda_{ijji} h_{ij} h_{ji},
\end{equation}

Apart from defining the scalar potential we need to consider different implementations which arise due to several possible vacuum configurations. Since the scalar potential is highly symmetric (the form of the scalar potential is left intact under any of the $S_3$ permutations among the doublets) we do not need to consider different permutations of vevs. For example, the scalar content and interactions of $(v,\,0,\,0)$ is identical to those of $(0,\,v,\,0)$ and $(0,\,0,\,v)$. Moreover, the absence of complex couplings in the $U(1)\times U(1)$ scalar potential indicates that the phases of the vacuum can be rotated away and hence are unphysical. We can exploit this property and restrict the discussion to real vacua.  Table~\ref{Table:U1_U1_Cases} summarises all possible cases.

{{\renewcommand{\arraystretch}{1.3}
\begin{table}[htb]
\caption{Different DM candidate implementations within the $U(1)\times U(1)$-symmetric 3HDM. Cases with a dash in the second column stand for the ones where the underlying symmetry is spontaneously broken, while the ``$\checkmark$" indicates that the underlying symmetry is preserved by the vacuum. In the third column the scalar potential is provided with additional, \textit{e.g.}, soft terms. In the fourth column patterns of mixing of the neutral states are presented; fields within braces mix to form a block-diagonal structure. Additional information is presented in the last column. There is no CP violation in any of the $U(1)\times U(1)$-symmetric 3HDM implementations.}
\label{Table:U1_U1_Cases}
\begin{center}
\begin{tabular}{|c|c|c|c|c|} \hline\hline
Vacuum & SYM & $V$ & \begin{tabular}[l]{@{}c@{}} Mixing of the \\ neutral states\end{tabular} & Comments \\ \hline
$(\hat{v}_1,\, \hat{v}_2,\, 0)$ & - & $V_0$ & \footnotesize\begin{tabular}[l]{@{}c@{}} $\{\eta_1,\, \eta_2\}{-}\{\eta_3\}$\\${-}\{\chi_1\}{-}\{\chi_2\}{-}\{\chi_3\}$ \end{tabular} & \footnotesize\begin{tabular}[l]{@{}c@{}} $m_{\eta_3} = m_{\chi_3}$ \\ $m_{\chi_1} = m_{\chi_2} = 0$ \end{tabular} \\ \hline
$(\hat{v}_1,\, \hat{v}_2,\, 0)$ & - & $V_0 + (\mu_{12}^2)^\mathrm{R}$ & \footnotesize\begin{tabular}[l]{@{}c@{}} $\{\eta_1,\, \eta_2\}{-}\{\eta_3\}$ \\ ${-}\{\chi_1,\, \chi_2\}{-}\{\chi_3\}$ \end{tabular} & $m_{\eta_3} = m_{\chi_3}$ \\ \hline
$(v,\, 0,\, 0)$  & $\checkmark$ & $V_0$ & diagonal & \footnotesize\begin{tabular}[l]{@{}c@{}} $m_{\eta_2} = m_{\chi_2},~m_{\eta_3} = m_{\chi_3}$ \end{tabular} \\ \hline \hline
\end{tabular}\vspace*{-9pt}
\end{center}
\end{table}}

\subsection{One vanishing vev}

First, let us consider the vacuum given by $(\hat{v}_1,\, \hat{v}_2,\, 0)$. The minimisation conditions are:
\begin{subequations}\label{Eq:U1U1_mincon1}
\begin{align}
\mu_{11}^2 ={}& - \lambda_{1111} \hat v_1^2 - \frac{1}{2}\left( \lambda_{1122} + \lambda_{1221} \right) \hat v_2^2,\\
\mu_{22}^2 ={}& - \lambda_{2222} \hat v_2^2 - \frac{1}{2}\left( \lambda_{1122} + \lambda_{1221} \right) \hat v_1^2.
\end{align}
\end{subequations}
There is a pair of neutral mass-degenerate states associated with the $h_3$ doublet. A vacuum with only one vanishing vev will always break the $U(1) \times U(1)$ symmetry.

It is known that spontaneously broken continuous symmetries lead to unwanted Goldstone bosons. A possible solution would be to introduce soft symmetry-breaking terms,
\begin{equation}
V_\mathrm{soft} = \mu_{12}^2 h_{12} + \mu_{13}^2 h_{13} + \mu_{23}^2 h_{23} + \mathrm{h.c.},
\end{equation}
where the coefficients of the bilinear terms can be \textit{complex}. Some of the soft symmetry-breaking terms will introduce mixing between $h_3$ and the other doublets $(h_1,\, h_2)$. This would allow for the decay of the DM candidate (associated with $h_3$) into SM particles. A possible approach would be to introduce only those soft symmetry-breaking terms which do not result in unwanted mixing, \textit{i.e.}, $\mu_{13}^2= \mu_{23}^2=0$, leaving only $\mu_{12}^2$ as a soft symmetry-breaking term. With the introduction of the soft symmetry-breaking term one must be careful about the vacuum phase, $\sigma$, not to miss any additional implementations. Let us consider $(\hat{v}_1 e^{i \sigma},\, \hat{v}_2,\, 0)$ with a complex $\mu_{12}^2$ bilinear term. The minimisation conditions are:
\vspace*{-15pt}\begin{subequations}
\begin{align}
(\mu_{12}^2)^\mathrm{I} ={}& -\sin \sigma\left[ 2 \mu_{11}^2 + 2 \lambda_{1111} \hat{v}_1^2 + \left( \lambda_{1122} + \lambda_{1221} \right) \hat{v}_2^2 \right]\frac{ \hat{v}_1 }{2 \hat{v}_2},\\
(\mu_{12}^2)^\mathrm{R} ={}& -\cos \sigma\left[ 2 \mu_{11}^2 + 2 \lambda_{1111} \hat{v}_1^2 + \left( \lambda_{1122} + \lambda_{1221} \right) \hat{v}_2^2 \right]\frac{ \hat{v}_1 }{2 \hat{v}_2},\\
\mu_{22}^2 ={}& \left( \mu_{11}^2 \hat{v}_1^2 + \lambda_{1111} \hat{v}_1^4 - \lambda_{2222} \hat{v}_2^4 \right)\frac{1}{\hat{v}_2^2}.
\end{align}
\end{subequations}

It does not make sense to consider simultaneously both a $\mu_{12}^2$ phase and a non-zero $\sigma$ since those  only appear together. One can see this from the minimisation conditions:
\begin{equation}
(\mu_{12}^2)^\mathrm{R} + i (\mu_{12}^2)^\mathrm{I} = -e^{i \sigma}\left[ 2 \mu_{11}^2 + 2 \lambda_{1111} \hat{v}_1^2 + \left( \lambda_{1122} + \lambda_{1221} \right) \hat{v}_2^2 \right]\frac{ \hat{v}_1 }{2 \hat{v}_2}.
\end{equation}
By setting either of the phases to zero, the phase of $\mu_{12}^2$ or $\sigma$, the other phase is also required to be zero due to the minimisation conditions, or else $\mu_{12}^2=0$. The $(\hat{v}_1,\, \hat{v}_2,\, 0)$ vacuum with a soft $\mu_{12}^2$ term yields:
\begin{subequations} \label{Eq:U1U1_MinCond2}
\begin{align}
\mu_{12}^2 ={}& -\left[ 2 \mu_{11}^2 + 2 \lambda_{1111} \hat{v}_1^2 + \left( \lambda_{1122} + \lambda_{1221} \right) \hat{v}_2^2 \right]\frac{\hat{v}_1 }{2 \hat{v}_2},\\
\mu_{22}^2 ={}&  \left(\mu_{11}^2 \hat{v}_1^2 + \lambda_{1111} \hat{v}_1^4 - \lambda_{2222} \hat{v}_2^4 \right)\frac{1}{\hat{v}_2^2}.
\end{align}
\end{subequations}
As a result, no complex parameter survives in the case of the discussed implementation.

In the case $(\hat{v}_1,\, \hat{v}_2,\, 0)$ with the soft symmetry-breaking term $\mu_{12}^2$ the entries of the mass-squared matrix associated with the $h_3$ doublet are left intact. We get $m_{\eta_3} = m_{\chi_3}$ as in the case without the soft symmetry-breaking term. However, due to the mixing caused by the soft symmetry-breaking term there are no longer unwanted massless states present.

In this implementation the overall $U(1) \times U(1)$ is broken down to a single $U(1)$, which remains unbroken in the $h_3$ sector even after the introduction of the soft breaking term. Therefore, a potentially interesting DM candidate might appear. In the following, we shall only discuss cases where the full symmetry is left unbroken.

\subsection{Two vanishing vevs}

The other option is to consider the vacuum given by $(v,\, 0,\, 0)$. In this case there is a single minimisation condition:
\begin{equation} \label{Eq:U1U1_MinCond3}
\mu_{11}^2 = - \lambda_{1111} v^2.
\end{equation}
This is the only vacuum which does not spontaneously break the underlying $U(1) \times U(1)$ symmetry, and therefore is free of unwanted massless states. The $h_1$ is then associated with the SM-like Higgs doublet. There is no mixing in the mass-squared matrix among the three doublets, the matrix is diagonal. Such behaviour causes two mass-degenerate pairs to appear; the CP-even and CP-odd states coming from the same doublet, $h_2$ and $h_3$, are mass degenerate, $m_{\eta_2} = m_{\chi_2}$ and $m_{\eta_3} = m_{\chi_3}$.

\clearpage
\bigskip\textbf{Case of \boldmath$(v,\,0,\,0)$} \labeltext{Case of $(v,\,0,\,0)$}{Sec:U1U1_v00}

The charged mass-squared matrix in the basis $\{h_1^\pm,\, h_2^\pm,\, h_3^\pm \}$ is:
\begin{equation}
\mathcal{M}_\mathrm{Ch}^2 = \mathrm{diag}\left(0,\, \mu_{22}^2 + \frac{1}{2} \lambda_{1122} v^2,\, \mu_{33}^2 + \frac{1}{2} \lambda_{1133} v^2   \right).
\end{equation}

The neutral mass-squared matrix in the basis $\{\eta_1,\, \eta_2,\, \eta_3,\, \chi_1,\, \chi_2,\, \chi_3 \}$ is:
\begin{equation}
\mathcal{M}_\mathrm{N}^2 = \mathrm{diag} \bigg(m_h^2,\, m_{H_2}^2,\, m_{H_3}^2,\, 0,\,  m_{H_2}^2,\, m_{H_3}^2   \bigg),
\end{equation}
where
\begin{subequations}\label{Eq:MN2_U1_U1}
\begin{align}
m_h^2 ={}& 2 \lambda_{1111} v^2,\\
 m_{H_i}^2 ={}& \mu_{ii}^2 + \frac{1}{2} \left( \lambda_{11ii} + \lambda_{1ii1} \right) v^2, \text{ for } i=\{2,\,3\}.
\end{align}
\end{subequations}

In this implementation $\{ \mu_{22}^2,\, \mu_{33}^2,\, \lambda_{1111},\, \lambda_{1122},\, \lambda_{1133},\, \lambda_{1221},\, \lambda_{1331}\}$
enter in the mass-squared matrices, and are responsible for generating five different mass-squared parameters. Some couplings appear only in the scalar interactions: $\{\lambda_{2222},\, \lambda_{2233},\, \lambda_{2332},\,$ $\lambda_{3333} \}$.

\section{\texorpdfstring{\boldmath$U(1)_1$}{U(1)1}-symmetric 3HDM}\label{Sec:Pot_U11}

The $U(1)_1$-symmetric scalar potential, with charges defined in eq.~\eqref{Eq:U11_U12_charges_U1a}, allows for a single phase-dependent coupling,
\begin{equation}\label{Eq:Vph_U11}
V^\mathrm{ph}_{U(1)_1} = \lambda_{1323} h_{13} h_{23} + \mathrm{h.c.},
\end{equation}
with the full scalar potential given by 
\begin{equation}
\begin{aligned}
V_{U(1)_1} ={}& V_0 + V^\mathrm{ph}_{U(1)_1}\\
={}& \sum_i \mu_{ii}^2 h_{ii} + \sum_i \lambda_{iiii} h_{ii}^2 + \sum_{i<j} \lambda_{iijj} h_{ii} h_{jj} + \sum_{i<j} \lambda_{ijji} h_{ij} h_{ji}\\
& + \left\lbrace \lambda_{1323} h_{13} h_{23} + \mathrm{h.c.} \right\rbrace.
\end{aligned}
\end{equation}
This potential also follows from imposing a $\mathbb{Z}_2 \times \mathbb{Z}_3$ symmetry as illustrated in Figure~\ref{Fig:Symmetry_breaking}.

The scalar potential is not symmetric under all permutations of vevs, and hence we shall consider different options. However, the form of the potential is left intact after permutations between $h_1$ and $h_2$. Since there is a single complex coupling $\lambda_{1323}$ sensitive to the phases of all three doublets, it is always possible to absorb all phases by a doublet with a vanishing vev. Even if one were to assume an implementation without a vanishing vev, the minimisation conditions would force the complex quartic coupling to become real. Therefore, we shall assume that $\lambda_{1323} \in \mathbb{R}$, except when soft symmetry-breaking terms are introduced.

The discussion of the $U(1)_1$-symmetric 3HDM is summarised in Table~\ref{Table:U11_Cases}. The only interesting cases are those with two vanishing vevs since all other cases result in SSB of $U(1)_1$. All discussed cases are CP conserving.

{\renewcommand{\arraystretch}{1.3}
\begin{table}[htb]
\caption{Similar to Table~\ref{Table:U1_U1_Cases}, but now for $U(1)_1$. In one case the minimisation conditions can lead to a higher symmetry. None of these cases violates CP.}
\label{Table:U11_Cases}
\begin{center}
\begin{tabular}{|c|c|c|c|c|} \hline\hline
Vacuum & SYM &$V$ & \begin{tabular}[l]{@{}c@{}} Mixing of the \\ neutral states\end{tabular} & Comments \\ \hline
$(\hat{v}_1,\, \hat{v}_2,\, 0)$ & $-$ &  $V_{U(1)_1}$ & \footnotesize\begin{tabular}[l]{@{}c@{}} $\{\eta_1,\, \eta_2\}{-}\{\eta_3\}$\\${-}\{\chi_1\}{-}\{\chi_2\}{-}\{\chi_3\}$ \end{tabular} & $m_{\chi_1} = m_{\chi_2} = 0$ \\ \hline
$(\hat{v}_1,\, \hat{v}_2,\, 0)$ & $-$ & $V_{U(1)_1} + (\mu_{12}^2)^\mathrm{R}$ & \footnotesize\begin{tabular}[l]{@{}c@{}} $\{\eta_1,\, \eta_2\}{-}\{\eta_3\}$\\ ${-}\{\chi_1,\, \chi_2\}{-}\{\chi_3\}$ \end{tabular} & - \\ \hline
$(0,\, \hat v_2,\, \hat v_3)$  & $-$ & $V_0$ & \footnotesize\begin{tabular}[l]{@{}c@{}} $\{\eta_1\}{-}\{\eta_2,\, \eta_3\}$ \\ ${-}\{\chi_1\}{-}\{\chi_2\}{-}\{\chi_3\}$ \end{tabular} & \footnotesize\begin{tabular}[l]{@{}c@{}} $m_{\eta_1} = m_{\chi_1}$ \\ $m_{\chi_2} = m_{\chi_3} = 0$ \end{tabular} \\ \hline
$(v,\,0,\,0)$ & $\checkmark$ & $V_{U(1)_1}$  & diagonal & \footnotesize\begin{tabular}[l]{@{}c@{}} $m_{\eta_2} = m_{\chi_2},~m_{\eta_3} = m_{\chi_3}$ \end{tabular} \\ \hline
$(0,\,0,\,v)$ & $\checkmark$ & $V_{U(1)_1}$ & \footnotesize\begin{tabular}[l]{@{}c@{}} $\{\eta_1,\, \eta_2\}{-}\{\eta_3\}$ \\ ${-}\{\chi_1,\, \chi_2\}{-}\{\chi_3\}$ \end{tabular} & \footnotesize\begin{tabular}[l]{@{}c@{}} Two pairs of \\ mass-degenerate states \end{tabular} \\ \hline \hline
\end{tabular} 
\end{center}
\end{table}

\subsection{One vanishing vev}

We start by considering vacuum configurations with a single vanishing vev. The first case is $(\hat{v}_1,\, \hat{v}_2,\, 0)$ with the minimisation conditions:
\begin{subequations}
\begin{align}
\mu_{11}^2 ={}& - \lambda_{1111} \hat v_1^2 - \frac{1}{2}\left( \lambda_{1122} + \lambda_{1221} \right) \hat v_2^2,\\
\mu_{22}^2 ={}& - \lambda_{2222} \hat v_2^2 - \frac{1}{2}\left( \lambda_{1122} + \lambda_{1221} \right) \hat v_1^2,
\end{align}
\end{subequations}
There is a massless state due to the broken $U(1)_1$ symmetry. A way around is to allow soft symmetry breaking. Not to introduce unwanted mixing, the only considered soft symmetry-breaking term is then $\mu_{12}^2$. Importantly, the minimisation conditions do not depend on the $\lambda_{1323}$ term. Although a new soft symmetry-breaking term is introduced, we can still consider $\lambda_{1323}$ to be real by re-phasing $h_3$. It can be seen that this case was already covered by the $U(1) \times U(1)$ minimisation conditions of eq.~\eqref{Eq:U1U1_MinCond2}. Due to the $(\mu_{12}^2)^\mathrm{I}=0$ condition we conclude that there is no CP violation in this case.

Another possibility is to consider the vacuum configuration $(0,\, \hat{v}_2,\, \hat{v}_3)$. In this implementation the minimisation conditions are:
\begin{subequations}
\begin{align}
\mu_{22}^2 ={}& - \lambda_{2222} \hat v_2^2 - \frac{1}{2}\left( \lambda_{2233} + \lambda_{2332} \right) \hat v_3^2,\\
\mu_{33}^2 ={}& - \lambda_{3333} \hat v_3^2 - \frac{1}{2}\left( \lambda_{2233} + \lambda_{2332} \right) \hat v_2^2,\\
\lambda_{1323}={}&0.
\end{align}
\end{subequations}
As a result, due to the vanishing of the quartic coupling $\lambda_{1323}$ this case is equivalent to $(\hat v_1,\, \hat v_2,\, 0)$ for $U(1) \times U(1)$ of Table~\ref{Table:U1_U1_Cases}.

\subsection{Two vanishing vevs}

Finally, we cover cases with two vanishing vevs. In the case $(v,\,0,\,0)$ there is no mixing among the mass eigenstates. The minimisation condition coincides with eq.~\eqref{Eq:U1U1_MinCond3}
\begin{equation*}
\mu_{11}^2 = - \lambda_{1111} v^2.
\end{equation*}
There are two pairs of neutral mass-degenerate states. In the case $(0,\,0,\,v)$ with the corresponding minimisation condition,
\begin{equation}
\mu_{33}^2 = - \lambda_{3333} v^2,
\end{equation}
there is mixing since $\mathcal{M}^2$ contains terms coming from $v^2 \lambda_{1323} h_1 h_2$. Nevertheless, there are still two pairs of neutral mass-degenerate states present.

\bigskip\textbf{Case of \boldmath$(v,\,0,\,0)$}

This case is identical, in terms of both the charged and neutral mass-squared matrices, to $(v,\,0,\,0)$ of $U(1) \times U(1)$, see \ref{Sec:U1U1_v00} in Section~\ref{Sec:Pot_U1U1}. However, it is physically different from $U(1) \times U(1)$. In this implementation $\{ \mu_{22}^2,\, \mu_{33}^2,\, \lambda_{1111},\, \lambda_{1122},\, \lambda_{1133},\,$ $\lambda_{1221},\,  \lambda_{1331}\}$ enter in the mass-squared matrices, and are responsible for generating five different mass-squared parameters. Some couplings appear only in the scalar interactions: $ \{ \lambda_{2222},\, \lambda_{2233},\, \lambda_{2332},\, \lambda_{3333},\, \lambda_{1323} \} $.

\bigskip\textbf{Case of \boldmath$(0,\,0,\,v)$}

The charged mass-squared matrix in the basis $\{h_1^\pm,\, h_2^\pm,\, h_3^\pm \}$ is diagonal:
\begin{equation}
\mathcal{M}_\mathrm{Ch}^2 = \mathrm{diag}\left( \mu_{11}^2 + \frac{1}{2} \lambda_{1133} v^2 ,\, \mu_{22}^2 + \frac{1}{2} \lambda_{2233} v^2 ,\, 0   \right).
\end{equation}

The neutral mass-squared matrix in the basis $\{\eta_1,\, \eta_2,\, \eta_3,\, \chi_1,\, \chi_2,\, \chi_3 \}$ is:
\begin{equation}
\mathcal{M}_\mathrm{N}^2 = \mathrm{diag}\Bigg(  \begin{pmatrix}
(\mathcal{M}_\mathrm{N}^2)_{11} & (\mathcal{M}_\mathrm{N}^2)_{12} \\
(\mathcal{M}_\mathrm{N}^2)_{12}& (\mathcal{M}_\mathrm{N}^2)_{22} \end{pmatrix},\, (\mathcal{M}_\mathrm{N}^2)_{33},\, \begin{pmatrix}
(\mathcal{M}_\mathrm{N}^2)_{11} & -(\mathcal{M}_\mathrm{N}^2)_{12}\\
-(\mathcal{M}_\mathrm{N}^2)_{12} & (\mathcal{M}_\mathrm{N}^2)_{22}
\end{pmatrix},\, 0 \Bigg),
\end{equation}
where
\begin{subequations}
\begin{align}
(\mathcal{M}_\mathrm{N}^2)_{ii} ={}& \mu_{ii}^2 + \frac{1}{2} \left( \lambda_{ii33} + \lambda_{i33i} \right) v^2, \text{ for } i=\{1,\,2\},\\
(\mathcal{M}_\mathrm{N}^2)_{12} ={}& \frac{1}{2} \lambda_{1323} v^2,\\
(\mathcal{M}_\mathrm{N}^2)_{33} ={}& 2 \lambda_{3333} v^2.
\end{align}
\end{subequations}

The eigenvalues of the two-by-two matrices are identical:
\begin{equation}
m_{H_i}^2 =  \frac{1}{4} \left[ 2\left( \mu_{11}^2 + \mu_{22}^2 \right) +  v^2 \left(\lambda_{1133} + \lambda_{2233} + \lambda_{1331} + \lambda_{2332}\right) \pm \Delta \right],
\end{equation}
where
\begin{equation}
\begin{aligned}
\Delta^2 ={}& v^4 \left[  4 \left(\lambda_{1323}^\mathrm{R}\right)^2 + \left( \lambda_{1133} - \lambda_{2233} + \lambda_{1331} - \lambda_{2332} \right)^2 \right]\\
&  + 4 v^2 \left( \lambda_{1133} - \lambda_{2233} + \lambda_{1331} - \lambda_{2332} \right)\left( \mu_{11}^2 - \mu_{22}^2 \right) + 4 \left( \mu_{11}^2 - \mu_{22}^2 \right)^2.
\end{aligned}
\end{equation}

In this implementation $\{ \mu_{11}^2,\, \mu_{22}^2,\, \lambda_{1133},\, \lambda_{1331},\, \lambda_{2233},\, \lambda_{2332},\, \lambda_{1323}\}$
enter in the mass-squared matrices, and are responsible for generating five different mass-squared parameters. Some couplings appear only in the scalar interactions: $\{\lambda_{1111},\, \lambda_{1122},\, \lambda_{1221},\,$ $ \lambda_{2222}\}$.

\section{\texorpdfstring{\boldmath$U(1) \times \mathbb{Z}_2$}{U(1) x Z2}-symmetric 3HDM}\label{Sec:Pot_U1Z2}

As an introduction to the discussion of the $U(1)_2$ symmetry, we shall here discuss the $U(1)\times \mathbb{Z}_2$-symmetric 3HDM, as suggested by Figure~\ref{Fig:Symmetry_breaking}. The $U(1)_2$ symmetry given by eq.~\eqref{Eq:U11_U12_charges_U1b} can be extended by a $\mathbb{Z}_2$ symmetry, under which
\begin{equation}
h_1 \to -h_1, \quad h_2 \to h_2, \quad h_3 \to h_3.
\end{equation}
One could have chosen an equivalent implementation with $h_2$ being odd under $\mathbb{Z}_2$, rather than $h_1$. The potential would have the same form regardless of the chosen $\mathbb{Z}_2$ charges. The phase-dependent part is then
\begin{equation}\label{Eq:Vph_U1Z2}
V^\mathrm{ph}_{U(1) \times \mathbb{Z}_2} = \lambda_{1212} h_{12}^2 + \mathrm{h.c.},
\end{equation}
with the full scalar potential given by 
\begin{equation}\label{Eq:V_U1Z2}
\begin{aligned}
V_{U(1) \times \mathbb{Z}_2} ={}& V_0 + V^\mathrm{ph}_{U(1) \times \mathbb{Z}_2}\\
={}& \sum_i \mu_{ii}^2 h_{ii} + \sum_i \lambda_{iiii} h_{ii}^2 + \sum_{i<j} \lambda_{iijj} h_{ii} h_{jj} + \sum_{i<j} \lambda_{ijji} h_{ij} h_{ji}\\
& + \left\lbrace \lambda_{1212} h_{12}^2 + \mathrm{h.c.} \right\rbrace.
\end{aligned}
\end{equation}

The scalar potential is symmetric under the exchange of indices, $h_1 \leftrightarrow h_2$. This constraint reduces the number of distinct cases in terms of the vevs. There is only one term present in the phase-sensitive part, like in the $U(1)_1$-symmetric model. One might be tempted to conclude that there exists a basis transformation from one model to the other, but this is not the case. Furthermore, the phase coming from the single phase-dependent coupling $\lambda_{1212}$ is not physical and can be rotated away. Therefore, we shall assume that $\lambda_{1212} \in \mathbb{R}$, except when soft symmetry-breaking terms are introduced. All implementations are summarised in Table~\ref{Table:U1Z2_Cases}.

{{\renewcommand{\arraystretch}{1.3}
\begin{table}[htb]
\caption{Similar to Table~\ref{Table:U1_U1_Cases}, but now for $U(1) \times \mathbb{Z}_2$. None of these cases violates CP.}
\label{Table:U1Z2_Cases}
\begin{center}
\begin{tabular}{|c|c|c|c|c|} \hline\hline
Vacuum & SYM & $V$ & \begin{tabular}[l]{@{}c@{}} Mixing of the \\ neutral states\end{tabular} & Comments \\ \hline
$(\hat{v}_1,\, \hat{v}_2,\, 0)$ & $-$ & $V_{U(1) \times \mathbb{Z}_2}$ & \footnotesize\begin{tabular}[l]{@{}c@{}} $\{\eta_1,\, \eta_2\}{-}\{\eta_3\}$ \\ ${-}\{\chi_1,\, \chi_2\}{-}\{\chi_3\}$ \end{tabular} & $m_{\eta_3} = m_{\chi_3}$ \\ \hline
$(0,\, \hat{v}_2,\, \hat{v}_3)$ & $-$ & $V_{U(1) \times \mathbb{Z}_2}$ & \footnotesize\begin{tabular}[l]{@{}c@{}} $\{\eta_1\}{-}\{\eta_2,\, \eta_3\}$ \\ ${-}\{\chi_1\}{-}\{\chi_2\}{-}\{\chi_3\}$ \end{tabular} & $m_{\chi_2} = m_{\chi_3} = 0$ \\ \hline
$(0,\, \hat{v}_2,\, \hat{v}_3)$ & $-$ & $V_{U(1) \times \mathbb{Z}_2} + (\mu_{23}^2)^\mathrm{R}$ & \footnotesize\begin{tabular}[l]{@{}c@{}} $\{\eta_1\}{-}\{\eta_2,\, \eta_3\}$ \\ ${-}\{\chi_1\}{-}\{\chi_2, \,\chi_3\}$ \end{tabular} & - \\ \hline
$(v, \,0, \, 0)$ & $\checkmark$ & $V_{U(1) \times \mathbb{Z}_2}$ & diagonal & $m_{\eta_3} = m_{\chi_3}$ \\ \hline
$(0, \,0, \, v)$  & $\checkmark$ & $V_{U(1) \times \mathbb{Z}_2}$ & diagonal & \footnotesize\begin{tabular}[l]{@{}c@{}} $m_{\eta_1} = m_{\chi_1},~m_{\eta_2} = m_{\chi_2}$ \end{tabular} \\ \hline \hline
\end{tabular} 
\end{center}
\end{table}}

\subsection{One vanishing vev}

We start by discussing the vacuum $(\hat{v}_1,\, \hat{v}_2,\, 0)$. The minimisation conditions are:
\begin{subequations}
\begin{align}
\mu_{11}^2 ={}& - \lambda_{1111} \hat v_1^2 - \frac{1}{2} \left( \lambda_{1122} + 2 \lambda_{1212}^\mathrm{R} + \lambda_{1221} \right) \hat v_2^2 ,\\
\mu_{22}^2 ={}& - \lambda_{2222} \hat v_2^2 - \frac{1}{2} \left( \lambda_{1122} + 2 \lambda_{1212}^\mathrm{R} + \lambda_{1221} \right) \hat v_1^2 .
\end{align}
\end{subequations}
The case $(\hat{v}_1 e^{i \sigma},\, \hat{v}_2,\, 0)$ is not realisable since the minimisation conditions force either $\lambda_{1212}=0$, increasing the symmetry to $U(1) \times U(1)$, or $\sigma = \pi/2$.

The other possibility is to consider a vacuum of the type $(0, \, v_2, \, v_3)$. Without loss of generality one can always absorb the phase of $\lambda_{1212}$ and choose the vacuum real through the following steps: i) re-phase $h_1$ in order to absorb the phase of $\lambda_{1212}$ (this phase will not appear in the vacuum), ii) perform an overall phase rotation of the scalars to make $v_2$ real, iii) re-phase $h_3$ making use of the $U(1)_2$ symmetry of the potential to make the vacuum real. Therefore, the case reduces to $(0, \, \hat{v}_2, \, \hat{v}_3)$ with the minimisation conditions:
\begin{subequations}
\begin{align}
\mu_{22}^2 ={}& - \lambda_{2222} \hat v_2^2 - \frac{1}{2} \left( \lambda_{2233} + \lambda_{2332} \right) \hat v_3^2,\\
\mu_{33}^2 ={}& - \lambda_{3333} \hat v_3^2 - \frac{1}{2} \left( \lambda_{2233} + \lambda_{2332} \right) \hat v_2^2.
\end{align}
\end{subequations}
The $U(1)$ symmetry gets spontaneously broken, which yields a massless state. Assuming soft symmetry breaking, the only term which does not contribute to unwanted mixing with $h_1$ is $\mu_{23}^2$. In the case $(0, \, \hat{v}_2, \, \hat{v}_3)$ the minimisation conditions are then given by:
\begin{subequations}
\begin{align}
\mu_{23}^2 ={}& -\left[ 2 \mu_{22}^2 + 2 \lambda_{2222} \hat{v}_2^2 + \left( \lambda_{2233} + \lambda_{2332} \right) \hat{v}_3^2 \right]\frac{\hat{v}_2 }{2 \hat{v}_3},\\
\mu_{33}^2 ={}&  \left( \mu_{22}^2 \hat{v}_2^2 + \lambda_{2222} \hat{v}_2^4 - \lambda_{3333} \hat{v}_3^4  \right)\frac{1}{\hat{v}_3^2}.
\end{align}
\end{subequations}
Then, as before, the phase of the $\lambda_{1212}$ term can be absorbed by $h_1$; there are no physical phases present. The case $(0, \, \hat{v}_2 e^{i \sigma}, \, \hat{v}_3)$ is not realisable as it would require $\mu_{23}^2=0$ or else $\sigma=0$, to satisfy the new minimisation conditions.

\subsection{Two vanishing vevs}

The last two cases are those with two vanishing vevs. For $(v,\, 0,\, 0)$ there is a single minimisation condition,
\begin{equation}
\mu_{11}^2 =- \lambda_{1111} v^2.
\end{equation}
There is a pair of mass-degenerate states coming from $h_3$. The other possibility is to assume $(0,\, 0,\ v)$. In this case the minimisation condition becomes
\begin{equation}
\mu_{33}^2 =- \lambda_{3333} v^2.
\end{equation}
For this vacuum the mass-squared matrix is diagonal,  and there are two pairs of neutral mass-degenerate states.

Both cases preserve the underlying $U(1) \times \mathbb{Z}_2$ symmetry. We shall now discuss these two implementations.

\bigskip\textbf{Case of \boldmath$(v,\,0,\,0)$}

The charged mass-squared matrix in the basis $\{h_1^\pm,\, h_2^\pm,\, h_3^\pm \}$ is:
\begin{equation}
\mathcal{M}_\mathrm{Ch}^2 = \mathrm{diag}\left(0,\, \mu_{22}^2 + \frac{1}{2} \lambda_{1122} v^2 ,\, \mu_{33}^2 + \frac{1}{2} \lambda_{1133} v^2   \right).
\end{equation}

The neutral mass-squared matrix in the basis $\{\eta_1,\, \eta_2,\, \eta_3,\, \chi_1,\, \chi_2,\, \chi_3 \}$ is:
\begin{equation}
\mathcal{M}_\mathrm{N}^2 = \mathrm{diag} \bigg( m_h^2 ,\, m_{\eta_2}^2,\, m_{H_3}^2,\, 0 ,\, m_{\chi_2}^2,\, m_{H_3}^2  \bigg),
\end{equation}
where
\begin{subequations}
\begin{align}
m_h^2 ={}& 2 \lambda_{1111} v^2,\\
m_{\eta_2}^2 ={}& \mu_{22}^2 + \frac{1}{2} \left( \lambda_{1122} + 2 \lambda_{1212} + \lambda_{1221} \right)v^2,\\
m_{H_3}^2 ={}& \mu_{33}^2 + \frac{1}{2} \left( \lambda_{1133} + \lambda_{1331} \right)v^2,\\
m_{\chi_2}^2 ={}& \mu_{22}^2 + \frac{1}{2} \left( \lambda_{1122} - 2 \lambda_{1212} + \lambda_{1221} \right)v^2.
\end{align}
\end{subequations}

In this implementation $\{ \mu_{22}^2,\, \mu_{33}^2,\, \lambda_{1111},\, \lambda_{1122},\, \lambda_{1133},\, \lambda_{1221},\, \lambda_{1331},\,\lambda_{1212}\}$
enter in the mass-squared matrices, and are responsible for generating six different mass-squared parameters. The couplings $\{\lambda_{2222},\, \lambda_{2233},\,$ $ \lambda_{2332},\, \lambda_{3333}\}$ appear only in the scalar interactions.

\bigskip\textbf{Case of \boldmath$(0,\,0,\,v)$}

After re-labeling the indices this case becomes identical, in terms of both the charged and the neutral mass-squared matrices, to \ref{Sec:U1U1_v00} of $U(1) \times U(1)$ in Section~\ref{Sec:Pot_U1U1}.

The charged mass-squared matrix in the basis $\{h_1^\pm,\, h_2^\pm,\, h_3^\pm \}$ is:
\begin{equation}
\mathcal{M}_\mathrm{Ch}^2 = \mathrm{diag}\left( \mu_{11}^2 + \frac{1}{2} \lambda_{1133} v^2 ,\, \mu_{22}^2 + \frac{1}{2} \lambda_{2233} v^2 ,\,0   \right).
\end{equation}

The neutral mass-squared matrix in the basis $\{\eta_1,\, \eta_2,\, \eta_3,\, \chi_1,\, \chi_2,\, \chi_3 \}$ is:
\begin{equation}
\mathcal{M}_\mathrm{N}^2 = \mathrm{diag} \bigg(m_{H_1}^2,\, m_{H_2}^2,\,m_h^2 ,\, m_{H_1}^2,\,m_{H_2}^2,\,0   \bigg),
\end{equation}
where
\begin{subequations}
\begin{align}
m_{H_i}^2 ={}& \mu_{ii}^2 + \frac{1}{2} \left( \lambda_{ii33} + \lambda_{i33i} \right) v^2,\text{ for } i=\{1,\,2\},\\
m_h^2 ={}& 2 \lambda_{3333} v^2.
\end{align}
\end{subequations}

In this implementation $\{ \mu_{11}^2,\, \mu_{22}^2,\, \lambda_{1133},\, \lambda_{1331},\, \lambda_{2233},\, \lambda_{2332},\, \lambda_{3333}\}$
enter in the mass-squared matrices, and are responsible for generating five different mass-squared parameters. The couplings $\{\lambda_{1111},\, \lambda_{1122},\, \lambda_{1221},\,\lambda_{2222},$ $\lambda_{1212}\}$ appear only in the scalar interactions.

\section{\texorpdfstring{\boldmath$U(1)_2$}{U(1)2}-symmetric 3HDM}\label{Sec:Pot_U12}

In contrast to the $U(1)_1$-symmetric case, the $U(1)_2$-symmetric 3HDM allows for six phase-dependent couplings rather than only one,
\begin{equation}\label{Eq:Vph_U12}
\begin{split}
V_{U(1)_2}^\text{ph} ={}& \mu_{12}^2 h_{12} + \lambda_{1212} h_{12}^2 + \lambda_{1112} h_{11}h_{12}+ \lambda_{1222} h_{12}h_{22}\\
&  + \lambda_{1233} h_{12}h_{33} + \lambda_{1332} h_{13}h_{32} + \mathrm{h.c.},
\end{split}
\end{equation}
with the scalar potential given by
\begin{equation}
\begin{aligned}
V_{U(1)_2} ={}& V_0 + V_{U(1)_2}^\text{ph}\\
={}& \sum_i \mu_{ii}^2 h_{ii} + \sum_i \lambda_{iiii} h_{ii}^2 + \sum_{i<j} \lambda_{iijj} h_{ii} h_{jj} + \sum_{i<j} \lambda_{ijji} h_{ij} h_{ji}\\
&  +  \Big\{\mu_{12}^2 h_{12} + \lambda_{1212} h_{12}^2 + \lambda_{1112} h_{11}h_{12}+ \lambda_{1222} h_{12}h_{22}\\
& \hspace{20pt} + \lambda_{1233} h_{12}h_{33} + \lambda_{1332} h_{13}h_{32} + \mathrm{h.c.} \Big\}.
\end{aligned}
\end{equation}

Once again, we do not need to consider all possible vacuum configurations as interchanging the vacua of $h_1$ and $h_2$ leads to identical solutions. In contrast to the above studied cases, in general, there is explicit CP violation. Then, at least one of the complex phases could be rotated away due to some $\left\langle h_i\right\rangle = 0$. Hence, like in other models, we shall consider two cases---with complex couplings and real vevs, and with real couplings and complex vevs. The different cases are summarised in Table~\ref{Table:U12_Cases}.

{{\renewcommand{\arraystretch}{1.3}
\begin{table}[htb]
\caption{ Similar to Table~\ref{Table:U1_U1_Cases}, but now for $U(1)_2$. In all cases complex parameters are present, and may result in CP violation. In cases of two non-vanishing vevs without hats, $v_i$, the vevs can be real or complex.}
\label{Table:U12_Cases}
\begin{center}
\begin{tabular}{|c|c|c|c|c|} \hline\hline
Vacuum & SYM & $V$ & \begin{tabular}[l]{@{}c@{}} Mixing of the \\ neutral states\end{tabular} & Comments \\ \hline
$(\hat v_1 e^{i \sigma},\, \hat v_2,\, 0)$ & $\checkmark$ & $V_{U(1)_2}$ & \footnotesize\begin{tabular}[l]{@{}c@{}} $\{\eta_1,\, \eta_2,\, \chi_1,\, \chi_2\}$ \\ ${-}\{\eta_3\}{-}\{\chi_3\}$ \end{tabular} & $m_{\eta_3} = m_{\chi_3}$ \\ \hline
$(0,\, \hat v_2 e^{i\sigma},\, \hat v_3)$ & $-$ & $V_{U(1)_2}$ & \footnotesize\begin{tabular}[l]{@{}c@{}} $\{\eta_1,\,\eta_2,\, \eta_3, \chi_1\}$ \\ ${-}\{\chi_2\}{-}\{\chi_3\}$ \end{tabular} & \footnotesize\begin{tabular}[l]{@{}c@{}} $m_{\chi_2} = m_{\chi_3} = 0 $ \\ No obvious DM \end{tabular} \\ \hline
$(0,\, \hat{v}_2,\, \hat{v}_3)$ & $-$ & $V_{U(1)_2} + (\mu_{23}^2)^\mathrm{R}$ & \footnotesize\begin{tabular}[l]{@{}c@{}} $\{\eta_1,\,\eta_2,\,\eta_3,\,\chi_1\}$\\${-}\{\chi_2,\, \chi_3\}$ \end{tabular} & \footnotesize No obvious DM  \\ \hline 
$(v,\, 0,\, 0)$ & $\checkmark$ & $V_{U(1)_2}$ & \footnotesize\begin{tabular}[l]{@{}c@{}} $\{\eta_1,\, \eta_2,\, \chi_2\}$ \\ ${-}\{\eta_3\}{-}\{\chi_1\}{-}\{\chi_3\}$ \end{tabular} & $m_{\eta_3} = m_{\chi_3}$ \\ \hline
$(0,\, 0,\, v)$ & $\checkmark$ & $V_{U(1)_2}$ & \footnotesize\begin{tabular}[l]{@{}c@{}} $\{\eta_1,\, \eta_2,\, \chi_1,\, \chi_2\}$ \\ ${-}\{\eta_3\}{-}\{\chi_3\}$ \end{tabular} & \footnotesize\begin{tabular}[l]{@{}c@{}} Two pairs of \\ mass-degenerate \\ neutral states\end{tabular} \\ \hline \hline
\end{tabular} 
\end{center}
\end{table}}

\subsection{One vanishing vev}

We start by considering vacuum configurations with a single vanishing vev. For $(v_1,\, v_2,\, 0)$ the minimisation conditions are given by:
\begin{subequations}\label{Eq:U12_Min_Con_v1v20_Gen}
\begin{align}
\begin{split}
(\mu_{12}^2)^\mathrm{I} ={}& -\frac{1}{2}\Bigg( \left[ \lambda_{1112}^\mathrm{I}\left( 2 - \cos (2 \sigma) \right)  + \lambda_{1112}^\mathrm{R} \sin (2 \sigma) \right] \hat v_1^2 + 2 \sin \sigma \left( \mu_{11}^2 + \lambda_{1111} \hat v_1^2 \right) \frac{\hat v_1}{\hat v_2} \\
& \hspace{30pt} + \left[ 2 \lambda_{1212}^\mathrm{I} \cos \sigma + \left( \lambda_{1122} - 2 \lambda_{1212}^\mathrm{R} + \lambda_{1221} \right) \sin \sigma  \right] \hat v_1 \hat v_2 +  \lambda_{1222}^\mathrm{I} \hat v_2^2 \Bigg),
\end{split}\\
\begin{split}
(\mu_{12}^2)^\mathrm{R} ={}& -\frac{1}{2}\Bigg(\left[ \lambda_{1112}^\mathrm{R}\left( 2 + \cos (2 \sigma) \right)  + \lambda_{1112}^\mathrm{I} \sin (2 \sigma) \right] \hat v_1^2 + 2 \cos \sigma \left( \mu_{11}^2 + \lambda_{1111} \hat v_1^2 \right) \frac{\hat v_1}{\hat v_2} \\
& \hspace{30pt} +\left[ 2 \lambda_{1212}^\mathrm{I} \sin \sigma + \left( \lambda_{1122} + 2 \lambda_{1212}^\mathrm{R} + \lambda_{1221} \right) \cos \sigma  \right] \hat v_1 \hat v_2 +  \lambda_{1222}^\mathrm{R} \hat v_2^2 \Bigg),
\end{split}\\
\begin{split}
\mu_{22}^2 ={}& \frac{1}{\hat{v}_2^2}  \Bigg[ \mu_{11}^2 \hat{v}_1^2 + \lambda_{1111} \hat{v}_1^4 + \left( \lambda_{1112}^\mathrm{R} \cos \sigma + \lambda_{1112}^\mathrm{I} \sin \sigma \right) \hat v_1^3 \hat v_2\\
& \hspace{17pt} - \left( \lambda_{1222}^\mathrm{R} \cos \sigma + \lambda_{1222}^\mathrm{I} \sin \sigma \right) \hat v_1 \hat v_2^3 - \lambda_{2222} \hat{v}_2^4 \Bigg].
\end{split}
\end{align}
\end{subequations}
It is sufficient to allow for a single complex parameter, either the imaginary part $(\mu_{12}^2)^\mathrm{I}$ or the $\sigma$ phase. The minimisation conditions are explicitly written down for the real case in eq.~\eqref{Eq:U12_Min_Con_1} and for the complex case (with real couplings) in eq.~\eqref{Eq:U12_Min_Con_2}. In both cases there is a mass-degenerate pair present in the neutral sector associated with the $h_3$ doublet.

Next we consider the vacuum $(0,\, v_2,\, v_3)$. In this case the minimisation conditions are:
\begin{subequations}
\begin{align}
\mu_{12}^2 ={}& - \frac{1}{2} \left[ \lambda_{1222} \hat v_2^2 + \left( \lambda_{1233} + \lambda_{1332} \right) \hat v_3^2 \right], \label{Eq:mu122_U12_0v2v3}\\
\mu_{22}^2 ={}& - \lambda_{2222} \hat v_2^2 - \frac{1}{2} \left( \lambda_{2233} + \lambda_{2332} \right) \hat v_3^2,\\
\mu_{33}^2 ={}& - \lambda_{3333} \hat v_3^2 - \frac{1}{2} \left( \lambda_{2233} + \lambda_{2332} \right) \hat v_2^2.
\end{align}
\end{subequations}
These expressions are independent of $\sigma$, but we should stress that $\mu_{12}^2$ can be complex. There are two massless neutral states associated with $\chi_2$ and $\chi_3$. The other neutral states mix to produce the mass eigenstates. Therefore, since the SM-like Higgs boson\footnote{In some cases, we shall refer to the 125 GeV state as the SM-like Higgs boson, rather than the $\eta_i$ field transforming together with the would-be Goldstone boson associated with the $\chi_i$ field.} mixes with all other scalars, the would-be DM candidate would necessarily interact with fermions. One could argue that as long as the decay rate of such processes exceeds the age of the Universe, it is possible to have a semi-stable DM state. We shall not consider such fine-tuning. Another possibility would be to try to force the mixing terms in the mass-squared matrix to vanish by requiring
\begin{equation}\label{Eq:0v2v3_AddCond}
\lambda_{1332} = \lambda_{1222} = \lambda_{1233} = 0.
\end{equation}
The $V_{U(1)_2}$ scalar potential then reduces to 
\begin{equation}\label{Eq:VU12pr}
\begin{split}
V ={}& V_0 + \left\lbrace \mu_{12}^2 h_{12} + \lambda_{1212} h_{12}^2 + \lambda_{1112} h_{11}h_{12} + \mathrm{h.c.} \right\rbrace.
\end{split}
\end{equation}
Due to additional vanishing quartic couplings~\eqref{Eq:0v2v3_AddCond} the minimisation condition~\eqref{Eq:mu122_U12_0v2v3} requires $\mu_{12}^2 =0$ for both $(0,\, \hat{v}_2,\, \hat{v}_3)$ and $(0,\, \hat{v}_2 e^{i \sigma},\, \hat{v}_3)$. The $U(1)_2$ symmetry remains spontaneously broken. There are two phase-sensitive terms in the potential, the phase of one of them can be rotated away. 

It should be noted that at the one-loop level the unwanted mixing might once again take place since the assumed vanishing couplings are not protected by an underlying symmetry. We checked the basis-invariant conditions of Ref.~\cite{deMedeirosVarzielas:2019rrp} and found that the scalar potential of eq.~\eqref{Eq:VU12pr} does not exhibit any additional symmetries apart from that of $U(1)_2$. Furthermore, no entry in the list of accidental symmetries of Ref.~\cite{Darvishi:2019dbh} is compatible with the reduced potential.

One can try to promote massless states of the cases with $(0,\, v_2,\, v_3)$ to massive ones by introducing the soft symmetry-breaking term $\mu_{23}^2$. The minimisation conditions are then:
\begin{subequations}
\begin{align}
\mu_{12}^2 ={}& - \frac{1}{2} \left[ \lambda_{1222} \hat v_2^2 + \left( \lambda_{1233} + \lambda_{1332} \right) \hat v_3^2 \right],\\
\mu_{23}^2 ={}& - e^{i \sigma} \left[ 2 \mu_{22}^2 + 2 \lambda_{2222} \hat v_2^2 + \left( \lambda_{2233} + \lambda_{2332} \right) \hat v_3^2 \right] \frac{\hat v_2}{2 \hat v_3},\\
\mu_{33}^2 ={}& \left( \mu_{22}^2 \hat v_2^2 + \lambda_{2222} \hat v_2^4 - \lambda_{3333} \hat v_3^4 \right) \frac{1}{\hat v_3^2} .
\end{align}
\end{subequations}
We recall that $\mu_{12}^2$ along with $\{\lambda_{1222},\, \lambda_{1233},\, \lambda_{1332} \}$ are complex parameters. In the case $(0,\, \hat{v}_2 e^{i \sigma},\, \hat{v}_3)$, requiring $\mu_{23}^I=0$ would imply that the soft symmetry-breaking term does not survive or else $\sigma=0$. Therefore, the case $(0,\, \hat{v}_2 e^{i \sigma},\, \hat{v}_3)$ including $\mu_{23}^2$, where both $ v_2$ and $\mu_{23}^2$ are complex, is not realisable. Even in the case of real vevs, $(0,\, \hat{v}_2,\, \hat{v}_3)$ there will be mixing among all doublets via the $\eta_i$ fields. Then, one can enforce the conditions of eq.~\eqref{Eq:0v2v3_AddCond}. In this case we get $\mu_{12}^2=0$, implying that only one phase is physical.

Let us now discuss in some detail cases with one zero vev where there is no SSB. These could potentially accommodate good DM candidates.

\bigskip\textbf{Case of \boldmath$(\hat{v}_1,\, \hat{v}_2,\, 0)$}

We start by considering the case with real vevs and complex couplings. The minimisation conditions~\eqref{Eq:U12_Min_Con_v1v20_Gen} in the case $(\hat{v}_1,\, \hat{v}_2,\, 0)$ are:
\begin{subequations}\label{Eq:U12_Min_Con_1}
\begin{align}
(\mu_{12}^2)^\mathrm{I} ={}& -\frac{1}{2} \left[ \lambda_{1112}^\mathrm{I} \hat{v}_1^2 + \left( 2 \lambda_{1212}^\mathrm{I} \hat{v}_1 + \lambda_{1222}^\mathrm{I} \hat{v}_2 \right)\hat{v}_2  \right],\\
\begin{split}
(\mu_{12}^2)^\mathrm{R} ={}& -\frac{1}{2}\Bigg[ 3 \lambda_{1112}^\mathrm{R} \hat v_1^2 + 2\left( \mu_{11}^2 + \lambda_{1111} \hat v_1^2 \right) \frac{\hat v_1}{\hat v_2} \\
& \hspace{30pt} + \left( \lambda_{1122} + 2 \lambda_{1212}^\mathrm{R} + \lambda_{1221} \right)  \hat v_1 \hat v_2 +  \lambda_{1222}^\mathrm{R} \hat v_2^2 \Bigg],
\end{split}\\
\mu_{22}^2 ={}&   \left( \mu_{11}^2 \hat{v}_1^2 + \lambda_{1111} \hat{v}_1^4 + \lambda_{1112}^\mathrm{R} \hat v_1^3 \hat v_2 - \lambda_{1222}^\mathrm{R} \hat v_1 \hat v_2^3 - \lambda_{2222} \hat{v}_2^4 \right)\frac{1}{\hat{v}_2^2}.
\end{align}
\end{subequations}

To simplify the off-diagonal elements of the mass-squared matrices we shall define the ratio of two vevs as (\textit{cf.} $\tan \beta$ in the 2HDM):
\begin{equation}
\hat v_{i/j} \equiv \frac{\hat v_i}{\hat v_j}.
\end{equation}

The charged mass-squared matrix in the basis $\{h_1^\pm,\, h_2^\pm,\, h_3^\pm \}$ is:
\begin{equation}\label{Eq:MCh2_v1_v2_0}
\mathcal{M}_\mathrm{Ch}^2 = \begin{pmatrix}
\left( \mathcal{M}_\textrm{Ch}^2 \right)_{11} & -\left( \mathcal{M}_\textrm{Ch}^2 \right)_{11} \hat v_{1/2} & 0 \\
-\left( \mathcal{M}_\textrm{Ch}^2 \right)_{11} \hat v_{1/2} & \left( \mathcal{M}_\textrm{Ch}^2 \right)_{11} \hat v_{1/2}^2 & 0 \\
 0 & 0 & \left( \mathcal{M}_\textrm{Ch}^2 \right)_{33}
\end{pmatrix},
\end{equation}
with 
\begin{subequations}
\begin{align}
\left( \mathcal{M}_\textrm{Ch}^2 \right)_{11} ={}& \mu_{11}^2 + \lambda_{1111} \hat v_1^2 + \lambda_{1112}^\mathrm{R}\hat v_1 \hat v_2 + \frac{1}{2} \lambda_{1122} \hat v_2^2,\\
\left( \mathcal{M}_\textrm{Ch}^2 \right)_{33} ={}& \mu_{33}^2 + \frac{1}{2}\lambda_{1133} \hat v_1^2 + \lambda_{1233}^\mathrm{R} \hat v_1 \hat v_2 + \frac{1}{2} \lambda_{2233} \hat v_2^2.
\end{align}
\end{subequations}

The neutral mass-squared matrix in the basis $\{\eta_1,\, \eta_2,\, \chi_1,\, \chi_2\}{-}\{\eta_3,\,\chi_3\}$ is:
\begin{equation}\label{Eq:MN2_v1_v2_0}
\mathcal{M}_\mathrm{N}^2 = \scriptstyle \begin{pmatrix}
(\mathcal{M}_\mathrm{N}^2)_{11} & (\mathcal{M}_\mathrm{N}^2)_{12}  & (\mathcal{M}_\mathrm{N}^2)_{13} & -(\mathcal{M}_\mathrm{N}^2)_{13} \hat v_{1/2} & 0 & 0\\
(\mathcal{M}_\mathrm{N}^2)_{12} & (\mathcal{M}_\mathrm{N}^2)_{22}  & (\mathcal{M}_\mathrm{N}^2)_{23} & -(\mathcal{M}_\mathrm{N}^2)_{23} \hat v_{1/2} & 0 & 0\\
(\mathcal{M}_\mathrm{N}^2)_{13} & (\mathcal{M}_\mathrm{N}^2)_{23}  & (\mathcal{M}_\mathrm{N}^2)_{33} & -(\mathcal{M}_\mathrm{N}^2)_{33} \hat v_{1/2} & 0 & 0\\
-(\mathcal{M}_\mathrm{N}^2)_{13} \hat v_{1/2} & -(\mathcal{M}_\mathrm{N}^2)_{23} \hat v_{1/2}  & -(\mathcal{M}_\mathrm{N}^2)_{33} \hat v_{1/2} & (\mathcal{M}_\mathrm{N}^2)_{33} \hat v_{1/2}^2 & 0 & 0\\
0 & 0 & 0 &  0 & (\mathcal{M}_\mathrm{N}^2)_{44} & 0 \\
0 & 0 & 0 & 0 & 0 & (\mathcal{M}_\mathrm{N}^2)_{44}  \\
\end{pmatrix},
\end{equation}

where
\begin{subequations}
\begin{align}
\begin{split}
(\mathcal{M}_\mathrm{N}^2)_{11} ={}& \mu_{11}^2 + 3 \lambda_{1111} \hat v_1^2 + 3 \lambda_{1112}^\mathrm{R} \hat v_1 \hat v_2 + \frac{1}{2}\left( \lambda_{1122} + 2 \lambda_{1212}^\mathrm{R} + \lambda_{1221} \right) \hat v_2^2,
\end{split}\\ 
\begin{split}
(\mathcal{M}_\mathrm{N}^2)_{12} ={}& - \left( \mu_{11}^2 + \lambda_{1111} \hat v_1^2 \right)\hat v_{1/2}^2 + \lambda_{1222}^\mathrm{R} \hat v_2^2+ \frac{1}{2} \left(  \lambda_{1122} + 2\lambda_{1212}^\mathrm{R} + \lambda_{1221}\right) \hat v_1 \hat v_2,
\end{split}\\
\begin{split}
(\mathcal{M}_\mathrm{N}^2)_{13} ={}& \lambda_{1112}^\mathrm{I} \hat v_1 \hat v_2 + \lambda_{1212}^\mathrm{I}\hat v_2^2,
\end{split}\\
\begin{split}
(\mathcal{M}_\mathrm{N}^2)_{22} ={}& \left( \mu_{11}^2 + \lambda_{1111} \hat v_1^2 \right)\hat v_{1/2}^2  + 2 \lambda_{2222} \hat v_2^2 + \frac{1}{2} \left( \lambda_{1122} + \lambda_{1221} + 2 \lambda_{1212}^\mathrm{R} \right) \hat v_1^2 \\
& +  \lambda_{1112}^\mathrm{R} v_1^2 \hat v_{1/2} + 2 \lambda_{1222}^\mathrm{R} \hat v_1 \hat v_2,
\end{split}\\
\begin{split}
(\mathcal{M}_\mathrm{N}^2)_{23} ={}& \lambda_{1212}^\mathrm{I} \hat v_1 \hat v_2 + \lambda_{1222}^\mathrm{I} \hat v_2^2,
\end{split}\\
\begin{split}
(\mathcal{M}_\mathrm{N}^2)_{33} ={}& \mu_{11}^2 + \lambda_{1111} \hat v_1^2 + \lambda_{1112}^\mathrm{R} \hat v_1 \hat v_2 + \frac{1}{2}\left( \lambda_{1122} - 2 \lambda_{1212}^\mathrm{R} + \lambda_{1221} \right) \hat v_2^2,
\end{split}\\
\begin{split}
(\mathcal{M}_\mathrm{N}^2)_{44} ={}& \mu_{33}^2 + \frac{1}{2} \left( \lambda_{1133} + \lambda_{1331} \right) \hat v_1^2 + \frac{1}{2} \left( \lambda_{2233} + \lambda_{2332} \right) \hat v_2^2\\
& + \left( \lambda_{1233}^\mathrm{R} + \lambda_{1332}^\mathrm{R} \right) \hat v_1 \hat v_2.
\end{split}
\end{align}
\end{subequations}

Although $\left\langle h_3 \right\rangle = 0$ there is no freedom to rotate away one of the imaginary parts of the quartic couplings since those fields always come in pairs proportional to $h_{33}$. Due to the complexity of the analytical expressions of the mass-squared parameters we do not provide the mass eigenvalues explicitly.

In this implementation $\{ \mu_{11}^2,\, \mu_{33}^2,\, \lambda_{1111},\, \lambda_{1122},\, \lambda_{1133},\, \lambda_{1221},\, \lambda_{1331},\, \lambda_{2222},\, \lambda_{2233},\,$  $ \lambda_{2332},\\$ $\lambda_{1112},\, \lambda_{1212},\, \lambda_{1222},\, \lambda_{1233}^\mathrm{R},\, \lambda_{1332}^\mathrm{R}\}$
enter in the mass-squared matrices, and are responsible for generating six different mass-squared parameters. Some couplings appear only in the scalar interactions: $\{\lambda_{3333},\,$ $ \lambda_{1233}^\mathrm{I},\, \lambda_{1332}^\mathrm{I} \}$.

It should be noted that this case contains an implementation with all coefficients real. The two cases, with or without complex couplings and real vevs, are physically distinct since one does not allow for CP violation in the scalar sector, while the other does.

\bigskip\textbf{Case of \boldmath$(\hat{v}_1 e^{i \sigma},\, \hat{v}_2,\, 0)$ and real couplings}

Let us next consider the case with complex vevs and real couplings, $(\mu_{12}^2)^\mathrm{I}=\lambda_{1212}^\mathrm{I} = \lambda_{1112}^\mathrm{I} = \lambda_{1222}^\mathrm{I}= \lambda_{1233}^\mathrm{I}= \lambda_{1332}^\mathrm{I}=0$. The minimisation conditions are:
\begin{subequations}\label{Eq:U12_Min_Con_2}
\begin{align}
\mu_{11}^2 ={}& -\lambda_{1111} \hat v_1^2 - \cos \sigma \lambda_{1112}^\mathrm{R} \hat v_1 \hat v_2 - \frac{1}{2}\left( \lambda_{1122} - 2 \lambda_{1212}^\mathrm{R} + \lambda_{1221} \right)\hat v_2^2,\\
(\mu_{12}^2)^\mathrm{R} ={}& -\frac{1}{2} \left[  \lambda_{1112}^\mathrm{R} \hat v_1^2 + \left( 4 \cos \sigma \lambda_{1212}^\mathrm{R} \hat v_1 + \lambda_{1222}^\mathrm{R} \hat v_2\right) \hat v_2 \right],\\
\mu_{22}^2 ={}& - \left( \cos \sigma \lambda_{1222}^\mathrm{R} \hat v_1 + \lambda_{2222} \hat v_2 \right) \hat v_2 - \frac{1}{2}\left( \lambda_{1122} - 2 \lambda_{1212}^\mathrm{R} + \lambda_{1221} \right)\hat v_1^2.
\end{align}
\end{subequations}

The structure of the charged mass-squared matrix is identical to eq.~\eqref{Eq:MCh2_v1_v2_0} with 
\begin{subequations}
\begin{align}
\left( \mathcal{M}_\textrm{Ch}^2 \right)_{11} ={}& \frac{1}{2} \left( 2 \lambda_{1212}^\mathrm{R} - \lambda_{1221} \right) \hat v_2^2,\\
\left( \mathcal{M}_\textrm{Ch}^2 \right)_{33} ={}& \mu_{33}^2 + \frac{1}{2}\lambda_{1133} \hat v_1^2 + \cos\sigma\lambda_{1233}^\mathrm{R} \hat v_1 \hat v_2 + \frac{1}{2} \lambda_{2233} \hat v_2^2.
\end{align}
\end{subequations}

The structure of the neutral mass-squared matrix is identical to eq.~\eqref{Eq:MN2_v1_v2_0} where
\begin{subequations}
\begin{align}
(\mathcal{M}_\mathrm{N}^2)_{11} ={}& 2 \left[ \lambda_{1111} \hat v_1^2 + \cos \sigma \left( \lambda_{1112}^\mathrm{R} \hat v_1 + \cos \sigma \lambda_{1212}^\mathrm{R} \hat v_2 \right) \hat v_2^2 \right],\\ 
\begin{split}
(\mathcal{M}_\mathrm{N}^2)_{12} ={}& \left(  \lambda_{1122} - 2\sin^2 \sigma\lambda_{1212}^\mathrm{R}  + \lambda_{1221}\right) \hat v_1 \hat v_2 + \cos \sigma \left( \lambda_{1112}^\mathrm{R} \hat v_1^2 + \lambda_{1222}^\mathrm{R} \hat v_2^2 \right),
\end{split}\\
\begin{split}
(\mathcal{M}_
\mathrm{N}^2)_{13} ={}& -\sin \sigma \left( \lambda_{1112}^\mathrm{R} \hat v_1 + 2 \cos \sigma \lambda_{1212}^\mathrm{R} \hat v_2 \right) \hat v_2,
\end{split}\\
(\mathcal{M}_\mathrm{N}^2)_{22} ={}& 2 \left( \cos^2 \sigma \lambda_{1212}^\mathrm{R} \hat v_1^2 + \cos \sigma \lambda_{1222}^\mathrm{R} \hat v_1 \hat v_2 + \lambda_{2222} \hat v_2^2  \right),\\
\begin{split}
(\mathcal{M}_\mathrm{N}^2)_{23} ={}& -\sin \sigma \left( \lambda_{1222}^\mathrm{R} \hat v_2 + 2 \cos \sigma \lambda_{1212}^\mathrm{R} \hat v_1 \right) \hat v_2,
\end{split}\\
(\mathcal{M}_\mathrm{N}^2)_{33} ={}& 2 \sin^2 \sigma \lambda_{1212}^\mathrm{R} \hat v_2^2,\\
\begin{split}
(\mathcal{M}_\mathrm{N}^2)_{44} ={}& \mu_{33}^2 + \frac{1}{2} \left( \lambda_{1133} + \lambda_{1331} \right) \hat v_1^2 + \frac{1}{2} \left( \lambda_{2233} + \lambda_{2332} \right) \hat v_2^2\\
& + \cos \sigma \left( \lambda_{1233}^\mathrm{R} + \lambda_{1332}^\mathrm{R} \right) \hat v_1 \hat v_2.
\end{split}
\end{align}
\end{subequations}

The counting of couplings is identical to the case $(\hat{v}_1,\, \hat{v}_2,\, 0)$, except that $\lambda_{ijkl}^\mathrm{I}=0$. In this case, CP can be spontaneously violated in the scalar sector.

\clearpage
\subsection{Two vanishing vevs}

In the case $(v,\, 0,\, 0)$, the minimisation conditions are:
\begin{subequations}
\begin{align}
\mu_{11}^2 ={}& - \lambda_{1111} v^2,\\
\mu_{12}^2 ={}& - \frac{1}{2} \lambda_{1112} v^2,
\end{align}
\end{subequations}
where $\mu_{12}^2$ is complex. There is a pair of neutral mass-degenerate states, $m_{\eta_3} = m_{\chi_3}$. An interesting observation is that the scalar potential is written in a Higgs-like basis. Here, the SM-like Higgs field, $\eta_1$, is not physical, it mixes with the $\{\eta_2,\, \chi_2\}$ fields, and the neutral would-be Goldstone boson is identified as $\chi_1$. This feature is caused by the presence of the $\lambda_{1112}$ term.

For $(0,\, 0,\, v)$ there is a single minimisation condition:
\begin{equation}
\mu_{33}^2 = - \lambda_{3333} v^2.
\end{equation}
There are two pairs of neutral mass-degenerate states. 

These cases do not spontaneously break the underlying symmetry and can violate CP.

\bigskip\textbf{Case of \boldmath$(v,\,0,\,0)$}

This case is obtained by starting with $(v_1,\, v_2,\, 0)$ and going into the Higgs basis.

The charged mass-squared matrix in the basis $\{h_1^\pm,\, h_2^\pm,\, h_3^\pm \}$ is:
\begin{equation}
\mathcal{M}_\mathrm{Ch}^2 = \mathrm{diag}\left(0,\, \mu_{22}^2 + \frac{1}{2} \lambda_{1122} v^2 ,\, \mu_{33}^2 + \frac{1}{2} \lambda_{1133} v^2   \right).
\end{equation}

The neutral mass-squared matrix in the basis $\{\eta_1,\, \eta_2,\, \chi_1,\, \chi_2 \}{-}\{\eta_3,\, \chi_3\}$ is:
\begin{equation}
\mathcal{M}_\mathrm{N}^2 = \begin{pmatrix}
(\mathcal{M}_\mathrm{N}^2)_{11} & \mathbb{R}\mathrm{e}\big((\mathcal{M}_\mathrm{N}^2)_{12} \big) & 0 & -\mathbb{I}\mathrm{m}\big((\mathcal{M}_\mathrm{N}^2)_{12} \big) & 0  & 0\\
\mathbb{R}\mathrm{e}\big((\mathcal{M}_\mathrm{N}^2)_{12} \big) & (\mathcal{M}_\mathrm{N}^2)_{22}  & 0 & (\mathcal{M}_\mathrm{N}^2)_{24} & 0  & 0\\
0 & 0 & 0 & 0 & 0 & 0 \\
-\mathbb{I}\mathrm{m}\big((\mathcal{M}_\mathrm{N}^2)_{12} \big) & (\mathcal{M}_\mathrm{N}^2)_{24}  & 0 & (\mathcal{M}_\mathrm{N}^2)_{44} & 0 & 0\\
0 & 0 & 0 & 0 & (\mathcal{M}_\mathrm{N}^2)_{55} & 0 \\
0 & 0 & 0 & 0 & 0 & (\mathcal{M}_\mathrm{N}^2)_{55} 
\end{pmatrix},
\end{equation}
where
\begin{subequations}
\begin{align}
(\mathcal{M}_\mathrm{N}^2)_{11} ={}& 2 \lambda_{1111} v^2,\\
(\mathcal{M}_\mathrm{N}^2)_{12} ={}& \lambda_{1112} v^2,\\
(\mathcal{M}_\mathrm{N}^2)_{22} ={}& \mu_{22}^2 + \frac{1}{2} \left( \lambda_{1122} + 2 \lambda_{1212}^\mathrm{R} + \lambda_{1221} \right)v^2,\\
(\mathcal{M}_\mathrm{N}^2)_{24} ={}& -\lambda_{1212}^\mathrm{I} v^2,\\
(\mathcal{M}_\mathrm{N}^2)_{44} ={}& \mu_{22}^2 + \frac{1}{2} \left( \lambda_{1122} - 2 \lambda_{1212}^\mathrm{R} + \lambda_{1221} \right)v^2,\\
(\mathcal{M}_\mathrm{N}^2)_{55} ={}& \mu_{33}^2 + \frac{1}{2} \left( \lambda_{1133} + \lambda_{1331} \right)v^2.
\end{align}
\end{subequations}

There is mixing among the $\{\eta_1,\, \eta_2, \chi_2\}$ fields due to the imaginary parts of the quartic couplings, $\{\lambda_{1112}^\mathrm{I},\, \lambda_{1212}^\mathrm{I}\}$. Since $\left\langle h_2 \right\rangle = 0$ there is freedom to re-phase the $h_2$ doublet and absorb one of the imaginary quartic coupling parts. However, there exists no rotation to split the $\{\eta_1,\, \eta_2\}$ and $\chi_2$ parts into a block-diagonal form.

In this implementation $\{ \mu_{22}^2,\, \mu_{33}^2,\, \lambda_{1111},\, \lambda_{1122},\, \lambda_{1133},\, \lambda_{1221},\, \lambda_{1331},\, \lambda_{1112},\, \lambda_{1212} \}$
enter in the mass-squared matrices, and are responsible for generating six different mass-squared parameters. Some couplings appear only in the scalar interactions: $\{\lambda_{2222},\, \lambda_{2233},\, \lambda_{2332},\,$ $\lambda_{3333},\, \lambda_{1222},\, \lambda_{1233},\, \lambda_{1332}\}$.

\bigskip\textbf{Case of \boldmath$(0,\,0,\,v)$}

The charged mass-squared matrix in the basis $\{h_1^\pm,\, h_2^\pm,\, h_3^\pm \}$ is:
\begin{equation}
\mathcal{M}_\mathrm{Ch}^2 = \begin{pmatrix}
\left( \mathcal{M}_\textrm{Ch}^2 \right)_{11}  & \left( \mathcal{M}_\textrm{Ch}^2 \right)_{12}^\ast  & 0 \\
\left( \mathcal{M}_\textrm{Ch}^2 \right)_{12}  & \left( \mathcal{M}_\textrm{Ch}^2 \right)_{22} & 0 \\
 0 & 0 & 0
\end{pmatrix},
\end{equation}
with 
\begin{equation}
\left( \mathcal{M}_\textrm{Ch}^2 \right)_{ij} = \mu_{ii}^2 + \frac{1}{2} \lambda_{ij33} v^2.
\end{equation}

The neutral mass-squared matrix in the basis $\{\eta_1,\, \eta_2,\, \chi_1,\, \chi_2\}{-}\{\eta_3\}{-}\{\chi_3\}$ is:
\begin{equation}
\mathcal{M}_\mathrm{N}^2 = \begin{pmatrix}
(\mathcal{M}_\mathrm{N}^2)_{11} & \mathbb{R}\mathrm{e}\left((\mathcal{M}_\mathrm{N}^2)_{12} \right) & 0 & -\mathbb{I}\mathrm{m}\left((\mathcal{M}_\mathrm{N}^2)_{12} \right) & 0  & 0\\
\mathbb{R}\mathrm{e}\left((\mathcal{M}_\mathrm{N}^2)_{12} \right) & (\mathcal{M}_\mathrm{N}^2)_{22} & \mathbb{I}\mathrm{m}\left((\mathcal{M}_\mathrm{N}^2)_{12} \right) & 0 & 0  & 0 \\
0 & \mathbb{I}\mathrm{m}\left((\mathcal{M}_\mathrm{N}^2)_{12} \right) & (\mathcal{M}_\mathrm{N}^2)_{11} & \mathbb{R}\mathrm{e}\left((\mathcal{M}_\mathrm{N}^2)_{12} \right) & 0  & 0\\
-\mathbb{I}\mathrm{m}\left((\mathcal{M}_\mathrm{N}^2)_{12} \right) & 0 & \mathbb{R}\mathrm{e}\left((\mathcal{M}_\mathrm{N}^2)_{12} \right) & (\mathcal{M}_\mathrm{N}^2)_{22} & 0  & 0\\
0 & 0 & 0 & 0 & (\mathcal{M}_\mathrm{N}^2)_{33} & 0 \\
0 & 0 & 0 & 0 & 0 & 0  \\
\end{pmatrix},
\end{equation}
where
\begin{subequations}
\begin{align}
(\mathcal{M}_\mathrm{N}^2)_{ij} ={}& \mu_{ij}^2 + \frac{1}{2} \left( \lambda_{ij33} + \lambda_{i33j} \right) v^2,\text{ for } \{i,\, j\}=\{1,\,2\},\\
(\mathcal{M}_\mathrm{N}^2)_{33} ={}& 2 \lambda_{3333} v^2.
\end{align}
\end{subequations}

There is mixing among the $\{\eta_1,\, \eta_2,\, \chi_1,\, \chi_2\}$ states due to the imaginary parts of the couplings, $\{\left(\mu_{12}^2 \right)^\mathrm{I},\, \lambda_{1233}^\mathrm{I},\, \lambda_{1332}^\mathrm{I}\}$. Since only the $h_3$ doublet develops a non-zero vev, it is possible to rotate away two out of three of the imaginary couplings causing mixing between the pairs of fields $\{\eta_1,\, \eta_2\}$ and $\{\chi_1,\, \chi_2\}$.

The mass eigenvalues of the neutral states associated with the $h_1$ and $h_2$ doublets, the upper-left four-by-four block of $\mathcal{M}_\mathrm{N}^2$, are pairwise degenerate:
\begin{equation}
m_{H_i}^2 =  \frac{1}{4} \left[ 2\left( \mu_{11}^2 + \mu_{22}^2 \right) +  v^2 \left(\lambda_{1133} + \lambda_{2233} + \lambda_{1331} + \lambda_{2332}\right) \pm \Delta \right],~i=1..4,
\end{equation}
where
\begin{equation}
\begin{aligned}
\Delta^2 ={}&  \left[ v^2 \left( \lambda_{1133} - \lambda_{2233} + \lambda_{1331} - \lambda_{2332} \right) + 2 \left( \mu_{11}^2 - \mu_{22}^2\right)\right]^2\\
&  + 4  \left| v^2 \left( \lambda_{1233} + \lambda_{1332}\right) + 2 \mu_{12}^2\right|^2.
\end{aligned}
\end{equation}

In this implementation $\{ \mu_{11}^2,\, \mu_{12}^2,\, \mu_{22}^2,\, \lambda_{1133},\, \lambda_{1331},\, \lambda_{2233},\, \lambda_{2332},\, \lambda_{3333},\, \lambda_{1233},\, \lambda_{1332} \}$ enter in the mass-squared matrices, and are responsible for generating six different mass-squared parameters. Some couplings appear only in the scalar interactions: $\{\lambda_{1111},\,  \lambda_{1122},\, $ $\lambda_{1221},\, \lambda_{2222},\, \lambda_{1112},\, \lambda_{1212},\, \lambda_{1222}\}$.

\section{Identifying more constrained potentials}\label{Sec:Pot_U1_additonal}

This section is devoted to a discussion of potentials that can be constructed by imposing additional constraints on the $U(1)$-based potentials discussed above. Some of these can be identified in terms of symmetries that have been considered in the 3HDM literature (we shall mainly focus on Refs.~\cite{deMedeirosVarzielas:2019rrp,Darvishi:2019dbh})\footnote{In Table II of Ref.~\cite{Darvishi:2019dbh}, an overall factor of ``1/2" multiples the $h_{ij}^2$ terms, see their eq.~(A.4).}. Three of the models discussed here seem not to have been investigated in the literature. We do not consider the $Sp(n)$ group, and its products, classified in Ref.~\cite{Darvishi:2019dbh}, since they do not apply directly to the $SU(2)$ doublets. Moreover, in some instances of this section the notation $``\times"$ is used instead of  $``\rtimes"$, where the latter might be more appropriate.

In the context of applying symmetries to potentials it is important to keep in mind that, since the scalar potential is restricted to bilinear and quartic terms, two different symmetries, $\mathcal{G}_a$ and $\mathcal{G}_b$ may lead to the same potential. This may even happen if one group is continuous and the other discrete, and also if a complex conjugation is involved.

In this section we shall predominantly focus on the $\{O(2) \times U(1),\, \left[ U(1) \times U(1)\right] \rtimes S_3,$ $U(1) \times D_4\}$ groups, which we did not find discussed in the literature. It was pointed out that the last symmetry, $U(1) \times D_4$, due to a non-trivial intersection of the groups might better be denoted in terms of a quotient group~\cite{Ivanonv_pr}. Even with these three additional cases our list might still be incomplete.

We found that the scalar potentials obtained from these three symmetries coincide with those obtained from particular generators of the $\{\mathbb{Z}_4,\, \mathbb{Z}_3,\, Q_8\}$ symmetries respectively. It should be noted that these symmetries are not maximal, in the sense that the scalar potentials are symmetric under further additional generators. The specific generators of $\{\mathbb{Z}_4,\, \mathbb{Z}_3\}$ were discussed in Refs.~\cite{Ivanov:2012ry,Ivanov:2012fp}, being a part of a larger group, but not as single generators. Two of the three models mentioned above, $\{O(2) \times U(1),\, \left[ U(1) \times U(1)\right] \rtimes S_3\}$, were denoted as $\{\left[ U(1) \times U(1)\right] \rtimes \mathbb{Z}_2,\, \left[ U(1) \times U(1)\right] \rtimes \mathbb{Z}_3\}$ in Ref.~\cite{BreeMT}. There, they were encountered through a different path, not related to $U(1)$-based symmetries.

\subsection{\texorpdfstring{$\left[ U(1) \times U(1) \right] \rtimes S_2 \sim O(2) \times U(1)$}{U(1) x U(1) x S2 ~ O(2) x U(1)}}\label{Sec:Ident_O2_U1}

In this section we start with the $U(1) \times U(1)$-symmetric potential and try to further constrain the scalar potential by applying discrete symmetries. Let us consider permutations of two doublets, \textit{i.e.}, $h_i \leftrightarrow h_j$, where $\{i,\,j\} \in \{1,\,2,\,3\}$ and $i \neq j$. Without loss of generality we can assume that additional symmetries act on any pair of the $SU(2)$ doublets; due to the underlying $U(1) \times U(1)$ symmetry. Let us enforce an additional $S_2$ permutation symmetry acting on two doublets, $\{h_1,\, h_2\}$, which we shall denote as $S_2(h_1,h_2)$, with the symmetry given by $\left[ U(1) \times U(1) \right] \rtimes S_2(h_1,h_2)$; we shall show that the corresponding scalar potential coincides with that of the $O(2) \times U(1)$-symmetric 3HDM. We indicate such coincidence of potentials by the ``$\sim$" symbol, \textit{e.g.}, $\left[ U(1) \times U(1) \right] \rtimes S_2 \sim O(2) \times U(1)$. 

The scalar potential invariant under $\left[ U(1) \times U(1) \right] \rtimes S_2$ can be expressed in terms of the transformation
\begin{equation}\label{Eq:U1_U1_S2_Rgen}
\begin{pmatrix}
h_1 \\
h_2 \\
h_3
\end{pmatrix} = \begin{pmatrix}
0 & e^{i \theta_1} & 0 \\
e^{i \theta_2} & 0 & 0 \\
0 & 0 & 1
\end{pmatrix} \begin{pmatrix}
h_1 ^\prime \\
h_2 ^\prime \\
h_3 ^\prime
\end{pmatrix},
\end{equation}
with $V(h_i^\prime) = V(h_i)$. The $h_3$ doublet plays a special role due to the $S_2$ group not acting on it. The scalar potential, invariant under this symmetry, can be written as
\begin{equation}\label{Eq:V_U1_U1_S2}
\begin{aligned}
V ={}& \mu_{11}^2 (h_{11} + h_{22}) + \mu_{33}^2 h_{33} + \lambda_{1111} (h_{11}^2 + h_{22}^2) + \lambda_{3333} h_{33}^2 + \lambda_{1122} h_{11} h_{22}\\
& + \lambda_{1133} (h_{11} h_{33} + h_{22} h_{33}) + \lambda_{1221} h_{12} h_{21} + \lambda_{1331} (h_{13} h_{31} + h_{23} h_{32}).
\end{aligned}
\end{equation}
Such scalar potential, removing one of the $U(1)$ symmetries, remains the same if we alternatively were to impose one of the following symmetries: $U(1)_1 \rtimes S_2(h_1,h_3)$ (or equivalently $U(1)_1 \rtimes S_2(h_2,h_3)$) or else $U(1)_2 \rtimes S_2(h_1,h_3)$ (or equivalently $U(1)_2 \rtimes S_2(h_2,h_3)$). It may seem that the potential of eq.~\eqref{Eq:V_U1_U1_S2} corresponds to $U(1) \times \mathbb{Z}_2$, different from the one provided in eq.~\eqref{Eq:V_U1Z2}, which is not the case. The crucial difference is that eq.~\eqref{Eq:V_U1_U1_S2} allows for two independent $U(1)$ phases, while $U(1) \times \mathbb{Z}_2$ of eq.~\eqref{Eq:V_U1Z2} allows for only one. The resulting potential~\eqref{Eq:V_U1_U1_S2} can also be seen as an $O(2) \times U(1)$-symmetric potential, as explained below. After identifying this underlying symmetry the case will be further on studied in Section~\ref{Sec:Pot_O2_U1}.

The generating set of the underlying symmetry, $\left[ U(1) \times U(1) \right] \rtimes S_2$, can be presented by three generators acting on the scalar fields:
\begin{subequations}
\begin{align}
g_1 ={}& \mathrm{diag}(e^{i \theta_1},\, 1,\, 1), \\
g_2 ={}& \mathrm{diag}(1,\, e^{i \theta_2},\, 1), \\
g_3 ={}& \begin{pmatrix}
0 & 1 & 0\\
1 & 0 & 0\\
0 & 0 & 1
\end{pmatrix},
\end{align}
\end{subequations}
with $\{g_1,\, g_2 \}$ not unique due to the overall $U(1)$ re-phasing symmetry of the potential.

It is known that $S_2$ and $\mathbb{Z}_2$ are isomorphic, $S_2 \cong \mathbb{Z}_2$. These two symmetries are related by the following basis transformation
\begin{equation}\label{Eq:Rot_tr_S2_to_Z2}
\begin{pmatrix}
h_1 \\
h_2 \\
h_3
\end{pmatrix} = \begin{pmatrix}
\cos \theta & \sin \theta & 0\\
-\sin \theta & \cos \theta & 0\\
0 & 0 & 1
\end{pmatrix} \begin{pmatrix}
h_1^\prime \\
h_2^\prime \\
h_3^\prime
\end{pmatrix},
\end{equation}
with $\theta= \pm \pi/4$. Starting from the $S_2$ symmetry, the $\mathbb{Z}_2$ symmetry becomes explicit (in terms of changes of signs and not cyclic permutations),
\begin{equation}\label{Eq:Rot_S2_to_Z2}
\begin{pmatrix}
\cos (\pm \frac{\pi}{4}) & \sin (\pm \frac{\pi}{4}) & 0\\
-\sin (\pm \frac{\pi}{4}) & \cos (\pm \frac{\pi}{4}) & 0\\
0 & 0 & 1
\end{pmatrix}
\begin{pmatrix}
0 & 1 & 0\\
1 & 0 & 0\\
0 & 0 & 1
\end{pmatrix}
\begin{pmatrix}
\cos (\pm \frac{\pi}{4}) & \sin (\pm \frac{\pi}{4}) & 0\\
-\sin (\pm \frac{\pi}{4}) & \cos (\pm \frac{\pi}{4}) & 0\\
0 & 0 & 1
\end{pmatrix}^\dagger = \begin{pmatrix}
\pm 1 & 0 & 0 \\
0 & \mp 1 & 0 \\
0 & 0 & 1
\end{pmatrix}.
\end{equation}
Regardless of the sign of the rotation angle  $\theta$ particularised in eq.~\eqref{Eq:Rot_S2_to_Z2}, this basis transformation (from an explicit $S_2$ to a $\mathbb{Z}_2$) takes the scalar potential of eq.~\eqref{Eq:V_U1_U1_S2} into
\begin{equation}\label{Eq:V_U1_U1_Z2}
\begin{aligned}
V ={}& \mu_{11}^2 (h_{11} + h_{22}) + \mu_{33}^2 h_{33} + \lambda_{a} (h_{11}^2 + h_{22}^2) + \lambda_b h_{12}h_{21}  + \lambda_c h_{11} h_{22} + \lambda_{3333} h_{33}^2\\
&  +\lambda_{1133} (h_{11} h_{33} + h_{22} h_{33})   + \lambda_{1331} (h_{13} h_{31} + h_{23} h_{32})  + \frac{1}{2}\Lambda\left( h_{12}^2 + h_{21}^2 \right),
\end{aligned}
\end{equation}
where
\begin{subequations}\label{Eq:la_lb_lc}
\begin{align}
\lambda_a ={}& \frac{1}{2} \lambda_{1111} + \frac{1}{4} \lambda_{1122} + \frac{1}{4} \lambda_{1221},\\
\lambda_b ={}& \lambda_{1111} - \frac{1}{2} \lambda_{1122} + \frac{1}{2} \lambda_{1221},\\
\lambda_c ={}& \lambda_{1111} + \frac{1}{2} \lambda_{1122} - \frac{1}{2} \lambda_{1221},
\end{align}
\end{subequations}
as well as
\begin{equation}\label{Eq:Incr_sym}
\Lambda \equiv -2 \lambda_a + \lambda_b + \lambda_c = \frac{1}{2}\left( 2\lambda_{1111} - \lambda_{1122}  - \lambda_{1221}\right).
\end{equation}
Setting $\Lambda=0$ will enlarge the underlying symmetry, as will be discussed in the sequel.

The particular transformation of going from the basis of $S_2$ to the basis of $\mathbb{Z}_2$ should be interpreted with caution. First of all, in the example above, eq.~\eqref{Eq:Rot_S2_to_Z2} only takes into account the $g_3$ generator, while the scalar potential is invariant under eq.~\eqref{Eq:U1_U1_S2_Rgen}, which includes two more generators. Moreover, the $U(1) \times U(1)$ symmetric potential does not allow for any phase-sensitive couplings in a particular basis, while as seen from eq.~\eqref{Eq:V_U1_U1_Z2} this underlying symmetry is not explicit any longer, as can be seen by the presence of the $\Lambda$ term. To be more explicit, the rotation into the $\mathbb{Z}_2$ basis, combining all three generators, see eq.~\eqref{Eq:U1_U1_S2_Rgen}, would produce,
\begin{equation}
\begin{aligned}
&\begin{pmatrix}
\cos ( \pm\frac{\pi}{4}) & \sin (\pm \frac{\pi}{4}) & 0\\
-\sin ( \pm\frac{\pi}{4}) & \cos ( \pm\frac{\pi}{4}) & 0\\
0 & 0 & 1
\end{pmatrix}
\begin{pmatrix}
0 & e^{i \theta_1} & 0\\
e^{i \theta_2} & 0 & 0\\
0 & 0 & 1
\end{pmatrix}
\begin{pmatrix}
\cos (\pm \frac{\pi}{4}) & \sin (\pm \frac{\pi}{4}) & 0\\
-\sin (\pm \frac{\pi}{4}) & \cos ( \pm\frac{\pi}{4}) & 0\\
0 & 0 & 1
\end{pmatrix}^\dagger= \\
&\hspace{120pt}= \frac{1}{2}\begin{pmatrix}
\pm\left( e^{i \theta_1} + e^{i \theta_2} \right) & \left( e^{i \theta_1} - e^{i \theta_2} \right) & 0 \\
-\left( e^{i \theta_1} - e^{i \theta_2} \right) & \mp\left( e^{i \theta_1} + e^{i \theta_2} \right) & 0 \\
0 & 0 & 2
\end{pmatrix} \subset U(2).
\end{aligned} 
\end{equation}
The most general $U(2)$-invariant matrix allows for four independent parameters while in our case there are only two.

We could not associate the scalar potential of eq.~\eqref{Eq:V_U1_U1_S2} (or equivalently of eq.~\eqref{Eq:V_U1_U1_Z2}) with any of the symmetries provided in the basis-independent way of Ref.~\cite{deMedeirosVarzielas:2019rrp}, by also requiring that the number of independent couplings would match. We would like to note that if one constructs a scalar potential by demanding invariance under some transformation, not all allowed couplings may be independent. Therefore we try to minimise the number of free couplings, if possible, by rotating into a new basis, assuming at most an $SU(3)$ transformation. In total, there are two independent bilinear couplings and six independent quartic couplings. In Ref.~\cite{Darvishi:2019dbh} the scalar potential invariant under $D_4 \times Sp(2)_{h_3}$:
\begin{equation}\label{Eq:V_D4_Sp2_h3}
\begin{aligned}
V_{D_4 \times Sp(2)_{h_3}} ={}& \mu_{11}^2 (h_{11} + h_{22}) + \mu_{33}^2 h_{33} + \lambda_{1111} (h_{11}^2 + h_{22}^2) + \lambda_{3333} h_{33}^2  + \lambda_{1221} h_{12}h_{21}\\
&  + \lambda_{1122} h_{11} h_{22} +\lambda_{1133} (h_{11} h_{33} + h_{22} h_{33}) + \lambda_{1212}\left( h_{12}^2 + h_{21}^2 \right),
\end{aligned}
\end{equation}
is listed as one with the structure of two plus six independent couplings. We checked for possible basis transformations, of the type $U(2)$, from the scalar potential of eq.~\eqref{Eq:V_U1_U1_S2} into the one of $D_4 \times Sp(2)_{h_3}$ given in eq.~\eqref{Eq:V_D4_Sp2_h3}, and concluded that there is no unitary transformation of the type $(h_{13} h_{31} + h_{23} h_{32}) \to \left( h_{12}^{\prime\,2} + h_{21}^{\prime\,2} \right)$.

One could try to guess which other groups the scalar potential~\eqref{Eq:V_U1_U1_S2} could correspond to. There are several potentials with underlying symmetries $\{O(2)\text{ or }SO(2),\, S_3\text{ or }D_3,\,$ $ D_4,\, \mathrm{CP4} \}$, which can be written in terms of phase-independent plus phase-sensitive parts. With the removal of the phase-sensitive part, \textit{e.g.}, via $U(1)$ symmetries, the truncated potential would correspond to that of eq.~\eqref{Eq:V_U1_U1_S2}. Let us discuss several of these cases.

In the case of $S_3$ one can impose two different symmetries to cancel the phase-sensitive part: $\mathbb{Z}_2$ or $U(1)$. It is known that $S_3 \times \mathbb{Z}_2 \cong D_6$, the generators of which can be written as phases acting on the scalar fields and a permutation symmetry acting on two doublets, with some freedom of signs. The largest realisable cyclic group acting on 3HDM is $\mathbb{Z}_4$. The 3HDM scalar potentials with $\mathbb{Z}_n$ symmetries, with $n \geq 5$, automatically lead to a $U(1)_1$-symmetric potential. As a result, $D_n$, with $n \geq 5$, would result in a scalar potential in the form of eq.~\eqref{Eq:V_U1_U1_S2}. As for the remaining symmetries one would need to impose an additional $\mathbb{Z}_3$ symmetry, acting on $\{O(2)\text{ or }SO(2),\, D_4,\, \mathrm{CP4} \}$ to get rid of the phase-sensitive part since all these cases allow for $h_{ij}^2$.

The case of $O(2)$~\cite{deMedeirosVarzielas:2019rrp} or $SO(2)$~\cite{Darvishi:2019dbh}\footnote{Note that the functional forms of the $SO(2)$-stabilised scalar potential in Ref.~\cite{Darvishi:2019dbh} suggest that the symmetry is presented as $\mathrm{CP1} \times SO(2)$, finding its origin in the 2HDM. Therefore, in Ref.~\cite{Darvishi:2019dbh}, all couplings are assumed to be real.} requires further discussion since applying different $U(1)$ symmetries enlarges the underlying symmetry to different ones. As $O(2) \supset SO(2)$, we shall assume the underlying symmetry to be $O(2)$. Following the work of Ref.~\cite{deMedeirosVarzielas:2019rrp}, we consider the following two bases\footnote{Although both transformations of eqs.~\eqref{Eq:Rep_O2_SO2_Z2} and \eqref{Eq:Rep_O2_U1_Z2} have determinant -1, taking similar transformations with determinant +1 leads to the same potentials.}:
\begin{itemize}
\item Given by orthogonal rotations of the scalar fields
\begin{equation}\label{Eq:Rep_O2_SO2_Z2}
\begin{pmatrix}
h_1 \\
h_2 \\
h_3
\end{pmatrix} = \begin{pmatrix}
\cos \theta & \sin \theta & 0\\
\sin \theta & -\cos \theta & 0\\
0 & 0 & 1
\end{pmatrix} \begin{pmatrix}
h_1 ^\prime \\
h_2 ^\prime \\
h_3 ^\prime
\end{pmatrix}.
\end{equation}

This orthogonal rotation combines $SO(2)$ and $\mathbb{Z}_2$ transformations, leading to the following scalar potential
\begin{equation}\label{Eq:V_O2_SO2_Z2}
\begin{aligned}
V_{SO(2) \rtimes \mathbb{Z}_2} ={}& \mu_{11}^2 (h_{11} + h_{22}) + \mu_{33}^2 h_{33} + \lambda_{1111} (h_{11}^2 + h_{22}^2) + \lambda_{3333} h_{33}^2 + \lambda_{1122} h_{11} h_{22}\\
& + \lambda_{1133} (h_{11} h_{33} + h_{22} h_{33}) + \lambda_{1221} h_{12} h_{21} + \lambda_{1331} (h_{13} h_{31} + h_{23} h_{32})\\
& + \Lambda \left( h_{12}^2 + h_{21}^2 \right) + \{\lambda_{1313} \left( h_{13}^2 + h_{23}^2 \right) + \mathrm{h.c.} \},
\end{aligned}
\end{equation}
or renaming $\Lambda$ as $\lambda_{1212}$ and writing eq.~\eqref{Eq:Incr_sym} as $\lambda_{1122} = 2 (\lambda_{1111} - \frac{1}{2} \lambda_{1221} - \lambda_{1212})$,
\begin{equation}\label{Eq:V_O2_SO2_Z2_II_cons}
\begin{aligned}
V_{SO(2) \rtimes \mathbb{Z}_2} ={}& \mu_{11}^2 (h_{11} + h_{22}) + \mu_{33}^2 h_{33} + \lambda_{1111} (h_{11} + h_{22})^2 + \lambda_{3333} h_{33}^2\\
& + \lambda_{1133} (h_{11} h_{33} + h_{22} h_{33}) + \lambda_{1221} (h_{12} h_{21} - h_{11}h_{22})\\
& + \lambda_{1331} (h_{13} h_{31} + h_{23} h_{32}) +  \lambda_{1212} \left( h_{12}^2 + h_{21}^2  - 2 h_{11} h_{22}\right)\\
& + \{\lambda_{1313} \left( h_{13}^2 + h_{23}^2 \right) + \mathrm{h.c.} \},
\end{aligned}
\end{equation}
presented in a form equivalent to that of Ref.~\cite{deMedeirosVarzielas:2019rrp}.

\clearpage
\item Given in the re-phasing basis of $U(1)_1$ by a transformation,
\begin{equation}\label{Eq:Rep_O2_U1_Z2}
\begin{pmatrix}
h_1 \\
h_2 \\
h_3
\end{pmatrix} = \begin{pmatrix}
0 & e^{- i \theta} & 0 \\
e^{i \theta} & 0 & 0 \\
0 & 0 & 1
\end{pmatrix} \begin{pmatrix}
h_1 ^\prime \\
h_2 ^\prime \\
h_3 ^\prime
\end{pmatrix},
\end{equation}
This re-phasing combines $U(1)_1$ and $\mathbb{Z}_2$ transformations, leading to the following scalar potential
\begin{equation}\label{Eq:V_O2_U1_Z2}
\begin{aligned}
V_{U(1)_1 \rtimes \mathbb{Z}_2} ={}& \mu_{11}^2 (h_{11} + h_{22}) + \mu_{33}^2 h_{33} + \lambda_{1111} (h_{11}^2 + h_{22}^2) + \lambda_{3333} h_{33}^2 + \lambda_{1122} h_{11} h_{22}\\
& + \lambda_{1133} (h_{11} h_{33} + h_{22} h_{33}) + \lambda_{1221} h_{12} h_{21} + \lambda_{1331} (h_{13} h_{31} + h_{23} h_{32})\\
& + \left\lbrace \lambda_{1323} h_{13} h_{23} + \mathrm{h.c.} \right\rbrace.
\end{aligned}
\end{equation}
To be more precise, in this case $\mathbb{Z}_2$ is expressed in terms of permutations, $\mathbb{Z}_2 \cong S_2$.
\end{itemize}

The two representations, eqs.~(\ref{Eq:Rep_O2_SO2_Z2}) and (\ref{Eq:Rep_O2_U1_Z2}), are connected via the transformation
\begin{equation}\label{Eq:Rot_O2_SO2_U1}
\begin{pmatrix}
\cos \theta & \sin \theta & 0 \\
\sin \theta & -\cos \theta & 0 \\
0 & 0 & 1
\end{pmatrix} = \frac{1}{\sqrt{2}} \begin{pmatrix}
1 & 1 & 0 \\
i & -i & 0 \\
0 & 0 & \sqrt{2}
\end{pmatrix} \begin{pmatrix}
0 & e^{- i \theta} & 0 \\
e^{i \theta} & 0 & 0 \\
0 & 0 & 1
\end{pmatrix}
\frac{1}{\sqrt{2}} \begin{pmatrix}
1 & -i & 0 \\
1 & i & 0 \\
0 & 0 & \sqrt{2}
\end{pmatrix}.
\end{equation}

Next, we note that there are two ways to remove the phase-sensitive part from the $O(2)$-symmetric scalar potentials by applying additional $U(1)$ symmetries:
\begin{itemize}
\item By applying $U(1)_2$ or $U(1)_{h_3}$ to the $SO(2) \rtimes \mathbb{Z}_2$ scalar potential of eq.~\eqref{Eq:V_O2_SO2_Z2}:
\begin{equation}\label{Eq:V_O2_U1_expl_O2}
\begin{aligned}
V ={}& \mu_{11}^2 (h_{11} + h_{22}) + \mu_{33}^2 h_{33} + \lambda_{1111} (h_{11}^2 + h_{22}^2) + \lambda_{3333} h_{33}^2 + \lambda_{1122} h_{11} h_{22}\\
& + \lambda_{1133} (h_{11} h_{33} + h_{22} h_{33}) + \lambda_{1221} h_{12} h_{21} + \lambda_{1331} (h_{13} h_{31} + h_{23} h_{32})\\
& + \Lambda \left( h_{12}^2 + h_{21}^2 \right),
\end{aligned}
\end{equation}
which is symmetric under 
\begin{equation}
\begin{pmatrix}
h_1 \\
h_2 \\
h_3
\end{pmatrix} = \begin{pmatrix}
e^{i \beta} & 0 & 0 \\
0 & e^{i \beta} & 0 \\
0 & 0 & 1
\end{pmatrix} \begin{pmatrix}
 \cos \alpha &  \sin \alpha & 0\\
 \sin \alpha &  -\cos \alpha & 0 \\
 0 & 0 & 1
\end{pmatrix} \begin{pmatrix}
h_1 ^\prime \\
h_2 ^\prime \\
h_3 ^\prime
\end{pmatrix}.
\end{equation}
In this case the underlying symmetry is increased to $O(2) \times U(1)$.

\item The application of $U(1)_1$ or $U(1)_{h_1}$ or $U(1)_{h_2}$ to the $SO(2) \rtimes \mathbb{Z}_2$ scalar potential of eq.~\eqref{Eq:V_O2_SO2_Z2} or equivalently applying $U(1)_2$ or $U(1)_{h_i}$ to the $U(1) \rtimes \mathbb{Z}_2$ scalar potential of eq.~\eqref{Eq:V_O2_U1_Z2} forces the phase-sensitive couplings to vanish. In this case the underlying symmetry is increased to $U(2)$~\cite{deMedeirosVarzielas:2019rrp} or an equivalent $SU(2)$~\cite{Darvishi:2019dbh}, with the scalar potential given by
\begin{equation}\label{Eq:V_U2}
\begin{aligned}
V_{U(2)} ={}& \mu_{11}^2 (h_{11} + h_{22}) + \mu_{33}^2 h_{33} + \lambda_{1111} (h_{11}^2 + h_{22}^2) + \lambda_{3333} h_{33}^2\\
& + \left( 2 \lambda_{1111} - \lambda_{1221} \right) h_{11} h_{22} + \lambda_{1133} (h_{11} h_{33} + h_{22} h_{33}) \\
& + \lambda_{1221} h_{12} h_{21}+ \lambda_{1331} (h_{13} h_{31} + h_{23} h_{32}).
\end{aligned}
\end{equation}
\end{itemize}
It is not possible to construct a scalar potential in such a way that only the $(h_{13}^2 + h_{23}^2)$ phase-sensitive part of the $SO(2) \rtimes \mathbb{Z}_2$-symmetric potential would survive. This can be understood from the bilinear terms, which would force the scalar fields transformation to span the subspace of $\mathrm{diag}(U(2),\, U(1))$.

We can perform a basis rotation applied on the scalar potential of eq.~\eqref{Eq:V_U1_U1_S2}
\begin{equation}\label{Eq:h_Rot1_h}
\begin{pmatrix}
h_1 \\
h_2 \\
h_3
\end{pmatrix} = \frac{1}{\sqrt{2}}\begin{pmatrix}
-e^{-i \frac{\pi}{4}} & -e^{i \frac{\pi}{4}} & 0 \\
e^{-i \frac{\pi}{4}} & -e^{i \frac{\pi}{4}} & 0 \\
0 & 0 & \sqrt{2}
\end{pmatrix} \begin{pmatrix}
h_1 ^\prime \\
h_2 ^\prime \\
h_3 ^\prime
\end{pmatrix},
\end{equation}
so that 
\begin{equation}\label{Eq:V_O2_U1_expl_O2_2}
\begin{aligned}
V ={}& \mu_{11}^2 (h_{11} + h_{22}) + \mu_{33}^2 h_{33} + \lambda_a (h_{11}^2 + h_{22}^2) + \lambda_b h_{12} h_{21} + \lambda_c h_{11} h_{22} + \lambda_{3333} h_{33}^2\\
&  + \lambda_{1331} (h_{13} h_{31} + h_{23} h_{32})  + \lambda_{1133} (h_{11} h_{33} + h_{22} h_{33}) - \frac{1}{2} \Lambda (h_{12}^2 + h_{21}^2),
\end{aligned}
\end{equation}
where $\{\lambda_a,\, \lambda_b,\, \lambda_c,\,\Lambda\}$ were provided in eqs.~\eqref{Eq:la_lb_lc} and \eqref{Eq:Incr_sym}. This scalar potential differs from the one provided in eq.~\eqref{Eq:V_U1_U1_Z2} by the minus sign next to the $\Lambda$ term. The scalar potential of eq.~\eqref{Eq:V_O2_U1_expl_O2_2} exhibits an explicit $O(2)$ symmetry, which resembles the potential $V_{SO(2) \rtimes \mathbb{Z}_2} $ of eq.~\eqref{Eq:V_O2_SO2_Z2}. However, note that the scalar potential invariant under $O(2)$ has two independent bilinear couplings and seven independent quartic couplings, while in this case the number of independent couplings is two plus six. This results from the fact that the scalar potential with the $\left[ U(1) \times U(1) \right] \rtimes S_2$ symmetry is more symmetric than just $O(2)$. It amounts to an $O(2) \times U(1)$ symmetry. 

After identifying the scalar potential of eq.~\eqref{Eq:V_U1_U1_S2} to be $O(2) \times U(1)$-symmetric in a certain basis, we would like to reiterate that there are two different $O(2)$ representations, which were given by eqs.~\eqref{Eq:Rep_O2_SO2_Z2} and \eqref{Eq:Rep_O2_U1_Z2}. One could start with either of the scalar potentials of eqs.~\eqref{Eq:V_O2_SO2_Z2} or \eqref{Eq:V_O2_U1_Z2}. If the starting point is the $SO(2) \rtimes \mathbb{Z}_2$-symmetric 3HDM of eq.~\eqref{Eq:V_O2_SO2_Z2}, then it is sufficient to apply any of the $U(1)$ symmetries to remove the $(h_{13}^2 + h_{23}^2)$ term. In this case, we end up with the scalar potential of eq.~\eqref{Eq:V_O2_U1_expl_O2}. On the other hand, if one considered the starting point to be the $U(1)_1 \rtimes \mathbb{Z}_2$-symmetric 3HDM given by eq.~\eqref{Eq:V_O2_U1_Z2}, then one would have to remove the $h_{13}h_{23}$ term. Since there is an underlying $U(1)_1$ symmetry present in the scalar potential, it is possible to remove the term by applying $U(1)_2$ or any of the $U(1)_{h_i}$ symmetries, which were specified in eq.~\eqref{Eq:U1i_charges}. In this case, the scalar potential is given by eq.~\eqref{Eq:V_U1_U1_S2}. As noted earlier, the scalar potentials of eqs.~\eqref{Eq:V_U1_U1_S2} and \eqref{Eq:V_O2_U1_expl_O2_2}, the latter resembling the one of eq.~\eqref{Eq:V_O2_U1_expl_O2},  are connected via the basis transformation of eq.~\eqref{Eq:h_Rot1_h}.

Another interesting observation is that the scalar potential of eq.~\eqref{Eq:V_U1_U1_S2} can be obtained from the most general 3HDM by imposing invariance under the transformation given by
\begin{equation}\label{Eq:O2_U1_ex_pr1}
g= \begin{pmatrix}
0 & e^{i \alpha} & 0 \\
-e^{-i \alpha} & 0 & 0\\
0 & 0 & 1
\end{pmatrix}.
\end{equation}
Applying the generator twice we get $g^2 = \mathrm{diag}(-1,\,-1,\,1)$, which, in turn, shows that ${g^4=\mathcal{I}_3}$. The presence of a free phase $\alpha$ implies that there is a family of groups parameterised by $\alpha$. This generator spans a $\mathbb{Z}_4$ symmetry. In Refs.~\cite{Ivanov:2012ry,Ivanov:2012fp} there was an attempt\footnote{It was concluded that $Q_8$ was not realisable since the potential acquired a higher continuous symmetry.} to construct the $Q_8$-symmetric 3HDM with the presentation of the group given by
\begin{equation}\label{Eq:O2_U1_ex_pr2}
Q_8 = \left\langle a=\mathrm{diag}(i,\,-i,\,1),~b= \begin{pmatrix}
0 & e^{i \alpha} & 0 \\
-e^{-i \alpha} & 0 & 0\\
0 & 0 & 1
\end{pmatrix} \Bigg| ~ a^4=\mathcal{I}_4,~a^2=b^2,~aba=b\right\rangle.
\end{equation}
As can be observed, $b$ of eq.~\eqref{Eq:O2_U1_ex_pr2} is identical to $g$ of eq.~\eqref{Eq:O2_U1_ex_pr1}. The scalar potential invariant under $b$ is also invariant under $a$. Therefore, we can conclude that a specific representation of $Q_8$ yields the $O(2) \times U(1)$-symmetric 3HDM.

The quartic part of the scalar potential of eq.~\eqref{Eq:V_U1_U1_S2} is presented in the bilinear formalism in Appendix~\ref{App:Bilinear_3symm}. This facilities comparison with Ref.~\cite{deMedeirosVarzielas:2019rrp}.

\subsection{\texorpdfstring{$\left[ U(1) \times U(1)\right] \rtimes S_3$}{U(1) x U(1) x S3}}\label{Sec:Ident_U1_U1_S3}

In this section we will study $\left[U(1) \times U(1)\right] \rtimes S_3$. Implementations based on different vacuum configurations will be covered in Section~\ref{Sec:Pot_U1_U1_S3}.

Here, in addition to the $U(1) \times U(1)$ transformations we start by considering a cyclic permutation acting on all three doublets, $h_1 \rightarrow h_2 \rightarrow h_3 \rightarrow h_1$, which corresponds to a representation of $\mathbb{Z}_3$. Then the scalar potential is invariant under the transformation
\begin{equation}\label{Eq:U1_U1_S3_Rgen}
\begin{pmatrix}
h_1 \\
h_2 \\
h_3
\end{pmatrix} = \begin{pmatrix}
0 & e^{i \theta_1} & 0 \\
0 & 0 & e^{i \theta_2} \\
1 & 0 & 0
\end{pmatrix} \begin{pmatrix}
h_1 ^\prime \\
h_2 ^\prime \\
h_3 ^\prime
\end{pmatrix}.
\end{equation}
We write the scalar potential as
\begin{equation}\label{Eq:V_U1_U1_S3}
\begin{aligned}
V ={}& \mu_{11}^2 (h_{11} + h_{22} + h_{33}) + \lambda_{1111} (h_{11}^2 + h_{22}^2 + h_{33}^2)\\
& + \lambda_{1122} (h_{11} h_{22} + h_{22} h_{33} + h_{33} h_{11}) + \lambda_{1221} (h_{12} h_{21} + h_{23} h_{32} + h_{31} h_{13})\\
={}& \mu_{11}^2 \sum_i h_{ii} + \lambda_{1111} \sum_i h_{ii}^2  + \lambda_{1122} \sum_{i<j} h_{ii} h_{jj} + \lambda_{1221} \sum_{i<j} h_{ij} h_{ji}.
\end{aligned}
\end{equation}
Actually, since the scalar potential can be written in terms of sums, the cyclic $\mathbb{Z}_3$ symmetry gets enlarged to an $S_3$ symmetry. 

The generating set acting on the scalar fields is given by
\begin{subequations}
\begin{align}
g_1 ={}& \mathrm{diag}(e^{i \theta_1},\, 1,\, 1), \\
g_2 ={}& \mathrm{diag}(1,\, e^{i \theta_2},\, 1), \\
g_3 ={}& \begin{pmatrix}
0 & 1 & 0\\
0 & 0 & 1\\
1 & 0 & 0
\end{pmatrix}.
\end{align}
\end{subequations}

We note that applying an additional $S_2(h_1,\, h_3)$ or $S_2(h_2,\, h_3)$ to the potential of eq.~\eqref{Eq:V_U1_U1_S2} would also result in eq.~\eqref{Eq:V_U1_U1_S3}. Then, $\left[ U(1) \times U(1)\right] \rtimes S_3 \sim \left[ O(2) \times U(1) \right] \rtimes S_2$. We shall abuse the notation of the semi-direct group to indicate that the two groups are not normal to each other, without specifying the automorphism mapping function.

The total number of free couplings in the scalar potential of eq.~\eqref{Eq:V_U1_U1_S3} coincides with that of the $SO(3)$ symmetry,
\begin{equation}\label{Eq:V_SO3}
\begin{aligned}
V_{SO(3)} ={}& \mu_{11}^2 (h_{11} + h_{22} + h_{33}) + \lambda_{1111} (h_{11}^2 + h_{22}^2 + h_{33}^2)\\
& + \lambda_{1122} (h_{11} h_{22} + h_{22} h_{33} + h_{33} h_{11}) + \lambda_{1221} (h_{12} h_{21} + h_{23} h_{32} + h_{31} h_{13})\\
& + \Lambda (h_{12}^2 + h_{21}^2 + h_{23}^2 + h_{32}^2 + h_{31}^2 + h_{13}^2)\\
={}& \mu_{11}^2 \sum_i h_{ii} + \lambda_{1111} \sum_i h_{ii}^2  + \lambda_{1122} \sum_{i<j} h_{ii} h_{jj} + \lambda_{1221} \sum_{i<j} h_{ij} h_{ji}\\
& + \Lambda \sum_{i<j} (h_{ij}^2 + h_{ji}^2),
\end{aligned}
\end{equation}
one bilinear and three (independent) quartic couplings, however the invariance under the $SO(3)$-symmetric potential strictly requires the presence of the $\Lambda(h_{ij}^2 + h_{ji}^2)$ term. We note that by imposing $\Lambda=0$ (see eq.~\eqref{Eq:Incr_sym}) the underlying symmetry is enlarged to $SU(3)$ (or equivalently $SU(3) \times U(1)$-symmetric in the notation of Ref.~\cite{Darvishi:2019dbh}),
\begin{equation}\label{Eq:V_SU3}
\begin{aligned}
V_{SU(3)} ={}& \mu_{11}^2 (h_{11} + h_{22} + h_{33}) + \lambda_{1111} (h_{11} + h_{22} + h_{33})^2\\
& +\lambda_{1221} (h_{12} h_{21} + h_{23} h_{32} + h_{31} h_{13} - h_{11} h_{22} - h_{22} h_{33} - h_{33} h_{11})\\
={}& \mu_{11}^2 \sum_i h_{ii} + \lambda_{1111} \left( \sum_i h_{ii} \right) ^2 + \lambda_{1221} \sum_{i<j} (h_{ij} h_{ji} -h_{ii} h_{jj} ).
\end{aligned}
\end{equation}

We checked the basis-invariant conditions of Ref.~\cite{deMedeirosVarzielas:2019rrp}, but failed to associate the scalar potential of eq.~\eqref{Eq:V_U1_U1_S3} with any of the symmetries listed there, requiring presence of all couplings allowed by the symmetry. The potential of eq.~\eqref{Eq:V_U1_U1_S3} coincides with that of several other cases, $\{A_4,\, S_4,\, \Delta(54),\, \Sigma(36)\}$, when the phase-sensitive part of these is removed.  Imposing cyclic groups to these parts, $\{A_4,\, S_4\} \times \mathbb{Z}_3$ or $\{\Delta(54),\, \Sigma(36)\} \times \mathbb{Z}_2$, leads to the scalar potential of eq.~\eqref{Eq:V_U1_U1_S3}. It is also possible to get an identical scalar potential for $\{U(1),\, \mathbb{Z}_3 \times U(1),\, \mathbb{Z}_4,$ $ U(1) \times \mathbb{Z}_2,\, O(2) \times U(1),\, S_3,\, D_4\} \times S_3$.

In the previous subsection we mentioned that a specific representation of $\mathbb{Z}_4$ generates a scalar potential identical to that of $O(2) \times U(1)$. In the case of $\left[ U(1) \times U(1)\right] \rtimes S_3$ there also exists such a special transformation. Consider the generator
\begin{equation}
g= \begin{pmatrix}
0 & e^{i \theta_1} & 0 \\
0 & 0 & e^{i \theta_2}\\
e^{-i (\theta_1 + \theta_2)} & 0 & 0
\end{pmatrix},
\end{equation}
which corresponds to a family of $\mathbb{Z}_3$ representations parameterised by $\theta_i$. This generator was considered in Refs.~\cite{Ivanov:2012ry,Ivanov:2012fp}, being part of a larger extension $\left[ \mathbb{Z}_2 \times \mathbb{Z}_2\right] \rtimes \mathbb{Z}_3$ identified as a symmetry group of the tetrahedron, which is of order twenty-four. Fixing the $\theta_i$ phases led the authors of Ref.~\cite{Ivanov:2012fp} to $A_4$, which is of order twelve, and would allow for additional terms, $h_{ij}^2$. With the most general phases this symmetry results in the  potential of eq.~\eqref{Eq:V_U1_U1_S3}, which is symmetric under $\left[ U(1) \times U(1)\right] \rtimes S_3$.

The quartic part of the scalar potential of eq.~\eqref{Eq:V_U1_U1_S3} is presented in the bilinear formalism in Appendix~\ref{App:Bilinear_3symm}, for an easy comparison with Ref.~\cite{deMedeirosVarzielas:2019rrp}.

\subsection{\texorpdfstring{$U(1)_1 \rtimes S_2 \sim O(2)$}{U(1)1 x S2 ~ O(2)}}\label{Sec:Ident_O2}

In this subsection we will identify the $O(2)$-symmetric 3HDM written in terms of $U(1)_1$ transformations; to be more explicit the studied symmetry is $U(1)_1 \rtimes \mathbb{Z}_2$, exploiting $S_2 \cong \mathbb{Z}_2$. This case has already been mentioned in Section~\ref{Sec:Ident_O2_U1} and the possibility of accommodating a DM candidate will be covered in Section~\ref{Sec:Pot_O2}.

For potentials with phase-sensitive terms, $V_\mathcal{G}^\mathrm{ph} \neq 0$, we need to consider only these terms of the scalar potential by applying symmetries in a way not to cancel or introduce any additional ones. The $U(1)_1$ phase-sensitive part given by eq.~\eqref{Eq:Vph_U11},
\begin{equation*}
V^\mathrm{ph}_{U(1)_1} = \lambda_{1323} h_{13} h_{23} + \mathrm{h.c.},
\end{equation*}
is only invariant under an additional transformation of the scalar fields given by eq.~\eqref{Eq:Rep_O2_U1_Z2}. The potential then respects $O(2) \cong U(1) \rtimes \mathbb{Z}_2$, see eq.~\eqref{Eq:V_O2_U1_Z2}. Since there is a single phase-sensitive coupling we can assume that $\lambda_{1323} \in \mathbb{R}$, unless stated otherwise.

\subsection{\texorpdfstring{$\left[U(1) \times \mathbb{Z}_2\right] \rtimes S_2 \sim U(1) \times D_4$}{U(1) x Z2 x S2 ~ U1 x D4}}\label{Sec:Ident_U1_Z2_S2}

In this section we will study $\left[U(1) \times \mathbb{Z}_2\right] \rtimes S_2$ and conclude that it may be written as $U(1) \times D_4$ (this might better be referred to as a quotient group). We believe this case was not previously discussed in the literature. Different vacuum configurations will be discussed in Section~\ref{Sec:Pot_U1_Z2_S2}.

For the $U(1) \times \mathbb{Z}_2$-symmetric 3HDM the phase-sensitive part is given by eq.~\eqref{Eq:Vph_U1Z2},
\begin{equation*}
V^\mathrm{ph}_{U(1) \times \mathbb{Z}_2} = \lambda_{1212} h_{12}^2 + \mathrm{h.c.},
\end{equation*}
which is invariant under $S_2(h_1, h_2)$. In this case the generating set acting on the scalar fields is given by:
\begin{subequations}\label{Eq:Gen_U1_Z2_S2}
\begin{align}
g_1 ={}& \mathrm{diag}(1,\, 1,\, e^{i \alpha}), \\
g_2 ={}& \mathrm{diag}(-1,\, 1,\, 1), \\
g_3 ={}& \begin{pmatrix}
0 & 1 & 0\\
1 & 0 & 0\\
0 & 0 & 1
\end{pmatrix}.
\end{align}
\end{subequations}
Then, the potential invariant under such transformations is given by
\begin{equation}\label{Eq:V_U1_Z2_S2}
\begin{aligned}
V ={}& \mu_{11}^2 (h_{11} + h_{22}) + \mu_{33}^2 h_{33} + \lambda_{1111} (h_{11}^2 + h_{22}^2) + \lambda_{3333} h_{33}^2 + \lambda_{1122} h_{11} h_{22}\\
& + \lambda_{1133} (h_{11} h_{33} + h_{22} h_{33}) + \lambda_{1221} h_{12} h_{21} + \lambda_{1331} (h_{13} h_{31} + h_{23} h_{32})\\
& + \lambda_{1212} (h_{12}^2 + h_{21}^2).
\end{aligned}
\end{equation}
The scalar potential looks identical to that of eq.~\eqref{Eq:V_U1_U1_Z2},
\begin{equation}
\lambda_a = \lambda_{1111},~ \lambda_{b} = \lambda_{1221},~ \lambda_{c} = \lambda_{1122},~\Lambda = \lambda_{1212},
\end{equation}
while $\Lambda$, or rather $\lambda_{1212}$, is an independent coupling in the scalar potential of eq.~\eqref{Eq:V_U1_Z2_S2}.

It should be noted that invariance of the potential under
\begin{equation}
\begin{pmatrix}
h_1 \\
h_2 \\
h_3
\end{pmatrix} = \begin{pmatrix}
0 & -e^{i \alpha} & 0 \\
e^{i \alpha} & 0 & 0 \\
0 & 0 & 1
\end{pmatrix} \begin{pmatrix}
h_1 ^\prime \\
h_2 ^\prime \\
h_3 ^\prime
\end{pmatrix}
\end{equation}
suggests that an additional term $\left\lbrace \lambda_{1112} \left( h_{11}h_{12} - h_{21}h_{22}\right) + \mathrm{h.c.} \right\rbrace$ is allowed. However, this term can be rotated away by going to a new basis,
\begin{equation}
\begin{pmatrix}
h_1 \\
h_2 \\
h_3
\end{pmatrix} = \frac{1}{\sqrt{2}}\begin{pmatrix}
e^{\frac{i \pi}{4}} & e^{\frac{i \pi}{4}} & 0 \\
-e^{-\frac{i \pi}{4}} & e^{-\frac{i \pi}{4}} & 0 \\
0 & 0 & \sqrt{2}
\end{pmatrix} \begin{pmatrix}
h_1 ^\prime \\
h_2 ^\prime \\
h_3 ^\prime
\end{pmatrix}.
\end{equation}
In the new basis $\lambda_{1212}$ becomes complex, but a further re-phasing can make it real, in agreement with eq.~\eqref{Eq:V_U1_Z2_S2}.

It might seem that the scalar potential of eq.~\eqref{Eq:V_U1_Z2_S2} corresponds to the $O(2)$-symmetric potential of eq.~\eqref{Eq:V_O2_SO2_Z2} since there is a common structure apart from the phase-independent term. For this scalar potential to become $O(2)$-symmetric, the condition $\lambda_{1212}= \Lambda$ should be satisfied. Using the basis-invariant conditions of Ref.~\cite{deMedeirosVarzielas:2019rrp} we verified that the symmetry is not $D_4$ either, which has an identical number of free parameters in the potential.

Now, consider a group generated by $\{g_2,\, g_3\}$, with a closed set of elements:
\begin{equation}
\{g_{2}^2,\, g_{2},\, g_{3},\, g_{2} g_{3},\, g_{3} g_{2},\, g_{3} g_{2} g_{3},\, g_{2} g_{3} g_{2},\, (g_{2} g_{3})^2\},
\end{equation}
which is of order eight. In total there are five groups of order eight, one of which is $D_4$. This set of generators can be identified as those of $D_4 \cong \mathbb{Z}_4 \rtimes \mathbb{Z}_2$ by going into another basis with the generators given by
\begin{subequations}\label{Eq:Gen_U1_Z2_S2_2}
\begin{align}
a ={}& g_2 g_3 = \begin{pmatrix}
0 & -1 & 0 \\
1 & 0 & 0\\
0 & 0 & 1
\end{pmatrix},\\
b ={}& g_3,
\end{align}
\end{subequations}
where the $\mathbb{Z}_4$ and $\mathbb{Z}_2$ symmetries are explicit.

Let us discuss the remaining order-eight groups, $\{\mathbb{Z}_8,\, \mathbb{Z}_4 \times \mathbb{Z}_2,\, \mathbb{Z}_2^3,\, Q_8\}$. Both for the $\mathbb{Z}_8$-symmetric and the $\mathbb{Z}_4 \times \mathbb{Z}_2$-symmetric scalar potentials there is the possibility of choosing the generators in such a way that the corresponding potential coincides with one of the $\{U(1)_1,\, U(1) \times U(1),\, O(2) \times U(1) \}$-sym\-metric ones. Next, the $\mathbb{Z}_2^3$-symmetric potential coincides with one of the $\{O(2),\, S_3,\, A_4\}$-symmetric potentials, depending on the basis. The last group of order eight is the quaternion group $Q_8$. In a two-dimensional matrix representation the $Q_8$ group is realisable over the complex numbers, $Q_8 \to \mathrm{SL}(2,\mathbb{C})$,
\begin{equation}
Q_8 = \left\langle \mathcal{I}_3,\, \begin{pmatrix}
 i &  0 &  0  \\
 0 & -i &  0 \\
 0 &  0 &  1
\end{pmatrix},\, \begin{pmatrix}
 0 &  1 &  0  \\
-1 &  0 &  0 \\
 0 &  0 &  1
\end{pmatrix},\, \begin{pmatrix}
 0 &  i &  0  \\
 i &  0 &  0 \\
 0 &  0 &  1
\end{pmatrix} \right\rangle.
\end{equation}
Actually, a larger group of order sixteen, the Pauli group $\mathcal{P}_1$, is also realised. Regardless of the choice of the group, the scalar potential of these representations coincides with that of eq.~\eqref{Eq:V_U1_Z2_S2}.

Although the $Q_8$ group seems like a good match, it is not a continuous group, and the starting point was the $U(1) \times \mathbb{Z}_2$-symmetric 3HDM. We shall identify the group generated by $\{g_2,\, g_3\}$ of eq.~\eqref{Eq:Gen_U1_Z2_S2} to be $D_4$. We are identifying the full set of generators of eq.~\eqref{Eq:Gen_U1_Z2_S2} as those of $U(1) \times D_4$; however, this should probably be denoted more correctly as a quotient group due to a non-trivial intersection of the groups~\cite{Ivanonv_pr}.

The scalar potential corresponding to the $D_4$ symmetry can be written as
\begin{equation}\label{Eq:V_D4pr}
\begin{aligned}
V_{D_4}^\prime ={}& \mu_{11}^2 (h_{11} + h_{22}) + \mu_{33}^2 h_{33} + \lambda_{1111} (h_{11}^2 + h_{22}^2) + \lambda_{3333} h_{33}^2 + \lambda_{1122} h_{11} h_{22}\\
& + \lambda_{1133} (h_{11} h_{33} + h_{22} h_{33}) + \lambda_{1221} h_{12} h_{21} + \lambda_{1331} (h_{13} h_{31} + h_{23} h_{32})\\
& + \lambda_{1212} (h_{12}^2 + h_{21}^2)  + \left\lbrace \lambda_{1313} (h_{13}^2 + h_{23}^2) + \mathrm{h.c.} \right\rbrace . 
\end{aligned}
\end{equation}
In Ref.~\cite{Darvishi:2019dbh} the number of free couplings for the $D_4$-symmetric 3HDM is two plus seven, rather than the two plus eight of eq.~\eqref{Eq:V_D4pr}. If one considers singlets under $D_4$ it is trivial to realise that the condition $\lambda_{1331} = \lambda_{2332} = \mathbb{R}\mathrm{e}\left( \lambda_{3231} \right)$ provided in Ref.~\cite{Darvishi:2019dbh} over-constrains the scalar potential. We find that it should be listed in their table as $\lambda_{1331} = \lambda_{2332}$ and $\mathbb{R}\mathrm{e}\left( \lambda_{3231} \right)$ should appear as an independent parameter\footnote{This point has been clarified with the authors of Ref.~\cite{Darvishi:2019dbh}.}. In Ref.~\cite{deMedeirosVarzielas:2019rrp} the $D_4$-symmetric scalar potential is presented as having two bilinear and eight quartic terms, in agreement with eq.~\eqref{Eq:V_D4pr}. We can perform a basis transformation
\begin{equation}\label{Eq:V_D4pr_to_D4}
\begin{pmatrix}
h_1 \\
h_2 \\
h_3
\end{pmatrix} = \frac{1}{\sqrt{2}}\begin{pmatrix}
e^{i \left( \beta +\frac{\pi}{4}\right)} & e^{-i \left( \beta - \frac{\pi}{4}\right)} & 0 \\
-e^{i \left( \beta -\frac{\pi}{4}\right)} & e^{-i \left( \beta +\frac{\pi}{4}\right)} & 0 \\
0 & 0 & \sqrt{2}
\end{pmatrix} \begin{pmatrix}
h_1^\prime \\
h_2^\prime \\
h_3^\prime
\end{pmatrix},
\end{equation}
so that the scalar potential becomes
\begin{equation}\label{Eq:V_D4}
\begin{aligned}
V_{D_4} ={}& \mu_{11}^2 (h_{11} + h_{22}) + \mu_{33}^2 h_{33} + \lambda_{a} (h_{11}^2 + h_{22}^2) + \lambda_{3333} h_{33}^2 + \lambda_{b} h_{11} h_{22}\\
& + \lambda_{1133} (h_{11} h_{33} + h_{22} h_{33}) + \lambda_{c} h_{12} h_{21} + \lambda_{1331} (h_{13} h_{31} + h_{23} h_{32})\\
& + \left\lbrace \lambda_{d}h_{12}^2  + \lambda_{e} h_{13}h_{23} + \mathrm{h.c.} \right\rbrace,
\end{aligned}
\end{equation}
where
\begin{subequations}
\begin{align}
\lambda_a ={}& \frac{1}{4}\left( 2  \lambda_{1111} + \lambda_{1122} + \lambda_{1221}- 2  \lambda_{1212}  \right),\\
\lambda_b ={}& \lambda_{1111} + \frac{1}{2} \lambda_{1122} - \frac{1}{2} \lambda_{1221}+ \lambda_{1212} ,\\
\lambda_c ={}& \lambda_{1111} - \frac{1}{2} \lambda_{1122} + \frac{1}{2} \lambda_{1221}+ \lambda_{1212} ,\\
\lambda_d ={}& \frac{1}{4}  e^{-4 i \beta}  \left( 2 \lambda_{1111} - \lambda_{1122}  - \lambda_{1221}- 2  \lambda_{1212} \right),\\
\lambda_e ={}& - 2 i \lambda_{1313}.
\end{align}
\end{subequations}
In the basis transformation of eq.~\eqref{Eq:V_D4pr_to_D4} we introduced a redundant $\beta$ phase which induces a mismatch between the number of phases in the scalar potentials of eqs.~\eqref{Eq:V_D4pr} and \eqref{Eq:V_D4}, in order to reproduce the notation of the scalar potential of Ref.~\cite{deMedeirosVarzielas:2019rrp}. On the other hand, in Ref.~\cite{Darvishi:2019dbh} there is no such redundancy in the number of phases.

We pointed out that $a$ of eq.~\eqref{Eq:Gen_U1_Z2_S2_2} generates the $\mathbb{Z}_4$ group. We note that such scalar potential coincides with that of the CP4 model, $\{h_1 \to h_2^\ast,\, h_2 \to -h_1^\ast,\, h_3 \to h_3^\ast\}$, (see Refs.~\cite{Ferreira:2017tvy,Haber:2018iwr} ),
\begin{equation}\label{Eq:V_Z4_alt}
\begin{aligned}
V_{\mathrm{CP4}} ={}& \mu_{11}^2 (h_{11} + h_{22}) +\mu_{33}^2 h_{33} + \lambda_{1111} (h_{11}^2 + h_{22}^2) + \lambda_{3333} h_{33}^2\\
& + \lambda_{1122} h_{11} h_{22} + \lambda_{1133} (h_{11} h_{33} + h_{22} h_{33}) + \lambda_{1221} h_{12} h_{21}\\
& + \lambda_{1331} (h_{13} h_{31} + h_{23} h_{32})+ \lambda_{1212} (h_{12}^2 +h_{21}^2 ) + \lambda_{1313}\left\lbrace  (h_{13}^2 + h_{23}^2) + \mathrm{h.c.} \right\rbrace\\
& + \left\lbrace \lambda_{1112} (h_{11}h_{12} - h_{12}h_{22}) + \mathrm{h.c.}\right\rbrace,
\end{aligned}
\end{equation}
while the specific $\mathbb{Z}_4$ one would allow for a complex $\lambda_{1313}$ coupling. Suppose we perform a basis rotation of the scalar potential of eq.~\eqref{Eq:V_Z4_alt} given by
\begin{equation}
\begin{pmatrix}
h_1 \\
h_2 \\
h_3
\end{pmatrix} = \frac{1}{\sqrt{2}}\begin{pmatrix}
i & 1 & 0 \\
1 & i & 0 \\
0 & 0 & \sqrt{2}
\end{pmatrix} \begin{pmatrix}
h_1^\prime \\
h_2^\prime \\
h_3^\prime
\end{pmatrix}.
\end{equation}
Then in the new basis the scalar potential is given by eq.~\eqref{Eq:V_D4} ($D_4$-symmetric 3HDM), provided that $\lambda_{1112} \in \mathbb{R}$. Furthermore, such scalar potential, with $\lambda_{1112} \in \mathbb{R}$, has a redundant quartic coupling.

The scalar potential of eq.~\eqref{Eq:V_U1_Z2_S2} is also identical to those of $\{{\mathbb{Z}_4 \times S_2 \times U(1)},\,$ $  \mathrm{CP4} \times S_2 \times U(1)\}$, where the $S_2$ group is represented by the $b$ generator of eq.~\eqref{Eq:Gen_U1_Z2_S2_2} and $U(1)$ by $g_1$. We would like to point out that we consider the above mentioned $\mathbb{Z}_4$ to be generated by $\mathrm{diag}(i,\,-i,\,1)$ and not by $a$ of eq.~\eqref{Eq:Gen_U1_Z2_S2_2}.

The quartic part of the scalar potential of eq.~\eqref{Eq:V_U1_Z2_S2} is presented in the bilinear formalism in Appendix~\ref{App:Bilinear_3symm}. This facilities comparison with Ref.~\cite{deMedeirosVarzielas:2019rrp}.

\subsection{\texorpdfstring{$U(1)_2 \rtimes S_2 \sim U(1) \times \mathbb{Z}_2$}{U(1)2 x S2 ~ U(1) x Z2}}

Here, we shall relate the $U(1)_2 \rtimes S_2$-symmetric case to the $U(1) \times \mathbb{Z}_2$-symmetric one. For the $U(1)_2$-symmetric 3HDM the phase-sensitive part was given in eq.~\eqref{Eq:Vph_U12}:
\begin{equation*}
\begin{split}
V_{U(1)_2}^\text{ph} ={}& \mu_{12}^2 h_{12} + \lambda_{1212} h_{12}^2 + \lambda_{1112} h_{11}h_{12}+ \lambda_{1222} h_{12}h_{22}\\
&  + \lambda_{1233} h_{12}h_{33} + \lambda_{1332} h_{13}h_{32} + \mathrm{h.c.}
\end{split}
\end{equation*}
The only possible permutation symmetry, which would not remove all phase-sensitive terms, is $S_2(h_1, h_2)$. Combined with the $U(1)_2$ symmetry, the full  transformation is
\begin{equation}
\begin{pmatrix}
h_1 \\
h_2 \\
h_3
\end{pmatrix} = \begin{pmatrix}
0 & e^{i \alpha} & 0 \\
e^{i \alpha} & 0 & 0 \\
0 & 0 & 1
\end{pmatrix} \begin{pmatrix}
h_1 ^\prime \\
h_2 ^\prime \\
h_3 ^\prime
\end{pmatrix}.
\end{equation}
The scalar potential in this case is 
\begin{equation}\label{Eq:V_U12_S2}
\begin{aligned}
V ={}& \mu_{11}^2 (h_{11} + h_{22}) + \mu_{12}^2 (h_{12}+ h_{21})+\mu_{33}^2 h_{33} + \lambda_{1111} (h_{11}^2 + h_{22}^2) + \lambda_{3333} h_{33}^2\\
& + \lambda_{1122} h_{11} h_{22} + \lambda_{1133} (h_{11} h_{33} + h_{22} h_{33}) + \lambda_{1221} h_{12} h_{21} + \lambda_{1331} (h_{13} h_{31} + h_{23} h_{32})\\
& + \lambda_{1212} (h_{12}^2 +h_{21}^2 ) + \lambda_{1233} (h_{12}h_{33} + h_{21}h_{33}) + \lambda_{1332} (h_{13}h_{32} + h_{23}h_{31})\\
&+ \left\lbrace \lambda_{1112} (h_{11}h_{12} + h_{22}h_{21}) + \mathrm{h.c.} \right\rbrace.
\end{aligned}
\end{equation}
The $U(1)_2 \rtimes S_2(h_1, h_2)$-symmetric scalar potential is related to the $U(1) \times \mathbb{Z}_2$ one through the basis rotation of eq.~\eqref{Eq:Rot_S2_to_Z2}. For further discussion of the $U(1) \times \mathbb{Z}_2$ model see Section~\ref{Sec:Pot_U1Z2}.

\section{\texorpdfstring{\boldmath$ O(2) \times U(1)$}{O(2) x U(1)}-symmetric 3HDM }\label{Sec:Pot_O2_U1}

The $O(2) \times U(1)$ symmetry was identified in Section~\ref{Sec:Ident_O2_U1}. Note that there are two different implementations of the $O(2)$ symmetry, given by eqs.~\eqref{Eq:Rep_O2_SO2_Z2} and \eqref{Eq:Rep_O2_U1_Z2}. Hence, we shall explicitly specify which of the $O(2)$ representations is considered in the subscript of the scalar potential, $V_x$. In the case of the $O(2)$ representation given in terms of $SO(2) \rtimes \mathbb{Z}_2$, with the scalar potential provided in eq.~\eqref{Eq:V_O2_SO2_Z2}, the $O(2) \times U(1)$ scalar potential can be written as the one given by eq.~\eqref{Eq:V_O2_U1_expl_O2},
\begin{equation*}
\begin{aligned}
V_{\left[ SO(2) \rtimes \mathbb{Z}_2 \right] \times U(1) } ={}& \mu_{11}^2 (h_{11} + h_{22}) + \mu_{33}^2 h_{33} + \lambda_{1111} (h_{11}^2 + h_{22}^2) + \lambda_{3333} h_{33}^2 + \lambda_{1122} h_{11} h_{22}\\
& + \lambda_{1133} (h_{11} h_{33} + h_{22} h_{33}) + \lambda_{1221} h_{12} h_{21} + \lambda_{1331} (h_{13} h_{31} + h_{23} h_{32})\\
& + \Lambda \left( h_{12}^2 + h_{21}^2 \right),
\end{aligned}
\end{equation*}
where $\Lambda$  was defined in eq.~\eqref{Eq:Incr_sym}. If instead we start with the $O(2)$ representation given by the $U(1)_1 \rtimes \mathbb{Z}_2$ symmetry, corresponding to the scalar potential provided in eq.~\eqref{Eq:V_O2_U1_Z2}, the $O(2) \times U(1)$ scalar potential can be written as the one given by eq.~\eqref{Eq:V_U1_U1_S2},
\begin{equation*}
\begin{aligned}
V_{\left[ U(1)_1 \rtimes \mathbb{Z}_2 \right] \times U(1) } ={}& \mu_{11}^2 (h_{11} + h_{22}) + \mu_{33}^2 h_{33} + \lambda_{1111} (h_{11}^2 + h_{22}^2) + \lambda_{3333} h_{33}^2 + \lambda_{1122} h_{11} h_{22}\\
& + \lambda_{1133} (h_{11} h_{33} + h_{22} h_{33}) + \lambda_{1221} h_{12} h_{21} + \lambda_{1331} (h_{13} h_{31} + h_{23} h_{32}).
\end{aligned}
\end{equation*}
These  $O(2) \times U(1)$ potentials are connected via the basis transformation of eq.~\eqref{Eq:h_Rot1_h}.

In this section we consider different possible vacua that may, in principle, accommodate DM. In the case of the $U(1) \times U(1)$-symmetric potential it was sufficient to analyse two vacua, $(\hat v_1,\, \hat v_2,\, 0)$ and $(v,\,0,\,0)$. Although the $O(2) \times U(1)$-symmetric potential contains the $U(1) \times U(1)$-symmetric potential, there are more implementations based on different vacua, \textit{i.e.}, one has to consider permutations of different vacuum configurations. To be more precise, in the $O(2) \times U(1)$-symmetric potential there are more ways to implement SSB, while in the $U(1) \times U(1)$ case these were equivalent under the permutation of indices of the coefficients of the potential. All discussed cases are summarised in Tables~\ref{Table:U1U1S2_Cases_1}~and~\ref{Table:U1U1S2_Cases_2}. 

{{\renewcommand{\arraystretch}{1.3}
\begin{table}[htb]
\caption{Similar to Table~\ref{Table:U1_U1_Cases}, but now for $\mathcal{G}_1 \equiv \left[ SO(2) \rtimes \mathbb{Z}_2 \right] \times U(1)$, with the scalar potential given by eq.~\eqref{Eq:V_O2_U1_expl_O2}. None of these cases violates CP.}
\label{Table:U1U1S2_Cases_1}
\begin{center}
\begin{tabular}{|c|c|c|c|c|} \hline\hline
Vacuum & SYM & $V$ & \begin{tabular}[l]{@{}c@{}} Mixing of the \\ neutral states\end{tabular} & Comments \\ \hline
$(\frac{v}{\sqrt{2}} e^{i \sigma},\, \pm \frac{v}{\sqrt{2}},\, 0)$ & $-$& $V_{\mathcal{G}_1}  + \mu_{12}^2$ & \footnotesize\begin{tabular}[l]{@{}c@{}} $\{\eta_1,\, \eta_2,\, \chi_1,\, \chi_2\}$ \\ ${-}\{\eta_3\}{-}\{\chi_3\}$ \end{tabular} & \begin{tabular}[l]{@{}c@{}} $m_{\eta_3} = m_{\chi_3}$\end{tabular} \\ \hline
$(i\frac{v}{\sqrt{2}},\, \pm \frac{v}{\sqrt{2}},\, 0)$ & $-$& $V_{\mathcal{G}_1}$ & \footnotesize\begin{tabular}[l]{@{}c@{}} $\{\eta_1,\, \eta_2\}{-}\{\eta_3\}$\\${-}\{\chi_1,\, \chi_2\}{-}\{\chi_3\}$ \end{tabular} & \footnotesize\begin{tabular}[l]{@{}c@{}} $m_{\eta_{(1,2)}}$ = $m_{\chi_{(1,2)}},~ m_{\eta_3} = m_{\chi_3}$ \end{tabular} \\ \hline
$(\hat{v}_1,\, \hat{v}_2,\, 0)$ & $-$& $V_{\mathcal{G}_1}$ & \footnotesize\begin{tabular}[l]{@{}c@{}} $\{\eta_1,\, \eta_2\}{-}\{\eta_3\}$\\${-}\{\chi_1,\, \chi_2\}{-}\{\chi_3\}$ \end{tabular} & \footnotesize\begin{tabular}[l]{@{}c@{}} $m_{\eta_3} = m_{\chi_3}$ \\ $m_{\eta_{(1,2)}}= m_{\chi_{(1,2)}} = 0$ \end{tabular} \\ \hline
$(0,\, \hat{v}_2,\, \hat{v}_3)$ & $-$& $V_{\mathcal{G}_1}$ & \footnotesize\begin{tabular}[l]{@{}c@{}} $\{\eta_1\}{-}\{\eta_2,\, \eta_3\}$\\${-}\{\chi_1\}{-}\{\chi_2\}{-}\{\chi_3\}$ \end{tabular} & \begin{tabular}[l]{@{}c@{}} $ m_{\eta_1} = m_{\chi_2} = m_{\chi_3} = 0$ \end{tabular} \\ \hline
$(0,\, \hat{v}_2,\, \hat{v}_3)$ & $-$& $V_{\mathcal{G}_1} + (\mu_{23}^2)^\mathrm{R} $ & \footnotesize\begin{tabular}[l]{@{}c@{}} $\{\eta_1\}{-}\{\eta_2,\, \eta_3\}$ \\ ${-}\{\chi_1\}{-}\{\chi_2,\, \chi_3\}$ \end{tabular} & - \\ \hline
$(v,\, 0,\, 0)$  & $-$ & $V_{\mathcal{G}_1}$ & diagonal & \footnotesize\begin{tabular}[l]{@{}c@{}} $m_{\eta_3} = m_{\chi_3}$ \\ $m_{\eta_2}= m_{\chi_1} = 0$ \end{tabular} \\ \hline
$(v,\, 0,\, 0)$  & $-$ & $V_{\mathcal{G}_1} + (\mu_{23}^2)^\mathrm{R} $ & \footnotesize\begin{tabular}[l]{@{}c@{}} $\{\eta_1\}{-}\{\eta_2,\, \eta_3\}$ \\ ${-}\{\chi_1\}{-}\{\chi_2,\, \chi_3\}$ \end{tabular} &  - \\ \hline
$(0,\, 0,\, v)$  & $\checkmark$ & $V_{\mathcal{G}_1}$ & diagonal & \footnotesize\begin{tabular}[l]{@{}c@{}} $ m_{h^+_1} = m_{h^+_2} $, \\ $m_{\eta_1} = m_{\eta_2} = m_{\chi_1} = m_{\chi_2}$ \end{tabular} \\ \hline \hline
\end{tabular}
\end{center}
\end{table}}

{{\renewcommand{\arraystretch}{1.3}
\begin{table}[htb]
\caption{Similar to Table~\ref{Table:U1_U1_Cases}, but now for $\mathcal{G}_2 \equiv \left[ U(1)_1 \rtimes \mathbb{Z}_2 \right] \times U(1)$, with the scalar potential given by eq.~\eqref{Eq:V_U1_U1_S2}. In one case the minimisation conditions can lead to a higher symmetry. None of these cases violates CP.}
\label{Table:U1U1S2_Cases_2}
\begin{center}
\begin{tabular}{|c|c|c|c|c|} \hline\hline
Vacuum & SYM & $V$ & \begin{tabular}[l]{@{}c@{}} Mixing of the \\ neutral states\end{tabular} & Comments \\ \hline
$(\hat{v}_1,\, \hat{v}_2,\, 0)$ & $-$& $V_{U(2)}$ & \footnotesize\begin{tabular}[l]{@{}c@{}} $\{\eta_1,\, \eta_2\}{-}\{\eta_3\}$\\${-}\{\chi_1\}{-}\{\chi_2\}{-}\{\chi_3\}$ \end{tabular} & \footnotesize\begin{tabular}[l]{@{}c@{}} $m_{\eta_3} = m_{\chi_3}$ \\ $m_{\eta_{(1,2)}}= m_{\chi_1} = m_{\chi_2} = 0$ \end{tabular} \\ \hline
$(\hat{v}_1,\, \hat{v}_2,\, 0)$ & $-$& $V_{\mathcal{G}_2}  + (\mu_{12}^2)^\mathrm{R}$ & \footnotesize\begin{tabular}[l]{@{}c@{}} $\{\eta_1,\, \eta_2\}{-}\{\eta_3\}$ \\ ${-}\{\chi_1,\, \chi_2\}{-}\{\chi_3\}$ \end{tabular} & \begin{tabular}[l]{@{}c@{}} $m_{\eta_3} = m_{\chi_3}$ \end{tabular} \\ \hline
$(\frac{v}{\sqrt{2}},\, \pm \frac{v}{\sqrt{2}},\, 0)$ & $-$ & $V_{\mathcal{G}_2}$ & \footnotesize\begin{tabular}[l]{@{}c@{}} $\{\eta_1,\, \eta_2\}{-}\{\eta_3\}$\\${-}\{\chi_1\}{-}\{\chi_2\}{-}\{\chi_3\}$ \end{tabular} & \footnotesize\begin{tabular}[l]{@{}c@{}} $m_{\eta_3} = m_{\chi_3}$ \\ $ m_{\chi_1} = m_{\chi_2} = 0$ \end{tabular} \\ \hline
$(\frac{v}{\sqrt{2}},\, \pm \frac{v}{\sqrt{2}},\, 0)$ & $-$ & $V_{\mathcal{G}_2} + (\mu_{12}^2)^\mathrm{R}$ & \footnotesize\begin{tabular}[l]{@{}c@{}} $\{\eta_1,\, \eta_2\}{-}\{\eta_3\}$ \\ ${-}\{\chi_1,\, \chi_2\}{-}\{\chi_3\}$ \end{tabular} & \begin{tabular}[l]{@{}c@{}} $m_{\eta_3} = m_{\chi_3}$ \end{tabular} \\ \hline
$(0,\, \hat{v}_2,\, \hat{v}_3)$ & $-$& $V_{\mathcal{G}_2}$ & \footnotesize\begin{tabular}[l]{@{}c@{}} $\{\eta_1\}{-}\{\eta_2,\, \eta_3\}$\\${-}\{\chi_1\}{-}\{\chi_2\}{-}\{\chi_3\}$ \end{tabular} & \footnotesize\begin{tabular}[l]{@{}c@{}} $m_{\eta_1} = m_{\chi_1}$ \\ $ m_{\chi_2} = m_{\chi_3} = 0$ \end{tabular} \\ \hline
$(0,\, \hat{v}_2,\, \hat{v}_3)$ & $-$& $V_{\mathcal{G}_2} + (\mu_{23}^2)^\mathrm{R} $ & \footnotesize\begin{tabular}[l]{@{}c@{}} $\{\eta_1\}{-}\{\eta_2,\, \eta_3\}$ \\ ${-}\{\chi_1\}{-}\{\chi_2,\, \chi_3\}$ \end{tabular} & \begin{tabular}[l]{@{}c@{}} $m_{\eta_1} = m_{\chi_1}$ \end{tabular} \\ \hline
$(v,\, 0,\, 0)$  & $-$ & $V_{\mathcal{G}_2}$ & diagonal & \footnotesize\begin{tabular}[l]{@{}c@{}} $m_{\eta_2} = m_{\chi_2},~m_{\eta_3} = m_{\chi_3}$ \end{tabular} \\ \hline
$(0,\, 0,\, v)$  & $\checkmark$ & $V_{\mathcal{G}_2}$ & diagonal & \footnotesize\begin{tabular}[l]{@{}c@{}} $ m_{h^+_1} = m_{h^+_2} $, \\ $m_{\eta_1} = m_{\eta_2} = m_{\chi_1} = m_{\chi_2}$ \end{tabular} \\ \hline \hline
\end{tabular}
\end{center}
\end{table}}

\subsection(A){\texorpdfstring{Case of $\left[ SO(2) \rtimes \mathbb{Z}_2 \right] \times U(1)$}{SO(2) x Z2 x U(1)}}\label{Sec:SO2_Z2_U1}

\subsubsection{One vanishing vev}

Although the scalar potential in the basis of $\left[ SO(2) \rtimes \mathbb{Z}_2 \right] \times U(1)$ is real, the $\Lambda$ term is phase-sensitive. Therefore we need to consider vacuum given by $(v_1,\, v_2,\, 0)$. Then,
\begin{equation}\label{Eq:SO2_Z2_U1_DiffCond_CompVev}
\Lambda \sin^2 \sigma \left( \hat v_1^2 - \hat v_2^2\right) = 0,
\end{equation}
as required by the minimisation conditions, needs to be satisfied. There are three solutions. First, let us consider $\sigma \neq 0$. For $(\hat v_1 e^{i \sigma},\, \hat v_2,\, 0)$ the minimisation conditions are:
\begin{subequations}
\begin{align}
\mu_{11}^2 ={}& - \lambda_{1111} \left(\hat v_1^2 + \hat v_2^2 \right),\\
\Lambda  ={}& 0,
\end{align}
\end{subequations}
and the overall symmetry is increased to $U(2)$. However, once the underlying symmetry is increased to $U(2)$, the $\sigma$ phase is unphysical and can be rotated away. This case is inconsistent  since $\sigma=0$ would result in a completely different implementation, as will be discussed below. We consider this case not to be realisable.

This case also suffers from two additional unwanted massless states. The usual approach would be to introduce a soft term $\mu_{12}^2$. Then, there are two possible solutions of the minimisation conditions. One solution would promote only one massless state to a massive one. Another solution forces vevs to be related, resulting in $ (\frac{v}{\sqrt{2}} e^{i \sigma},\, \pm \frac{v}{\sqrt{2}} ,\, 0)$. In this case the minimisation conditions are given by:
\begin{subequations}
\begin{align}
\left( \mu_{12}^2 \right)^\mathrm{R} ={}& \mp \cos \sigma \left( \mu_{11}^2 + \lambda_{1111} v^2 \right),\\
\left( \mu_{12}^2 \right)^\mathrm{I} ={}& \mp \sin \sigma \left( \mu_{11}^2 + \frac{1}{2} \left( \lambda_{1122} + \lambda_{1221} \right) v^2 \right).
\end{align}
\end{subequations}
Notice that the phase of $\mu_{12}^2$ is not equal to  $\sigma$. Although there are two independent phases, it is possible to perform a transformation into a basis with real coefficients. 

Another solution of eq.~\eqref{Eq:SO2_Z2_U1_DiffCond_CompVev}, with $\Lambda \neq 0$, forces the vevs to be related, $ (\frac{v}{\sqrt{2}} e^{i \sigma},\, \pm \frac{v}{\sqrt{2}} ,\, 0)$. For this vacuum, from the other minimisation conditions, we need to impose $\sigma = \pi/2$. In the case $ (i\frac{v}{\sqrt{2}},\, \pm \frac{v}{\sqrt{2}} ,\, 0)$, there is a single minimisation condition given by
\begin{equation}
\mu_{11}^2 = - \frac{1}{2} \left( \lambda_{1122} + \lambda_{1221} \right) v^2.
\end{equation}
There are no massless states and there are two pairs of mass-degenerate states.

Finally, we consider the real vacuum $(\hat v_1,\, \hat v_2,\, 0)$. There is a minimisation condition:
\begin{equation}
\mu_{11}^2 = - \lambda_{1111} \left( \hat v_1^2 + \hat v_2^2 \right).
\end{equation}
In this case the underlying continuous $SO(2)$ group is broken and an unwanted neutral Goldstone boson appears. This can be avoided by introducing a soft symmetry-breaking term $\mu_{12}^2$. For the soft term to survive, the new minimisation conditions require $\hat v_1^2 = \hat v_2^2$. This case is contained within the more general $ (\frac{v}{\sqrt{2}} e^{i \sigma},\, \pm \frac{v}{\sqrt{2}} ,\, 0)$ with the $\mu_{12}^2$ term.

Yet another possibility is to consider the vacuum given by $(0,\,  v_2,\,  v_3)$. While the $\Lambda$ term is present, it involves the $h_1$ doublet, which is vev-less. Therefore, without loss of generality we can re-define the $h_1$ and $h_2$ doublets simultaneously to absorb the phase appearing in the vev. As a consequence, it is sufficient to consider a real vacuum. Due to the $S_2(h_1,\, h_2)$ symmetry we do not need to consider separately the case $(\hat v_1,\, 0,\, \hat v_3)$. In the case $(0,\, \hat v_2,\, \hat v_3)$, the minimisation conditions are:
\begin{subequations}\label{Eq:MinCon_O2Z2U1_0v2v3}
\begin{align}
\mu_{11}^2 &= - \lambda_{1111} \hat v_2^2 - \frac{1}{2}\left( \lambda_{1133} + \lambda_{1331} \right) \hat v_3^2,\\
\mu_{33}^2 &= - \frac{1}{2} \left( \lambda_{1133} + \lambda_{1331} \right) \hat v_2^2 - \lambda_{3333} \hat v_3^2.
\end{align}
\end{subequations}
In this case there are two unwanted massless states. As before, one could introduce a soft symmetry-breaking term $\mu_{23}^2$. Then, the minimisation conditions are:
\begin{subequations}\label{Eq:MinCon_O2Z2U1_0v2v3_SBT}
\begin{align}
\mu_{23}^2 &= - \left[ 2 \mu_{11}^2 + 2 \lambda_{1111} \hat v_2^2 + \left( \lambda_{1133} + \lambda_{1331} \right) \hat v_3^2 \right]\frac{\hat v_2}{2 \hat v_3},\\
\mu_{33}^2 &= \frac{1}{\hat v_3^2} \left( \mu_{11}^2 \hat v_2^2 + \lambda_{1111} \hat v_2^4 - \lambda_{3333} \hat v_3^4 \right),
\end{align}
\end{subequations}
and there are no unwanted massless states present.

\subsubsection{Two vanishing vevs}\label{Sec:SO2_Z2_U1_2van_vevs}

One should consider two different vacuum configurations. In the case $\left(v,\,0,\,0\right)$, there is a single minimisation condition,
\begin{equation}
\mu_{11}^2 = - \lambda_{1111} v^2.
\end{equation}
There is an unwanted massless state. This is the first occurrence where an additional massless state arises for a case with two vanishing vevs. Next, we can introduce a soft symmetry-breaking term. There are several possibilities to implement soft symmetry-breaking terms, however neither $\mu_{12}^2$ nor $\mu_{13}^2$ would survive the minimisation conditions. The only option left is to introduce the $\mu_{23}^2$ term. This term does not alter the minimisation conditions and the unwanted massless states get promoted to massive ones. The phase of $\mu_{23}^2$ can be absorbed via a re-definition of the $h_3$ doublet, so that $\left( \mu_{23}^2 \right)^\mathrm{I} = 0$.

Let us next consider $\left(0,\,0,\,v\right)$. The minimisation condition is given by
\begin{equation}
\mu_{33}^2 = - \lambda_{3333} v^2.
\end{equation}
In this case there is mass degeneracy between the charged physical scalars. Whenever there is a pair of charged mass-degenerate scalars present, as a consequence of applying a symmetry, the neutral states of the corresponding doublets will also experience some degeneracy pattern. In this specific case all four additional neutral states become mass degenerate.

The only case which does not spontaneously break the underlying symmetry is given by $\left(0,\,0,\,v\right)$, and will be discussed in some more detail below.

\bigskip\textbf{Case of \boldmath$(0,\,0,\,v)$}

The charged mass-squared matrix in the basis $\{h_1^\pm,\, h_2^\pm,\, h_3^\pm \}$ is:
\begin{equation}
\mathcal{M}_\mathrm{Ch}^2 = \mathrm{diag} \left( \mu_{11}^2 + \frac{1}{2} \lambda_{1133}v^2,\, \mu_{11}^2 + \frac{1}{2} \lambda_{1133}v^2,\, 0 \right).
\end{equation}

The neutral mass-squared matrix in the basis $\{\eta_1,\, \eta_2,\, \eta_3,\, \chi_1,\, \chi_2,\, \chi_3 \}$ is:
\begin{equation} \label{Eq:masses}
\mathcal{M}_\mathrm{N}^2 = \mathrm{diag} \left( m_H^2,\, m_H^2,\, m_h^2,\, m_H^2,\, m_H^2,\, 0 \right),
\end{equation}
where
\begin{subequations}\label{Eq:MN2_U1_U1_S2}
\begin{align}
m_H^2 ={}& \mu_{11}^2 + \frac{1}{2} \left( \lambda_{1133} + \lambda_{1331} \right) v^2,\\
m_h^2 ={}& 2 \lambda_{3333} v^2,
\end{align}
\end{subequations}
and the $h$ state could be associated with the SM-like Higgs boson.

In this implementation $\{ \mu_{11}^2,\, \lambda_{1133},\, \lambda_{1331},\, \lambda_{3333} \}$ enter in the mass-squared matrices, and are responsible for generating three different mass-squared parameters. Some couplings appear only in the scalar interactions: $\{\lambda_{1111},\, \lambda_{1122},\, \lambda_{1221}\}$.

\subsection(B){\texorpdfstring{Case of $\left[ U(1)_1 \rtimes \mathbb{Z}_2 \right] \times U(1)$}{U(1) x Z2 x U(1)}}\label{Sec:U11_Z2_U1}

\subsubsection{One vanishing vev}\label{Sec:U1U1S2_v1v20}

In the previously discussed representation of $O(2) \times U(1)$ we had to consider vacua with a non-vanishing phase. In the representation given by $\left[ U(1)_1 \rtimes \mathbb{Z}_2 \right] \times U(1)$ this is no longer the case since the scalar potential does not have any phase-sensitive couplings. Even with the introduction of (relevant) soft symmetry-breaking terms, the minimisation conditions would force the phase of a complex symmetry-breaking term to be related to the phase of the vacuum, $\arg( \mu_{ij}^2) = \sigma$.

Let us consider the vacuum given by $(\hat v_1,\, \hat v_2,\, 0)$. The minimisation conditions are:
\begin{subequations}\label{Eq:U1U1S2_MinCon_v1v20_gen}
\begin{align}
& \mu_{11}^2 = - \lambda_{1111} \hat v_1^2 - \frac{1}{2} \left( \lambda_{1122} + \lambda_{1221} \right) \hat v_2^2,\\
& \left( 2\lambda_{1111} - \lambda_{1122}  - \lambda_{1221}\right) \left( \hat v_1^2 - \hat v_2^2 \right) = 0.
\end{align}
\end{subequations}
There are two solutions to the second minimisation condition. The first solution is given by relating the quartic couplings. This solution increases the overall symmetry to $U(2)$ and together with this vacuum leads to an additional Goldstone boson associated with the SSB of this symmetry. The second solution requires a vacuum given by $ (\frac{v}{\sqrt{2}},\, \pm \frac{v}{\sqrt{2}} ,\, 0)$. In this solution there is one fewer massless state. For both solutions there is mass degeneracy between the neutral states associated with the $h_3$ doublet.

A soft symmetry-breaking term $\mu_{12}^2$ can be introduced to promote the massless states to massive ones. In the case $(\hat v_1,\, \hat v_2,\, 0)$, the minimisation conditions become:
\begin{subequations}\label{Eq:MinCon_O2_U1_soft_v1v20}
\begin{align}
\mu_{11}^2 &= - \lambda_{1111} \left( \hat v_1^2 + \hat v_2^2 \right),\\
\mu_{12}^2 &= \frac{1}{2} \left( 2 \lambda_{1111} - \lambda_{1122} - \lambda_{1221} \right) \hat v_1 \hat v_2,
\end{align}
\end{subequations}
and for the case $ (\frac{v}{\sqrt{2}},\, \pm \frac{v}{\sqrt{2}} ,\, 0)$ it is:
\begin{equation}\label{Eq:MinCon_O2_U1_soft_v1v10}
\mu_{12}^2  = \mp \mu_{11}^2 \mp \frac{1}{4} \left( 2 \lambda_{1111} + \lambda_{1122} + \lambda_{1221} \right) v^2.
\end{equation}

Yet another possibility is to consider the vacuum given by $(0,\, \hat v_2,\, \hat v_3)$. In this case the minimisation conditions are identical to those of eq.~\eqref{Eq:MinCon_O2Z2U1_0v2v3}. There will be an additional massless state and a pair of mass-degenerate states associated with the $h_1$ doublet. One could introduce a soft symmetry-breaking term $\mu_{23}^2$. The minimisation conditions are identical to those of eq.~\eqref{Eq:MinCon_O2Z2U1_0v2v3_SBT}.

\subsubsection{Two vanishing vevs}

The minimisation conditions of the implementations with a single non-vanishing vev are identical to those of the $\left[ SO(2) \rtimes \mathbb{Z}_2 \right] \times U(1)$ case.

In the case $\left(v,\,0,\,0\right)$ the neutral mass-squared matrix is diagonal and  there are two pairs of mass-degenerate states. For the $\left[ SO(2) \rtimes \mathbb{Z}_2 \right] \times U(1)$-symmetric 3HDM with an identical vacuum configuration there was an additional massless state, see Section~\ref{Sec:SO2_Z2_U1_2van_vevs}.

In the case $\left(0,\,0,\,v\right)$ there is a mass degeneracy between the charged states and all four neutral states associated with $h_1$ and $h_2$ are mass-degenerate. As before, this is the only case that does not spontaneously break the underlying symmetry. The discussion is identical to that of $\left[ SO(2) \rtimes \mathbb{Z}_2 \right] \times U(1)$.

The two implementations of the $O(2)\times U(1)$ symmetry considered in Sections~\ref{Sec:SO2_Z2_U1} and \ref{Sec:U11_Z2_U1} are related by a basis transformation. Several features (but not all) are therefore common to these two implementations and can be mapped between Tables~\ref{Table:U1U1S2_Cases_1} and~\ref{Table:U1U1S2_Cases_2}:
\begin{itemize}
\item Unbroken symmetry for $(0,0,v)$ in both cases since $O(2) \times U(1)$ acts on the first two doublets, and hence these vacuum configurations coincide;
\item The vacuum $(v,0,0)$ leads to two pairs of degenerate neutral states under the ${\cal G}_2$ implementation, but not under ${\cal G}_1$, since the basis transformation of eq.~\eqref{Eq:h_Rot1_h} acts on the first two doublets.
\end{itemize}

\section{\texorpdfstring{\boldmath$[ U(1) \times U(1)] \rtimes S_3$}{U(1) x U(1) x S3}-symmetric 3HDM}\label{Sec:Pot_U1_U1_S3}

The $[ U(1) \times U(1)] \rtimes S_3$ symmetry was identified in Section~\ref{Sec:Ident_U1_U1_S3}. Now we consider specific examples with possible DM implementations. Since there is an $S_3$ symmetry present we do not have to consider permutations of different vacuum configurations. The scalar potential was provided in eq.~\eqref{Eq:V_U1_U1_S3},
\begin{equation*}
\begin{aligned}
V_{\mathcal{G}_3} ={}& \mu_{11}^2 \sum_i h_{ii} + \lambda_{1111} \sum_i h_{ii}^2  + \lambda_{1122} \sum_{i<j} h_{ii} h_{jj} + \lambda_{1221} \sum_{i<j} h_{ij} h_{ji},
\end{aligned}
\end{equation*}
where, for simplicity, we denote $\mathcal{G}_3 \equiv \left[U(1) \times U(1)\right] \rtimes S_3$.

All vacuum configurations with at least one vanishing vev are summarised in Table~\ref{Table:U11S3_Cases}. There are no vacuum configurations which would preserve the underlying $\mathcal{G}_3$ symmetry.

\subsection{One vanishing vev}

For a single vanishing vev the minimisation conditions are identical to those of the $O(2) \times U(1)$ 3HDM, given in Section~\ref{Sec:U1U1S2_v1v20}. There are two possible vacuum configurations: $(\hat v_1,\, \hat v_2,\, 0)$ and $ (\frac{v}{\sqrt{2}},\, \pm \frac{v}{\sqrt{2}} ,\, 0)$. In the case $(\hat v_1,\, \hat v_2,\, 0)$ the minimisation conditions force $\Lambda=0$, see eq.~\eqref{Eq:Incr_sym}, and hence the scalar potential becomes $SU(3)$-symmetric. For $ (\frac{v}{\sqrt{2}},\, \pm \frac{v}{\sqrt{2}} ,\, 0)$ without the soft symmetry-breaking terms the mass-squared parameters are:
\begin{subequations}
\begin{align}
m_{H_1}^2 ={}& \frac{1}{2} \left( 2 \lambda_{1111} - \lambda_{1122} - \lambda_{1221} \right) v^2,\\
m_{H_2}^2 ={}& \frac{1}{2} \left( 2 \lambda_{1111} + \lambda_{1122} + \lambda_{1221} \right) v^2,\\
m_{\eta_3}^2 = m_{\chi_3}^2 ={}& -\frac{1}{4} \left( 2 \lambda_{1111} - \lambda_{1122} - \lambda_{1221} \right) v^2.
\end{align}
\end{subequations}
It is not possible to have simultaneously all mass-squared parameters positive definite. Note that the simultaneous existence of mass-squared parameters of opposite signs represents a saddle point. This case is thus unphysical. A way around is to introduce a soft $\mu_{12}^2$ term.

{{\renewcommand{\arraystretch}{1.3}
\begin{table}[htb]
\caption{Similar to Table~\ref{Table:U1_U1_Cases}, but now for $\mathcal{G}_3 \equiv \left[ U(1) \times U(1) \right] \rtimes S_3 $. In one case the minimisation conditions lead to a higher symmetry. None of these cases violates~CP.}
\label{Table:U11S3_Cases}
\begin{center}
\begin{tabular}{|c|c|c|c|c|} \hline\hline
Vacuum & SYM & $V$ & \begin{tabular}[l]{@{}c@{}} Mixing of the \\ neutral states\end{tabular} & Comments \\ \hline
$(\hat{v}_1,\, \hat{v}_2,\, 0)$ & $-$& $V_{SU(3)}$ & \footnotesize\begin{tabular}[l]{@{}c@{}} $\{\eta_1,\, \eta_2\}{-}\{\eta_3\}$\\${-}\{\chi_1\}{-}\{\chi_2\}{-}\{\chi_3\}$ \end{tabular} & \footnotesize\begin{tabular}[l]{@{}c@{}} $m_{h^+_{(1,2)}} = m_{h^+_3}$ \\ Five neutral massless states \end{tabular} \\ \hline
$(\hat{v}_1,\, \hat{v}_2,\, 0)$ & $-$& $V_{\mathcal{G}_3}  + (\mu_{12}^2)^\mathrm{R}$ & \footnotesize\begin{tabular}[l]{@{}c@{}} $\{\eta_1,\, \eta_2\} {-}\{\eta_3\}$ \\ ${-}\{\chi_1,\, \chi_2\}{-}\{\chi_3\}$ \end{tabular} & \begin{tabular}[l]{@{}c@{}} $m_{h^+_{(1,2)}} = m_{h^+_3},$ \\ $ m_{\eta_3} = m_{\chi_3} = m_{\chi_{(2,3)}}$ \end{tabular} \\ \hline
$(\frac{v}{\sqrt{2}},\, \pm \frac{v}{\sqrt{2}},\, 0)$ & $-$ & $V_{\mathcal{G}_3}$ & \footnotesize\begin{tabular}[l]{@{}c@{}} $\{\eta_1,\, \eta_2\}{-}\{\eta_3\}$\\${-}\{\chi_1\}{-}\{\chi_2\}{-}\{\chi_3\}$ \end{tabular} & \footnotesize\begin{tabular}[l]{@{}c@{}} $m_{\eta_3} = m_{\chi_3}$ \\ $ m_{\chi_1} = m_{\chi_2} = 0$ \\ A negative $m_\mathrm{neutral}^2$  \end{tabular} \\ \hline
$(\frac{v}{\sqrt{2}},\, \pm \frac{v}{\sqrt{2}},\, 0)$ & $-$ & $V_{\mathcal{G}_3} + (\mu_{12}^2)^\mathrm{R}$ & \footnotesize\begin{tabular}[l]{@{}c@{}} $\{\eta_1,\, \eta_2\} {-}\{\eta_3\}$ \\ ${-}\{\chi_1,\, \chi_2\}{-}\{\chi_3\}$ \end{tabular} & \begin{tabular}[l]{@{}c@{}} $m_{\eta_3} = m_{\chi_3}$ \end{tabular} \\ \hline
$(v,\, 0,\, 0)$  & $-$ & $V_{\mathcal{G}_3}$ & diagonal & \footnotesize\begin{tabular}[l]{@{}c@{}} $ m_{h^+_2} = m_{h^+_3}, $ \\ $m_{\eta_2} = m_{\eta_3} = m_{\chi_2} = m_{\chi_3}$ \end{tabular} \\ \hline\hline
\end{tabular}
\end{center}
\end{table}}

\subsection{Two vanishing vevs}

Due to the superimposed $S_3$ it is sufficient to consider a single vacuum configuration with two vanishing vevs, $\left( v,\, 0,\, 0\right)$. There is a single minimisation condition given by:
\begin{equation}
\mu_{11}^2 = - \lambda_{1111} v^2.
\end{equation}
Then there are several mass-degeneracies among the inert scalars present in this model. First of all, two of the charged states become mass degenerate. Furthermore, due to the $S_3$ symmetry, all four neutral states associated with $h_2$ and $h_3$ are mass degenerate. Then, the entries of the mass-squared matrix are independent of the bilinear terms and thus one can put a limit on the heavy states by adopting the perturbativity constraint, $|\lambda_{ijkl}| \leq 4 \pi$.

\section{\texorpdfstring{\boldmath$O(2)$}{O(2)}-symmetric 3HDM}\label{Sec:Pot_O2}

For completeness, we shall consider two representations of the $O(2)$ symmetry, in accordance with Ref.~\cite{deMedeirosVarzielas:2019rrp}. These were provided in Section~\ref{Sec:Ident_O2_U1}. The $O(2)$ group can be presented by orthogonal rotations. In this case we consider the symmetry group $SO(2) \rtimes \mathbb{Z}_2$. The scalar potential was presented in eq.~\eqref{Eq:V_O2_SO2_Z2},
\begin{equation*}
\begin{aligned}
V_{SO(2) \rtimes \mathbb{Z}_2} ={}& \mu_{11}^2 (h_{11} + h_{22}) + \mu_{33}^2 h_{33} + \lambda_{1111} (h_{11}^2 + h_{22}^2) + \lambda_{3333} h_{33}^2 + \lambda_{1122} h_{11} h_{22}\\
& + \lambda_{1133} (h_{11} h_{33} + h_{22} h_{33}) + \lambda_{1221} h_{12} h_{21} + \lambda_{1331} (h_{13} h_{31} + h_{23} h_{32})\\
& + \Lambda \left( h_{12}^2 + h_{21}^2 \right) + \{\lambda_{1313} \left( h_{13}^2 + h_{23}^2 \right) + \mathrm{h.c.} \}.
\end{aligned}
\end{equation*}
Another representation of the $O(2)$ group, which we shall also consider, is given in the re-phasing basis of $U(1)_1$. In this case, the symmetry group can be expressed in terms of $U(1)_1 \rtimes \mathbb{Z}_2$, with the scalar potential of eq.~\eqref{Eq:V_O2_U1_Z2},
\begin{equation*}
\begin{aligned}
V_{U(1)_1 \rtimes \mathbb{Z}_2} ={}& \mu_{11}^2 (h_{11} + h_{22}) + \mu_{33}^2 h_{33} + \lambda_{1111} (h_{11}^2 + h_{22}^2) + \lambda_{3333} h_{33}^2 + \lambda_{1122} h_{11} h_{22}\\
& + \lambda_{1133} (h_{11} h_{33} + h_{22} h_{33}) + \lambda_{1221} h_{12} h_{21} + \lambda_{1331} (h_{13} h_{31} + h_{23} h_{32})\\
& + \left\lbrace \lambda_{1323} h_{13} h_{23} + \mathrm{h.c.} \right\rbrace.
\end{aligned}
\end{equation*}

These two $O(2)$-symmetric scalar potentials differ by the $h_{ij}^2 \leftrightarrow h_{13}h_{23}$ terms. Since the above scalar potentials exhibit an $O(2)$ symmetry, it is expected that both $V_{SO(2) \rtimes \mathbb{Z}_2}$ and $V_{U(1)_1 \rtimes \mathbb{Z}_2}$ are connected via a basis transformation given by  eq.~\eqref{Eq:Rot_O2_SO2_U1}.

Different vacuum configurations are summarised in Tables~\ref{Table:O2_Cases_1} and~\ref{Table:O2_Cases_2}.

{\renewcommand{\arraystretch}{1.3}
\begin{table}[htb]
\caption{Similar to Table~\ref{Table:U1_U1_Cases}, but now for $SO(2) \rtimes \mathbb{Z}_2$, with the scalar potential given by eq.~\eqref{Eq:V_O2_SO2_Z2}. There may be CP violation in the $(0,\, \hat{v}_2,\, \hat{v}_3)$ implementation with $\mu_{23}^2$.}
\label{Table:O2_Cases_1}
\begin{center}
\begin{tabular}{|c|c|c|c|c|} \hline\hline
Vacuum & SYM &$V$ & \begin{tabular}[l]{@{}c@{}} Mixing of the \\ neutral states\end{tabular} & Comments \\ \hline
$(\hat{v}_1 e^{i \sigma},\, \hat{v}_2,\, 0)$ & $-$ &  $V_{SO(2) \rtimes \mathbb{Z}_2},\,\Lambda=0$ & \footnotesize\begin{tabular}[l]{@{}c@{}} $\{\eta_1,\, \eta_2\}{-}\{\eta_3,\, \chi_3\}$\\${-}\{\chi_1\}{-}\{\chi_2\}$ \end{tabular} & \footnotesize$m_{\eta_{(1,2)}} = m_{\chi_1} = m_{\chi_2} = 0$ \\ \hline
$(\frac{v}{\sqrt{2}} e^{i \sigma},\, \pm \frac{v}{\sqrt{2}},\, 0)$ & $-$ &  $V_{SO(2) \rtimes \mathbb{Z}_2} + \mu_{12}^2$ & \footnotesize\begin{tabular}[l]{@{}c@{}} $\{\eta_1,\, \eta_2,\, \chi_1,\, \chi_2\}$\\${-}\{\eta_3,\, \chi_3\}$ \end{tabular} & - \\ \hline
$(i\frac{v}{\sqrt{2}},\, \pm \frac{v}{\sqrt{2}},\, 0)$ & $-$ &  $V_{SO(2) \rtimes \mathbb{Z}_2}$ & \footnotesize\begin{tabular}[l]{@{}c@{}} $\{\eta_1,\, \eta_2\}{-}\{\eta_3\}$\\${-}\{\chi_1,\,\chi_2\}{-}\{\chi_3\}$ \end{tabular} & $m_{\eta_{(1,2)}} = m_{\chi_{(1,2)}}$  \\ \hline
$(\hat{v}_1,\, \hat{v}_2,\, 0)$ & $-$ &  $V_{SO(2) \rtimes \mathbb{Z}_2}$ & \footnotesize\begin{tabular}[l]{@{}c@{}} $\{\eta_1,\, \eta_2\}{-}\{\eta_3\}$\\${-}\{\chi_1,\,\chi_2\}{-}\{\chi_3\}$ \end{tabular} & $m_{\eta_{(1,2)}} = m_{\chi_{(1,2)}} = 0$ \\ \hline
$(0,\, \hat{v}_2,\, \hat{v}_3)$ & $-$& $V_{SO(2) \rtimes \mathbb{Z}_2}$ & \footnotesize\begin{tabular}[l]{@{}c@{}} $\{\eta_1\}{-}\{\eta_2,\, \eta_3\}$\\${-}\{\chi_1\}{-}\{\chi_2,\,\chi_3\}$ \end{tabular} & $ m_{\eta_1} = m_{\chi_{(2,3)}} = 0$ \\ \hline
$(0,\, \hat{v}_2,\, \hat{v}_3)$ & $-$& $V_{SO(2) \rtimes \mathbb{Z}_2} + \mu_{23}^2$ & \footnotesize\begin{tabular}[l]{@{}c@{}} $\{\eta_1,\, \chi_1\}$\\${-}\{\eta_2,\, \eta_3,\,\chi_2,\,\chi_3\}$ \end{tabular} & - \\ \hline
$(v,\,0,\,0)$ & $-$ & $V_{SO(2) \rtimes \mathbb{Z}_2}$  & diagonal & $m_{\eta_2} = m_{\chi_1} = 0$\\ \hline
$(v,\,0,\,0)$ & $-$ & $V_{SO(2) \rtimes \mathbb{Z}_2} + (\mu_{23}^2)^\mathrm{R}$  & \footnotesize\begin{tabular}[l]{@{}c@{}} $\{\eta_1\}{-}\{\chi_1\}$ \\  ${-}\{\eta_2,\, \eta_3,\, \chi_2,\, \chi_3\}$ \end{tabular} & -\\ \hline
$(0,\,0,\,v)$ & $\checkmark$ & $V_{SO(2) \rtimes \mathbb{Z}_2}$ & diagonal & \footnotesize\begin{tabular}[l]{@{}c@{}} $m_{h_1^+} = m_{h_2^+}$, \\ $m_{\eta_1} = m_{\eta_2},~m_{\chi_1} = m_{\chi_2}$ \end{tabular} \\ \hline \hline
\end{tabular} 
\end{center}
\end{table}}

{\renewcommand{\arraystretch}{1.3}
\begin{table}[htb]
\caption{Similar to Table~\ref{Table:U1_U1_Cases}, but now for $U(1)_1 \rtimes \mathbb{Z}_2$, with the scalar potential given by eq.~\eqref{Eq:V_O2_U1_Z2}. In one case the minimisation conditions can lead to a higher $O(2) \times U(1)$ symmetry. None of these cases violates CP.}
\label{Table:O2_Cases_2}
\begin{center}
\begin{tabular}{|c|c|c|c|c|} \hline\hline
Vacuum & SYM &$V$ & \begin{tabular}[l]{@{}c@{}} Mixing of the \\ neutral states\end{tabular} & Comments \\ \hline
$(\hat{v}_1 ,\, \hat{v}_2,\, 0)$ & $-$ &  $V_{U(1)_1 \rtimes \mathbb{Z}_2},\,\Lambda=0$ & \footnotesize\begin{tabular}[l]{@{}c@{}} $\{\eta_1,\, \eta_2\}{-}\{\eta_3,\,\chi_3\}$\\${-}\{\chi_1\}{-}\{\chi_2\}$ \end{tabular} & \footnotesize\begin{tabular}[l]{@{}c@{}} $m_{\eta_{(1,2)}} = m_{\chi_1}= m_{\chi_2} = 0$ \end{tabular} \\ \hline
$(\hat{v}_1 ,\, \hat{v}_2,\, 0)$ & $-$ &  $V_{U(1)_1 \rtimes \mathbb{Z}_2} + (\mu_{12}^2)^\mathrm{R}$ & \footnotesize\begin{tabular}[l]{@{}c@{}} $\{\eta_1,\, \eta_2\}{-}\{\eta_3,\,\chi_3\}$\\${-}\{\chi_1,\,\chi_2\}$ \end{tabular} & - \\ \hline
$(\frac{v}{\sqrt{2}} ,\, \pm \frac{v}{\sqrt{2}},\, 0)$ & $-$ &  $V_{U(1)_1 \rtimes \mathbb{Z}_2}$ & \footnotesize\begin{tabular}[l]{@{}c@{}} $\{\eta_1,\, \eta_2\}{-}\{\eta_3,\,\chi_3\}$\\${-}\{\chi_1\}{-}\{\chi_2\}$ \end{tabular} & \footnotesize $m_{\chi_1} = m_{\chi_2} = 0$ \\ \hline
$(\frac{v}{\sqrt{2}} ,\, \pm \frac{v}{\sqrt{2}},\, 0)$ & $-$ &  $V_{U(1)_1 \rtimes \mathbb{Z}_2} + (\mu_{12}^2)^\mathrm{R}$ & \footnotesize\begin{tabular}[l]{@{}c@{}} $\{\eta_1,\, \eta_2\}{-}\{\eta_3,\,\chi_3\}$\\${-}\{\chi_1,\,\chi_2\}$ \end{tabular} & - \\ \hline
$(0,\, \hat v_2,\, \hat v_3)$ & $-$& $V_{O(2) \times U(1)}$ & \footnotesize\begin{tabular}[l]{@{}c@{}} $\{\eta_1\}{-}\{\eta_2,\, \eta_3\}$\\${-}\{\chi_1\}{-}\{\chi_2\}{-}\{\chi_3\}$ \end{tabular} & - \\ \hline
$(v,\,0,\,0)$ & $-$ & $V_{U(1)_1 \rtimes \mathbb{Z}_2}$  & diagonal & \footnotesize\begin{tabular}[l]{@{}c@{}} $m_{\eta_2} = m_{\chi_2},\,m_{\eta_3} = m_{\chi_3}$ \end{tabular} \\ \hline
$(0,\,0,\,v)$ & $\checkmark$ & $V_{U(1)_1 \rtimes \mathbb{Z}_2}$ & \footnotesize\begin{tabular}[l]{@{}c@{}} $\{\eta_1,\, \eta_2\}{-}\{\eta_3\}$\\${-}\{\chi_1,\,\chi_2\}{-}\{\chi_3\}$ \end{tabular} & \footnotesize\begin{tabular}[l]{@{}c@{}} $m_{h_1^+} = m_{h_2^+}$ \\ Two pairs of neutral \\ mass-degenerate states \end{tabular} \\ \hline \hline
\end{tabular} 
\end{center}
\end{table}}

\subsection(A){\texorpdfstring{Case of $ SO(2) \rtimes \mathbb{Z}_2$}{SO(2) x Z2}}

\subsubsection{One vanishing vev}

We start by analysing cases with a single vanishing vev. For the $O(2)$-symmetric scalar potential we need to consider complex vacuum configurations. We shall begin with the case $(v_1,\, v_2,\, 0)$.  As in the case of the $O(2) \times U(1)$-symmetric scalar potential there are several solutions to the minimisation conditions which would correspond to different  vacuum configurations; the condition of eq.~\eqref{Eq:SO2_Z2_U1_DiffCond_CompVev} must also be satisfied here. Since both $h_1$ and $h_2$ are stabilised by a continuous symmetry, all implementations with only a single vanishing vev will result in a SSB.

First, let us consider the $(\hat v_1 e^{i \sigma},\, \hat v_2,\, 0)$ vacuum. The minimisation conditions are:
\begin{subequations} \label{Eq:SO2_Z2_Incr_symm}
\begin{align}
\mu_{11}^2 ={}& - \lambda_{1111} \left( \hat v_1^2 + \hat v_2^2 \right),\\
\Lambda ={}& 0.
\end{align}
\end{subequations}
The latter condition presumably signals an increase of the overall symmetry, but we have not been able to identify it (actually, in the bilinear formalism presented in Appendix~\ref{App:Bilinear_3symm} the degeneracy pattern is seen to change.) In total there are three massless states. By introducing the soft symmetry-breaking term $\mu_{12}^2$, the minimisation conditions require
\begin{equation}
\left[ \mu_{11}^2 + \lambda_{1111} \left( \hat v_1^2 + \hat v_2^2  \right) \right]\left( \hat v_1^2 - \hat v_2^2 \right) = 0.
\end{equation}
Putting the first factor to zero would lead to a massless state. Therefore, we need to assume $v_1^2 = v_2^2$, leading to the vacuum $ (\frac{v}{\sqrt{2}} e^{i \sigma},\, \pm \frac{v}{\sqrt{2}} ,\, 0)$, with the minimisation conditions:
\begin{subequations}
\begin{align}
\left( \mu_{12}^2 \right)^\mathrm{R} ={}& \mp \cos \sigma \left( \mu_{11}^2 + \lambda_{1111} v^2 \right),\\
\left( \mu_{12}^2 \right)^\mathrm{I} ={}& \mp \sin \sigma \left( \mu_{11}^2 + \frac{1}{2} \left( \lambda_{1122} + \lambda_{1221} \right) v^2 \right).
\end{align}
\end{subequations}
In this case there are no additional massless states.

Another possibility would be to choose the vacuum given by $ (i\frac{v}{\sqrt{2}},\, \pm \frac{v}{\sqrt{2}} ,\, 0)$. There is one minimisation condition,
\begin{equation}
\mu_{11}^2 = - \frac{1}{2}\left( \lambda_{1122} + \lambda_{1221} \right) v^2.
\end{equation}
There are then no additional massless states and there are two pairs of neutral mass-degenerate states.

Let us now consider the vacuum given by $(\hat v_1,\, \hat v_2,\, 0)$. The minimisation condition is
\begin{equation}
\mu_{11}^2 = - \lambda_{1111} \left( \hat v_1^2 + \hat v_2^2 \right).
\end{equation}
The underlying $O(2)$ symmetry is broken, and as a result an unwanted massless state is present. A soft symmetry-breaking term $\mu_{12}^2$ can be introduced. One needs to assume ${\hat v_1 = \hat v_2}$ since otherwise the minimisation conditions would require $\mu_{12}^2=0$. This case is contained within the more general case $ (\frac{v}{\sqrt{2}} e^{i \sigma},\, \pm \frac{v}{\sqrt{2}} ,\, 0)$, where the $\eta_i$ decouple from the $\chi_i$, and with the $\mu_{12}^2$ term forced to be real.

Another possible vacuum configuration is given by $(0,\, v_2,\, v_3)$. In this case the minimisation conditions do not relate parameters of the vacuum. For $(0,\, \hat v_2 e^{i \sigma},\, \hat v_3)$ the minimisation conditions would force $\lambda_{1313}=0$. This would increase the underlying symmetry to $O(2) \times U(1)$ and the $\sigma$ phase would become redundant. Thus, this case is inconsistent.

For the vev $(0,\, \hat v_2 ,\, \hat v_3)$, the minimisation conditions are:
\begin{subequations}
\begin{align}
\mu_{11}^2 ={}& - \lambda_{1111} \hat v_2^2 - \frac{1}{2}\left( \lambda_{1133} + 2 \lambda_{1313}^\mathrm{R} +  \lambda_{1331} \right) \hat v_3^2,\\
\mu_{33}^2 ={}& - \frac{1}{2} \left( \lambda_{1133} + 2 \lambda_{1313}^\mathrm{R} + \lambda_{1331} \right) \hat v_2^2 - \lambda_{3333} \hat v_3^2,\\
\lambda_{1313}^\mathrm{I} ={}0.
\end{align}
\end{subequations}

In this case there is one unwanted massless state. 
With the soft symmetry-breaking term $\mu_{23}^2$ the minimisation conditions become:
\begin{subequations}
\begin{align}
\mu_{33}^2 ={}&  \left( \mu_{11}^2 \hat v_2^2 + \lambda_{1111} \hat v_2^4 - \lambda_{3333} \hat v_3^4 \right)\frac{1}{\hat v_3^2} ,\\
\left( \mu_{23}^2 \right)^\mathrm{R} ={}&   -\bigg[ 2 \mu_{11}^2 + 2 \lambda_{1111} \hat v_2^2 + \left( \lambda_{1133} + 2 \lambda_{1313}^\mathrm{R} + \lambda_{1331} \right) \hat v_3^2 \bigg]\frac{ \hat v_2}{2 \hat v_3},\\
\begin{split}
\left( \mu_{23}^2 \right)^\mathrm{I} ={}&  -\lambda_{1313}^\mathrm{I} \hat v_2 \hat v_3,
\end{split}
\end{align}
\end{subequations}
with no unwanted massless states. There is now the possibility for explicit CP violation.

\subsubsection{Two vanishing vevs}

Next, we consider cases with two vanishing vevs. The first case is given by the vacuum $(v,\,0,\,0)$ with a single minimisation condition:
\begin{equation}
\mu_{11}^2 = - \lambda_{1111} v^2.
\end{equation}
An additional massless state is present. There are several possibilities to implement soft symmetry-breaking terms, however neither $\mu_{12}^2$ nor $\mu_{13}^2$ would survive the minimisation conditions. The only option is to introduce the $\mu_{23}^2$ term, which we can choose, without loss of generality, to be real. This term does not alter the minimisation conditions and the unwanted massless states get promoted to massive ones.

The final vacuum configuration which we have to consider is $(0,\,0,\,v)$. In this case there is a single minimisation condition,
\begin{equation}
\mu_{33}^2 = - \lambda_{3333} v^2,
\end{equation}
leading to a pair of mass-degenerate charged scalars and two pairs of neutral mass-degenerate states. This is the only case which does not result in SSB, and therefore we shall discuss it in some detail below.

\bigskip\textbf{Case of \boldmath$(0,\,0,\,v)$}

The charged mass-squared matrix in the basis $\{h_1^\pm,\, h_2^\pm,\, h_3^\pm \}$ is:
\begin{equation}
\mathcal{M}_\mathrm{Ch}^2 = \mathrm{diag} \left( \mu_{11}^2 + \frac{1}{2} \lambda_{1133}v^2 ,\, \mu_{11}^2 + \frac{1}{2} \lambda_{1133}v^2 ,\, 0 \right).
\end{equation}

The neutral mass-squared matrix in the basis $\{\eta_1,\, \eta_2,\, \eta_3,\, \chi_1,\, \chi_2,\, \chi_3 \}$ is:
\begin{equation}
\mathcal{M}_\mathrm{N}^2 = \mathrm{diag} \left( m_\eta^2,\, m_\eta^2,\, m_h^2,\, m_\chi^2,\, m_\chi^2,\, 0 \right),
\end{equation}
where
\begin{subequations}
\begin{align}
m_\eta^2 ={}& \mu_{11}^2 + \frac{1}{2} \left( \lambda_{1133} + 2 \lambda_{1313}^\mathrm{R} + \lambda_{1331} \right) v^2,\\
m_h^2 ={}& 2 \lambda_{3333} v^2,\\
m_\chi^2 ={}& \mu_{11}^2 + \frac{1}{2} \left( \lambda_{1133} - 2 \lambda_{1313}^\mathrm{R} + \lambda_{1331} \right) v^2,
\end{align}
\end{subequations}
and the $h$ state is associated with the SM-like Higgs boson. 

In this implementation $\{ \mu_{11}^2,\, \lambda_{1133},\, \lambda_{1331},\, \lambda_{3333}, \lambda_{1313}^\mathrm{R} \}$ enter in the mass-squared matrices, and are responsible for generating four different mass-squared parameters. Some couplings appear only in the scalar interactions: $\{\lambda_{1111},\, \lambda_{1122},\, \lambda_{1221}\}$.

\subsection(B){\texorpdfstring{Case of $ U(1)_1 \rtimes \mathbb{Z}_2$}{U(1)1 x Z2}}

\subsubsection{One vanishing vev}

Since there is a single phase-sensitive term $\lambda_{1323}$ and the soft symmetry-breaking terms are chosen to be complex, without loss of generality, we can assume that the vacuum is real. Consider the vacuum given by $( \hat v_1,\,  \hat v_2,\, 0)$. The minimisation conditions require:
\begin{equation}
\left( 2\lambda_{1111} - \lambda_{1122}  - \lambda_{1221}\right) \left(\hat v_1^2 - \hat v_2^2  \right) = 0,
\end{equation}
to be satisfied, as in the case of $O(2) \times U(1)$, see Section~\ref{Sec:U1U1S2_v1v20}. Requiring no constraints on the vacuum, the minimisation conditions are:
\begin{subequations}
\begin{align}
& \mu_{11}^2 = - \lambda_{1111} \left( \hat v_1^2 + \hat v_2^2 \right),\\
& \left( 2\lambda_{1111} - \lambda_{1122}  - \lambda_{1221}\right) = 0.
\end{align}
\end{subequations}
The latter condition (which could be expressed as $\Lambda=0$) presumably increases the overall symmetry, see the comment following eq.~\eqref{Eq:SO2_Z2_Incr_symm}. There are three massless states. After introducing the soft $\mu_{12}^2$ term, the new minimisation conditions are:
\begin{subequations}
\begin{align}
\mu_{11}^2 ={}& - \lambda_{1111} \left( \hat v_1^2 + \hat v_2^2 \right),\\
\mu_{12}^2 ={}& \frac{1}{2} \left( 2 \lambda_{1111} - \lambda_{1122} - \lambda_{1221} \right) \hat v_1 \hat v_2,
\end{align}
\end{subequations}
and there are no additional Goldstone bosons.

Another vacuum is given by $ (\frac{v}{\sqrt{2}},\, \pm \frac{v}{\sqrt{2}} ,\, 0)$ with the minimisation condition:
\begin{equation}
\mu_{11}^2 = -  \frac{1}{4}\left( 2\lambda_{1111} + \lambda_{1221} + \lambda_{1221} \right) v^2.
\end{equation}
In this case there is an unwanted massless state. With the soft $\mu_{12}^2$ term the minimisation conditions become:
\begin{equation}
\mu_{12}^2 = \mp \left[ \mu_{11}^2 + \frac{1}{4} \left( 2 \lambda_{1111} + \lambda_{1122} + \lambda_{1221} \right) v^2 \right].
\end{equation}

Another possibility would be to assume the vacuum given by $(0,\, \hat v_2,\, \hat v_3)$. In the case of exact symmetry and also when a soft symmetry-breaking term is introduced, the minimisation conditions require $\lambda_{1323}=0$ to be satisfied. Hence, we are led to the $O(2) \times U(1)$-symmetric 3HDM, discussed in Section~\ref{Sec:U1U1S2_v1v20}.

We note that the CP-violating case $(0,\, \hat v_2,\, \hat v_3)$ of $SO(2) \rtimes \mathbb{Z}_2$ in Table~\ref{Table:O2_Cases_1} corresponds to the case $(\hat v_2/\sqrt{2},\, \hat v_2/\sqrt{2},\, \hat v_3)$ of $U(1)_1 \rtimes \mathbb{Z}_2$ and therefore does not appear in Table~\ref{Table:O2_Cases_2}.

\subsubsection{Two vanishing vevs}

There are two ways to implement vacuum configurations with two vanishing vevs. For the case $(v,\, 0,\, 0)$ there is a single minimisation condition given by
\begin{equation}
\mu_{11}^2 = - \lambda_{1111} v^2.
\end{equation}
The neutral mass-squared matrix is diagonal and as a result of the $U(1)$ charges there are two pairs of neutral mass-degenerate states. The $\mu_{33}^2$ bilinear term will contribute to masses of $h_3$. Due to $\left\langle h_1 \right\rangle \neq 0$ the vacuum does not preserve the underlying symmetry.

Another case is given by $\left(0,\,0,\,v\right)$. In this case the minimisation condition is
\begin{equation}
\mu_{33}^2 = - \lambda_{3333} v^2.
\end{equation}
Like in the previous case, there are two pairs of neutral mass-degenerate states. In contrast, the underlying $O(2)$ is preserved by the vacuum. We shall now consider this case.

\clearpage
\bigskip\textbf{Case of \boldmath$(0,\,0,\,v)$}

The charged mass-squared matrix in the basis $\{h_1^\pm,\, h_2^\pm,\, h_3^\pm \}$ is:
\begin{equation}
\mathcal{M}_\mathrm{Ch}^2 = \mathrm{diag} \left( \mu_{11}^2 + \frac{1}{2} \lambda_{1133}v^2 ,\, \mu_{11}^2 + \frac{1}{2} \lambda_{1133}v^2 ,\, 0 \right).
\end{equation}

The neutral mass-squared matrix in the basis $\{\eta_1,\, \eta_2,\, \eta_3,\, \chi_1,\, \chi_2,\, \chi_3 \}$ is:
\begin{equation}
\mathcal{M}_\mathrm{N}^2 = \mathrm{diag}\Bigg( \begin{pmatrix}
(\mathcal{M}_\mathrm{N}^2)_{11} & (\mathcal{M}_\mathrm{N}^2)_{12}\\
(\mathcal{M}_\mathrm{N}^2)_{12} & (\mathcal{M}_\mathrm{N}^2)_{11}
\end{pmatrix},\, (\mathcal{M}_\mathrm{N}^2)_{33},\,
\begin{pmatrix}
(\mathcal{M}_\mathrm{N}^2)_{11} & -(\mathcal{M}_\mathrm{N}^2)_{12}\\
-(\mathcal{M}_\mathrm{N}^2)_{12}  & (\mathcal{M}_\mathrm{N}^2)_{11}
\end{pmatrix},\, 0 \Bigg),
\end{equation}
where
\begin{subequations}
\begin{align}
(\mathcal{M}_\mathrm{N}^2)_{11} ={}& \mu_{11}^2 + \frac{1}{2} \left( \lambda_{1133} + \lambda_{1331} \right) v^2,\\
(\mathcal{M}_\mathrm{N}^2)_{12} ={}& \frac{1}{2} \lambda_{1323} v^2,\\
(\mathcal{M}_\mathrm{N}^2)_{33} ={}& 2 \lambda_{3333} v^2.
\end{align}
\end{subequations}
The mass-squared parameters of the mixed states are:
\begin{equation}
m_{H_i}^2 = \mu_{11}^2 + \frac{1}{2} v^2 \left( \lambda_{1133} + \lambda_{1331} \pm \lambda_{1323}^\mathrm{R} \right).
\end{equation}

In this implementation $\{ \mu_{11}^2,\, \lambda_{1133},\, \lambda_{1331},\, \lambda_{3333},\, \lambda_{1323} \}$ enter in the mass-squared matrices, and are responsible for generating four different mass-squared parameters. Some couplings appear only in the scalar interactions: $\{\lambda_{1111},\, \lambda_{1122},\, \lambda_{1221}\}$.

\section{\texorpdfstring{\boldmath$U(1) \times D_4$}{U(1) x D4}-symmetric 3HDM }\label{Sec:Pot_U1_Z2_S2}

The $U(1) \times D_4$ symmetry was identified in Section~\ref{Sec:Ident_U1_Z2_S2}. Here, we turn our attention to different cases realisable within the symmetric scalar potential, as provided in eq.~\eqref{Eq:V_U1_Z2_S2},
\begin{equation*}
\begin{aligned}
V_{U(1) \times D_4} ={}& \mu_{11}^2 (h_{11} + h_{22}) + \mu_{33}^2 h_{33} + \lambda_{1111} (h_{11}^2 + h_{22}^2) + \lambda_{3333} h_{33}^2 + \lambda_{1122} h_{11} h_{22}\\
& + \lambda_{1133} (h_{11} h_{33} + h_{22} h_{33}) + \lambda_{1221} h_{12} h_{21} + \lambda_{1331} (h_{13} h_{31} + h_{23} h_{32})\\
& + \lambda_{1212} (h_{12}^2 + h_{21}^2).
\end{aligned}
\end{equation*}

Different vacua are summarised in Table~\ref{Table:U1Z2_S2_Cases}.

{{\renewcommand{\arraystretch}{1.3}
\begin{table}[htb]
\caption{Similar to Table~\ref{Table:U1_U1_Cases}, but now for $U(1) \times D_4$. In one case the minimisation conditions can lead to a higher symmetry. None of these cases violates CP.}
\label{Table:U1Z2_S2_Cases}
\begin{center}
\begin{tabular}{|c|c|c|c|c|} \hline\hline
Vacuum & SYM & $V$ & \begin{tabular}[l]{@{}c@{}} Mixing of the \\ neutral states\end{tabular} & Comments \\ \hline
$(\hat{v}_1,\, \hat{v}_2,\, 0)$ & $-$ & $V_{O(2)\times U(1)}$ & \footnotesize\begin{tabular}[l]{@{}c@{}} $\{\eta_1,\, \eta_2\}{-}\{\eta_3\}$  \\ ${-}\{\chi_1,\, \chi_2\}{-}\{\chi_3\}$ \end{tabular} & \footnotesize\begin{tabular}[l]{@{}c@{}}  $m_{\eta_3} = m_{\chi_3}$ \\ $m_{\eta_{(1,2)}} = m_{\chi_{(1,2)}}=0$ \\\end{tabular} \\ \hline
$(\hat{v}_1,\, \hat{v}_2,\, 0)$ & $-$ & $V_{U(1) \times D_4} + (\mu_{12}^2)^\mathrm{R}$ & \footnotesize\begin{tabular}[l]{@{}c@{}} $\{\eta_1,\, \eta_2\}{-}\{\eta_3\}$ \\ ${-}\{\chi_1,\, \chi_2\}{-}\{\chi_3\}$ \end{tabular} & \begin{tabular}[l]{@{}c@{}} $m_{\eta_3} = m_{\chi_3}$ \end{tabular} \\ \hline
$ (\frac{v}{\sqrt{2}},\, \pm \frac{v}{\sqrt{2}} ,\, 0)$ & $-$ & $V_{U(1) \times D_4}$ & \footnotesize\begin{tabular}[l]{@{}c@{}} $\{\eta_1,\, \eta_2\}{-}\{\eta_3\}$ \\ ${-}\{\chi_1,\, \chi_2\}{-}\{\chi_3\}$ \end{tabular} & $m_{\eta_3} = m_{\chi_3}$  \\ \hline
$(0,\, \hat{v}_2,\, \hat{v}_3)$ & $-$ & $V_{U(1) \times D_4}$ & \footnotesize\begin{tabular}[l]{@{}c@{}} $\{\eta_1\}{-}\{\eta_2,\, \eta_3\}$ \\ ${-}\{\chi_1\}{-}\{\chi_2\}{-}\{\chi_3\}$ \end{tabular} & $m_{\chi_2} = m_{\chi_3}=0$ \\ \hline
$(0,\, \hat{v}_2,\, \hat{v}_3)$ & $-$ & $V_{U(1) \times D_4} + (\mu_{23}^2)^\mathrm{R}$ & \footnotesize\begin{tabular}[l]{@{}c@{}} $\{\eta_1\}{-}\{\eta_2,\, \eta_3\}$ \\ ${-}\{\chi_1\}{-}\{\chi_2, \,\chi_3\}$ \end{tabular} & - \\ \hline
$(v, \,0, \, 0)$ & $-$ & $V_{U(1) \times D_4}$ & diagonal & $m_{\eta_3} = m_{\chi_3}$ \\ \hline
$(0, \,0, \, v)$  & $\checkmark$ & $V_{U(1) \times D_4}$ & diagonal & \footnotesize\begin{tabular}[l]{@{}c@{}}  $m_{h_1^+} = m_{h_2^+}$, \\ $m_{\eta_1} = m_{\eta_2}=$ \\ $= m_{\chi_1}= m_{\chi_2}$ \end{tabular} \\ \hline \hline
\end{tabular} 
\end{center}
\end{table}}

\subsection{One vanishing vev}

Due to the presence of the $\lambda_{1212}$ term, which is forced to be real by the symmetry, we need to consider vacuum configurations with a non-vanishing phase. The presence of $\sigma$ leads to several possible vacuum configurations. We start by discussing the case when $\sigma \neq 0$.

For the case $(\hat{v}_1 e^{i \sigma},\, \hat{v}_2,\, 0)$ the minimisation conditions are given by:
\begin{subequations}
\begin{align}
\mu_{11}^2 ={}& - \lambda_{1111} \left( \hat v_1^2 + \hat v_2^2 \right),\\
\lambda_{1221} ={}&  2 \lambda_{1111} - \lambda_{1122},\\
\lambda_{1212} ={}&  0.
\end{align}
\end{subequations}
Due to the last two conditions the overall symmetry is increased to $U(2)$. However, in the $U(2)$-symmetric 3HDM there are no phase-sensitive couplings. As a result, we can rotate the $\sigma$ phase away. This, in turn, is inconsistent with the vacuum configuration. Therefore, we conclude that this case is not realisable. In addition, there are three massless states. Allowing for the soft $\mu_{12}^2$ symmetry-breaking term the minimisation conditions become:
\begin{subequations}
\begin{align}
\mu_{11}^2 ={}& - \lambda_{1111} \left( \hat v_1^2 + \hat v_2^2 \right),\\
\left( \mu_{12}^2 \right) ^\mathrm{R} ={}& \frac{1}{2} \cos \sigma \left( 2 \lambda_{1111} - \lambda_{1122} - \lambda_{1221} - 2 \lambda_{1212} \right)  \hat v_1 \hat v_2,\\
\left( \mu_{12}^2 \right) ^\mathrm{I} ={}& \frac{1}{2} \sin \sigma  \left( 2 \lambda_{1111} - \lambda_{1122} - \lambda_{1221} + 2 \lambda_{1212} \right) \hat v_1 \hat v_2.
\end{align}
\end{subequations}
We note that $\sigma$ is independent of the $\mu_{12}^2$ phase. However, it is possible to go into a basis where the scalar potential becomes real. Due to this, we shall not explicitly consider this case, and instead cover the real case below.

Another case $ (\frac{v}{\sqrt{2}} e^{i \sigma},\, \pm \frac{v}{\sqrt{2}} ,\, 0)$ results in the minimisation conditions given by:
\begin{subequations}
\begin{align}
\mu_{11}^2 ={}& - \frac{1}{4} \left( 2\lambda_{1111} + \lambda_{1122} + \lambda_{1221}  \right) \hat v^2,\\
\lambda_{1212} ={}& 0.
\end{align}
\end{subequations}
The latter condition increases the overall symmetry to $O(2) \times U(1)$. As in the case $(\hat{v}_1 e^{i \sigma},\, \hat{v}_2,\, 0)$, we conclude that the vacuum is not realisable. There is also an additional neutral massless state present. With the soft symmetry-breaking term $\mu_{12}^2$ the discussion is identical to that of $(\hat{v}_1 e^{i \sigma},\, \hat{v}_2,\, 0)$.

Finally, the vacuum can be given by $ (i \frac{v}{\sqrt{2}},\, \pm \frac{v}{\sqrt{2}} ,\, 0)$. Re-phasing of $h_1$ by ``$i$" forces the $\lambda_{1212}$ coupling to change by an overall sign. Then, the vacuum becomes real, $ (\frac{v}{\sqrt{2}},\, \pm \frac{v}{\sqrt{2}} ,\, 0)$.

Next, we consider the real vacuum $(\hat{v}_1,\, \hat{v}_2,\, 0)$ with the minimisation conditions:
\begin{subequations}
\begin{align}
& \mu_{11}^2 = - \lambda_{1111} \hat v_1^2 - \frac{1}{2} \left( \lambda_{1122} + \lambda_{1221} + 2 \lambda_{1212}  \right) \hat v_2^2,\\
& \left( 2\lambda_{1111} -\lambda_{1122} - \lambda_{1221} - 2 \lambda_{1212} \right) \left( \hat v_1^2 - \hat v_2^2 \right) = 0,
\end{align}
\end{subequations}
As in some of the previous cases, \textit{e.g.}, the $O(2) \times U(1)$-symmetric 3HDM, see Section~\ref{Sec:U1U1S2_v1v20}, the minimisation conditions can be solved in two different ways: relating the quartic couplings or assuming $\hat v_1^2 = \hat v_2^2$. 

By requiring $ 2\lambda_{1111} -\lambda_{1122} - \lambda_{1221} - 2 \lambda_{1212} = 0$, the scalar potential takes the form of the $O(2) \times U(1)$-symmetric 3HDM in a specific basis given by eq.~\eqref{Eq:V_O2_U1_expl_O2}. This case does not preserve the underlying $O(2) \times U(1)$ symmetry, leading to an unwanted Goldstone boson. One can introduce a soft symmetry-breaking term $\mu_{12}^2$ to eliminate this additional massless state. The minimisation conditions in this case are:
\begin{subequations}
\begin{align}
\mu_{11}^2 &= - \lambda_{1111} \left( \hat v_1^2 + \hat v_2^2 \right),\\
\mu_{12}^2  &= \frac{1}{2} \left( 2 \lambda_{1111} - \lambda_{1122} - \lambda_{1221} - 2 \lambda_{1212} \right) \hat v_1 \hat v_2.
\end{align}
\end{subequations}

Another possibility is to consider the vacuum given by $ (\frac{v}{\sqrt{2}},\, \pm \frac{v}{\sqrt{2}} ,\, 0)$. In this case there is a single minimisation condition:
\begin{equation}
\mu_{11}^2 = - \frac{1}{4}  \left( 2 \lambda_{1111} + \lambda_{1122} + \lambda_{1221} + 2 \lambda_{1212} \right)v^2,
\end{equation}
and no unwanted massless state is present.

Next, we consider the vacuum $(0,\, \hat{v}_2,\, \hat{v}_3)$. The minimisation conditions are:
\begin{subequations}
\begin{align}
\mu_{11}^2 &= - \lambda_{1111} \hat v_2^2 - \frac{1}{2}\left( \lambda_{1133} + \lambda_{1331} \right) \hat v_3^2,\\
\mu_{33}^2 &= - \frac{1}{2} \left( \lambda_{1133} + \lambda_{1331} \right) \hat v_2^2 - \lambda_{3333} \hat v_3^2.
\end{align}
\end{subequations}
There is now an additional massless state. Again, one could introduce a soft symmetry-breaking term, this time $\mu_{23}^2$. Then, the minimisation conditions are:
\begin{subequations}
\begin{align}
\mu_{23}^2 &= - \left[ 2 \mu_{11}^2 + 2 \lambda_{1111} \hat v_2^2 + \left( \lambda_{1133} + \lambda_{1331} \right) \hat v_3^2 \right]\frac{\hat v_2}{2 \hat v_3},\\
\mu_{33}^2 &=  \left( \mu_{11}^2 \hat v_2^2 + \lambda_{1111} \hat v_2^4 - \lambda_{3333} \hat v_3^4 \right)\frac{1}{\hat v_3^2},
\end{align}
\end{subequations}
and there are no unwanted massless states or any mass-degenerate states.

\subsection{Two vanishing vevs}

There are two cases when two vanishing vevs are assumed. For the case $(v,\, 0,\, 0)$ there is a single minimisation condition given by
\begin{equation}
\mu_{11}^2 = - \lambda_{1111} v^2.
\end{equation}
The neutral mass-squared matrix is diagonal and there is a pair of neutral mass-degenerate states associated with the $h_3$ doublet. 

Another case is given by $\left(0,\,0,\,v\right)$. This time the minimisation condition is
\begin{equation}
\mu_{33}^2 = - \lambda_{3333} v^2.
\end{equation}
In contrast to the previous case, there is now a pair of mass-degenerate charged states as well as four neutral mass-degenerate states, since the vacuum preserves the underlying symmetry.

\bigskip\textbf{Case of \boldmath$(0,\,0,\,v)$}\labeltext{Case of $(0,\,0,\,v)$}{Sec:U1Z2S2_00v}

The charged mass-squared matrix in the basis $\{h_1^\pm,\, h_2^\pm,\, h_3^\pm \}$ is:
\begin{equation}
\mathcal{M}_\mathrm{Ch}^2 = \mathrm{diag} \left( \mu_{11}^2 + \frac{1}{2} \lambda_{1133}v^2,\, \mu_{11}^2 + \frac{1}{2} \lambda_{1133}v^2,\, 0 \right).
\end{equation}

The neutral mass-squared matrix in the basis $\{\eta_1,\, \eta_2,\, \eta_3,\, \chi_1,\, \chi_2,\, \chi_3 \}$ is:
\begin{equation}
\mathcal{M}_\mathrm{N}^2 = \mathrm{diag} \left( m_H^2,\, m_H^2,\, m_h^2,\, m_H^2,\, m_H^2,\, 0 \right),
\end{equation}
where
\begin{subequations}
\begin{align}
m_H^2 ={}& \mu_{11}^2 + \frac{1}{2} \left( \lambda_{1133} + \lambda_{1331} \right) v^2,\\
m_h^2 ={}& 2 \lambda_{3333} v^2,
\end{align}
\end{subequations}
and the $h$ state is associated with the SM-like Higgs boson.

In this implementation $\{ \mu_{11}^2,\, \lambda_{1133},\, \lambda_{1331},\, \lambda_{3333} \}$ enter in the mass-squared matrices, and are responsible for generating three different mass-squared parameters. Some couplings appear only in the scalar interactions: $\{\lambda_{1111},\, \lambda_{1122},\, \lambda_{1221},\, \lambda_{1212}\}$.

\section{\texorpdfstring{\boldmath$U(2)$}{U(2)}-symmetric 3HDM}\label{Sec:Pot_U2}

We shall now proceed to other continuous symmetries, which were mentioned in Section~\ref{Sec:Pot_U1_additonal}. The first case is the $U(2)$-symmetric 3HDM, with the scalar potential of eq.~\eqref{Eq:V_U2},
\begin{equation*}
\begin{aligned}
V_{U(2)} ={}& \mu_{11}^2 (h_{11} + h_{22}) + \mu_{33}^2 h_{33} + \lambda_{1111} (h_{11}^2 + h_{22}^2) + \lambda_{3333} h_{33}^2 + \left( 2 \lambda_{1111} - \lambda_{1221} \right) h_{11} h_{22}\\
&  + \lambda_{1133} (h_{11} h_{33} + h_{22} h_{33}) + \lambda_{1221} h_{12} h_{21}+ \lambda_{1331} (h_{13} h_{31} + h_{23} h_{32}).
\end{aligned}
\end{equation*}

All the cases with different vacuum configurations are summarised in Table~\ref{Table:U2_Cases}. 

{{\renewcommand{\arraystretch}{1.3}
\begin{table}[htb]
\caption{Similar to Table~\ref{Table:U1_U1_Cases}, but now for $U(2)$. None of these cases violates CP.}
\label{Table:U2_Cases}
\begin{center}
\begin{tabular}{|c|c|c|c|c|} \hline\hline
Vacuum & SYM & $V$ & \begin{tabular}[l]{@{}c@{}} Mixing of the \\ neutral states\end{tabular} & Comments \\ \hline
$(\hat{v}_1,\, \hat{v}_2,\, 0)$ & $-$ & $V_{U(2)}$ & \footnotesize\begin{tabular}[l]{@{}c@{}} $\{\eta_1,\, \eta_2\}{-}\{\eta_3\}$ \\ ${-}\{\chi_1\}{-}\{\chi_2\}{-}\{\chi_3\}$ \end{tabular} & \footnotesize\begin{tabular}[l]{@{}c@{}} $m_{\eta_3} = m_{\chi_3}$ \\ $m_{\eta_{(1,2)}}=m_{\chi_1} = m_{\chi_2} = 0$ \\  \end{tabular} \\ \hline
$(\frac{v}{\sqrt{2}},\, \pm \frac{v}{\sqrt{2}} ,\, 0)$ & $-$ & $V_{U(2)} + \left( \mu_{12}^2 \right)^\mathrm{R}$ & \footnotesize\begin{tabular}[l]{@{}c@{}} $\{\eta_1,\, \eta_2\}{-}\{\eta_3\}$ \\ ${-}\{\chi_1,\, \chi_2\}{-}\{\chi_3\}$ \end{tabular} & \footnotesize\begin{tabular}[l]{@{}c@{}} $m_{\eta_{(1,2)}}=m_{\chi_{(1,2)}},~m_{\eta_3} = m_{\chi_3}$ \end{tabular}\\ \hline
$(0,\, \hat{v}_2,\, \hat{v}_3)$ & $-$ & $V_{U(2)}$ & \footnotesize\begin{tabular}[l]{@{}c@{}} $\{\eta_1\}{-}\{\eta_2,\,\eta_3\}$ \\ ${-}\{\chi_1\}{-}\{\chi_2\}{-}\{\chi_3\}$ \end{tabular} & \footnotesize\begin{tabular}[l]{@{}c@{}} $m_{\eta_1} = m_{\chi_i}=0$ \end{tabular} \\ \hline
$(0,\, \hat{v}_2,\, \hat{v}_3)$ & $-$ & $V_{U(2)} + \left( \mu_{23}^2 \right)^\mathrm{R}$ & \footnotesize\begin{tabular}[l]{@{}c@{}} $\{\eta_1\}{-}\{\eta_2,\,\eta_3\}$ \\ ${-}\{\chi_1\}{-}\{\chi_2,\,\chi_3\}$ \end{tabular} &  $m_{\eta_1} = m_{\chi_1}$  \\ \hline
$(v,\, 0,\, 0)$ & $-$ & $V_{U(2)}$ & diagonal &  \footnotesize\begin{tabular}[l]{@{}c@{}} $m_{\eta_3} = m_{\chi_3}$ \\ $m_{\eta_2}=m_{\chi_1} = m_{\chi_2} = 0$ \end{tabular} \\ \hline
$(v,\, 0,\, 0)$ & $-$ & $V_{U(2)} + \left( \mu_{23}^2 \right)^\mathrm{R}$ & \footnotesize\begin{tabular}[l]{@{}c@{}} $\{\eta_1\}{-}\{\eta_2,\, \eta_3\}$ \\ ${-}\{\chi_1\}{-}\{\chi_2,\,\chi_3\}$ \end{tabular} &  \footnotesize\begin{tabular}[l]{@{}c@{}} Two pairs of neutral \\ mass-degenerate states \end{tabular}\\ \hline
$(0,\, 0,\, v)$ & $\checkmark$ & $V_{U(2)}$ & diagonal & \footnotesize\begin{tabular}[l]{@{}c@{}} $m_{h_1^+} = m_{h_2^+}$, \\ $m_{\eta_1}  = m_{\eta_2} = m_{\chi_1} = m_{\chi_2}$ \end{tabular} \\ \hline
\end{tabular}  
\end{center}
\end{table}}

\subsection{One vanishing vev}

First, we consider different vacuum configurations with a single vanishing vev. In the case $(\hat{v}_1,\, \hat{v}_2,\, 0)$ there is a single minimisation condition:
\begin{equation}
\mu_{11}^2 = - \lambda_{1111} \left( \hat v_1^2 + \hat v_2^2 \right).
\end{equation}
This implementation leads to two massless states. We can introduce a soft symmetry-breaking term, $\mu_{12}^2$. There are two possibilities. Solving the minimisation conditions for independent vevs forces $\mu_{12}^2=0$. On the other hand, fixing $\hat v_2 = \pm \hat v_1$ leads to a real $\mu_{12}^2$,
\begin{equation}
\mu_{12}^2 = \mp \left( \mu_{11}^2 + 2 \lambda_{1111} \hat v_1^2 \right).
\end{equation}

Next we consider the vacuum given by $(0,\, \hat v_2,\, \hat v_3)$. In this case the minimisation conditions are:
\begin{subequations}
\begin{align}
\mu_{11}^2 &= - \lambda_{1111} \hat v_2^2 - \frac{1}{2}\left( \lambda_{1133} + \lambda_{1331} \right) \hat v_3^2,\\
\mu_{33}^2 &= - \frac{1}{2} \left( \lambda_{1133} + \lambda_{1331} \right) \hat v_2^2 - \lambda_{3333} \hat v_3^2.
\end{align}
\end{subequations}
In this implementation there is an additional massless state, four in total, due to the fact that $\left\langle h_3 \right\rangle \neq 0$. As before, one can introduce a soft symmetry-breaking term with the $\mu_{23}^2$ coefficient. The new minimisation conditions are:
\begin{subequations}
\begin{align}
\mu_{23}^2 &= - \left[ 2 \mu_{11}^2 + 2 \lambda_{1111} \hat v_2^2 + \left( \lambda_{1133} + \lambda_{1331} \right) \hat v_3^2 \right]\frac{\hat v_2}{2 \hat v_3},\\
\mu_{33}^2 &=  \left( \mu_{11}^2 \hat v_2^2 + \lambda_{1111} \hat v_2^4 - \lambda_{3333} \hat v_3^4 \right)\frac{1}{\hat v_3^2},
\end{align}
\end{subequations}
and all states become massive.

\subsection{Two vanishing vevs}

We shall now consider cases where two vevs vanish. The first case is $(v,\,0,\,0)$ with the minimisation condition:
\begin{equation}
\mu_{11}^2 = - \lambda_{1111} v^2.
\end{equation}
There are two additional massless states, both of them associated with the $h_2$ doublet. There are several possibilities to implement soft symmetry-breaking terms, however neither $\mu_{12}^2$ nor $\mu_{13}^2$ would survive the minimisation conditions. Therefore, the only appropriate soft symmetry-breaking term is $\mu_{23}^2$. The minimisation conditions do not change since the additional mixing occurs between the vev-less doublets. There is freedom to rotate away the imaginary part of $\mu_{23}^2$. Also, there will be two pairs of mass-degenerate states.

Another implementation is given by $\left(0,\,0,\,v\right)$. This time the minimisation condition is
\begin{equation}
\mu_{33}^2 = - \lambda_{3333} v^2.
\end{equation}
The neutral states of the vev-less doublets acquire a fourfold mass degeneracy.

\bigskip\textbf{Case of \boldmath$(0,\,0,\,v)$}

This case is identical, in terms of both the charged and the neutral mass-squared matrices, to \ref{Sec:U1Z2S2_00v} of $U(1) \times D_4$ in Section~\ref{Sec:Pot_U1_Z2_S2}. The two cases are physically different since the numbers of independent scalar couplings differ, \textit{e.g.}, in the $U(1) \times D_4$ model there are vertices $\eta_1 \eta_2 \chi_1 \chi_2 $ and $h_1^\pm h_1^\pm h_2^\mp h_2^\mp $ originating from the $\lambda_{1212}$ coupling, while in the $U(2)$-symmetric model we have $\lambda_{1212}=0$.

In this implementation $\{ \mu_{11}^2,\, \lambda_{1133},\, \lambda_{1331},\, \lambda_{3333} \}$ enter in the mass-squared matrices, and are responsible for generating three different mass-squared parameters. The couplings $\{\lambda_{1111},\, \lambda_{1122}\}$ appear only in the scalar interactions.

\section{\texorpdfstring{\boldmath$SO(3)$}{O(3)}-symmetric 3HDM }\label{Sec:Pot_SO3}

We shall now consider models stabilised by continuous symmetries of dimension three. In such cases it is sufficient to consider just a single option with different vanishing vevs, without permutations of the vevs. We start by exploring the $SO(3)$-symmetric 3HDM. The scalar potential was provided in eq.~\eqref{Eq:V_SO3} (with $\Lambda$ defined in eq.~\eqref{Eq:Incr_sym}),
\begin{equation*}
\begin{aligned}
V_{SO(3)} ={}& \mu_{11}^2 \sum_i h_{ii} + \lambda_{1111} \sum_i h_{ii}^2  + \lambda_{1122} \sum_{i<j} h_{ii} h_{jj} + \lambda_{1221} \sum_{i<j} h_{ij} h_{ji}+ \Lambda \sum_{i<j} (h_{ij}^2 + h_{ji}^2).
\end{aligned}
\end{equation*}

All cases with at least one zero vev are summarised in Table~\ref{Table:SO3_Cases}. These vacuum configurations, apart from a vev with all zeros, will break the underlying symmetry.

{{\renewcommand{\arraystretch}{1.3}
\begin{table}[htb]
\caption{Similar to Table~\ref{Table:U1_U1_Cases}, but now for $SO(3)$. None of these cases violates CP.}
\label{Table:SO3_Cases}
\begin{center}
\begin{tabular}{|c|c|c|c|c|} \hline\hline
Vacuum & SYM & $V$ & \begin{tabular}[l]{@{}c@{}} Mixing of the \\ neutral states\end{tabular} & Comments \\ \hline
$(\frac{v}{\sqrt{2}} e^{i \sigma},\, \pm \frac{v}{\sqrt{2}} ,\, 0)$ & $-$ & $V_{SO(3)} + \mu_{12}^2 $ & \footnotesize\begin{tabular}[l]{@{}c@{}} $\{\eta_1,\, \eta_2,\,\chi_1,\, \chi_2\}$ \\ ${-}\{\eta_3,\, \chi_3\}$ \end{tabular} & - \\ \hline
$(i \frac{v}{\sqrt{2}},\, \pm \frac{v}{\sqrt{2}} ,\, 0)$ & $-$ &  $V_{SO(3)} + \left( \mu_{12}^2 \right)^\mathrm{I}$  & \footnotesize\begin{tabular}[l]{@{}c@{}} $\{\eta_1,\, \eta_2\}{-}\{\eta_3\}$ \\ ${-}\{\chi_1,\,\chi_2\}{-}\{\chi_3\}$ \end{tabular} & \footnotesize\begin{tabular}[l]{@{}c@{}} $m_{\eta_{(1,2)}} = m_{\chi_{(1,2)}},~m_{\eta_3} = m_{\chi_3}$ \end{tabular} \\ \hline
$(\hat{v}_1,\, \hat{v}_2,\, 0)$ & $-$ & $V_{SO(3)}$ & \footnotesize\begin{tabular}[l]{@{}c@{}} $\{\eta_1,\, \eta_2\}{-}\{\eta_3\}$ \\ ${-}\{\chi_1,\,\chi_2\}{-}\{\chi_3\}$ \end{tabular} & \footnotesize\begin{tabular}[l]{@{}c@{}} $m_{h^+_{(1,2)}} = m_{h_3^+},~m_{\chi_{(1,2)}}^\prime = m_{\chi_3}$ \\ $m_{\eta_{(1,2)}} = m_{\eta_3} = m_{\chi_{(1,2)}}=0$ \end{tabular} \\ \hline
$(v,\, 0,\, 0)$ & $-$ & $V_{SO(3)}$ & diagonal &  \footnotesize\begin{tabular}[l]{@{}c@{}} $m_{h_2^+} = m_{h_3^+},~m_{\chi_2} = m_{\chi_3}$ \\ $m_{\eta_2} = m_{\eta_3} = m_{\chi_1}=0$ \end{tabular}\\ \hline
$(v,\, 0,\, 0)$ & $-$ & $V_{SO(3)} + \left( \mu_{23}^2 \right)^\mathrm{R}$ & \footnotesize\begin{tabular}[l]{@{}c@{}} $\{\eta_1\}{-}\{\eta_2,\, \eta_3\}$ \\ ${-}\{\chi_1\}{-}\{\chi_2,\, \chi_3\}$ \end{tabular} &  \footnotesize\begin{tabular}[l]{@{}c@{}} Two pairs of neutral \\ mass-degenerate states \end{tabular}\\ \hline
\end{tabular}  
\end{center}
\end{table}}

\subsection{One vanishing vev}

As mentioned before, due to the underlying $SO(3)$ symmetry it is sufficient to discuss a single vacuum configuration. However, due to the presence of $\Lambda$ it is not sufficient to consider only real vacua. In the case of $(\hat v_1 e^{i \sigma},\, \hat v_2,\, 0)$ the minimisation conditions are:
\begin{subequations}
\begin{align}
\mu_{11}^2 ={}& - \lambda_{1111} \left( \hat v_1^2 + \hat v_2^2 \right),\\
\Lambda ={}& 0.
\end{align}
\end{subequations}
The $\Lambda=0$ conditions indicates increase of the underlying symmetry. This vacuum configuration forces the scalar potential to become $SU(3)$-symmetric. It should be noted that there are no phase-sensitive terms present in the $SU(3)$ scalar potential (which we shall discuss in Section~\ref{Sec:Pot_SU3}) and the $\sigma$ phase becomes redundant. We conclude that this case is not realisable. Furthermore, there are five massless neutral states along with two mass-degenerate charged states. One could introduce the soft symmetry-breaking term $\mu_{12}^2$. The new minimisation conditions are:
\begin{subequations}
\begin{align}
\mu_{11}^2 ={}& - \lambda_{1111} \left( \hat v_1^2 + \hat v_2^2 \right),\\
\mu_{12}^2  ={}& 2 i  \Lambda \sin \sigma \hat v_1 \hat v_2.
\end{align}
\end{subequations}
However, in this case there is still one unwanted massless state present.

The vacuum configuration $(\frac{v}{\sqrt{2}} e^{i \sigma},\, \pm \frac{v}{\sqrt{2}} ,\, 0)$ yields the minimisation conditions:
\begin{subequations}
\begin{align}
\mu_{11}^2 ={}& - \lambda_{1111} v^2,\\
\Lambda ={}& 0,
\end{align}
\end{subequations}
where due to the latter condition the overall symmetry is increased to that of $SU(3)$. The $\sigma$ phase is redundant in the $SU(3)$-symmetric 3HDM. This implementation has two additional unwanted massless states along with a pair of neutral mass-degenerate states. With the soft symmetry-breaking term $\mu_{12}^2$ the minimisation conditions become:
\begin{subequations}
\begin{align}
\left( \mu_{12}^2 \right)^\mathrm{R} ={}& \mp \cos \sigma \left( \mu_{11}^2 + \lambda_{1111} v^2 \right),\\
\left( \mu_{12}^2 \right)^\mathrm{I} ={}& \mp \sin \sigma \left( \mu_{11}^2 + \frac{1}{2} \left( \lambda_{1122} + \lambda_{1221} \right) v^2 \right).
\end{align}
\end{subequations}
This implementation does not exhibit any particular mass pattern. Another possible vacuum configuration with the soft symmetry-breaking term is given by $(i\frac{v}{\sqrt{2}},\, \pm \frac{v}{\sqrt{2}} ,\, 0)$. The minimisation condition is
\begin{equation}
 \mu_{12}^2 = \mp i\left[ \mu_{11}^2 + \frac{1}{2}\left( \lambda_{1122} + \lambda_{1221} \right) v^2\right].
\end{equation}
In this case there are two pairs of mass-degenerate neutral states.

After going through different complex vacua, we can turn out attention to the real vacuum. In the case $(\hat v_1,\, \hat v_2,\, 0)$, the minimisation condition is:
\begin{equation}
\mu_{11}^2 = - \lambda_{1111} \left( \hat v_1^2 + \hat v_2^2 \right).
\end{equation}
There are two unwanted massless states. These states can be promoted to massive ones via the introduction of a $\mu_{12}^2$ soft symmetry-breaking term. The soft symmetry-breaking term survives the minimisation conditions only when the vevs are related as $\hat v_1^2= \hat v_2^2$. The case $ (\frac{v}{\sqrt{2}},\, \pm \frac{v}{\sqrt{2}} ,\, 0)$ is contained within a more general implementation $ (\frac{v}{\sqrt{2}} e^{i \sigma},\, \pm \frac{v}{\sqrt{2}} ,\, 0)$.

\subsection{Two vanishing vevs}

For the vacuum configurations with a single non-zero vev we are free to consider a single case. In the case $(v,\,0,\,0)$, the minimisation condition is:
\begin{equation}
\mu_{11}^2 = - \lambda_{1111} v^2.
\end{equation}
In total there will be three neutral massless states. The two unwanted ones can be removed via the introduction of a soft symmetry-breaking term. There are several possibilities to implement soft symmetry-breaking terms, however neither $\mu_{12}^2$ nor $\mu_{13}^2$ would survive the minimisation conditions. Therefore, the only option left is to introduce the $\mu_{23}^2$ term. This term does not alter the minimisation conditions and both unwanted massless states get promoted to massive ones. There is freedom to rotate away the imaginary part of the $\mu_{23}^2$ term since the phase can be absorbed into the $h_2$ or the $h_3$ doublet.

\section{\texorpdfstring{\boldmath$SU(3)$}{SU(3)}-symmetric 3HDM}\label{Sec:Pot_SU3}

Finally, we consider the most symmetric model, $SU(3)$-symmetric 3HDM. As in the previous case, the procedure of identifying different vacuum configurations is simplified. The scalar potential was provided in eq.~\eqref{Eq:V_SU3},
\begin{equation*}
\begin{aligned}
V_{SU(3)} ={}& \mu_{11}^2 \sum_i h_{ii} + \lambda_{1111} \left( \sum_i h_{ii} \right) ^2 + \lambda_{1221} \sum_{i<j} (h_{ij} h_{ji} -h_{ii} h_{jj} ).
\end{aligned}
\end{equation*}
Vacuum configurations with at least a single vanishing vev are summarised in Table~\ref{Table:SU3_Cases}. 

{{\renewcommand{\arraystretch}{1.21}
\begin{table}[htb]
\caption{Similar to Table~\ref{Table:U1_U1_Cases}, but now for $SU(3)$. None of these cases violates CP.}
\label{Table:SU3_Cases}
\begin{center}
\begin{tabular}{|c|c|c|c|c|} \hline\hline
Vacuum & SYM & $V$ & \begin{tabular}[l]{@{}c@{}} Mixing of the \\ neutral states\end{tabular} & Comments \\ \hline
$(\hat{v}_1,\, \hat{v}_2,\, 0)$ & $-$ & $V_{SU(3)}$ & \footnotesize\begin{tabular}[l]{@{}c@{}} $\{\eta_1,\, \eta_2\}{-}\{\eta_3\}$ \\ ${-}\{\chi_1\}{-}\{\chi_2\}{-}\{\chi_3\}$ \end{tabular} & \footnotesize\begin{tabular}[l]{@{}c@{}} $m_{h^+_{(1,2)}} = m_{h_3^+}$ \\ $m_{\eta_{(1,2)}} = m_{\eta_3} = m_{\chi_i}=0$ \end{tabular} \\ \hline
$(\frac{v}{\sqrt{2}},\, \pm \frac{v}{\sqrt{2}} ,\, 0)$ & $-$ & $V_{SU(3)} + \left( \mu_{12}^2 \right)^\mathrm{R}$ & \footnotesize\begin{tabular}[l]{@{}c@{}} $\{\eta_1,\, \eta_2\}{-}\{\eta_3\}$ \\ ${-}\{\chi_1,\, \chi_2\}{-}\{\chi_3\}$ \end{tabular} & \footnotesize\begin{tabular}[l]{@{}c@{}} $m_{\eta_{(1,2)}} = m_{\chi_{(1,2)}},~m_{\eta_3} = m_{\chi_3}$\end{tabular}\\ \hline
$(v,\, 0,\, 0)$ & $-$ & $V_{SU(3)}$ & diagonal &  \footnotesize\begin{tabular}[l]{@{}c@{}} $m_{h_2^+} = m_{h_3^+}$ \\ $m_{\eta_2} = m_{\eta_3} = m_{\chi_i}=0$ \end{tabular}\\ \hline
$(v,\, 0,\, 0)$ & $-$ & $V_{SU(3)} + \left( \mu_{23}^2 \right)^\mathrm{R}$ & \footnotesize\begin{tabular}[l]{@{}c@{}} $\{\eta_1\}{-}\{\eta_2,\, \eta_3\}$ \\ ${-}\{\chi_1\}{-}\{\chi_2,\, \chi_3\}$ \end{tabular} &  \footnotesize\begin{tabular}[l]{@{}c@{}} $m_{H_1}^2 = - m_{H_2}^2$ \end{tabular}\\ \hline
\end{tabular}  
\end{center}\vspace*{-10pt}
\end{table}}

\subsection{One vanishing vev}

For the vacuum configuration $(\hat v_1,\, \hat v_2,\, 0 )$ the minimisation condition is:
\begin{equation}
\mu_{11}^2 = - \lambda_{1111} \left( \hat v_1^2 + \hat v_2^2 \right).
\end{equation}
Since all $SU(3)$ generators are broken, we get five massless states. We can introduce the $\mu_{12}^2$ soft symmetry-breaking term to promote unwanted massless states to massive ones. Requiring the $\mu_{12}^2$  term to survive we need to take $\hat v_1^2 = \hat v_2^2$, so that the vacuum configuration becomes $ (\frac{v}{\sqrt{2}},\, \pm \frac{v}{\sqrt{2}} ,\, 0)$. The new minimisation conditions is
\begin{equation}
\mu_{12}^2 = \mp \mu_{11}^2 \mp \lambda_{1111} v^2.
\end{equation}
and there will be two pairs of neutral mass-degenerate states.

\subsection{Two vanishing vevs}

For the case $(v,\,0,\,0)$ the minimisation condition is:
\begin{equation}
\mu_{11}^2 = - \lambda_{1111} v^2.
\end{equation}
Once again the $SU(3)$ symmetry will be maximally broken, leading to five neutral massless states. There are several possibilities to implement soft symmetry-breaking terms, however neither $\mu_{12}^2$ nor $\mu_{13}^2$ would survive the minimisation conditions. To remove the unwanted massless states one can introduce a $\mu_{23}^2$ soft symmetry-breaking term which, without loss of generality, can be taken to be real. In this case the minimisation conditions do not change. For the exact $SU(3)$ symmetry, only one entry of the neutral mass-squared matrix is non-zero. With the soft symmetry-breaking term, four entries will become equal to $\left(\mu_{23}^2\right)^\mathrm{R}$. These terms will form two anti-diagonal sub-matrices of $\mathcal{M}_\mathrm{N}^2$, resulting in two pairs of neutral states with squared masses of opposite signs (a saddle point), $m_{H_1}^2 = - m_{H_2}^2 =  \left(\mu_{23}^2\right)^\mathrm{R}.$

\section{The Dark Matter candidates}\label{Sec:Different_DM_cases}

In the previous sections we identified cases {\it without} SSB:
{\quad\begin{longtable}{llll}
\hyperref[Table:U1_U1_Cases]{$\bullet \quad U(1) \times U(1)$}\hspace{20pt} & $(v,\,0,\,0)$ &  & \\
\hyperref[Table:U11_Cases]{$\bullet \quad U(1)_1$} & $(v,\,0,\,0)$ & $(0,\,0,\,v)$ & \\
\hyperref[Table:U1Z2_Cases]{$\bullet \quad U(1) \times \mathbb{Z}_2$} & $(v,\,0,\,0)$ & $(0,\,0,\,v)$ & \\
\hyperref[Table:U12_Cases]{$\bullet \quad U(1)_2$} & $(v,\,0,\,0)$ & $(0,\,0,\,v)$ &  $(v_1,\,v_2,\,0)$ \\
\hyperref[Table:U1U1S2_Cases_1]{$\bullet \quad O(2) \times U(1)$} & & $(0,\,0,\,v)$ & \\
\hyperref[Table:O2_Cases_1]{$\bullet \quad O(2)$} & & $(0,\,0,\,v)$ & \\
\hyperref[Table:U1Z2_S2_Cases]{$\bullet \quad U(1) \times D_4$} & & $(0,\,0,\,v)$ & \\
\hyperref[Table:U2_Cases]{$\bullet \quad U(2)$} & & $(0,\,0,\,v)$ & \\
\addtocounter{table}{-1}
\end{longtable}}
\vspace{-32pt}\begin{flushleft}We recall that the \hyperref[Table:U11S3_Cases]{$\left[ U(1) \times U(1) \right] \rtimes S_3$}, \hyperref[Table:SO3_Cases]{$SO(3)$}, \hyperref[Table:SU3_Cases]{$SU(3)$} cases always result in SSB.\end{flushleft}

\clearpage

An important generic feature of the $U(1)$-stabilised DM candidates is the existence of a mass-degeneracy pattern in the neutral sector. This can be understood by studying the allowed patterns of the $SU(2)$ singlets, $h_{ij}$. Let us consider a generic case where the vacuum of the $h_3$ doublet is stabilised, $\left\langle h_3 \right\rangle =0$, by an underlying symmetry. Contributions to the mass-squared parameters of the $h_3$ doublet come only from the quartic terms $h_{ij} h_{33}$ and $h_{i3}h_{3j}$, and the bilinear term $h_{33}$. Since $h_3$ is stabilised by $U(1)$ there is no contribution from the quartic $\lambda_{3333}$ term. Then, the entries of the mass-squared matrix associated with $h_3$ will be invariant under an interchange of the two neutral fields, $h_3^\dagger h_3 \supset \eta_3^2 + \chi_3^2$. An obvious consequence can be observed---when the DM candidate, in $n$HDMs, is stabilised by continuous symmetries and the underlying symmetry is not broken, there will be at least one pair of mass-degenerate DM candidates, $m_{\eta_3}^2 = m_{\chi_3}^2$.

If two of the vev-less doublets are stabilised by a $U(1)$ symmetry, this will cause two separate degeneracy patterns to appear in the neutral sector. In some cases there might be two inert sectors with no mixing in the mass matrix. This, then, results in two pairs of degenerate scalar fields.  However, these sectors could still have mixed interactions. Different cases with mass-degenerate pairs are listed in Table~\ref{Table:Mass_degen_patterns}. We need to determine if any of these models are physically equivalent. We begin by comparing different implementations within a given symmetry.

First, we consider the $U(1)_1$-symmetric 3HDM. There are two cases which preserve the underlying symmetry, both given by implementations with two vanishing vevs. There are two pairs of mass-degenerate states, which we see from Table~\ref{Table:Mass_degen_patterns}. In the case $(v,\, 0,\, 0)$ the neutral mass-squared matrix is diagonal while for the $(0,\, 0,\, v)$ case there is mixing between the $h_1$ and $h_2$ doublets caused by the $\lambda_{1323}$ term, \textit{e.g.}, in the latter implementation there will be a trilinear coupling between the SM-like Higgs boson, residing in $h_3$, and the states associated with $h_1$ and $h_2$. Such trilinear coupling is absent in the other case. Therefore, these cases are physically different.

In the $U(1) \times \mathbb{Z}_2$ model we observe different mass-degeneracy patterns, see Table~\ref{Table:Mass_degen_patterns}. Based on this observation we conclude that these implementations are physically different.

Then, there are four different $U(1)_2$ implementations. By examining the mass-squared parameters of $(v,\, 0,\, 0)$ and $(0,\, 0,\, v)$ one can see that the number of mass-degenerate states is different for these two cases, see Table~\ref{Table:Mass_degen_patterns}. Next, there is the case $(v_1,\, v_2,\, 0)$, connected via a Higgs basis transformation to $(v,\,0,\,0)$, with either explicit or spontaneous CP violation, or CP conservation. Due to the CP indefinite states the implementations with and without CP violation yield different physical cases. We need to check for the cases $(\hat{v}_1,\,\hat{v}_2,\,0)$ and $(\hat{v}_1 e^{i \sigma},\,\hat{v}_2,\,0)$ to verify if the latter case is a more constrained one. We stress that the spontaneous CP violating cases can be represented by a complex potential and real vevs. However, the phases of the potential terms are now interrelated. It could be that the explicit CP violating case might introduce redundant phases. To sum up, the distinction among different implementations can not only be based on the mass-degeneracy patterns. 

In the $U(1)_2$ model, with explicit CP violation, there are five complex quartic couplings, while in the implementation with spontaneous CP violation there is a basis where all these couplings become real. In the general case it is a difficult task to discriminate experimentally between different types of CP violation. By performing a naive numerical analysis of the trilinear and quartic couplings in the physical basis (a random scan of the physical couplings without implementing any theoretical or experimental constraints on the parameter space) we found that these two implementations are physically different.

To sum up, different implementations within the same underlying symmetry correspond to a different physical parameter space. However, we also need to check if the same is true across different underlying symmetries. We shall consider the $U(1) \times U(1)$-symmetric 3HDM to be the reference model since it corresponds to the most general real 3HDM. The only vacuum respecting the $U(1) \times U(1)$ symmetry is $(v,\,0,\,0)$, with all possible permutations of the non-zero element.

{{\renewcommand{\arraystretch}{1.25}
\setlength\LTcapwidth{\linewidth}
\begin{table}[htb]
\caption{Mass degeneracy patterns of implementations without SSB. Some shared features of different blocks are indicated above the block. In the last column a combination of fields, or a single field, responsible for the mass eigenstate of the SM-like Higgs boson is provided.}
\begin{center}
\begin{threeparttable}
\begin{tabular}{|c|c|c|c|c|}\hline\hline
Model & Vacuum & Symmetry & Reference & $h_\text{SM}$ \\ \hline\hline
\multicolumn{5}{c}{$m_{\eta_3}=m_{\chi_3}$} {\rule{0pt}{15pt}} \\ \hline
I-a & $(v,\, 0,\, 0)$ & $U(1) \times \mathbb{Z}_2$ & Table~\ref{Table:U1Z2_Cases} & $\eta_1$ \\ \hline
I-b & $(v,\, 0,\, 0)$ & $U(1)_2$ & Table~\ref{Table:U12_Cases} & $\{\eta_1,\, \eta_2,\, \chi_2\}$ \\ \hline
I-c & $(\hat{v}_1,\, \hat{v}_2,\, 0)$ & $U(1)_2$ & Table~\ref{Table:U12_Cases} & $\{\eta_1,\, \eta_2 ,\, \chi_1,\, \chi_2\}$\\ \hline
I-d & $(\hat{v}_1 e^{i\sigma},\, \hat{v}_2,\, 0)$ & $U(1)_2$ & Table~\ref{Table:U12_Cases} &  $\{\eta_1,\, \eta_2 ,\, \chi_1,\, \chi_2\}$ \\ \hline
\multicolumn{5}{c}{$m_{\eta_i}=m_{\chi_i}, \text{ for } \left\langle h_i\right\rangle = 0$} {\rule{0pt}{15pt}} \\ \hline
II-a &$(v,\, 0,\, 0)$ & $U(1)\times U(1)$ & Table~\ref{Table:U1_U1_Cases} &  $\eta_1$\\ \hline
II-b & $(v,\, 0,\, 0)$ & $U(1)_1$ & Table~\ref{Table:U11_Cases} &  $\eta_1$\\ \hline
II-c & $(0,\, 0,\, v)$ & $U(1) \times \mathbb{Z}_2$ & Table~\ref{Table:U1Z2_Cases} & $\eta_3$ \\ \hline \hline
\multicolumn{5}{c}{$m_{\{\eta_1,\eta_2\}} = m_{\{\chi_1,\chi_2\}}$, $\eta_i$ and $\chi_i$ mix separately} {\rule{0pt}{15pt}} \\ \hline
II-d & $(0,\, 0,\,v)$ & $U(1)_1$ & Table~\ref{Table:U11_Cases} & $\eta_3$\\ \hline
II-e & $(0,\, 0,\, v)$ & $O(2)$ & Tables~\ref{Table:O2_Cases_1} and~\ref{Table:O2_Cases_2} & $\eta_3$ \\ \hline
\multicolumn{5}{c}{$m_{\{\eta_1,\eta_2,\chi_1,\chi_2\}}$, all states mix together} {\rule{0pt}{15pt}} \\ \hline
II-f & $(0,\, 0,\, v)$ & $U(1)_2$ & Table~\ref{Table:U12_Cases} & $\eta_3$ \\ \hline\hline
\multicolumn{5}{c}{$m_{\eta_1}=m_{\eta_2}=m_{\chi_1}=m_{\chi_2}$ $^\alpha$} {\rule{0pt}{15pt}} \\ \hline
II-g & $(0,\, 0,\, v)$ & $O(2) \times U(1)$ & Tables~\ref{Table:U1U1S2_Cases_1} and~\ref{Table:U1U1S2_Cases_2} & $\eta_3$ \\ \hline
II-h & $(0,\, 0,\, v)$ & $U(1) \times D_4$ & Table~\ref{Table:U1Z2_S2_Cases} & $\eta_3$ \\ \hline \hline
II-i & $(0,\, 0,\, v)$ & $U(2)$ & Table~\ref{Table:U2_Cases} & $\eta_3$ \\ \hline
\end{tabular}
\begin{tablenotes}
\item [$\alpha$] In these cases the potential is completely symmetric under interchange $h_1\leftrightarrow h_2$. This symmetry remains unbroken in these implementations. There is also an unbroken symmetry for $\eta_i\leftrightarrow\chi_i$, CP is conserved. We have two pairs of identical fields with different names.\end{tablenotes}
\end{threeparttable}
\end{center}
\label{Table:Mass_degen_patterns}
\end{table}}

Consider the $U(1)_1$-symmetric 3HDM. There is a single phase-sensitive coupling $\lambda_{1323}$, the phase of which may be freely rotated away. For the $(v,\,0,\,0)$ implementation the new trilinear couplings, with respect to the $U(1) \times U(1)$-symmetric model, are
\begin{equation}
\frac{1}{2} \lambda_{1323} \left( \eta_2 \eta_3^2 + 2 \eta_3 \chi_2 \chi_3 - \eta_2 \chi_3^2 \right)v +  \frac{1}{2} \lambda_{1323} \left( \eta_3  h_2^+ h_3^- - i \chi_3 h_2^+ h_3^- + \mathrm{h.c.} \right)v,
\end{equation}
and the quartic ones are
\begin{equation}
\frac{1}{2} \lambda_{1323} h \left( \eta_2 \eta_3^2 + 2 \eta_3 \chi_2 \chi_3 - \eta_2 \chi_3^2 \right) + \frac{1}{2} \lambda_{1323} h \left( \eta_3 h_2^+ h_3^- -i \chi_3 h_2^+ h_3^- + \mathrm{h.c.} \right),
\end{equation}
having removed the Goldstone bosons, and where $h$ is the SM-like Higgs boson.

Let us now consider $(0,\,0,\,v)$. Increasing the $U(1) \times U(1)$ symmetry by $S_2$, the couplings of the two doublets, symmetric under $S_2$, get related and therefore two of the mass-degenerate pairs present in $U(1) \times U(1)$ become degenerate among themselves, forcing four of the neutral states to have identical masses, as seen in Table~\ref{Table:Mass_degen_patterns}. Two charged scalars also become mass degenerate. In both cases the Higgs portal couplings are:
\begin{equation}
\left( \lambda_{jj33} + \lambda_{j33j} \right)\left( v + h \right) h H_i H_i  = 2 \frac{m_{H_i}^2 - \mu_{jj}^2}{v^2} \left( v + h \right) h H_i H_i,
\end{equation} 
not accounting for the symmetry factor. Here, the $j=\{1,2\}$ index is associated with the $SU(2)$ doublet, either $h_1$ or $h_2$, in which the mass eigenstate $H_i$ ($i=1..4$) resides. By tuning the value of the bilinear term, which is a free parameter of the implementation, it is possible to cancel the portal couplings at tree level. Then, the dark sector would interact with the visible sector via the gauge couplings or charged scalars,
\begin{equation}
\lambda_{jj33} \left( v + h \right) h h_i^+ h_i^- = 2 \frac{m_{h^+_i}^2 - \mu_{jj}^2}{v^2} \left( v + h \right) h h_i^+ h_i^-.
\end{equation}

In Table~\ref{Table:Couplings_tri_quart} we present the total numbers of trilinear and quartic couplings, not counting vertices with Goldstone bosons. The purpose of this is to identify if some of the implementations of the different models could possibly result in identical physical models. There is a repeating pattern of ``7 + 32" couplings among several models: $U(1) \times U(1)$, $U(1) \times \mathbb{Z}_2$ (there are two distinct implementations), $O(2) \times U(1)$ and $U(2)$. One can observe that the mass-degeneracy patterns, which were provided in Table~\ref{Table:Mass_degen_patterns}, are different, except for the two models, $ O(2) \times U(1)$ and $U(2)$, which share the same pattern. Since these cases share identical numbers of couplings, but not the scalar potentials, one could check if some of the physical couplings are related by a constant within these two symmetries. We checked explicitly that the patterns of the couplings are different for these four models, and therefore conclude that they are physically different. 

Apart from the scalar interactions there are interactions involving scalars with gauge bosons and with fermions. There exist several ways in which one could construct the Yukawa Lagrangian. So far we were only dealing with the scalar sector, and thus there was some freedom in assigning charges of the underlying symmetry to different $SU(2)$ doublets. For instance, in eq.~\eqref{Eq:U11_U12_charges}, introducing Yukawa couplings would remove the equivalence among cases.

The $Z$-scalar interactions can be written as:
\begin{equation}
\mathcal{L}_{VHH} = -\frac{ g}{2 \cos \theta_W}Z^\mu \eta_i \overset\leftrightarrow{\partial_\mu} \chi_i.
\end{equation}
The DM candidates in the discussed models will be associated with the fields $\eta_i$ and $\chi_i$, for some $i$. We recall that two DM candidates will be mass degenerates as a result of the underlying symmetry. This, in turn, indicates that processes involving the $Z$ boson will be kinematically suppressed. This feature together with the mass degeneracy of the DM candidates are notable differences from the IDM and other studied 3HDMs listed in Section~\ref{Sec:3HDM-review}. However, there will be contributions to the relic density via the trilinear interactions involving the $W^\pm$ bosons and the charged scalars.

\begin{table}[htb]
\caption{Physical scalar couplings in different cases which preserve the underlying symmetry. Entries correspond to the number of non-vanishing trilinear plus quartic couplings. The number of independent couplings (not related by simple constants, \textit{e.g.}, $g(X) = C g(Y)$) are given in parentheses. The counting does not include the Goldstone bosons, nor the Hermitian conjugated couplings, \textit{i.e.}, $g(H_i H_j^+ H_k^-) = g(H_i H_j^- H_k^+) ^\ast$ are counted as a single entry in the parentheses. The two $U(1)\times U(1)$ implementations have identical structures under a re-labeling of the fields. The $U(1)_2$ model is the only one with CP violation.}
\label{Table:Couplings_tri_quart}
\begin{center}
\begin{tabular}{|c|c|c|c|c|} \hline\hline
Symmetry & $(v,\, 0,\, 0)$ & $(0,\, 0,\, v)$ & $(v_1,\, v_2,\, 0)$ \\ \hline
$ \quad U(1) \times U(1)$&	\multicolumn{2}{c|}{7 + 32 (5 + 10)} &\\ \hline 
$ \quad U(1)_1$ &	15 + 40 (8 + 13) &	9 + 53 (6 + 24) &	\\ \hline  
$ \quad U(1) \times \mathbb{Z}_2$ &	7 + 32 (6 + 11)&	 7 + 38 (5 + 16) &	\\ \hline 
$ \quad U(1)_2$ &  \begin{tabular}[l]{@{}c@{}} $\mathbb{R}:16 + 41\,(13 + 24)$ \\ $\mathbb{C}:24 + 55\,(20 + 40)$  \end{tabular} &	\begin{tabular}[l]{@{}c@{}} $\mathbb{R}:10 + 55\,(7 + 30)$ \\ $\mathbb{C}:12 + 73\,(8 + 36)$  \end{tabular} &	\begin{tabular}[l]{@{}c@{}} $\mathbb{R}:16 + 41\,(13 + 24)$ \\ $\mathbb{C}:24 + 55\,(20 + 40)$  \end{tabular}\\ \hline
$ \quad O(2) \times U(1)$ &	& 7 + 32 (3 + 7)	&	\\ \hline   
$ \quad O(2)$ &	& 7 + 39 (4 + 12)	&	\\ \hline   
$ \quad U(1) \times D_4$ &	&	7 + 34 (3 + 11) &	\\ \hline   
$ \quad U(2)$ &	&	7 + 32 (3 + 6) &	\\ \hline   
\end{tabular} 
\end{center}
\end{table}}

In summary, we verified that all analysed models and implementations, which were provided as a list at the beginning of this section, are physically distinct. In this section, our goal was to identify cases without SSB which could be of interest for model builders. In total, we identified thirteen different cases capable of accommodating DM. The full study of different DM models would require an involved numerical analysis. 

\section{Comparison with the \texorpdfstring{\boldmath$\mathbb{Z}_2\text{-3HDM and}\; \mathbb{Z}_3\text{-3HDM}$}{Z2-3HDM and Z3-3HDM}}\label{Sec:Pot_Zn}

For completeness we consider both $\mathbb{Z}_2$-symmetric and $\mathbb{Z}_3$-symmetric 3HDMs as both are the least symmetric models and can serve as a basis for the discussion of the more symmetric $U(1)$-based models by imposing conditions on the couplings. We are interested in cases with CP violation and therefore shall consider two possibilities: explicit CP violation and spontaneous CP violation. Mixing in the neutral sectors of these two symmetries as well as different scalar potentials are summarised in Tables~\ref{Table:Z2_cases} and \ref{Table:Z3_Cases}. Further discussion of these two symmetries can be found in Appendix~\ref{App:Pot_Z2} and Appendix~\ref{App:Pot_Z3}.

{{\renewcommand{\arraystretch}{1.295}
\begin{table}[htb]
\caption{ Similar to Table~\ref{Table:U1_U1_Cases}, but now for $\mathbb{Z}_2$. In all cases complex parameters are present, they may result in CP violation.}
\label{Table:Z2_cases}
\begin{center}
\begin{tabular}{|c|c|c|c|c|} \hline\hline
Vacuum & SYM & $V$ & \begin{tabular}[l]{@{}c@{}} Mixing of the \\ neutral states\end{tabular} & Comments \\ \hline
$(\hat v_1 e^{i \sigma},\, \hat v_2,\, 0)$ & $\checkmark$ & $V_{\mathbb{Z}_2}$ & \footnotesize\begin{tabular}[l]{@{}c@{}} $\{\eta_1,\, \eta_2,\, \chi_1,\, \chi_2\}$ \\ ${-}\{\eta_3,\,\chi_3\}$ \end{tabular} & -\\ \hline
$(0,\, \hat v_2,\, \hat v_3)$ & $-$ & $V_{\mathbb{Z}_2}$ &  total mixing & No obvious DM \\ \hline
$(0,\, \hat v_2 e^{i \sigma},\, \hat v_3)$ & $-$ & $V_{\mathbb{Z}_2}$  & \footnotesize\begin{tabular}[l]{@{}c@{}} $\{\eta_1,\,\eta_2,\, \eta_3,\, \chi_1\}$ \\ ${-}\{\chi_2\}{-}\{\chi_3\}$ \end{tabular} & \footnotesize\begin{tabular}[l]{@{}c@{}} $m_{\chi_2} = m_{\chi_3}=0$ \\ No obvious DM  \end{tabular} \\ \hline 
$(0,\, \hat v_2 e^{i \sigma},\, \hat v_3)$ & $-$ & $V_{\mathbb{Z}_2} + (\mu_{23}^2)^\mathrm{R}$  & total mixing & No obvious DM \\ \hline 
$(v,\, 0,\, 0)$ & $\checkmark$ & $V_{\mathbb{Z}_2}$ & \footnotesize\begin{tabular}[l]{@{}c@{}} $\{\eta_1,\, \eta_2,\, \chi_2\}{-}\{\chi_1\}$ \\ ${-}\{\eta_3,\,\chi_3\}$ \end{tabular} & - \\ \hline
$(0,\, 0,\, v)$ & $\checkmark$ & $V_{\mathbb{Z}_2}$ & \footnotesize\begin{tabular}[l]{@{}c@{}} $\{\eta_1,\, \eta_2,\, \chi_1,\, \chi_2\}$ \\ ${-}\{\eta_3\}{-}\{\chi_3\}$ \end{tabular} & - \\ \hline \hline
\end{tabular} 
\end{center}
\end{table}}

{{\renewcommand{\arraystretch}{1.295}
\begin{table}[htb]
\caption{Similar to Table~\ref{Table:U1_U1_Cases}, but now for $\mathbb{Z}_3$. It is not possible to have simultaneously CP violation and a DM candidate.}
\label{Table:Z3_Cases}
\begin{center}
\begin{tabular}{|c|c|c|c|c|} \hline\hline
Vacuum & SYM &$V$ & \begin{tabular}[l]{@{}c@{}} Mixing of the \\ neutral states\end{tabular} & Comments \\ \hline
$(\hat{v}_1,\, \hat{v}_2,\, 0)$ & $-$ &  $V_{\mathbb{Z}_3}$ & total mixing & No obvious DM \\ \hline
$(\hat{v}_1 e^{i \sigma},\, \hat{v}_2,\, 0)$ & $-$ &  $V_{U(1)_1}$ & \footnotesize\begin{tabular}[l]{@{}c@{}} $\{\eta_1,\, \eta_2\}{-}\{\eta_3\}$\\${-}\{\chi_1\}{-}\{\chi_2\}{-}\{\chi_3\}$ \end{tabular} & $m_{\chi_1} = m_{\chi_2} = 0$ \\ \hline
$(\hat{v}_1 e^{i \pi/3},\, \hat{v}_2,\, 0)$  & $-$ &  $V_{\mathbb{Z}_3}$ & total mixing & No obvious DM \\ \hline
$(\hat v_1 e^{i \sigma},\, \hat v_2,\, 0)$ & $-$ & $V_{U(1)_1} + (\mu_{12}^2)^\mathrm{R}$ & \footnotesize\begin{tabular}[l]{@{}c@{}} $\{\eta_1,\, \eta_2\}{-}\{\eta_3\}$\\ ${-}\{\chi_1,\, \chi_2\}{-}\{\chi_3\}$ \end{tabular} & - \\ \hline
$(v,\,0,\,0)$ & $\checkmark$ & $V_{\mathbb{Z}_3}$ & \footnotesize\begin{tabular}[l]{@{}c@{}} $\{\eta_1\}{-}\{\chi_1\}$ \\ ${-}\{\eta_2,\ \eta_3,\, \chi_2,\, \chi_3\}$ \end{tabular} & \footnotesize\begin{tabular}[l]{@{}c@{}} Two pairs of \\ mass-degenerate states \end{tabular} \\ \hline \hline
\end{tabular} 
\end{center}
\end{table}}

All the models with continuous symmetries considered in the previous sections can be obtained from the $\mathbb{Z}_2$-symmetric 3HDM by requiring some couplings to vanish or imposing specific relations among them. Likewise, some of those models can be obtained from the $\mathbb{Z}_3$-symmetric 3HDM. We recall that the symmetry-breaking patterns were presented in Figure~\ref{Fig:Symmetry_breaking}. To illustrate the point, let us consider several examples. 

For example, consider the $U(1)_1$-symmetric 3HDM. Starting from the $\mathbb{Z}_2$-symmetric model we need to require $\mu_{12}^2 = 0$ and $\lambda_{ijij} = \lambda_{1112} = \lambda_{1222} = \lambda_{1233} = \lambda_{1332} = 0$ to be satisfied. Or, going from the $\mathbb{Z}_3$-symmetric model, the conditions are $\lambda_{1213} = \lambda_{1232} = 0$. Then, one can plug these conditions into the mass-squared matrices of Appendices~\ref{App:Pot_Z2} or \ref{App:Pot_Z3} to get the results presented in Section~\ref{Sec:Pot_U11}. We checked explicitly that such procedure always holds for all directions of the symmetry-breaking patterns presented in Figure~\ref{Fig:Symmetry_breaking}.

\section{Dark Matter in the \texorpdfstring{\boldmath$U(1)\times U(1)$}{U(1) x U(1)} 3HDM }\label{Sec:U1U1_analysis}

After identifying and classifying different models, and different implementations of these models, we turn our attention to a numerical scan of a particular model. We decided to take a closer look at the $U(1) \times U(1)$-symmetric 3HDM. This choice is based on the fact that this is the most general phase-independent scalar potential. In the discussion of different models in the earlier sections we took the $U(1) \times U(1)$-symmetric 3HDM as the starting point. However, from the discussion in Section~\ref{Sec:Different_DM_cases} we already know that different models will exhibit different physical properties.

The case of the CP4-symmetric 3HDM, see Section~\ref{Sec:3HDM-review}, shares one of the main aspects of the DM candidates considered here---there is  a pair of mass-degenerate DM candidates also in that model. In Ref.~\cite{Ivanov:2018srm} it was assumed that the thermal evolution of the inert sector takes place in the asymmetric regime. We shall instead assume a conventional freeze-out mechanism, analogous to that usually adopted for the IDM-like scenario, but with four (two pairs of mass-degenerate states) DM candidates. Another possibility, not considered here, would be to associate the mass-degenerate states with complex scalars. In our approach these two possibilities would yield identical results. We would also like to point out that the study of the $U(1) \times U(1)$-symmetric 3HDM is just for illustrative purposes. We consider it a toy model. 

\subsection{Mass-squared parameters}

There is a single vev, $(v,\,0,\,0)$, which does not spontaneously break the $U(1) \times U(1)$ symmetry. We shall focus on this case. Permutations of the vacuum configurations amount to re-labeling of the indices. Both the charged and the neutral mass-squared matrices are diagonal. The mass of the SM-like Higgs boson is given by:
\begin{equation}
m_h^2 = 2 \lambda_{1111} v^2.
\end{equation}
There is a charged and two neutral states coming from each of $h_2$ and $h_3$. The mass-squared parameters of the charged states are:
\begin{equation}
m_{h_i^+}^2 = \mu_{ii}^2 + \frac{1}{2} \lambda_{11ii}v^2,\text{ for }i=\{2,3\}.
\end{equation}
The index $i$ identifies which doublet the state is associated with. The neutral states of each doublet are mass-degenerate, see eq.~\eqref{Eq:MN2_U1_U1},
\begin{equation}
m_{\eta_i}^2 = m_{\chi_i}^2 \equiv m_{H_i}^2 = \mu_{ii}^2 + \frac{1}{2} \left( \lambda_{11ii} + \lambda_{1ii1} \right) v^2, \text{ for } i=\{2,\,3\}.
\end{equation}
We can express the neutral masses, $m_{H_i}$, in term of the charged ones, $m_{h_i^+}$, as:
\begin{equation}\label{Eq:U1U1_mHNi}
m_{H_i}^2 = m_{h_i^+}^2 + \frac{v^2}{2} \lambda_{1ii1}, \text{ for } i=\{2,\,3\}.
\end{equation} 
Since the DM candidates are assumed to be neutral, we require $\lambda_{1ii1}<0$.

Without loss of generality, it may be assumed that the lighter DM candidate is associated with $h_2$ and the heavier one with the $h_3$ doublet. However, there is still freedom in the ordering of the charged scalar masses. 

\subsection{The kinetic Lagrangian}

The kinetic Lagrangian of the $U(1) \times U(1)$ 3HDM resembles the one of the IDM, the only difference being that there is an additional copy of the vev-less doublet. Not considering the Goldstone bosons, the kinetic Lagrangian is:
\begin{subequations}
\begin{align}
\begin{split}
\mathcal{L}_{VVH} =& \left[ \frac{g}{2 \cos \theta_W}m_ZZ_\mu Z^\mu + g m_W W_\mu^+ W^{\mu-} \right] h,
 \end{split}\\
\begin{split}
\mathcal{L}_{VHH} =& -\frac{ g}{2 \cos \theta_W}Z^\mu \eta_i \overset\leftrightarrow{\partial_\mu} \chi_i  - \frac{g}{2}\bigg\{ i W_\mu^+ \left[ ih_i^- \overset\leftrightarrow{\partial^\mu} \chi_i + h_i^- \overset\leftrightarrow{\partial^\mu} \eta_i  \right] + \mathrm{h.c.} \bigg\}\\
& + \left[ i e A^\mu + \frac{i g}{2} \frac{\cos (2\theta_W)}{\cos \theta_W} Z^\mu \right]  h_i^+ \overset\leftrightarrow{\partial_\mu} h_i^-  ,
\end{split}\\
\begin{split}
\mathcal{L}_{VVHH} =& \left[ \frac{g^2}{8 \cos^2\theta_W}Z_\mu Z^\mu + \frac{g^2}{4} W_\mu^+ W^{\mu-} \right] \left( \eta_i^2 + \chi_i^2 + h^2\right)\\
& + \bigg\{ \left[ \frac{e g}{2} A^\mu W_\mu^+ - \frac{g^2}{2} \frac{\sin^2\theta_W}{\cos \theta_W}Z^\mu W_\mu^+ \right] \left[ \eta_i h_i^- + i \chi_i h_i^- \right] + \mathrm{h.c.} \bigg\}\\
&+ \left[ e^2 A_\mu A^\mu + e g \frac{\cos (2\theta_W)}{\cos \theta_W}A_\mu Z^\mu + \frac{g^2}{4} \frac{\cos^2(2\theta_W)}{\cos^2\theta_W}Z_\mu Z^\mu + \frac{g^2}{2} W_\mu^- W^{\mu +} \right] \\
&\quad~\times\left( h_i^-h_i^+ \right),
\end{split}
\end{align}
\end{subequations}
where the index $``i"$  is summed over the doublets $\{h_2,\, h_3\}$.

Note that in the $U(1) \times U(1)$ model there are two pairs of neutral mass-degenerate states. Thus, in the early Universe there were no processes $\eta_i \rightarrow Z \chi_i$ or $\chi_i \rightarrow Z \eta_i$. Furthermore, for masses $m_{H_i} < m_Z/2 $ the decay rate of $Z \rightarrow \eta_i \chi_i$ will be constrained by the invisible width of the $Z$ boson.

\subsection{The scalar interactions}\label{Sec:U1xU1_SSS_SSSS}

The scalar interactions (accounting for the symmetry factors, but not the overall ``$-i$") involving three scalars are:
\begin{subequations}
\begin{align}
g (\eta_i^2 h) = g(\chi_i^2h) ={}& v \left( \lambda_{11ii} + \lambda_{1ii1} \right),\label{Eq:U1xU1_portal_3}\\
g(h^3) ={}& 6 v \lambda_{1111},\\
g(h_i^+ h_i^- h) ={}& v \lambda_{11ii},
\end{align}
\end{subequations}
where the  $\{h_i^\pm,\, \eta_i,\, \chi_i\}$ fields are associated with the $h_2$ and $h_3$ doublets.

Interactions involving four scalars with $i \neq j$ are:
\begin{subequations}
\begin{align}
g(\eta_i^4) = g(\chi_i^4) ={}& 6 \lambda_{iiii},\\
g(h^4) ={}& 6 \lambda_{1111},\\
g(\eta_i^2 \chi_i^2) = g(\eta_i^2 h^+_i h^-_i) = g(\chi_i^2 h^+_i h_i^-) ={}& 2 \lambda_{iiii},\\
g(\eta_i^2 \eta_j^2) = g(\chi_i^2 \eta_j^2) = g(\chi_i^2 \chi_j^2) = g(h_i^+ h_i^- h_j^+ h_j^-) ={}& \lambda_{iijj} + \lambda_{ijji},\\
g(\eta_i^2 h^2) = g(\chi_i^2 h^2) ={}& \lambda_{11ii} + \lambda_{1ii1}, \label{Eq:U1xU1_portal_4}\\
g(\eta^2_i h_j^+ h_j^-) = g(\chi^2_i h_j^+ h_j^-) ={}& \lambda_{iijj},\\
g(h_i^+ h_i^- h_i^+ h_i^-) ={}& 4 \lambda_{iiii},\\
g(\eta_i \eta_j h_i^+ h_j^-) = g(\chi_i \chi_j h_i^+ h_j^-) ={}& \frac{1}{2} \lambda_{ijji},\\
g(h^2 h_i^+ h_i^-) ={}& \lambda_{11ii},\label{Eq:U1xU1_portal_Ch}\\
g(\chi_i \eta_j h^+_j h^-_i) ={}& \frac{i}{2} \lambda_{ijji},\\
g(\eta_i \chi_j h^+_j h^-_i) ={}& -\frac{i}{2} \lambda_{ijji}.   
\end{align}
\end{subequations}
All these couplings, apart from the last two, $\chi_i \eta_j h^+_j h^-_i$ and $\eta_i \chi_j h^+_j h^-_i$, are symmetric under the interchange $\eta_i \leftrightarrow \chi_i$. The last two, involving charged scalars, are related by complex conjugation. In addition, the scalar interactions for fields associated with $h_2$ are identical to those of $h_3$. Also, there are no trilinear vertices which could cause decays of the neutral states between $h_2 \leftrightarrow h_3$. The spectrum of DM particles is given by the $\{ \eta_2,\, \chi_2,\, \eta_3,\, \chi_3 \}$ fields.

It should be noted that there exists a vertex $h^+_i h^-_i h$. Such interactions will modify the Higgs di-photon decay rate, which should be compared against the experimental results.

\subsection{The Yukawa Lagrangian}

We would like to emphasise that the (unbroken) $U(1) \times U(1)$ symmetry acts only on the $h_2$ and $h_3$ doublets. There are several possibilities to construct the Yukawa Lagrangian. An intriguing possibility would be to couple one of the inert scalar doublets to additional (non-SM) fermions. In this case different fermions are assigned different $U(1)$ charges of the underlying $U(1) \times U(1)$ symmetry. We do not consider such cases, but rather restrict ourselves to the SM-like fermionic sector coupling to $h_1$. In this case the generic SM Yukawa Lagrangian is reproduced. As a result, we do not impose any direct constraints on the Yukawa couplings.

\subsection{Comparison with the IDM}

In the IDM, with $h_1$ and $h_2$ transforming under $\mathbb{Z}_2$ as $\{h_1 \rightarrow h_1,~h_2 \rightarrow -h_2\}$, the scalar potential is given by:
\begin{equation}
\begin{aligned}
V_\mathrm{IDM} ={}& \mu_{11}^2 h_{11} + \mu_{22}^2 h_{22} + \lambda_{1111} h_{11}^2 + \lambda_{2222} h_{22}^2 + \lambda_{1122} h_{11} h_{22} + \lambda_{1221} h_{12} h_{21}\\
&  + \lambda_{1212} \left( h_{12}^2 + h_{21}^2 \right),
\end{aligned}
\end{equation}
where all parameters are taken to be real. Even if one starts with a complex $\lambda_{1212}$ (known in the literature as $\lambda_5$) there is the possibility to re-phase the second doublet, $h_2$, and make $\lambda_{1212}$ real, since $h_2$ is vev-less.

For simplicity, let us split the $U(1) \times U(1)$ scalar potential into:
\begin{equation}\label{Eq:V_U1xU1_expl}
\begin{aligned}
V_{U(1) \times U(1)} &= \frac{1}{2}\mu_{11}^2 h_{11} + \mu_{22}^2 h_{22} + \frac{1}{2}\lambda_{1111} h_{11}^2 + \lambda_{2222} h_{22}^2 + \lambda_{1122} h_{11} h_{22} + \lambda_{1221} h_{12} h_{21}\\
& + \frac{1}{2}\mu_{11}^2 h_{11} + \mu_{33}^2 h_{33} + \frac{1}{2}\lambda_{1111} h_{11}^2 + \lambda_{3333} h_{33}^2 + \lambda_{1133} h_{11} h_{33} + \lambda_{1331} h_{13} h_{31}\\
& + \lambda_{2233} h_{22} h_{33} + \lambda_{2332} h_{23} h_{32}.
\end{aligned}
\end{equation}
so that it is easy to compare the two scalar potentials. Each of the first two lines corresponds to an IDM-like scalar potential, but with $\lambda_{1212}=0$. Removal of the $\lambda_{1212}$ term in the case of the 2HDM, increases the $\mathbb{Z}_2$ symmetry to a $U(1)$ symmetry. Also, this coupling generates the mass splitting of the neutral states associated with the inert sector,
\begin{equation}
\begin{aligned}
m_{H_1}^2 ={}& \mu_{22}^2 + \frac{1}{2}\left( \lambda_{1122} + \lambda_{1221} - 2\lambda_{1212} \right)v^2,\\
m_{H_2}^2 ={}& \mu_{22}^2 + \frac{1}{2}\left( \lambda_{1122} + \lambda_{1221} + 2\lambda_{1212} \right)v^2.
\end{aligned}
\end{equation}

Disregarding the couplings among the inert sector fields, there are fewer parameters in the $U(1) \times U(1)$ model (comparing either of the first two lines against the IDM), however it is still possible to adjust the relevant channels, like the Higgs portal, the strength of which could be taken as an input parameter. Some of the relevant contributions to the cosmological and astrophysical parameters will come from the Higgs portal, given by $\{\lambda_{11ii},\, \lambda_{1ii1}\}$, see Section~\ref{Sec:U1xU1_SSS_SSSS}.

Contrary to the IDM, in the $U(1) \times U(1)$ model there are several new interactions between the doublets of the inert sector, see the third line of eq.~\eqref{Eq:V_U1xU1_expl}. These couplings will be relevant to control annihilations of the heavier inert doublet into the lighter one. Since neither $\lambda_{2233}$ nor $\lambda_{2332}$ is a coupling involving $h_1$, these terms will only contribute to the quartic interactions among the dark-sector fields. 

It should also be mentioned that there is another closely related model---the Replicated Inert Doublet Model (RIDM). The RIDM was presented in Ref.~\cite{Haber:2018iwr} with the scalar potential given by
\begin{equation}
\begin{aligned}
V_\mathrm{RIDM} ={}& \mu_{11}^2 h_{11} + \mu_{22}^2 \left( h_{22} + h_{33} \right) + \lambda_{1111} h_{11}^2 + \lambda_{2222} \left( h_{22} + h_{33} \right)^2\\
& + \lambda_{1122} \left( h_{11} h_{22} + h_{11} h_{33} \right) + \lambda_{1221} \left( h_{12} h_{21} + h_{13} h_{31} \right)\\
& + \lambda_{1212} \left( h_{12}^2 + h_{21}^2 + h_{13}^2 + h_{31}^2 \right).
\end{aligned}
\end{equation}
In the RIDM, for the vacuum $(v,\,0,\,0)$ the mass-squared parameters are given by:
\begin{subequations}
\begin{align}
m_{h_2^\pm}^2 = m_{h_3^\pm}^2 ={}& \mu_{22}^2 + \frac{1}{2} \lambda_{1122} v^2,\\
m_{\eta_2}^2 = m_{\eta_3}^2 ={}& \mu_{22}^2 + \frac{1}{2} \left( \lambda_{1122} + \lambda_{1221} + 2 \lambda_{1212} \right)v^2,\\
m_{\chi_2}^2 = m_{\chi_3}^2 ={}& \mu_{22}^2 + \frac{1}{2} \left( \lambda_{1122} + \lambda_{1221} - 2 \lambda_{1212} \right)v^2.
\end{align}
\end{subequations}
In the case of $\lambda_{1212}=0$ there is a mass degeneracy present among all the neutral states associated with the inert doublets; those would be $h_2$ and $h_3$. The degeneracy patterns can then be explained by the global underlying $O(4)$ symmetry.

\subsection{Numerical studies}

The parameter space of the $U(1)\times U(1)$-symmetric model  was scanned, accounting for:
\begin{itemize}
\item An upper bound of 2 TeV was imposed on all additional scalars;
\item The DM candidates are associated with the neutral states of the $h_2$ and $h_3$ doublets. Since the form of the scalar potential is invariant under re-naming of $h_2$ and $h_3$, without loss of generality we take $m_{H_2} \leq m_{H_3}$;
\item LEP constraints. A conservative bound on the light charged scalars is usually assumed to be $m_{H_i^\pm} \geq 80\text{ GeV}$~\cite{Pierce:2007ut,Arbey:2017gmh}. We allow for a more generous value of $m_{H_i^\pm} \geq 70\text{ GeV}$. Moreover, measurements of the $Z$ widths at LEP forbid decays of the $Z$ boson into lighter scalars. In the $U(1) \times U(1)$ model this condition puts the bound $m_{H_i} > \frac{1}{2}m_Z$ on the neutral scalar masses. 
\item Perturbativity, Unitarity ($8\pi$), Stability conditions. We verified the stability conditions presented in Ref.~\cite{Faro:2019vcd} and later analysed in Ref.~\cite{Boto:2022uwv}. We checked the unitarity conditions, finding them consistent with those provided in Ref.~\cite{Bento:2022vsb};
\item Electroweak precision observables $S$ and $T$~\cite{Peskin:1990zt,Peskin:1991sw} are evaluated using techniques of Refs.~\cite{Grimus:2007if,Grimus:2008nb};
\item The Higgs boson invisible decay is limited by $\text{Br} \left(h \to \mathrm{invisible} \right)\leq 0.1$~\cite{CMS:2022dwd,ATLAS:2022vkf,ATLAS:2023tkt};
\item LHC search results implemented in $\mathsf{HiggsTools}$~\cite{Bahl:2022igd}, utilising $\mathsf{HiggsBounds}$~\cite{Bechtle:2008jh,Bechtle:2011sb,Bechtle:2012lvg,Bechtle:2013wla,Bechtle:2015pma,Bechtle:2020pkv,Bahl:2021yhk} and $\mathsf{HiggsSignals}$~\cite{Bechtle:2013xfa,Stal:2013hwa,Bechtle:2014ewa,Bechtle:2020uwn}; 
\item DM relic density is assumed to be consistent with the Planck measurements~\cite{Planck:2018vyg} and is evaluated as $\Omega_\mathrm{DM}h^2 = 2 \Omega_2 h^2 + 2 \Omega_3 h^2$, where $\Omega_i$ are contributions of the $h_2$ and $h_3$ doublets;
\item Direct DM searches based on XENONnT~\cite{XENON:2020kmp,XENON:2022ltv,XENON:2023cxc} and LUX-ZEPLIN~\cite{LZ:2021xov,LZ:2022lsv};
\item Indirect DM searches based on Ref.~\cite{Hess:2021cdp}.
\end{itemize}
We allow for a 3-$\sigma$ tolerance with the numerical values taken from the PDG~\cite{ParticleDataGroup:2022pth}. Also, we take into account a 10\% tolerance, due to theoretical and computational uncertainties. For the evaluation of the DM observables we utilise the generic $\mathtt{micrOMEGAs}$~\cite{Belanger:2004yn,Belanger:2006is,Belanger:2008sj,Belanger:2010pz,Belanger:2013oya,Belanger:2014vza,Barducci:2016pcb,Belanger:2018ccd,Alguero:2023zol} code (version 6.0.5). We accumulated a total of $10^6$ parameter points simultaneously satisfying all constraints.

Separate components of the relic density are shown in Figure~\ref{Fig:Omega_mHi} as functions of the DM mass. Consider a situation where there is a single doublet, the lighter one, $h_2$, contributing to the DM relic density. The relic density plot is then comparable to that of the IDM. With the inclusion of the other inert doublet, there is now the possibility to have quartic interactions between the two doublets, $g(h_2 h_2 h_3 h_3)=f(\lambda_{2233},\, \lambda_{2332})$. As a consequence of such quartic interactions between the two inert sectors we may expect annihilations of the heavier doublet into the lighter one. This can be concluded from noticing that there is a region of low relic density in panel (b), in contrast to panel (a), of Figure~\ref{Fig:Omega_mHi}.

\begin{figure}[htb]
\begin{center}
\includegraphics[scale=0.255]{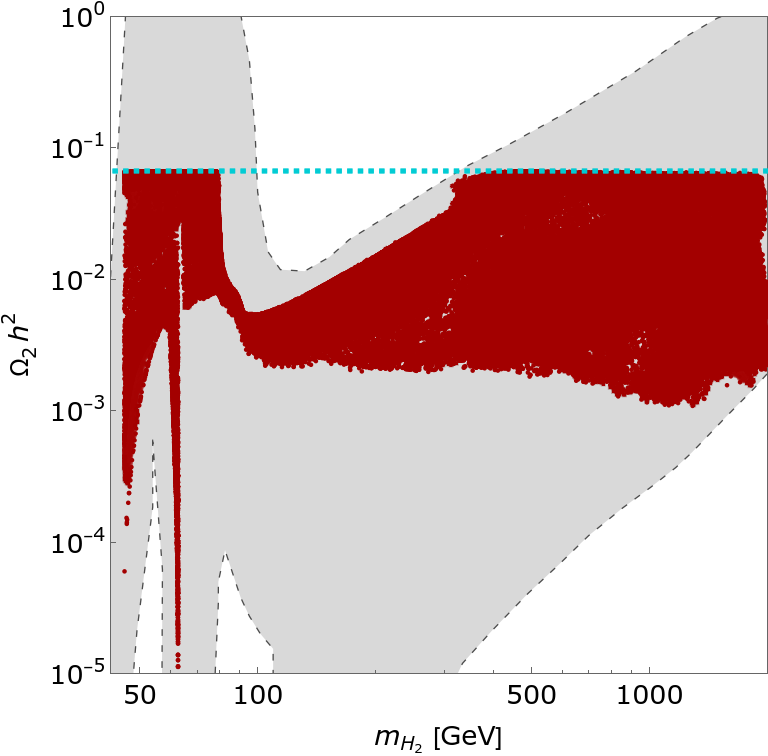}
\includegraphics[scale=0.255]{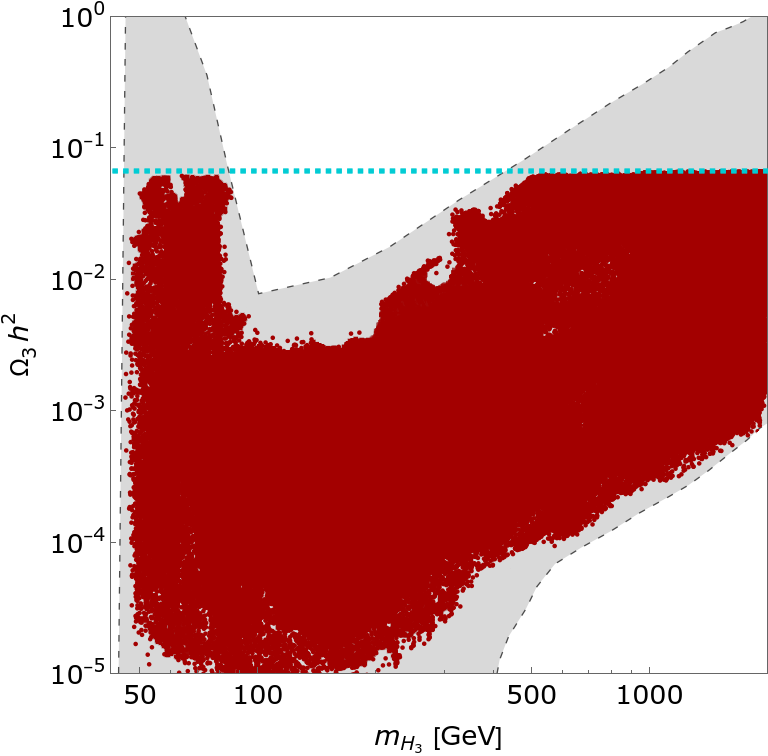}
\includegraphics[scale=0.255]{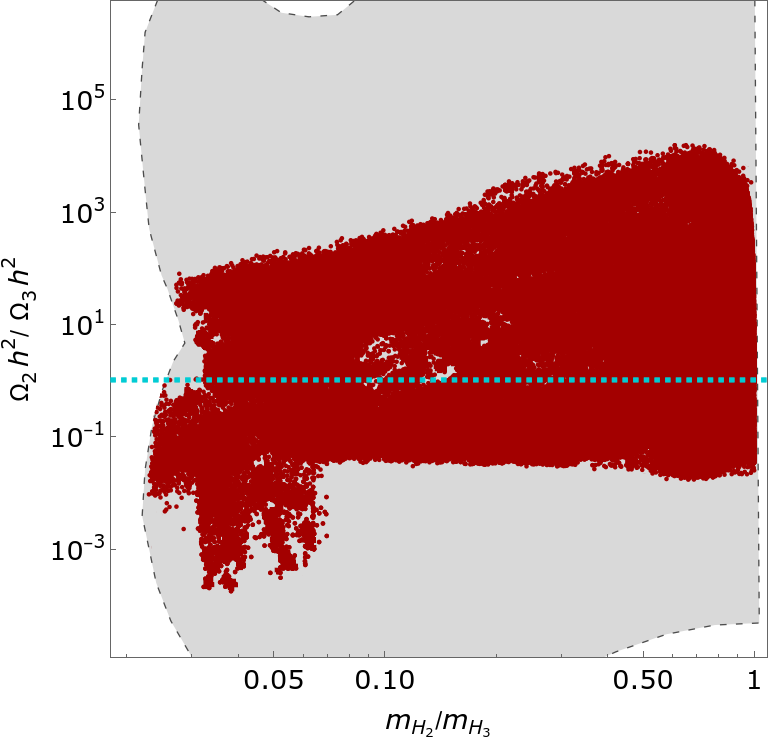}
\put (-370,132) {\small (a)}
\put (-220,132) {\small (b)}
\put (- 65,132) {\small (c)}
\end{center}
\vspace*{-3mm}
\caption{ Panels (a), (b): relic density as a function of mass for the two DM candidates. The  horizontal cyan line indicates the maximal relic density ($\Omega h^2 \approx 2 \Omega_i h^2$) which would be consistent with the Planck mission observations~\cite{Planck:2018vyg}. Panel (c): ratio of the relic density components, $\Omega_2 h^2$ and $\Omega_3 h^2$, as a function of the ratio of the DM masses. The cyan line indicates equal contribution of the two scalar sectors to the total relic density, \textit{i.e.}, $\Omega_2 h^2 = \Omega_3 h^2$. The grey blob represents parameter points satisfying only theoretical constraints. Red points correspond to parameters also satisfying experimental constraints. }
\label{Fig:Omega_mHi}
\end{figure}

Since two scalar sectors contribute to the total relic density, it is interesting to compare the individual contributions. Such a comparison is shown in panel (c) of Figure~\ref{Fig:Omega_mHi}. We also checked for cases where the two mass scales and the contributions to the total relic density are similar. For this check we required that both masses and relic densities of the two sectors are within 5\% of each other. This selection criterion led to the identification of two regions, $[50;\,80]$ GeV and $[300;\,2000]$ GeV. The lower bound of the high-mass region of the $U(1) \times U(1)$-symmetric 3HDM, $m_\mathrm{DM} \sim 300\text{ GeV}$, is significantly below the corresponding bound of the IDM, $m_\mathrm{DM} \sim 500\text{ GeV}$, while the discussed region of the $U(1) \times U(1)$-symmetric 3HDM is compatible with that of the truncated $\mathbb{Z}_2$-symmetric 3HDM, $m_\mathrm{DM} \sim 360\text{ GeV}$, which can be seen from Figure~\ref{Fig:DM_mass_ranges_different_models}.

A significant part of the parameter space is ruled out due to the direct DM detection constraints. In the low-mass region we need to keep the sum of the couplings $(\lambda_{11ii} + \lambda_{1ii1})$ small, this is the Higgs portal coupling, see eqs.~\eqref{Eq:U1xU1_portal_3} and \eqref{Eq:U1xU1_portal_4}. On the other hand, the $\lambda_{1ii1}$ coupling is responsible for the mass splitting between the neutral and the charged states, see eq.~\eqref{Eq:U1U1_mHNi}. In principle, the scalar potential couplings could be traded for the physical couplings and used as input parameters. The need to keep the portal couplings quite small is also due to the constraints coming from the Higgs boson invisible channel. 

Finally, we checked for the indirect DM detection constraints and found that those introduce no further significant cuts on the viable parameter space, while the direct DM detection constraints do. The main contribution to the Early Universe annihilation rate involves gauge bosons in the final state. While the current indirect DM searches~\cite{Hess:2021cdp} do not impose strict constraints, it should be noted that a considerable parameter space could be ruled out if no signals are observed in the near future and the experimental bounds from the indirect DM searches are improved by two orders of magnitude.

Let us summarise several aspects of the $U(1)\times U(1)$ model:
\begin{itemize}  \setlength\itemsep{3.2pt}
\item The main difference from the $\mathbb{Z}_2$ stabilisation of the DM candidate is the presence of pairs of mass-degenerate neutral states with opposite CP parities (or, equivalently, complex neutral scalars);
\item In comparison to the IDM and other 3HDMs, see Figure~\ref{Fig:DM_mass_ranges_different_models}, the model seems to be viable over the entire range of DM masses, see Figure~\ref{Fig:U1xU1_masses}. When the DM masses are in the intervals $m_{H_2} \in [100;\,300]$ GeV together with $m_{H_3} \in [100;\,450]$ GeV, the annihilation in the early Universe into gauge bosons is too fast, whereas annihilations into Higgs bosons could be controlled;
\item There are regions for the masses $m_{H_2}$ and $m_{H_3}$, $[50;\,80]$ GeV and $[300;\,2000]$ GeV respectively, with both masses close to each other where the contributions to the total relic density are comparable;
\item There are regions where the total relic density could be dominated by a single state. These are $[45;\,80]$ GeV and $[330;\,2000]$ GeV for $h_2$ and $[52;\,80]$ GeV and  $[470;\,2000]$~GeV for $h_3$;
\item There are instances where either the lighter or the heavier DM particle can account for the dominant relic density, see the previous bullet point and (c) in Figure~\ref{Fig:Omega_mHi}.
\end{itemize}

\begin{figure}[htb]
\begin{center}
\includegraphics[scale=0.3]{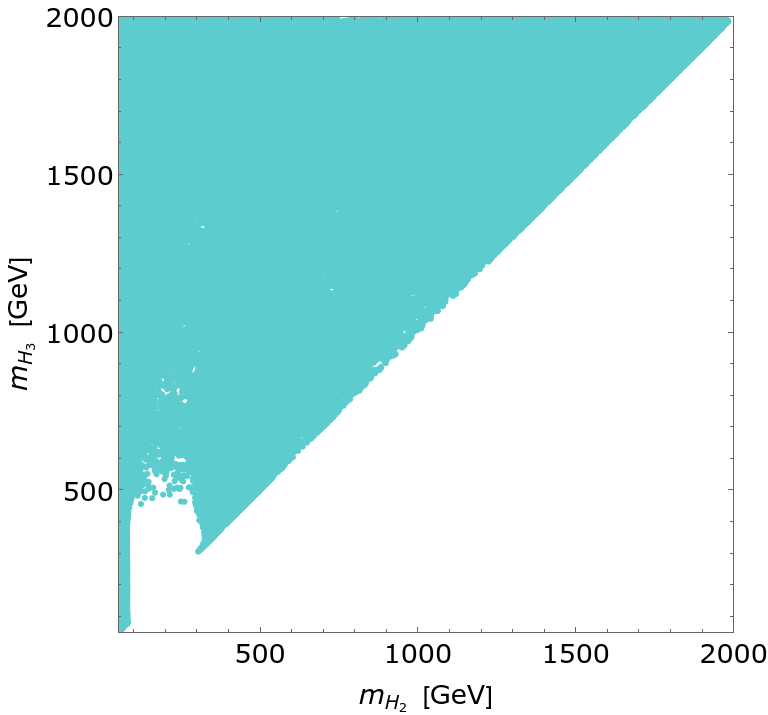}
\includegraphics[scale=0.3]{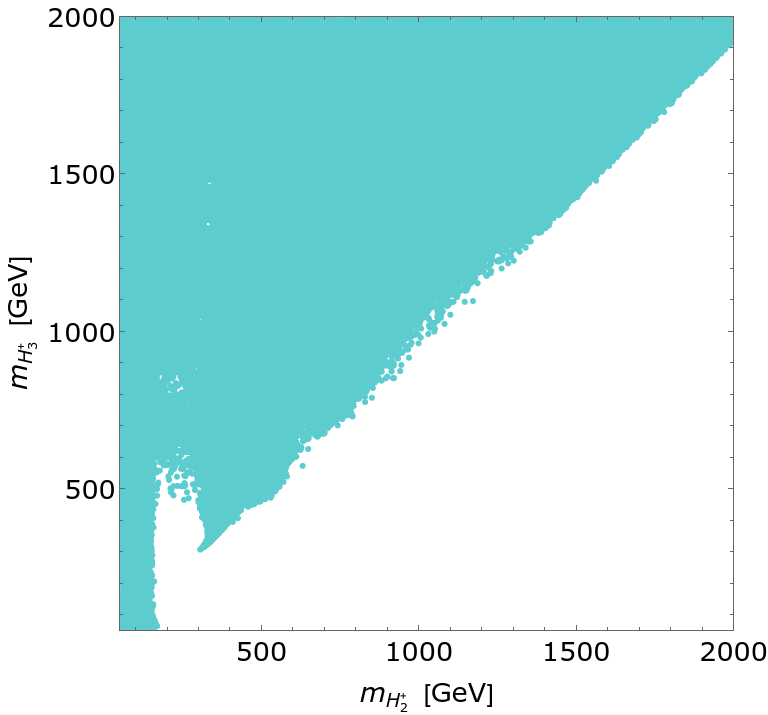}
\end{center}
\vspace*{-3mm}
\caption{ Mass scatter plots of parameters satisfying all constraints. Left: masses of the neutral inert sector. The labeling is $m_{H_2} \leq m_{H_3}$.  Right: masses of the charged inert sectors. While the bulk of the points satisfy $m_{H_2^\pm} \leq m_{H_3^\pm}$, this was not imposed by hand. However, there is a connection between charged and neutral masses, see eq.~\eqref{Eq:U1U1_mHNi}.}
\label{Fig:U1xU1_masses}
\end{figure}

As stated at the beginning of this section, our aim was not to scrutinise systematically the model but rather to consider it as a toy model. It may be of interest to consider different production channels at colliders since there are two dark doublets. While the model seems to be highly flexible, due to a broad parameter space available, it could well also be that the there are ``specific directions" in the scalar potential where couplings of the inert sectors are pairwise equal (through the interchange of the indices ``2" and ``3"), apart from bilinear terms,
\begin{equation}
\lambda_{2222} = \lambda_{3333},~ \lambda_{1122} = \lambda_{1133},~ \lambda_{1221} = \lambda_{1331},
\end{equation}
or that there is no tree-level Higgs portal to the neutral inert states, requiring
\begin{equation}
\lambda_{1221} = - \lambda_{1122},~ \lambda_{1331} = - \lambda_{1133},
\end{equation}
or else the two inert sectors do not ``see" each other,
\begin{equation}
\lambda_{2233} = \lambda_{2332}=0,
\end{equation}
and many more possibilities that may be viable.

Although this is a simple model, different implementations may lead to different observable signals. The DM candidates stabilised by continuous symmetries are mass degenerate, in contrast to those stabilised by $\mathbb{Z}_2$. However, it could be quite challenging to try to discriminate between the two classes of symmetries, if a candidate were found in a region compatible with both models. Normally, both the CP-even and the CP-odd components of the same sector equally contribute to the astrophysical and cosmological constraints. However, this does not imply that both components are equally abundant. There could be some mechanism that would lead to local variations in the relative abundance. It could also be that the components of the mass-degenerate states decouple at different temperatures.

This example illustrates the possibility of having potentially successful complex DM candidates. Beware of statements in the literature claiming that ``scenarios featuring a complex (non self-conjugate) dark matter field are excluded by cosmology and astrophysics alone", see Ref.~\cite{Arina:2023msd}, since these claims are made in model-dependent contexts.

\section{Summary}

We have identified 3HDMs that can accommodate DM, where the DM stability is due to some unbroken continuous symmetry, \textit{e.g.}, the $U(1)$ symmetry. Symmetries generally imply a reduction in the number of independent parameters. This can lead to some couplings or masses vanishing, or being related. Different symmetries may lead to the same form of the potential. Conversely, while two different symmetries may have the same number of independent couplings, see Table~\ref{Table:Diff_Cases} for a summary of the symmetries that were considered, they would have different structures in terms of the $SU(2)$ doublets. Experimentally, they could, in principle, be distinguished by having different patterns of couplings, see Table~\ref{Table:Couplings_tri_quart}, or by having different mass relations among the physical states, see Table~\ref{Table:Mass_degen_patterns}. The latter property is a generic feature of the models considered here. As it was shown in Figure~\ref{Fig:Symmetry_breaking}, the $U(1)$ symmetry can be implemented in many ways. Some of the symmetries $\{U(1) \times D_4,\, O(2) \times U(1),\, \left[ U(1) \times U(1) \right] \rtimes S_3\}$ have not been explicitly considered in the literature. 

In some cases, considering complex vacua with a vanishing vev, \textit{e.g.}, $(\hat v_1 e^{i\sigma},\, \hat v_2,\,0)$, where the potential has real (as required by the underlying symmetry) phase-sensitive terms, the minimisation conditions force such terms to vanish. As a result, a higher symmetry is reached, and the phase appearing in the vevs becomes redundant since it can be rotated away without appearing in the potential. However, had we started with real vevs, there would be no need to remove such terms from the potential. We considered those cases as inconsistent and did not put them in the tables. This can be illustrated by the absence in Table~\ref{Table:U1U1S2_Cases_1} of the vacuum $(\hat v_1 e^{i\sigma},\, \hat v_2,\,0)$ for the symmetry $[ SO(2) \rtimes \mathbb{Z}_2 ] \times U(1)$, where the solution of the minimisation conditions would increase the symmetry to that of $U(2)$ and the phase would become redundant. However, we can consistently start with $\sigma =0$ and get a completely different case without increasing the symmetry. The previous case is presented in Table~\ref{Table:U2_Cases}, in the context of the $U(2)$ symmetry. 

{{\renewcommand{\arraystretch}{1.2}
\setlength\LTcapwidth{\linewidth}
\begin{center}
\begin{longtable}[htb]{|c|c|c|l|} 
\caption{ Considered models. In the third column the numbers of independent bilinear plus quartic couplings are provided. The notation of $V_0$, see eq.~\eqref{V_U1xU1}, stands for allowed couplings $\{\mu_{ii}^2,~\lambda_{iiii},~ \lambda_{iijj},~ \lambda_{ijji}\}$, and $V_{O(2) \times U(1)}$, see eq.~\eqref{Eq:V_U1_U1_S2}, requiring $\{\mu_{11}^2 = \mu_{22}^2,~ \mu_{33}^2,~ \lambda_{1111} = \lambda_{2222},~\lambda_{3333},~\lambda_{1122},~\lambda_{1133} = \lambda_{2233},~\lambda_{1221},~\lambda_{1331} = \lambda_{2332}\}$. We recall that $\Lambda=\frac{1}{2}\left( 2 \lambda_{1111} - \lambda_{1221} - \lambda_{1122} \right)$. }
\label{Table:Diff_Cases} \\ \hline\hline
\begin{tabular}[l]{@{}c@{}} Underlying \\ symmetry \end{tabular} & Reference & \begin{tabular}[l]{@{}c@{}} Indep. \\ couplings \end{tabular} & \begin{tabular}[l]{@{}c@{}} Allowed couplings and \\ necessary relations \end{tabular} \\ \hline\hline
$\mathbb{Z}_2$ & Table~\ref{Table:Z2_cases}  & 4 + 17 & \begin{tabular}[l]{@{}l@{}} $V_0,~\mu_{12}^2,~\lambda_{ijij},~\lambda_{1112},~\lambda_{1222},~\lambda_{1233},~$ \\$\lambda_{1323},~\lambda_{1332}$ \end{tabular} \\ \hline 
$U(1)_2$ & Table~\ref{Table:U12_Cases}  &  4 + 14 & \begin{tabular}[l]{@{}l@{}} $V_0,~\mu_{12}^2,~\lambda_{1212},~\lambda_{1112},~\lambda_{1222},$\\$\lambda_{1233},~\lambda_{1332}$\end{tabular}  \\ \hline
$\mathbb{Z}_3$ & Table~\ref{Table:Z3_Cases}  &  3 + 12 & \begin{tabular}[l]{@{}l@{}} $V_0,~ \lambda_{1323},~\lambda_{1213},~\lambda_{1232}$ \end{tabular} \\ \hline
$U(1)_1$ & Table~\ref{Table:U11_Cases}  &  3 + 10 & $V_0,~\lambda_{1323}$\\ \hline
$U(1) \times \mathbb{Z}_2$ & Table~\ref{Table:U1Z2_Cases}  &  3 + 10 & $V_0,~\lambda_{1212}$\\ \hline
$U(1) \times U(1)$ & Table~\ref{Table:U1_U1_Cases}  &  3 + 9 & $V_0$ \\ \hline
$O(2)_{\scriptscriptstyle[SO(2) \rtimes \mathbb{Z}_2]}$ &   Table~\ref{Table:O2_Cases_1}  &  2 + 7 & \begin{tabular}[l]{@{}l@{}} $V_{O(2) \times U(1)},~\lambda_{1212}= \Lambda,~\lambda_{1313}=\lambda_{2323} $ \end{tabular}\\ \hline
$O(2)_{\scriptscriptstyle[U(1) \rtimes \mathbb{Z}_2]}$ & Table~\ref{Table:O2_Cases_2}  &  2 + 7 & \begin{tabular}[l]{@{}l@{}} $V_{O(2) \times U(1)},~\lambda_{1323} $ \end{tabular}\\ \hline 
$U(1) \times D_4$ & Table~\ref{Table:U1Z2_S2_Cases}  &  2 + 7 & \begin{tabular}[l]{@{}l@{}} $V_{O(2) \times U(1)},~ \lambda_{1212}$ \end{tabular}\\ \hline
$O(2)_{\scriptscriptstyle [SO(2) \rtimes \mathbb{Z}_2]} \times U(1)$\hspace*{-0.5pt} & Table~\ref{Table:U1U1S2_Cases_1}  &  2 + 6 & \begin{tabular}[l]{@{}l@{}} $V_{O(2) \times U(1)},~\lambda_{1212} = \Lambda$ \end{tabular} \\  \hline 
$O(2)_{\scriptscriptstyle [U(1) \rtimes \mathbb{Z}_2]} \times U(1)$\hspace*{-0.5pt} & Table~\ref{Table:U1U1S2_Cases_2}  &  2 + 6 & $V_{O(2) \times U(1)}$ \\  \hline 
$U(2)$ & Table~\ref{Table:U2_Cases}  &  2 + 5 & \begin{tabular}[l]{@{}l@{}} $\mu_{11}^2 = \mu_{22}^2,~ \mu_{33}^2,~ \lambda_{1111} = \lambda_{2222},~\lambda_{3333},$ \\ $\Lambda=0,~\lambda_{1133} = \lambda_{2233},~\lambda_{1221}$\\$\lambda_{1331} = \lambda_{2332}$ \end{tabular}\\ \hline
$\left[ U(1) \times U(1) \right] \rtimes S_3$ & Table~\ref{Table:U11S3_Cases}  &  1 + 3 & \begin{tabular}[l]{@{}l@{}}$\mu_{11}^2 = \mu_{22}^2 = \mu_{33}^2,~\lambda_{1111} = \lambda_{2222} = \lambda_{3333},$ \\ $\lambda_{1122} = \lambda_{1133} = \lambda_{2233},$ \\$\lambda_{1221} = \lambda_{1331} = \lambda_{2332}$ \end{tabular}\\ \hline
$SO(3)$ & Table~\ref{Table:SO3_Cases}  &  1 + 3 & \begin{tabular}[l]{@{}l@{}}$\mu_{11}^2 = \mu_{22}^2 = \mu_{33}^2,~\lambda_{1111} = \lambda_{2222} = \lambda_{3333},$ \\ $\lambda_{1122} = \lambda_{1133} = \lambda_{2233},$\\$\lambda_{1221} = \lambda_{1331} = \lambda_{2332}, ~\lambda_{ijij} = \Lambda$\end{tabular}\\ \hline
$SU(3)$ & Table~\ref{Table:SU3_Cases}  &  1 + 2 & \begin{tabular}[l]{@{}l@{}}$\mu_{11}^2 = \mu_{22}^2 = \mu_{33}^2,~\lambda_{1111} = \lambda_{2222} = \lambda_{3333},$ \\ $\lambda_{1122} = \lambda_{1133} = \lambda_{2233},$ \\ $\lambda_{ijji} = 2 \lambda_{1111} - \lambda_{1122}$\end{tabular}\\ \hline \hline
\end{longtable}
\end{center}}

If one requires that the vacuum should conserve the underlying symmetry, the number of possible DM models is reduced by three, see the comment at the beginning of Section~\ref{Sec:Different_DM_cases}. However, technically speaking, the cases with SSB could also be of interest. Such implementations might still be stabilised by a remnant symmetry, where the presence of a DM candidate could be hidden in a basis where none of the vevs vanishes.

We chose to analyse the parameter space of the $U(1) \times U(1)$-symmetric potential, since it is the most general real scalar 3HDM potential. Assuming the mainstream freeze-out scenario, with hypothetical weakly interacting massive particles, we discussed the parameter space compatible with constraints coming from Particle Physics, Astrophysics and Cosmology. In this model there are two DM mass scales, associated with two different inert doublets. While there seems to be some preference for the lighter doublet to give a dominant contribution to the relic density, we note that this is due to the allowed decay channels of the heavier states into the lighter ones, which, in principle, could be tweaked in a way to suppress such processes. Inside the region we scanned, 45--2000 GeV, we found that only the region where \textit{both} neutral dark scalars are in the mass range $\mathcal{O}(100)-\mathcal{O}(300)$~GeV is excluded. This analysis provides some insight into how to discriminate models with continuous symmetries, specifically $U(1)$, from discrete symmetries within the context of 3HDMs.

\section*{Acknowledgements}

It is a pleasure to thank Genevi\`eve Belanger, Howard Haber, Igor Ivanov and Odd Magne \O greid for discussions at different stages of this work. We thank Ivo de Medeiros Varzielas and Igor Ivanov, as well as Neda Darvishi and Apostolos Pilaftsis for clarifying discussions of their papers~\cite{deMedeirosVarzielas:2019rrp,Darvishi:2019dbh}, as well as Alexander Pukhov for discussions about $\mathtt{micrOMEGAs}$. 

The work of AK and MNR was partially supported by Funda\c c\~ ao para a  Ci\^ encia e a Tecnologia (FCT, Portugal) through the  projects CFTP-FCT Unit UIDB/00777/2020 and UIDP/00777/2020, CERN/FIS-PAR/0008/2019 and CERN/FIS-PAR/0002/2021, which are partially funded through POCTI (FEDER), COMPETE, QREN and  EU. Furthermore, the work of AK has been supported by the FCT PhD fellowship with reference UI/BD/150735/2020. PO~is supported in part by the Research Council of Norway. AK would like to thank Laboratoire d'Annecy-le-Vieux de Physique Théorique for hospitality. We also thank the University of Bergen and CFTP/IST/University of Lisbon as well as the CERN Theory Department, where collaboration visits took place.

\appendix

\section{\texorpdfstring{\boldmath$\mathbb{Z}_2$}{Z2}-symmetric 3HDM}\label{App:Pot_Z2}

Without loss of generality one can assign the following $\mathbb{Z}_2$ charges to the doublets:
\begin{equation}\label{Eq:Z2_charges}
\mathbb{Z}_2: ~ h_1\to h_1,~~ h_2\to h_2,~~ h_3\to -h_3.
\end{equation}
The phase-sensitive part of the scalar potential is given by\footnote{
In Ref.~\cite{Darvishi:2019dbh} there is a coupling missing in the $\mathbb{Z}_2$ classification. For $\mathbb{Z}_2$ it is $\lambda_{1223}$ with the $\mathbb{Z}_2$ charges given by $[(+)(-)(-)(+)]$ and for $\mathbb{Z}_2^\prime$ the relevant coupling is $\lambda_{1321}$ with charges $[(-)(+)(+)(-)]$, as clarified with the authors.}:
\begin{equation}\label{Eq:Z2_Vph}
\begin{split}
V_{\mathbb{Z}_2}^\text{ph} ={}& \mu_{12}^2 h_{12} + \sum_{i<j} \lambda_{ijij} h_{ij}^2 + \lambda_{1112} h_{11}h_{12} + \lambda_{1222} h_{12}h_{22}\\
&  + \lambda_{1233} h_{12}h_{33} + \lambda_{1323} h_{13}h_{23} + \lambda_{1332} h_{13}h_{32} + \mathrm{h.c.}
\end{split}
\end{equation} 

The first-order derivatives are provided in Appendix~\ref{App:Derivatives}. In order to cover the whole parameter space of a model, containing at least a single vanishing vev, specific vacuum directions should be analysed, possibly fixing the ratio of two vevs if that would minimise the scalar potential in a unique way. Since the scalar potential is symmetric under the interchange of $h_1 \leftrightarrow h_2$, some permutations of vevs result in identical physical models. One of the main conditions to stabilise a DM candidate is to prevent any mixing between the DM and the active scalars. This can be achieved by imposing an underlying symmetry.

\subsection{One vanishing vev}

\textbf{Case of \boldmath$(\hat v_1,\, \hat v_2,\,0)$}\labeltext{Case of $(\hat v_1,\, \hat v_2,\,0)$}{App:Pot_Z2_v1_v2_0}

We start by checking the structure of the mass eigenstates considering that one of the vevs vanishes, the other vevs are real. Requiring that the vacuum does not break the underlying symmetry imposes $\left\langle h_3 \right\rangle=0$. We shall assume a case of explicit CP violation. The minimisation conditions are given by:
\begin{subequations}
\begin{align}
\begin{split}
\mu_{11}^2 & = - \Bigg( 2 (\mu_{12}^2)^\mathrm{R} \hat v_2 + 2 \lambda_{1111} \hat v_1^3 + \lambda_{1122} \hat v_1 \hat v_2^2 + \lambda_{1221} \hat v_1 \hat v_2^2\\
&\hspace{50pt} + 3  \lambda_{1112}^\mathrm{R} \hat v_1^2 \hat v_2 +  \lambda_{1222}^\mathrm{R} \hat v_2^3 + 2 \lambda_{1212}^\mathrm{R} \hat v_1 \hat v_2^2 \Bigg)\frac{1}{2 \hat v_1},
\end{split}\\
\begin{split}
\mu_{22}^2 & = - \Bigg( 2 (\mu_{12}^2)^\mathrm{R} \hat v_1 + 2 \lambda_{2222} \hat v_2^3 + \lambda_{1122} \hat v_1^2 \hat v_2 + \lambda_{1221} \hat v_1^2 \hat v_2\\
&\hspace{50pt} + \lambda_{1112}^\mathrm{R} \hat v_1^3 +  3 \lambda_{1222}^\mathrm{R} \hat v_1 \hat v_2^2 + 2 \lambda_{1212}^\mathrm{R} \hat v_1^2 \hat v_2 \Bigg)\frac{1}{2 \hat v_2},
\end{split}\\
(\mu_{12}^2)^\mathrm{I} & = -\frac{1}{2} \Bigg( \lambda_{1112}^\mathrm{I} \hat v_1^2 + 2 \lambda_{1212}^\mathrm{I} \hat v_1 \hat v_2 + \lambda_{1222}^\mathrm{I} \hat v_2^2 \Bigg).
\end{align}
\end{subequations}

The charged mass-squared matrix is:
\begin{equation}
\mathcal{M}_\mathrm{Ch}^2 = \begin{pmatrix}
(\mathcal{M}_\mathrm{Ch}^2)_{11} & (\mathcal{M}_\mathrm{Ch}^2)_{11} \hat v_{1/2} & 0 \\
(\mathcal{M}_\mathrm{Ch}^2)_{11} \hat v_{1/2} & (\mathcal{M}_\mathrm{Ch}^2)_{11} \hat v_{1/2}^2 & 0 \\
0 & 0 & (\mathcal{M}_\mathrm{Ch}^2)_{33}   
\end{pmatrix},
\end{equation}
where
\begin{subequations}
\begin{align}
(\mathcal{M}_\mathrm{Ch}^2)_{11} ={}& -  \left[ 2(\mu_{12}^2)^\mathrm{R} + \lambda_{1112}^\mathrm{R} \hat v_1^2 + \left( 2 \lambda_{1212}^\mathrm{R} + \lambda_{1221} \right)\hat v_1 \hat v_2 + \lambda_{1222}^\mathrm{R}\hat v_2^2 \right] \frac{\hat v_2}{2 \hat v_1},\\
(\mathcal{M}_\mathrm{Ch}^2)_{33} ={}& \mu_{33}^2 + \frac{1}{2}\left( \lambda_{1133} \hat v_1^2 + 2 \lambda_{1233}^\mathrm{R} \hat v_1 \hat v_2 + \lambda_{2233} \hat v_2^2  \right).
\end{align}
\end{subequations}
The neutral mass-squared matrix in the basis $\{\eta_1,\, \eta_2,\, \chi_1,\, \chi_2\}{-}\{\eta_3,\,\chi_3\}$ is given by:
\begin{equation}
\begin{multlined}[c]
\mathcal{M}_\mathrm{N}^2 = { \left(\begin{matrix}
(\mathcal{M}_\mathrm{N}^2)_{11} & (\mathcal{M}_\mathrm{N}^2)_{12} & (\mathcal{M}_\mathrm{N}^2)_{13}\\
(\mathcal{M}_\mathrm{N}^2)_{12} & (\mathcal{M}_\mathrm{N}^2)_{22} & (\mathcal{M}_\mathrm{N}^2)_{23}\\
(\mathcal{M}_\mathrm{N}^2)_{13} & (\mathcal{M}_\mathrm{N}^2)_{23} & (\mathcal{M}_\mathrm{N}^2)_{33}\\
-(\mathcal{M}_\mathrm{N}^2)_{13} \hat v_{1/2} & - (\mathcal{M}_\mathrm{N}^2)_{23} \hat v_{1/2} & (\mathcal{M}_\mathrm{N}^2)_{34}\\
0 & 0 & 0 &\\
0 & 0 & 0 &
\end{matrix}\right.}{ \left.\begin{matrix}
 -(\mathcal{M}_\mathrm{N}^2)_{13} \hat v_{1/2} & 0 & 0\\
-(\mathcal{M}_\mathrm{N}^2)_{23} \hat v_{1/2} & 0 & 0\\
(\mathcal{M}_\mathrm{N}^2)_{34} & 0 & 0\\
(\mathcal{M}_\mathrm{N}^2)_{44} & 0 & 0\\
0 & (\mathcal{M}_\mathrm{N}^2)_{55} & (\mathcal{M}_\mathrm{N}^2)_{56}\\
0 & (\mathcal{M}_\mathrm{N}^2)_{56} & (\mathcal{M}_\mathrm{N}^2)_{66}
\end{matrix}\right)}, 
\end{multlined}
\end{equation}
where
\begin{subequations}
\begin{align}
(\mathcal{M}_\mathrm{N}^2)_{11}={}& -\frac{1}{2 \hat v_1} \left( 2 (\mu_{12}^2)^\mathrm{R} \hat v_2  -4 \lambda_{1111} \hat v_1^3 - 3 \lambda_{1112}^\mathrm{R} \hat v_1^2 \hat v_2 + \lambda_{1222}^\mathrm{R} \hat v_2^3   \right),\\
(\mathcal{M}_\mathrm{N}^2)_{12}={}& (\mu_{12}^2)^\mathrm{R} + \frac{3}{2}\left( \lambda_{1112}^\mathrm{R} \hat v_1^2 + \lambda_{1222}^\mathrm{R} \hat v_2^2 \right) + \left( \lambda_{1122} + \lambda_{1221} + 2 \lambda_{1212}^\mathrm{R} \right)\hat v_1 \hat v_2,\\
(\mathcal{M}_\mathrm{N}^2)_{13}={}& \left( \lambda_{1112}^\mathrm{I} \hat v_1 + \lambda_{1212}^\mathrm{I} \hat v_2 \right) \hat v_2,\\
(\mathcal{M}_\mathrm{N}^2)_{22}={}& -\frac{1}{2 \hat v_2} \left( 2 (\mu_{12}^2)^\mathrm{R} \hat v_1  -4 \lambda_{2222} \hat v_2^3 - 3 \lambda_{1222}^\mathrm{R} \hat v_1 \hat v_2^2 + \lambda_{1112}^\mathrm{R} \hat v_1^3   \right),\\
(\mathcal{M}_\mathrm{N}^2)_{23}={}& \left( \lambda_{1212}^\mathrm{I} \hat v_1 + \lambda_{1222}^\mathrm{I} \hat v_2 \right) \hat v_2,\\
(\mathcal{M}_\mathrm{N}^2)_{33}={}& -\frac{\hat v_2}{2 \hat v_1} \left( 2 (\mu_{12}^2)^\mathrm{R} + \lambda_{1112}^\mathrm{R} \hat v_1^2 + 4 \lambda_{1212}^\mathrm{R} \hat v_1 \hat v_2 + \lambda_{1222}^\mathrm{R} \hat v_2^2\right),\\
(\mathcal{M}_\mathrm{N}^2)_{34}={}& \frac{1}{2}\left( 2(\mu_{12}^2)^\mathrm{R} + \lambda_{1112}^\mathrm{R} \hat v_1^2 + 4 \lambda_{1212}^\mathrm{R} \hat v_1 \hat v_2 + \lambda_{1222}^\mathrm{R} \hat v_2^2 \right),\\
(\mathcal{M}_\mathrm{N}^2)_{44}={}& -\frac{\hat v_1}{2 \hat v_2} \left( 2 (\mu_{12}^2)^\mathrm{R} + \lambda_{1112}^\mathrm{R} \hat v_1^2 + 4 \lambda_{1212}^\mathrm{R} \hat v_1 \hat v_2 + \lambda_{1222}^\mathrm{R} \hat v_2^2\right),\\
\begin{split}
(\mathcal{M}_\mathrm{N}^2)_{55} = {}&  \mu_{33}^2 + \frac{1}{2}\left( \lambda_{1133} + 2 \lambda_{1313}^\mathrm{R} + \lambda_{1331}\right)\hat v_1^2  + \frac{1}{2}\left( \lambda_{2233} + 2 \lambda_{2323}^\mathrm{R} + \lambda_{2332}\right)\hat v_2^2\\
& + \left( \lambda_{1233}^\mathrm{R} + \lambda_{1323}^\mathrm{R} + \lambda_{1332}^\mathrm{R}\right)\hat v_1 \hat v_2 ,
\end{split}\\
(\mathcal{M}_\mathrm{N}^2)_{56} = {}&  -\lambda_{1313}^\mathrm{I} \hat v_1^2 -\left( \lambda_{1323}^\mathrm{I} \hat v_1 + \lambda_{2323}^\mathrm{I} \hat v_2 \right)\hat v_2,\\
\begin{split}
(\mathcal{M}_\mathrm{N}^2)_{66} = {}&  \mu_{33}^2 + \frac{1}{2}\left( \lambda_{1133} - 2 \lambda_{1313}^\mathrm{R} + \lambda_{1331}\right)\hat v_1^2  + \frac{1}{2}\left( \lambda_{2233} - 2 \lambda_{2323}^\mathrm{R} + \lambda_{2332}\right)\hat v_2^2\\
& + \left( \lambda_{1233}^\mathrm{R} - \lambda_{1323}^\mathrm{R} + \lambda_{1332}^\mathrm{R}\right)\hat v_1 \hat v_2.
\end{split}
\end{align}
\end{subequations}

The available parameter space of this case, in terms of the degrees of freedom coming from the couplings, ensures that there are no massless or mass-degenerate states. 

\bigskip\textbf{Case of \boldmath$(\hat v_1 e^{i \sigma},\, \hat v_2,\,0)$ and real couplings}

Next, we consider the case of spontaneous CP violation, when all  couplings are real. In this case the minimisation conditions are:
\begin{subequations}
\begin{align}
\begin{split}
\mu_{11}^2 ={}& -\lambda_{1111} \hat v_1^2 - \lambda_{1112}^\mathrm{R} \cos \sigma \hat v_1 \hat v_2   -\frac{1}{2}\left( \lambda_{1122} - 2 \lambda_{1212}^\mathrm{R} + \lambda_{1221} \right) \hat v_2^2,
\end{split}\\
\begin{split}
\mu_{22}^2 ={}& -\lambda_{2222} \hat v_2^2 - \lambda_{1222}^\mathrm{R} \cos \sigma \hat v_1 \hat v_2   -\frac{1}{2}\left( \lambda_{1122} - 2 \lambda_{1212}^\mathrm{R} + \lambda_{1221} \right) \hat v_1^2,
\end{split}\\
(\mu_{12}^2)^\mathrm{R} ={}& -\frac{1}{2}\lambda_{1112}^\mathrm{R} \hat v_1^2 - 2\lambda_{1212}^\mathrm{R} \cos \sigma \hat v_1 \hat v_2 -\frac{1}{2} \lambda_{1222}^\mathrm{R} \hat v_2^2,
\end{align}
\end{subequations}

The charged mass-squared matrix is:
\begin{equation}
\mathcal{M}_\mathrm{Ch}^2 = \begin{pmatrix}
(\mathcal{M}_\mathrm{Ch}^2)_{11} & -(\mathcal{M}_\mathrm{Ch}^2)_{11} \hat v_{1/2}& 0 \\
-(\mathcal{M}_\mathrm{Ch}^2)_{11} \hat v_{1/2} & (\mathcal{M}_\mathrm{Ch}^2)_{11} \hat v_{1/2}^2 & 0 \\
0 & 0 & (\mathcal{M}_\mathrm{Ch}^2)_{33}   
\end{pmatrix},
\end{equation}
where
\begin{subequations}
\begin{align}
(\mathcal{M}_\mathrm{Ch}^2)_{11} ={}& \frac{1}{2}\left( 2 \lambda_{1212}^\mathrm{R} - \lambda_{1221} \right) \hat v_2^2,\\
(\mathcal{M}_\mathrm{Ch}^2)_{33} ={}&\mu_{33}^2 + \frac{1}{2}\lambda_{1133}\hat v_1^2 + \lambda_{1233}^\mathrm{R} \cos \sigma \hat v_1 \hat v_2 + \frac{1}{2} \lambda_{2233} \hat v_2^2.
\end{align}
\end{subequations}

The neutral mass-squared matrix in the basis $\{\eta_1,\, \eta_2,\, \chi_1,\, \chi_2\}{-}\{\eta_3,\,\chi_3\}$ is given by:
\begin{equation}
\begin{multlined}[c]
\mathcal{M}_\mathrm{N}^2 = { \left(\begin{matrix}
(\mathcal{M}_\mathrm{N}^2)_{11} & (\mathcal{M}_\mathrm{N}^2)_{12} & (\mathcal{M}_\mathrm{N}^2)_{13}\\
(\mathcal{M}_\mathrm{N}^2)_{12} & (\mathcal{M}_\mathrm{N}^2)_{22} & (\mathcal{M}_\mathrm{N}^2)_{23}\\
(\mathcal{M}_\mathrm{N}^2)_{13} & (\mathcal{M}_\mathrm{N}^2)_{23} & (\mathcal{M}_\mathrm{N}^2)_{33}\\
-(\mathcal{M}_\mathrm{N}^2)_{13} \hat v_{1/2} & -(\mathcal{M}_\mathrm{N}^2)_{23} \hat v_{1/2} & -(\mathcal{M}_\mathrm{N}^2)_{33} \hat v_{1/2}\\
0 & 0 & 0\\
0 & 0 & 0
\end{matrix}\right.}{ \left.\begin{matrix}
-(\mathcal{M}_\mathrm{N}^2)_{13} \hat v_{1/2} & 0 & 0\\
-(\mathcal{M}_\mathrm{N}^2)_{23} \hat v_{1/2} & 0 & 0\\
-(\mathcal{M}_\mathrm{N}^2)_{33} \hat v_{1/2} & 0 & 0\\
(\mathcal{M}_\mathrm{N}^2)_{33} \hat v_{1/2}^2 & 0 & 0\\
0 & (\mathcal{M}_\mathrm{N}^2)_{55} & (\mathcal{M}_\mathrm{N}^2)_{56}\\
0 & (\mathcal{M}_\mathrm{N}^2)_{56} & (\mathcal{M}_\mathrm{N}^2)_{66}
\end{matrix}\right)}, 
\end{multlined}
\end{equation}
where
\begin{subequations}
\begin{align}
(\mathcal{M}_\mathrm{N}^2)_{11}={}& 2 \left( \lambda_{1111}  \hat v_1^2 + \lambda_{1112}^\mathrm{R} \cos \sigma \hat v_1 \hat v_2 + \lambda_{1212}^\mathrm{R} \cos^2 \sigma \hat v_2^2 \right),\\
(\mathcal{M}_\mathrm{N}^2)_{12}={}& \lambda_{1112}^\mathrm{R} \cos \sigma \hat v_1^2 + \left( \lambda_{1122} + \lambda_{1221} -2 \lambda_{1212}^\mathrm{R} \sin^2 \sigma \right) \hat v_1 \hat v_2 + \lambda_{1222}^\mathrm{R} \cos \sigma \hat v_2^2,\\
(\mathcal{M}_\mathrm{N}^2)_{13}={}& -\sin \sigma \left( 2\lambda_{1212}^\mathrm{R} \cos \sigma \hat v_2 + \lambda_{1112}^\mathrm{R} \hat v_1 \right) \hat v_2,\\
(\mathcal{M}_\mathrm{N}^2)_{22}={}& 2 \left( \lambda_{2222} \hat v_2^2 + \lambda_{1222}^\mathrm{R} \cos \sigma \hat v_1 \hat v_2 + \lambda_{1212}^\mathrm{R} \cos^2 \sigma \hat v_1^2 \right),\\
(\mathcal{M}_\mathrm{N}^2)_{23}={}& -\sin \sigma \left( 2\lambda_{1212}^\mathrm{R}\cos \sigma \hat v_1 + \lambda_{1222}^\mathrm{R} \hat v_2 \right) \hat v_2,\\
(\mathcal{M}_\mathrm{N}^2)_{33}={}& 2 \lambda_{1212}^\mathrm{R} \sin^2 \sigma \hat v_2^2,\\
\begin{split}
(\mathcal{M}_\mathrm{N}^2)_{55} = {}&  \mu_{33}^2 + \frac{1}{2}\left( \lambda_{1133} + 2 \lambda_{1313}^\mathrm{R}\cos 2\sigma + \lambda_{1331}\right)\hat v_1^2\\
& + \frac{1}{2}\left( \lambda_{2233} + 2 \lambda_{2323}^\mathrm{R} + \lambda_{2332}\right)\hat v_2^2 + \left( \lambda_{1233}^\mathrm{R} + \lambda_{1323}^\mathrm{R} + \lambda_{1332}^\mathrm{R}\right)\cos \sigma\hat v_1 \hat v_2,
\end{split}\\
(\mathcal{M}_\mathrm{N}^2)_{56} = {}&  \sin \sigma \left( 2 \lambda_{1313}^\mathrm{R} \cos \sigma \hat v_1 + \lambda_{1323}^\mathrm{R} \hat v_2\right) \hat v_1,\\
\begin{split}
(\mathcal{M}_\mathrm{N}^2)_{66} = {}&  \mu_{33}^2 + \frac{1}{2}\left( \lambda_{1133} - 2 \lambda_{1313}^\mathrm{R}\cos 2\sigma + \lambda_{1331}\right)\hat v_1^2\\
& + \frac{1}{2}\left( \lambda_{2233} - 2 \lambda_{2323}^\mathrm{R} + \lambda_{2332}\right)\hat v_2^2 + \left( \lambda_{1233}^\mathrm{R} - \lambda_{1323}^\mathrm{R} + \lambda_{1332}^\mathrm{R}\right)\cos \sigma\hat v_1 \hat v_2.
\end{split}
\end{align}
\end{subequations}

\clearpage
\bigskip\textbf{Case of \boldmath$(0,\,\hat v_2,\,\hat v_3)$} 

Next, we consider that the $h_3$ doublet develops a non-zero vev. In this case the $\mathbb{Z}_2$ symmetry gets broken by $\left\langle h_3 \right\rangle \neq 0$. The case $(\hat v_1,\,0,\,\hat v_3)$ is identical to the case $(0,\,\hat v_2,\,\hat v_3)$, corresponding to the interchange of indices ``1" and ``2".

The minimisation conditions are given by:
\begin{subequations}
\begin{align}
\mu_{22}^2 & = -\frac{1}{2} \left( 2 \lambda_{2222} \hat v_2^2 + \lambda_{2233} \hat v_3^2 + \lambda_{2332} \hat v_3^2 + 2 \lambda_{2323}^\mathrm{R} \hat v_3^2\right),\\
\mu_{33}^2 & = -\frac{1}{2} \left( 2 \lambda_{3333} \hat v_3^2 + \lambda_{2233} \hat v_2^2 + \lambda_{2332} \hat v_2^2 + 2 \lambda_{2323}^\mathrm{R} \hat v_2^2\right),\\
\mu_{12}^2 & = -\frac{1}{2} \left( \lambda_{1222} \hat v_2^2 + \lambda_{1233} \hat v_3^2 + \lambda_{1323} \hat v_3^2 + \lambda_{1332} \hat v_3^2 \right),\\
\lambda_{2323}^\mathrm{I} & = 0,
\end{align}
\end{subequations}
where $\mu_{12}^2$ is complex.

The charged mass-squared matrix is:
\begin{equation}
\mathcal{M}_\mathrm{Ch}^2 = \begin{pmatrix}
(\mathcal{M}_\mathrm{Ch}^2)_{11} & (\mathcal{M}_\mathrm{Ch}^2)_{12} & -(\mathcal{M}_\mathrm{Ch}^2)_{12} \hat v_{2/3} \\
(\mathcal{M}_\mathrm{Ch}^2)_{12}^\ast & (\mathcal{M}_\mathrm{Ch}^2)_{22} & -(\mathcal{M}_\mathrm{Ch}^2)_{22} \hat v_{2/3}  \\
-(\mathcal{M}_\mathrm{Ch}^2)_{12}^\ast \hat v_{2/3}  & - (\mathcal{M}_\mathrm{Ch}^2)_{23} \hat v_{2/3}  & (\mathcal{M}_\mathrm{Ch}^2)_{22}  \hat v_{2/3}^2  
\end{pmatrix},
\end{equation}
where
\begin{subequations}
\begin{align}
(\mathcal{M}_\mathrm{Ch}^2)_{11} ={}& \frac{1}{2} \left( 2 \mu_{11}^2 + \lambda_{1122} \hat v_2^2 + \lambda_{1133} \hat v_3^2 \right),\\
(\mathcal{M}_\mathrm{Ch}^2)_{12} ={}& \frac{i}{2} \left( \lambda_{1323}^\mathrm{I} + i \lambda_{1323}^\mathrm{R} + \lambda_{1332}^\mathrm{I} + i \lambda_{1332}^\mathrm{R} \right) \hat v_3^2,\\
(\mathcal{M}_\mathrm{Ch}^2)_{22} ={}& -\frac{1}{2}\left( 2 \lambda_{2323}^\mathrm{R} + \lambda_{2332} \right) \hat v_3^2.
\end{align}
\end{subequations}

The neutral mass-squared matrix in the basis $\{\eta_1,\, \eta_2,\, \eta_3,\, \chi_1,\, \chi_2,\, \chi_3\}$ is:
\begin{equation}
\begin{multlined}[c]
\mathcal{M}_\mathrm{N}^2 = { \left(\begin{matrix}
(\mathcal{M}_\mathrm{N}^2)_{11} & \mathbb{R}\mathrm{e}\big((\mathcal{M}_\mathrm{N}^2)_{12} \big) & \mathbb{R}\mathrm{e}\big((\mathcal{M}_\mathrm{N}^2)_{13} \big)\\
\mathbb{R}\mathrm{e}\big((\mathcal{M}_\mathrm{N}^2)_{12} \big) & (\mathcal{M}_\mathrm{N}^2)_{22} & (\mathcal{M}_\mathrm{N}^2)_{23}\\
\mathbb{R}\mathrm{e}\big((\mathcal{M}_\mathrm{N}^2)_{13} \big) & (\mathcal{M}_\mathrm{N}^2)_{23} & (\mathcal{M}_\mathrm{N}^2)_{33}\\
(\mathcal{M}_\mathrm{N}^2)_{14} & \mathbb{I}\mathrm{m}\big((\mathcal{M}_\mathrm{N}^2)_{12} \big) & \mathbb{I}\mathrm{m}\big((\mathcal{M}_\mathrm{N}^2)_{13} \big)\\
\mathbb{I}\mathrm{m}\big((\mathcal{M}_\mathrm{N}^2)_{15} \big) & 0 & 0\\
-\mathbb{I}\mathrm{m}\big((\mathcal{M}_\mathrm{N}^2)_{15} \big) \hat v_{2/3} & 0 & 0\\
\end{matrix}\right.}\quad\dots\hspace{50pt}\vspace*{10pt}\\
\dots\quad{ \left.\begin{matrix}
(\mathcal{M}_\mathrm{N}^2)_{14} & \mathbb{I}\mathrm{m}\big((\mathcal{M}_\mathrm{N}^2)_{15} \big) & -\mathbb{I}\mathrm{m}\big((\mathcal{M}_\mathrm{N}^2)_{15} \big) \hat v_{2/3} \\
\mathbb{I}\mathrm{m}\big((\mathcal{M}_\mathrm{N}^2)_{12} \big) & 0 & 0\\
\mathbb{I}\mathrm{m}\big((\mathcal{M}_\mathrm{N}^2)_{13} \big) & 0 & 0\\
(\mathcal{M}_\mathrm{N}^2)_{44} & -\mathbb{R}\mathrm{e}\big((\mathcal{M}_\mathrm{N}^2)_{15} \big) & \mathbb{R}\mathrm{e}\big((\mathcal{M}_\mathrm{N}^2)_{15}\big) \hat v_{2/3}  \\
-\mathbb{R}\mathrm{e}\big((\mathcal{M}_\mathrm{N}^2)_{15} \big) & (\mathcal{M}_\mathrm{N}^2)_{55} & -(\mathcal{M}_\mathrm{N}^2)_{55} \hat v_{2/3} \\
\mathbb{R}\mathrm{e}\big((\mathcal{M}_\mathrm{N}^2)_{15} \big) \hat v_{2/3}  & -(\mathcal{M}_\mathrm{N}^2)_{55} \hat v_{2/3}  & (\mathcal{M}_\mathrm{N}^2)_{55} \hat v_{2/3} ^2\\
\end{matrix}\right)}, 
\end{multlined}
\end{equation}
where
\begin{subequations}
\begin{align}
(\mathcal{M}_\mathrm{N}^2)_{11} ={}& \mu_{11}^2 + \frac{1}{2}\left[ \left( \lambda_{1122} + 2 \lambda_{1212}^\mathrm{R} + \lambda_{1221} \right)\hat v_2^2 + \left( \lambda_{1133} + 2\lambda_{1313}^\mathrm{R} + \lambda_{1331} \right)\hat v_3^2 \right],\\
(\mathcal{M}_\mathrm{N}^2)_{12} ={}& \lambda_{1222} \hat v_2^2,\\
(\mathcal{M}_\mathrm{N}^2)_{13} ={}& \left( \lambda_{1233} + \lambda_{1323} + \lambda_{1332} \right) \hat v_2 \hat v_3,\\
(\mathcal{M}_\mathrm{N}^2)_{14} ={}& \lambda_{1212}^\mathrm{I} \hat v_2^2 + \lambda_{1313}^\mathrm{I} \hat v_3^2,\\
(\mathcal{M}_\mathrm{N}^2)_{15} ={}& \lambda_{1323} \hat v_3^2,\\
(\mathcal{M}_\mathrm{N}^2)_{22} ={}& 2\lambda_{2222} \hat v_2^2,\\
(\mathcal{M}_\mathrm{N}^2)_{23} ={}& \left( \lambda_{2233} + 2 \lambda_{2323}^\mathrm{R} + \lambda_{2332} \right) \hat v_2 \hat v_3,\\
(\mathcal{M}_\mathrm{N}^2)_{33} ={}& 2\lambda_{3333} \hat v_3^2,\\
(\mathcal{M}_\mathrm{N}^2)_{44} ={}& \mu_{11}^2 + \frac{1}{2}\left[ \left( \lambda_{1122} - 2 \lambda_{1212}^\mathrm{R} + \lambda_{1221} \right)\hat v_2^2 + \left( \lambda_{1133} - 2\lambda_{1313}^\mathrm{R} + \lambda_{1331} \right)\hat v_3^2 \right],\\
(\mathcal{M}_\mathrm{N}^2)_{55} ={}& -2 \lambda_{2323}^\mathrm{R} \hat v_3^2.
\end{align}
\end{subequations}

There is mixing between $h_1$ and $\{h_2,\, h_3\}$. As a result, the mass eigenstates will be given in terms of all $\{\eta_i,\chi_i\}$ fields. Mixing between the neutral scalar fields would then suggest that a coupling $g(H_i Z Z)$ would arise from the kinetic terms, where $H_i$ is a neutral mass eigenstate. Therefore, the lightest state $H_1$ will not be stable, unless the rotation matrix is tuned in such a way that the vertex $g(H_1 Z Z)$ vanishes to all orders. Fixing the rotation angles would force additional constraints on the couplings. However, such tuning might not guarantee cancellations of the couplings to $H_1$ in the Yukawa Lagrangian. A possible solution to handle both issues simultaneously would be to prevent mixing between the doublets $h_1$ and $\{h_2,\, h_3\}$. In this case one would have to require: 
\begin{equation}\label{Eq:Z2_0v1v2_AddCond}
\lambda_{1323} = \lambda_{1332} = \lambda_{1222} = \lambda_{1233} = 0.
\end{equation}
Assuming these conditions, the phase-sensitive part~\eqref{Eq:Z2_Vph} reduces to:
\begin{equation}
V_\text{ph} = \sum_{i<j} \lambda_{ijij} h_{ij}^2 + \lambda_{1112} h_{11}h_{12} + \mathrm{h.c.},
\end{equation}
with $\lambda_{2323}^\mathrm{I}=0$ coming from the minimisation conditions. Then, the scalar potential can be written as $V_0 + V_\mathrm{ph} = V_{\mathbb{Z}_2 \times \mathbb{Z}_2} + \left\lbrace \lambda_{1112} h_{11}h_{12} + \mathrm{h.c.} \right\rbrace $, where
\begin{equation}
\begin{aligned}
V_{\mathbb{Z}_2 \times \mathbb{Z}_2} ={}& \sum_i \mu_{ii}^2 h_{ii} + \sum_i \lambda_{iiii} h_{ii}^2 + \sum_{i<j} \lambda_{iijj} h_{ii} h_{jj}\\
& + \sum_{i<j} \lambda_{ijji} h_{ij} h_{ji} + \left\lbrace \sum_{i<j} \lambda_{ijij} h_{ij}^2 + \mathrm{h.c.} \right\rbrace.
\end{aligned}
\end{equation}
This might not be a valid solution once the higher-order contributions are taken into consideration if the case is not stabilised by a symmetry. The lists of symmetries given in Refs.~\cite{deMedeirosVarzielas:2019rrp,Darvishi:2019dbh} do not contain such a symmetry.

The term $\lambda_{1112}$ allows for couplings of the type
\begin{equation}
\lambda_{1112} \left( \varphi_1^2 + \varphi_2^2 \right) \left[ \left( H_i + A_i \right)\left( \varphi_1 + \varphi_2 \right) + \hat v_2\left( \varphi_1 + \varphi_2 \right) \right].
\end{equation}
Above it is assumed that $\eta_1$ and $\chi_1$ mix to produce mass eigenstates denoted by $\varphi_i$. The doublet $h_2$ in terms of mass eigenstates is $h_2 \sim (\hat v_2 + H_i + i A_i)$. These forms of the mass eigenstates are suggested by the neutral mass-squared matrix, which in the basis $\{\eta_1,\, \chi_1\}{-}\{\eta_2,\, \eta_3\}{-}\{\chi_2,\, \chi_3\}$ can be split into three blocks,
\begin{equation}\label{Eq:Z2_Neutral_0_v1_v2}
\mathcal{M}_\mathrm{N}^2 = \mathrm{diag}  \left( \mathcal{M}^2_{\eta_1 \chi_1},\, \mathcal{M}_{\eta_{23}}^2,\, \mathcal{M}_{\chi_{23}}^2 \right).
\end{equation}

\bigskip\textbf{Case of \boldmath$(0,\,\hat v_2 e^{i \sigma},\,\hat v_3)$ and real couplings}

Next, we consider a case corresponding to the previous one but with spontaneous CP violation. The minimisation conditions are then given by:
\begin{subequations}
\begin{align}
\mu_{22}^2 ={}& - \lambda_{2222} \hat v_2^2 - \frac{1}{2} \left( \lambda_{2233} + \lambda_{2332} \right) \hat v_3^2,\\
\mu_{33}^2 ={}&  - \lambda_{3333} \hat v_3^2 - \frac{1}{2} \left( \lambda_{2233} + \lambda_{2332} \right) \hat v_2^2,\\
\left(\mu_{12}^2\right)^\mathrm{R} ={}&  -\frac{1}{2} \left( \lambda_{1222}^\mathrm{R} \hat v_2^2 + \lambda_{1233}^\mathrm{R} \hat v_3^2  + \lambda_{1332}^\mathrm{R} \hat v_3^2 \right),\\
\lambda_{2323}^\mathrm{R} ={}& 0,\\
\lambda_{1323}^\mathrm{R} ={}& 0.
\end{align}
\end{subequations}

The charged mass-squared matrix is
\begin{equation}
\mathcal{M}_\mathrm{Ch}^2 = \begin{pmatrix}
(\mathcal{M}_\mathrm{Ch}^2)_{11} & (\mathcal{M}_\mathrm{Ch}^2)_{21}^\ast & -(\mathcal{M}_\mathrm{Ch}^2)_{21}^\ast \hat v_{2/3}  \\
(\mathcal{M}_\mathrm{Ch}^2)_{21} & (\mathcal{M}_\mathrm{Ch}^2)_{22} & -(\mathcal{M}_\mathrm{Ch}^2)_{22} \hat v_{2/3}  \\
-(\mathcal{M}_\mathrm{Ch}^2)_{21} \hat v_{2/3}  & -(\mathcal{M}_\mathrm{Ch}^2)_{22} \hat v_{2/3}  & (\mathcal{M}_\mathrm{Ch}^2)_{22}   \hat v_{2/3}^2 
\end{pmatrix},
\end{equation}
where
\begin{subequations}
\begin{align}
(\mathcal{M}_\mathrm{Ch}^2)_{11} ={}& \frac{1}{2} \left( 2 \mu_{11}^2 + \lambda_{1122} \hat v_2^2 + \lambda_{1133} \hat v_3^2 \right),\\
(\mathcal{M}_\mathrm{Ch}^2)_{21} ={}& -\frac{1}{2}  e^{i \sigma} \lambda_{1332}^\mathrm{R} \hat v_3^2 ,\\
(\mathcal{M}_\mathrm{Ch}^2)_{22} ={}& -\frac{1}{2} \lambda_{2332} \hat v_3^2.
\end{align}
\end{subequations}

The neutral mass-squared matrix in the basis $\{\eta_1,\, \eta_2,\, \eta_3,\, \chi_1,\, \chi_2,\, \chi_3\}$ is
\begin{equation}
\mathcal{M}_\mathrm{N}^2 = \mathrm{diag} \left( \begin{pmatrix}
(\mathcal{M}_\mathrm{N}^2)_{11} & (\mathcal{M}_\mathrm{N}^2)_{12} & (\mathcal{M}_\mathrm{N}^2)_{13} & (\mathcal{M}_\mathrm{N}^2)_{14} \\
(\mathcal{M}_\mathrm{N}^2)_{12} & (\mathcal{M}_\mathrm{N}^2)_{22} & (\mathcal{M}_\mathrm{N}^2)_{23} & (\mathcal{M}_\mathrm{N}^2)_{24} \\
(\mathcal{M}_\mathrm{N}^2)_{13} & (\mathcal{M}_\mathrm{N}^2)_{23} & (\mathcal{M}_\mathrm{N}^2)_{33} & (\mathcal{M}_\mathrm{N}^2)_{34} \\
(\mathcal{M}_\mathrm{N}^2)_{14} & (\mathcal{M}_\mathrm{N}^2)_{24} & (\mathcal{M}_\mathrm{N}^2)_{34} & (\mathcal{M}_\mathrm{N}^2)_{44} \\
\end{pmatrix},\, 0,\, 0\right),
\end{equation}
where
\begin{subequations}
\begin{align}
\begin{split}
(\mathcal{M}_\mathrm{N}^2)_{11} & = \mu_{11}^2 + \frac{1}{2}\Big[ \left( \lambda_{1122} + 2 \lambda_{1212}^\mathrm{R} \cos (2\sigma) + \lambda_{1221} \right)\hat v_2^2\\
 &\hspace{60pt}+ \left( \lambda_{1133} + 2\lambda_{1313}^\mathrm{R} + \lambda_{1331} \right)\hat v_3^2 \Big],
\end{split}\\
(\mathcal{M}_\mathrm{N}^2)_{12} & = \lambda_{1222}^\mathrm{R} \cos\sigma \hat v_2^2,\\
(\mathcal{M}_\mathrm{N}^2)_{13} & = \cos \sigma \left( \lambda_{1233}^\mathrm{R} + \lambda_{1332}^\mathrm{R} \right) \hat v_2 \hat v_3,\\
(\mathcal{M}_\mathrm{N}^2)_{14} & = \lambda_{1212}^\mathrm{R} \sin  (2\sigma) \hat v_2^2,\\
(\mathcal{M}_\mathrm{N}^2)_{22} & = 2 \lambda_{2222} \hat v_2^2,\\
(\mathcal{M}_\mathrm{N}^2)_{23} & = \left( \lambda_{2233} + \lambda_{2332} \right) \hat v_2 \hat v_3,\\
(\mathcal{M}_\mathrm{N}^2)_{24} & = \lambda_{1222}^\mathrm{R} \sin \sigma \hat v_2^2,\\
(\mathcal{M}_\mathrm{N}^2)_{33} & = 2 \lambda_{3333} \hat v_3^2,\\
(\mathcal{M}_\mathrm{N}^2)_{34} & = \sin \sigma \left( \lambda_{1233}^\mathrm{R} + \lambda_{1332}^\mathrm{R} \right) \hat v_2 \hat v_3,\\
\begin{split}
(\mathcal{M}_\mathrm{N}^2)_{44} & = \mu_{11}^2 + \frac{1}{2}\Big[ \left( \lambda_{1122} - 2 \lambda_{1212}^\mathrm{R} \cos (2\sigma) + \lambda_{1221} \right)\hat v_2^2\\
 &\hspace{60pt}+ \left( \lambda_{1133} - 2\lambda_{1313}^\mathrm{R} + \lambda_{1331} \right)\hat v_3^2 \Big].
\end{split}
\end{align}
\end{subequations}

There are two neutral massless states present, $m_{\chi_2} = m_{\chi_3} = 0$. A possible way around is to introduce soft symmetry-breaking terms. Since we are interested in a DM candidate, we do not want to mix $h_1$ with $\{h_2,\, h_3\}$. This constraint leaves us with a single possibility---a $\mu_{23}^2$ soft symmetry-breaking term, which for consistency we take to be real since we consider spontaneous CP violation. The minimisation conditions then become:
\begin{subequations}
\begin{align}
\mu_{22}^2 ={}& - \lambda_{2222} \hat v_2^2 - \frac{1}{2} \left( \lambda_{2233} -2 \lambda_{2323}^\mathrm{R}+ \lambda_{2332} \right) \hat v_3^2,\\
\mu_{33}^2 ={}&  - \lambda_{3333} \hat v_3^2 - \frac{1}{2} \left( \lambda_{2233} -2 \lambda_{2323}^\mathrm{R}+ \lambda_{2332} \right) \hat v_2^2,\\
\left(\mu_{12}^2\right)^\mathrm{R} ={}&  -\frac{1}{2} \left( \lambda_{1222}^\mathrm{R} \hat v_2^2 + \lambda_{1233}^\mathrm{R} \hat v_3^2  + \lambda_{1332}^\mathrm{R} \hat v_3^2 \right),\\
\left(\mu_{23}^2\right)^\mathrm{R} ={}&  - 2\lambda_{2323}^\mathrm{R} \cos \sigma \hat v_2 \hat v_3,\\
\lambda_{1323}^\mathrm{R} ={}& 0.
\end{align}
\end{subequations}

The charged mass-squared matrix is:
\begin{equation}
\mathcal{M}_\mathrm{Ch}^2 = \begin{pmatrix}
(\mathcal{M}_\mathrm{Ch}^2)_{11} & (\mathcal{M}_\mathrm{Ch}^2)_{21}^\ast & -(\mathcal{M}_\mathrm{Ch}^2)_{21}^\ast \hat v_{2/3}  \\
(\mathcal{M}_\mathrm{Ch}^2)_{21} & (\mathcal{M}_\mathrm{Ch}^2)_{22} & -(\mathcal{M}_\mathrm{Ch}^2)_{22} \hat v_{2/3}  \\
- (\mathcal{M}_\mathrm{Ch}^2)_{21} \hat v_{2/3}  & -(\mathcal{M}_\mathrm{Ch}^2)_{22} \hat v_{2/3}  & (\mathcal{M}_\mathrm{Ch}^2)_{22} \hat v_{2/3}^2
\end{pmatrix},
\end{equation}
where
\begin{subequations}
\begin{align}
(\mathcal{M}_\mathrm{Ch}^2)_{11} ={}& \frac{1}{2} \left( 2 \mu_{11}^2 + \lambda_{1122} \hat v_2^2 + \lambda_{1133} \hat v_3^2 \right),\\
(\mathcal{M}_\mathrm{Ch}^2)_{21} ={}& -\frac{1}{2}  e^{i \sigma} \lambda_{1332}^\mathrm{R} \hat v_3^2 ,\\
(\mathcal{M}_\mathrm{Ch}^2)_{22} ={}& \frac{1}{2} \left( 2\lambda_{2323}^\mathrm{R} - \lambda_{2332} \right) \hat v_3^2.
\end{align}
\end{subequations}

The neutral mass-squared matrix in the basis $\{\eta_1,\, \eta_2,\, \eta_3,\, \chi_1,\, \chi_2,\, \chi_3\}$ is:
\begin{equation}
\begin{multlined}[c]
\mathcal{M}_\mathrm{N}^2 = { \left(\begin{matrix}
(\mathcal{M}_\mathrm{N}^2)_{11} & (\mathcal{M}_\mathrm{N}^2)_{12} & (\mathcal{M}_\mathrm{N}^2)_{13}\\
(\mathcal{M}_\mathrm{N}^2)_{12} & (\mathcal{M}_\mathrm{N}^2)_{22} & (\mathcal{M}_\mathrm{N}^2)_{23}\\
(\mathcal{M}_\mathrm{N}^2)_{13} & (\mathcal{M}_\mathrm{N}^2)_{23} & (\mathcal{M}_\mathrm{N}^2)_{33}\\
(\mathcal{M}_\mathrm{N}^2)_{14} & (\mathcal{M}_\mathrm{N}^2)_{24} & (\mathcal{M}_\mathrm{N}^2)_{34}\\
0 & (\mathcal{M}_\mathrm{N}^2)_{25} & (\mathcal{M}_\mathrm{N}^2)_{25} \hat v_{2/3} \\
0 & -(\mathcal{M}_\mathrm{N}^2)_{25} \hat v_{2/3}  & -(\mathcal{M}_\mathrm{N}^2)_{25} \hat v_{2/3}^2
\end{matrix}\right.}{ \left.\begin{matrix}
(\mathcal{M}_\mathrm{N}^2)_{14} & 0 & 0\\
(\mathcal{M}_\mathrm{N}^2)_{24} & (\mathcal{M}_\mathrm{N}^2)_{25} & -(\mathcal{M}_\mathrm{N}^2)_{25} \hat v_{2/3} \\
(\mathcal{M}_\mathrm{N}^2)_{34} & (\mathcal{M}_\mathrm{N}^2)_{25} \hat v_{2/3}  & -(\mathcal{M}_\mathrm{N}^2)_{25} \hat v_{2/3}^2\\
(\mathcal{M}_\mathrm{N}^2)_{44} & 0 & 0\\
0 & (\mathcal{M}_\mathrm{N}^2)_{55} & -(\mathcal{M}_\mathrm{N}^2)_{55} \hat v_{2/3} \\
0 & -(\mathcal{M}_\mathrm{N}^2)_{55} \hat v_{2/3}  & (\mathcal{M}_\mathrm{N}^2)_{55} \hat v_{2/3}^2
\end{matrix}\right)}, 
\end{multlined}
\end{equation}
where
\begin{subequations}
\begin{align}
\begin{split}
(\mathcal{M}_\mathrm{N}^2)_{11} & = \mu_{11}^2 + \frac{1}{2}\Big[ \left( \lambda_{1122} + 2 \lambda_{1212}^\mathrm{R} \cos (2\sigma) + \lambda_{1221} \right)\hat v_2^2\\
& \hspace{60pt}+\left( \lambda_{1133} + 2\lambda_{1313}^\mathrm{R} + \lambda_{1331} \right)\hat v_3^2 \Big],
\end{split}\\
(\mathcal{M}_\mathrm{N}^2)_{12} & = \lambda_{1222}^\mathrm{R} \cos\sigma \hat v_2^2,\\
(\mathcal{M}_\mathrm{N}^2)_{13} & = \cos \sigma \left( \lambda_{1233}^\mathrm{R} + \lambda_{1332}^\mathrm{R} \right) \hat v_2 \hat v_3,\\
(\mathcal{M}_\mathrm{N}^2)_{14} & = \lambda_{1212}^\mathrm{R} \sin  (2\sigma) \hat v_2^2,\\
(\mathcal{M}_\mathrm{N}^2)_{22} & = 2 \lambda_{2222} \hat v_2^2 + 2\lambda_{2323}^\mathrm{R} \cos ^2 \sigma \hat v_3^2,\\
(\mathcal{M}_\mathrm{N}^2)_{23} & = \left( \lambda_{2233} + \lambda_{2332} - 2\lambda_{2323}^\mathrm{R} \sin^2\sigma \right) \hat v_2 \hat v_3,\\
(\mathcal{M}_\mathrm{N}^2)_{24} & = \lambda_{1222}^\mathrm{R} \sin \sigma \hat v_2^2,\\
(\mathcal{M}_\mathrm{N}^2)_{25} & = - \lambda_{2323}^\mathrm{R} \sin (2\sigma) \hat v_3^2,\\
(\mathcal{M}_\mathrm{N}^2)_{33} & = 2 \lambda_{3333} \hat v_3^2 + 2\lambda_{2323}^\mathrm{R} \cos^2 \sigma \hat v_2^2,\\
(\mathcal{M}_\mathrm{N}^2)_{34} & = \sin \sigma \left( \lambda_{1233}^\mathrm{R} + \lambda_{1332}^\mathrm{R} \right) \hat v_2 \hat v_3,\\
\begin{split}
(\mathcal{M}_\mathrm{N}^2)_{44} & = \mu_{11}^2 + \frac{1}{2}\Big[ \left( \lambda_{1122} - 2 \lambda_{1212}^\mathrm{R} \cos (2\sigma) + \lambda_{1221} \right)\hat v_2^2 \\
&\hspace{60pt}+\left( \lambda_{1133} - 2\lambda_{1313}^\mathrm{R} + \lambda_{1331} \right)\hat v_3^2 \Big],
\end{split}\\
(\mathcal{M}_\mathrm{N}^2)_{55} & = 2 \lambda_{2323}^\mathrm{R} \sin^2 \sigma \hat v_3 ^2.
\end{align}
\end{subequations}

The neutral mass eigenstates will be given by a combination of all neutral fields, $\eta_i$ and $\chi_i$. As a result the DM candidate will not be stable. At tree level one can decouple $h_1$ from $\{h_2,\, h_3\}$ by requiring:
\begin{equation}
\lambda_{1332}^\mathrm{R} = \lambda_{1222}^\mathrm{R} = \lambda_{1233}^\mathrm{R} = 0.
\end{equation}
Then, the phase-sensitive part of the scalar potential becomes:
\begin{equation}
V_\text{ph} = \sum_{i<j} \lambda_{ijij}^\mathrm{R} (h_{ij}^2 + h_{ji}^2) +  \lambda_{1112}^\mathrm{R} (h_{11}h_{12} + h_{11} h_{21}),
\end{equation}
which can be written as $V_{\mathbb{Z}_2 \times \mathbb{Z}_2}^\mathrm{\;R} + \lambda_{1112}^\mathrm{R} h_{11}h_{12} + \mathrm{h.c.}$ and does not correspond to a symmetry.

\clearpage
\subsection{Two vanishing vevs}

\bigskip\textbf{Case of \boldmath$(0,\,0,\,v)$}

There are two possibilities to have a vacuum with only a single non-vanishing vev. The $\mathbb{Z}_2$ charges given by eq.~\eqref{Eq:Z2_charges} suggest that the considered vacuum would break the $\mathbb{Z}_2$ symmetry. This is not the case since without loss of generality we could assume that the charges are rotated by the overall $U(1)$ symmetry so that:
\begin{equation}\label{Eq:Z2_charges_00v}
\mathbb{Z}_2: ~ h_1\to -h_1,~~ h_2\to -h_2,~~ h_3\to h_3.
\end{equation}

In this case there is a single minimisation condition:
\begin{equation}
\mu_{33}^2 = - \lambda_{3333}v^2 .
\end{equation}

The charged mass-squared matrix is given by:
\begin{equation}
\mathcal{M}_\mathrm{Ch}^2 = \begin{pmatrix}
(\mathcal{M}_\mathrm{Ch}^2)_{11} & (\mathcal{M}_\mathrm{Ch}^2)^\ast_{12} & 0 \\
(\mathcal{M}_\mathrm{Ch}^2)_{12} & (\mathcal{M}_\mathrm{Ch}^2)_{22} & 0 \\
0 & 0 & 0   
\end{pmatrix},
\end{equation}
where
\begin{equation}
(\mathcal{M}_\mathrm{Ch}^2)_{ij} = \mu_{ij}^2 + \frac{v^2}{2} \lambda_{ij33}.
\end{equation}

The neutral mass-squared matrix in the basis $\{\eta_1,\, \eta_2,\, \chi_1,\, \chi_2\}{-}\{\eta_3\}{-}\{\chi_3\}$ is:
\begin{equation}\label{Eq:00v_Z2_Ninert}
\mathcal{M}_\mathrm{N}^2 = \begin{pmatrix}
(\mathcal{M}_\mathrm{N}^2)_{11} & \mathbb{R}\mathrm{e}\big((\mathcal{M}_\mathrm{N}^2)_{12} \big) & (\mathcal{M}_\mathrm{N}^2)_{13} & -\mathbb{I}\mathrm{m}\big((\mathcal{M}_\mathrm{N}^2)_{14} \big) & 0 & 0\\
\mathbb{R}\mathrm{e}\big((\mathcal{M}_\mathrm{N}^2)_{12} \big) & (\mathcal{M}_\mathrm{N}^2)_{22} & \mathbb{I}\mathrm{m}\big((\mathcal{M}_\mathrm{N}^2)_{12} \big) & (\mathcal{M}_\mathrm{N}^2)_{24} & 0 & 0\\
(\mathcal{M}_\mathrm{N}^2)_{13} & \mathbb{I}\mathrm{m}\big((\mathcal{M}_\mathrm{N}^2)_{12} \big) & (\mathcal{M}_\mathrm{N}^2)_{33} & \mathbb{R}\mathrm{e}\big((\mathcal{M}_\mathrm{N}^2)_{14} \big) & 0 & 0\\
-\mathbb{I}\mathrm{m}\big((\mathcal{M}_\mathrm{N}^2)_{14} \big) & (\mathcal{M}_\mathrm{N}^2)_{24} & \mathbb{R}\mathrm{e}\big((\mathcal{M}_\mathrm{N}^2)_{14} \big) & (\mathcal{M}_\mathrm{N}^2)_{44} & 0 & 0\\
0 & 0 & 0 & 0 & m_h^2 & 0 \\
0 & 0 & 0 & 0 & 0 & 0
\end{pmatrix},
\end{equation}
where
\begin{subequations}
\begin{align}
(\mathcal{M}_\mathrm{N}^2)_{11}={}& \mu_{11}^2 + \frac{v^2}{2} \left( \lambda_{1133} + 2 \lambda_{1313}^\mathrm{R} + \lambda_{1331} \right),\\
(\mathcal{M}_\mathrm{N}^2)_{12}={}& \mu_{12}^2 + \frac{v^2}{2} \left( \lambda_{1233} + \lambda_{1323} + \lambda_{1332}\right),\\
(\mathcal{M}_\mathrm{N}^2)_{13}={}&  \lambda_{1313}^\mathrm{I}v^2,\\
(\mathcal{M}_\mathrm{N}^2)_{14}={}& \mu_{12}^2 + \frac{v^2}{2}\left( \lambda_{1233} - \lambda_{1323} + \lambda_{1332}\right),\\
(\mathcal{M}_\mathrm{N}^2)_{22}={}& \mu_{22}^2 + \frac{v^2}{2} \left( \lambda_{2233} + 2 \lambda_{2323}^\mathrm{R} + \lambda_{2332} \right),\\
(\mathcal{M}_\mathrm{N}^2)_{24}={}&  \lambda_{2323}^\mathrm{I}v^2,\\
(\mathcal{M}_\mathrm{N}^2)_{33}={}& \mu_{11}^2 + \frac{v^2}{2} \left( \lambda_{1133} - 2 \lambda_{1313}^\mathrm{R} + \lambda_{1331} \right),\\
(\mathcal{M}_\mathrm{N}^2)_{44}={}& \mu_{22}^2 + \frac{v^2}{2} \left( \lambda_{2233} - 2 \lambda_{2323}^\mathrm{R} + \lambda_{2332} \right),\\
m_h^2 ={}& 2  \lambda_{3333} v^2.
\end{align}
\end{subequations}

\bigskip\textbf{Case of \boldmath$(v,\,0,\,0)$}

The other possibility is to consider a situation where out of the two doublets with identical $\mathbb{Z}_2$ charges only one develops a non-zero vev. In this case the $\mathbb{Z}_2$ symmetry~\eqref{Eq:Z2_charges} is not spontaneously broken by the vacuum. The minimisation conditions are given by:
\begin{subequations}
\begin{align}
\mu_{11}^2 & = - \lambda_{1111}v^2,\\
\mu_{12}^2 & = -\frac{v^2}{2} \lambda_{1112}.
\end{align}
\end{subequations}

There is no mixing between the charged scalars and the squared masses are given by:
\begin{subequations}
\begin{align}
m_{h_2^\pm}^2 & = \mu_{22}^2 + \frac{v^2}{2} \lambda_{1122},\\
m_{h_3^\pm}^2 & = \mu_{33}^2 + \frac{v^2}{2} \lambda_{1133}.
\end{align}
\end{subequations}

The neutral mass-squared matrix in the basis $\{\eta_1,\, \eta_2,\, \chi_1,\, \chi_2\}{-}\{\eta_3, \,\chi_3\}$ is:
\begin{equation}\label{Eq:v00_Z2_Ninert}
\mathcal{M}_\mathrm{N}^2 = \begin{pmatrix}
(\mathcal{M}_\mathrm{N}^2)_{11} & \mathbb{R}\mathrm{e}\big((\mathcal{M}_\mathrm{N}^2)_{12} \big) & 0 & -\mathbb{I}\mathrm{m}\big((\mathcal{M}_\mathrm{N}^2)_{12} \big) & 0 & 0\\
\mathbb{R}\mathrm{e}\big((\mathcal{M}_\mathrm{N}^2)_{12} \big) & (\mathcal{M}_\mathrm{N}^2)_{22} & 0 & (\mathcal{M}_\mathrm{N}^2)_{23} & 0 & 0\\
0 & 0 & 0 & 0 & 0 & 0\\
-\mathbb{I}\mathrm{m}\big((\mathcal{M}_\mathrm{N}^2)_{12} \big) & (\mathcal{M}_\mathrm{N}^2)_{23} & 0 & (\mathcal{M}_\mathrm{N}^2)_{33} & 0 & 0\\
0 & 0 & 0 & 0 & (\mathcal{M}_\mathrm{N}^2)_{55} & (\mathcal{M}_\mathrm{N}^2)_{56} \\
0 & 0 & 0 & 0 & (\mathcal{M}_\mathrm{N}^2)_{56} & (\mathcal{M}_\mathrm{N}^2)_{66}
\end{pmatrix},
\end{equation}
where
\begin{subequations}
\begin{align}
(\mathcal{M}_\mathrm{N}^2)_{11}={}& 2  \lambda_{1111}v^2,\\
(\mathcal{M}_\mathrm{N}^2)_{12}={}&  \lambda_{1112} v^2,\\
(\mathcal{M}_\mathrm{N}^2)_{22}={}& \mu_{22}^2 + \frac{v^2}{2}\left( \lambda_{1122} + 2 \lambda_{1212}^\mathrm{R} + \lambda_{1221} \right),\\
(\mathcal{M}_\mathrm{N}^2)_{23}={}& - \lambda_{1212}^\mathrm{I}v^2,\\
(\mathcal{M}_\mathrm{N}^2)_{33}={}& \mu_{22}^2 + \frac{v^2}{2}\left( \lambda_{1122} - 2 \lambda_{1212}^\mathrm{R} + \lambda_{1221} \right),\\
(\mathcal{M}_\mathrm{N}^2)_{55} ={}& \mu_{33}^2 + \frac{v^2}{2}\left( \lambda_{1133} + 2 \lambda_{1313}^\mathrm{R} + \lambda_{1331} \right),\\
(\mathcal{M}_\mathrm{N}^2)_{56} ={}&- \lambda_{1313}^\mathrm{I}v^2,\\
(\mathcal{M}_\mathrm{N}^2)_{66} ={}& \mu_{33}^2 + \frac{v^2}{2}\left( \lambda_{1133} - 2 \lambda_{1313}^\mathrm{R} + \lambda_{1331} \right).
\end{align}
\end{subequations}
Since there are two vanishing vevs it is possible to rotate away two phases of the quartic couplings. The most natural choice would be to set $\lambda_{1212}^\mathrm{I}=\lambda_{1313}^\mathrm{I}=0$ to reduce mixing between the neutral states in the mass-squared matrix. The doublets $h_1$ and $h_2$ would not mix in the limit $\lambda_{1112} = 0$, which does not correspond to a higher symmetry.

\section{\texorpdfstring{\boldmath$\mathbb{Z}_3$}{Z3}-symmetric 3HDM}\label{App:Pot_Z3}

The generators of order three are chosen to be re-phasing transformations with phases
\begin{equation}\label{Eq:Z3_gen_ph_tr}
\mathbb{Z}_3:~\mathrm{diag} \left( 1,\, e^{i \frac{2\pi}{3}},\, e^{i \frac{4\pi}{3}} \right).
\end{equation}
A scalar potential invariant under this choice of charges remains invariant for any other ordering of the $\mathbb{Z}_3$ charges. The phase-sensitive part of the scalar potential is given by:
\begin{equation}
V_{\mathbb{Z}_3}^\text{ph} = \lambda_{1323} h_{13} h_{23} + \lambda_{1213} h_{12} h_{13} + \lambda_{1232} h_{12} h_{32} + \mathrm{h.c.}
\end{equation}
There are two couplings $\{\lambda_{1213},\, \lambda_{1232}\}$ which were not present in the $\mathbb{Z}_2$-symmetric 3HDM, while some of the terms of the $\mathbb{Z}_2$ scalar potential are not allowed in the case of the $\mathbb{Z}_3$ symmetry. Since the form of the $\mathbb{Z}_3$ scalar potential is preserved under permutations of the scalar doublets, it is sufficient to consider only two vacuum configurations.

\subsection{One vanishing vev}

\bigskip\textbf{Case of \boldmath$(\hat v_1,\,\hat v_2,\,0)$}

First, we consider a vacuum with two non-vanishing vevs. Such vacuum spontaneously breaks the underlying $\mathbb{Z}_3$ symmetry. The minimisation conditions are given by:
\begin{subequations}\label{Eq:Z3_v1v20_MinCond}
\begin{align}
\mu_{11}^2 & = -\frac{1}{2} \left( 2 \lambda_{1111} \hat v_1^2 + \lambda_{1122} \hat v_2^2 + \lambda_{1221} \hat v_2^2 \right),\\
\mu_{22}^2 & = -\frac{1}{2} \left( 2 \lambda_{2222} \hat v_2^2 + \lambda_{1122} \hat v_1^2 + \lambda_{1221} \hat v_1^2 \right),\\
\lambda_{1213}^\mathrm{R} & = - \frac{\hat v_2}{\hat v_1} \lambda_{1232}^\mathrm{R},\\
\lambda_{1213}^\mathrm{I} & = \frac{\hat v_2}{\hat v_1} \lambda_{1232}^\mathrm{I}.
\end{align}
\end{subequations}

The charged mass-squared matrix is:
\begin{equation}
\mathcal{M}_\mathrm{Ch}^2 = \begin{pmatrix}
(\mathcal{M}_\mathrm{Ch}^2)_{11} & -(\mathcal{M}_\mathrm{Ch}^2)_{11} \hat v_{1/2} & (\mathcal{M}_\mathrm{Ch}^2)_{13} \\
-(\mathcal{M}_\mathrm{Ch}^2)_{11} \hat v_{1/2} & (\mathcal{M}_\mathrm{Ch}^2)_{11} \hat v_{1/2}^2 & -(\mathcal{M}_\mathrm{Ch}^2)_{13} \hat v_{1/2} \\
(\mathcal{M}_\mathrm{Ch}^2)_{13} & -(\mathcal{M}_\mathrm{Ch}^2)_{13} \hat v_{1/2} & (\mathcal{M}_\mathrm{Ch}^2)_{33}   
\end{pmatrix},
\end{equation}
where
\begin{subequations}
\begin{align}
(\mathcal{M}_\mathrm{Ch}^2)_{11} ={}& -\frac{1}{2}\lambda_{1221}\hat v_2^2,\\
(\mathcal{M}_\mathrm{Ch}^2)_{13} ={}& -\frac{1}{2}\lambda_{1232}\hat v_2^2,\\
(\mathcal{M}_\mathrm{Ch}^2)_{33} ={}& \mu_{33}^2 + \frac{1}{2} \left( \lambda_{1133} \hat v_1^2 + \lambda_{2233} \hat v_2^2 \right).
\end{align}
\end{subequations}

The neutral mass-squared matrix in the basis $\{\eta_1,\, \eta_2,\, \eta_3,\, \chi_1,\, \chi_2,\,\chi_3\}$ is given by
\begin{equation}
\begin{multlined}[c]
\mathcal{M}_\mathrm{N}^2 = { \left(\begin{matrix}
(\mathcal{M}_\mathrm{N}^2)_{11} & (\mathcal{M}_\mathrm{N}^2)_{12} & \mathbb{R}\mathrm{e}\big((\mathcal{M}_\mathrm{N}^2)_{13} \big)\\
(\mathcal{M}_\mathrm{N}^2)_{12} & (\mathcal{M}_\mathrm{N}^2)_{22} & -\mathbb{R}\mathrm{e}\big((\mathcal{M}_\mathrm{N}^2)_{13} \big) \hat v_{1/2}\\
\mathbb{R}\mathrm{e}\big((\mathcal{M}_\mathrm{N}^2)_{13} \big) & -\mathbb{R}\mathrm{e}\big((\mathcal{M}_\mathrm{N}^2)_{13} \big) \hat v_{1/2} & (\mathcal{M}_\mathrm{N}^2)_{33} &\\
0                               & 0                               &-3\mathbb{I}\mathrm{m}\big((\mathcal{M}_\mathrm{N}^2)_{13} \big)\\
0                               & 0                               &3\mathbb{I}\mathrm{m}\big((\mathcal{M}_\mathrm{N}^2)_{13} \big) \hat v_{1/2}\\
\mathbb{I}\mathrm{m}\big((\mathcal{M}_\mathrm{N}^2)_{13} \big) & -\mathbb{I}\mathrm{m}\big((\mathcal{M}_\mathrm{N}^2)_{13} \big) \hat v_{1/2} & (\mathcal{M}_\mathrm{N}^2)_{36}
\end{matrix}\right.}\quad\dots\quad\hspace{20pt}\vspace*{10pt}\\
\dots\quad{ \left.\begin{matrix}
0                               & 0                               & \mathbb{I}\mathrm{m}\big((\mathcal{M}_\mathrm{N}^2)_{13} \big) \\
0                               & 0                               & -\mathbb{I}\mathrm{m}\big((\mathcal{M}_\mathrm{N}^2)_{13} \big) \hat v_{1/2} \\
-3\mathbb{I}\mathrm{m}\big((\mathcal{M}_\mathrm{N}^2)_{13} \big)& 3\mathbb{I}\mathrm{m}\big((\mathcal{M}_\mathrm{N}^2)_{13} \big) \hat v_{1/2}& (\mathcal{M}_\mathrm{N}^2)_{36} \\
0                               & 0                               & 3 \mathbb{R}\mathrm{e}\big((\mathcal{M}_\mathrm{N}^2)_{13}\big) \\
0                               & 0                               &-3\mathbb{R}\mathrm{e}\big((\mathcal{M}_\mathrm{N}^2)_{13} \big) \hat v_{1/2}\\
3 \mathbb{R}\mathrm{e}\big((\mathcal{M}_\mathrm{N}^2)_{13}\big) &-3\mathbb{R}\mathrm{e}\big((\mathcal{M}_\mathrm{N}^2)_{13}  \big) \hat v_{1/2} & (\mathcal{M}_\mathrm{N}^2)_{66} 
\end{matrix}\right)}, 
\end{multlined}
\end{equation}
where
\begin{subequations}
\begin{align}
(\mathcal{M}_\mathrm{N}^2)_{11} ={}& 2 \lambda_{1111} \hat v_1^2, \\
(\mathcal{M}_\mathrm{N}^2)_{12} ={}& \left( \lambda_{1122} + \lambda_{1221}\right) \hat v_1 \hat v_2 , \\
(\mathcal{M}_\mathrm{N}^2)_{13} ={}& -\frac{1}{2} \lambda_{1232}\hat v_2^2, \\
(\mathcal{M}_\mathrm{N}^2)_{22} ={}& 2 \lambda_{2222} \hat v_2^2,\\
(\mathcal{M}_\mathrm{N}^2)_{33} ={}& \mu_{33}^2 + \frac{1}{2}\left[ \left( \lambda_{1133} + \lambda_{1331} \right) \hat v_1^2 + 2 \lambda_{1323}^\mathrm{R} \hat v_1 \hat v_2 + \left( \lambda_{2233} + \lambda_{2332} \right)\hat v_2^2 \right],\\
(\mathcal{M}_\mathrm{N}^2)_{36} ={}& - \lambda_{1323}^\mathrm{I} \hat v_1 \hat v_2,\\
(\mathcal{M}_\mathrm{N}^2)_{66} ={}& \mu_{33}^2 + \frac{1}{2}\left[ \left( \lambda_{1133} + \lambda_{1331} \right) \hat v_1^2 - 2 \lambda_{1323}^\mathrm{R} \hat v_1 \hat v_2 + \left( \lambda_{2233} + \lambda_{2332} \right)\hat v_2^2 \right].
\end{align}
\end{subequations}

Both in the charged and neutral sectors, the mass eigenstates will be given in terms of all fields. Such mixing is due to the $\lambda_{1232}$ term, the $\lambda_{1213}$ term is fixed by the $\lambda_{1232}$ coupling. Requiring no mixing between $\{h_1, \, h_2\}$ and $h_3$, we would need to take $\lambda_{1232}= 0$, resulting in $\lambda_{1213}=0$ due to the minimisation condition. As a result, the phase-sensitive part of the potential would consist of a single $\lambda_{1323}$ term, and the overall symmetry would be increased to $U(1)_1$. An unwanted Goldstone boson would be present due to the breaking of this symmetry. It is possible to promote the state to a massive one by soft symmetry breaking. The only possible term is $\mu_{12}^2$ since other terms would introduce unwanted mixing with the $h_3$ doublet. This leads to an additional minimisation condition forcing the $\mu_{12}^2$ term to be real. For this vacuum, all terms of the softly broken $\mathbb{Z}_3$ potential are contained within the $\mathbb{Z}_2$-symmetric 3HDM, see \ref{App:Pot_Z2_v1_v2_0} in Appendix~\ref{App:Pot_Z2}.

\bigskip\textbf{Case of \boldmath$(\hat v_1 e^{i \sigma},\,\hat v_2,\,0)$ and real couplings}

Next, we consider the case of spontaneous CP violation. Allowing for a free phase $\sigma$ the minimisation conditions force $\lambda_{1232}^\mathrm{R} = \lambda_{1213}^\mathrm{R} = 0$. This case reduces to the $U(1)_1$-symmetric potential, and the $\sigma$ phase can then be rotated away. 

An exception arises for the choice of $\sigma = \pm \pi/3$, which no longer requires the above conditions on the phase-sensitive couplings. In this case the minimisation conditions are:
\begin{subequations}
\begin{align}
\mu_{11}^2 & = -\frac{1}{2} \left( 2 \lambda_{1111} \hat v_1^2 + \lambda_{1122} \hat v_2^2 + \lambda_{1221} \hat v_2^2 \right),\\
\mu_{22}^2 & = -\frac{1}{2} \left( 2 \lambda_{2222} \hat v_2^2 + \lambda_{1122} \hat v_1^2 + \lambda_{1221} \hat v_1^2 \right),\\
\lambda_{1213}^\mathrm{R} & = \frac{\hat v_2}{\hat v_1} \lambda_{1232}^\mathrm{R}.
\end{align}
\end{subequations}

The charged mass-squared matrix is:
\begin{equation}
\mathcal{M}_\mathrm{Ch}^2 = \begin{pmatrix}
(\mathcal{M}_\mathrm{Ch}^2)_{11} & -(\mathcal{M}_\mathrm{Ch}^2)_{11} \hat v_{1/2} & (\mathcal{M}_\mathrm{Ch}^2)_{13} \\
-(\mathcal{M}_\mathrm{Ch}^2)_{11}\hat v_{1/2} & (\mathcal{M}_\mathrm{Ch}^2)_{11} \hat v_{1/2}^2 & -(\mathcal{M}_\mathrm{Ch}^2)_{13} \hat v_{1/2} \\
(\mathcal{M}_\mathrm{Ch}^2)_{13}^\ast & -(\mathcal{M}_\mathrm{Ch}^2)_{13}^\ast \hat v_{1/2} & (\mathcal{M}_\mathrm{Ch}^2)_{33}   
\end{pmatrix},
\end{equation}
where
\begin{subequations}
\begin{align}
(\mathcal{M}_\mathrm{Ch}^2)_{11} ={}& -\frac{1}{2}\lambda_{1221}\hat v_2^2,\\
(\mathcal{M}_\mathrm{Ch}^2)_{13} ={}& -\frac{(1-i\sqrt{3})}{4}\lambda_{1232}^\mathrm{R} \hat v_2^2,\\
(\mathcal{M}_\mathrm{Ch}^2)_{33} ={}& \mu_{33}^2 + \frac{1}{2} \left( \lambda_{1133} \hat v_1^2 + \lambda_{2233} \hat v_2^2 \right).
\end{align}
\end{subequations}

The neutral mass-squared matrix in the basis $\{\eta_1,\, \eta_2,\, \eta_3,\, \chi_1,\, \chi_2,\,\chi_3\}$ is given by:
\begin{equation}\label{Eq:Z3_MN2_v1sv20}
\begin{multlined}[c]
\mathcal{M}_\mathrm{N}^2 = { \left(\begin{matrix}
 (\mathcal{M}_\mathrm{N}^2)_{11} & (\mathcal{M}_\mathrm{N}^2)_{12} & -\frac{1}{4}(\mathcal{M}_\mathrm{N}^2)_{13}\\
 (\mathcal{M}_\mathrm{N}^2)_{12} & (\mathcal{M}_\mathrm{N}^2)_{22}  & \frac{1}{4}(\mathcal{M}_\mathrm{N}^2)_{13} \hat v_{1/2}\\
 -\frac{1}{4}(\mathcal{M}_\mathrm{N}^2)_{13} & \frac{1}{4}(\mathcal{M}_\mathrm{N}^2)_{13} \hat v_{1/2} & (\mathcal{M}_\mathrm{N}^2)_{33}\\
 0 & 0 & -\frac{3 \sqrt{3}}{4} (\mathcal{M}_\mathrm{N}^2)_{13}\\
 0 & 0 & \frac{3 \sqrt{3} }{4}(\mathcal{M}_\mathrm{N}^2)_{13} \hat v_{1/2}\\
 \frac{\sqrt{3}}{4}(\mathcal{M}_\mathrm{N}^2)_{13} & -\frac{\sqrt{3}}{4} (\mathcal{M}_\mathrm{N}^2)_{13} \hat v_{1/2} & (\mathcal{M}_\mathrm{N}^2)_{36}\\
\end{matrix}\right.}\quad\dots\quad\hspace{20pt}\vspace*{10pt}\\
\dots\quad{ \left.\begin{matrix}
0 & 0 & \frac{\sqrt{3}}{4}(\mathcal{M}_\mathrm{N}^2)_{13} \\
0 & 0 & -\frac{\sqrt{3}}{4}  (\mathcal{M}_\mathrm{N}^2)_{13} \hat v_{1/2} \\
-\frac{3 \sqrt{3}}{4} (\mathcal{M}_\mathrm{N}^2)_{13} & \frac{3 \sqrt{3}}{4}(\mathcal{M}_\mathrm{N}^2)_{13} \hat v_{1/2} & (\mathcal{M}_\mathrm{N}^2)_{36} \\
0 & 0 & -\frac{3}{4} (\mathcal{M}_\mathrm{N}^2)_{13} \\
0 & 0 & \frac{3 }{4}(\mathcal{M}_\mathrm{N}^2)_{13} \hat v_{1/2} \\
-\frac{3}{4} (\mathcal{M}_\mathrm{N}^2)_{13} & \frac{3 }{4}(\mathcal{M}_\mathrm{N}^2)_{13} \hat v_{1/2} & (\mathcal{M}_\mathrm{N}^2)_{66} \\
\end{matrix}\right)}, 
\end{multlined}
\end{equation}

\clearpage
where
\begin{subequations}
\begin{align}
(\mathcal{M}_\mathrm{N}^2)_{11} ={}& 2 \lambda_{1111} \hat v_1^2, \\
(\mathcal{M}_\mathrm{N}^2)_{12} ={}& \left( \lambda_{1122} + \lambda_{1221}\right) \hat v_1 \hat v_2 , \\
(\mathcal{M}_\mathrm{N}^2)_{13} ={}& \lambda_{1232}^\mathrm{R}\hat v_2^2, \\
(\mathcal{M}_\mathrm{N}^2)_{22} ={}& 2 \lambda_{2222} \hat v_2^2,\\
(\mathcal{M}_\mathrm{N}^2)_{33} ={}& \mu_{33}^2 + \frac{1}{2}\left[ \left( \lambda_{1133} + \lambda_{1331} \right) \hat v_1^2 + \lambda_{1323}^\mathrm{R} \hat v_1 \hat v_2 + \left( \lambda_{2233} + \lambda_{2332} \right)\hat v_2^2 \right],\\
(\mathcal{M}_\mathrm{N}^2)_{36} ={}& \frac{\sqrt{3}}{2} \lambda_{1323}^\mathrm{R} \hat v_1 \hat v_2,\\
(\mathcal{M}_\mathrm{N}^2)_{66} ={}& \mu_{33}^2 + \frac{1}{2}\left[ \left( \lambda_{1133} + \lambda_{1331} \right) \hat v_1^2 - \lambda_{1323}^\mathrm{R} \hat v_1 \hat v_2 + \left( \lambda_{2233} + \lambda_{2332} \right)\hat v_2^2 \right].
\end{align}
\end{subequations}

Since there is mixing between $\{h_1,\, h_2\}$ and $h_3$ one could force no mixing at tree level by requiring $\lambda_{1232}^\mathrm{R}=0$, which would also result in $\lambda_{1213}^\mathrm{R}=0$. Then, the symmetry would again be increased to $U(1)_1$. Inspection of $\mathcal{M}_\mathrm{N}^2 $ of eq.~\eqref{Eq:Z3_MN2_v1sv20} shows that $\chi_1$ and $\chi_2$ are massless. We cannot eliminate the unwanted massless state via the introduction of the soft symmetry-breaking term $\left( \mu_{12}^2 \right)^\mathrm{R}$ since it would not survive the minimisation conditions.

\subsection{Two vanishing vevs}

\bigskip\textbf{Case of \boldmath$(v,\,0,\,0)$}

In the case $(v,\,0,\,0)$ there is a single minimisation condition given by:
\begin{equation}
\mu_{11}^2 = - \lambda_{1111} v^2.
\end{equation}

The charged mass-squared matrix is diagonal with the mass-squared parameters:
\begin{subequations}
\begin{align}
m_{h_2^\pm}^2 & = \mu_{22}^2 + \frac{v^2}{2} \lambda_{1122},\\
m_{h_3^\pm}^2 & = \mu_{33}^2 + \frac{v^2}{2} \lambda_{1133}.
\end{align}
\end{subequations}

The neutral mass-squared matrix in the basis $\{\eta_1\}{-}\{\chi_1\}{-}\{\eta_2,\, \eta_3,\, \chi_2,\,\chi_3\}$ is:
\begin{equation}
\mathcal{M}_\mathrm{N}^2 = \begin{pmatrix}
m_h^2 & 0 & 0 & 0 & 0 & 0 \\
0 & 0 & 0 & 0 & 0 & 0 \\
0 & 0 & (\mathcal{M}_\mathrm{N}^2)_{11} & \mathbb{R}\mathrm{e}\big((\mathcal{M}_\mathrm{N}^2)_{12} \big) & 0 & -\mathbb{I}\mathrm{m}\big((\mathcal{M}_\mathrm{N}^2)_{12} \big),\\
0 & 0 & \mathbb{R}\mathrm{e}\big((\mathcal{M}_\mathrm{N}^2)_{12} \big) & (\mathcal{M}_\mathrm{N}^2)_{22} & -\mathbb{I}\mathrm{m}\big((\mathcal{M}_\mathrm{N}^2)_{12} \big) & 0 \\
0 & 0 & 0 & -\mathbb{I}\mathrm{m}\big((\mathcal{M}_\mathrm{N}^2)_{12} \big) & (\mathcal{M}_\mathrm{N}^2)_{11} & -\mathbb{R}\mathrm{e}\big((\mathcal{M}_\mathrm{N}^2)_{12} \big) \\
0 & 0 & -\mathbb{I}\mathrm{m}\big((\mathcal{M}_\mathrm{N}^2)_{12} \big) & 0 & -\mathbb{R}\mathrm{e}\big((\mathcal{M}_\mathrm{N}^2)_{12} \big) & (\mathcal{M}_\mathrm{N}^2)_{22}
\end{pmatrix},
\end{equation}
where
\begin{subequations}
\begin{align}
m_h^2 ={}& 2 \lambda_{1111} v^2,\\
(\mathcal{M}_\mathrm{N}^2)_{11}={}& \mu_{22}^2 + \frac{v^2}{2} \left( \lambda_{1122} + \lambda_{1221} \right),\\
(\mathcal{M}_\mathrm{N}^2)_{12}={}& \frac{1}{2} \lambda_{1213} v^2,\\
(\mathcal{M}_\mathrm{N}^2)_{22}={}& \mu_{33}^2 + \frac{v^2}{2} \left( \lambda_{1133} + \lambda_{1331} \right).
\end{align}
\end{subequations}
The neutral mass-squared matrix shows that all scalars associated with $h_2$ and $h_3$ mix. There is freedom to remove some of the complex phases of the scalar potential by re-phasing $h_2$ and $h_3$. An obvious choice would be to absorb the phase of $\lambda_{1213}$ to simplify the neutral mass-squared matrix by making $\mathbb{I}\mathrm{m}\big((\mathcal{M}_\mathrm{N}^2)_{12} \big)=0$. There are two pairs of mass-degenerate states.

\section{Minimisation conditions}\label{App:Derivatives}

We present the minimisation conditions of the most general scalar potential, which includes all the terms present in the  $\mathbb{Z}_2$-symmetric and the $\mathbb{Z}_3$-symmetric 3HDM potential,
\begin{equation}
\begin{aligned}
V ={}& \sum_i \mu_{ii}^2 h_{ii} + \sum_i \lambda_{iijj} h_{ii} h_{jj} + \sum_{i<j} \lambda_{ijji} h_{ij} h_{ji} + \sum_{i<j} \lambda_{ijij} h_{ij}^2\\
& + \Big\{  \mu_{12}^2 h_{12} +  \lambda_{1112} h_{11}h_{12} + \lambda_{1222} h_{12}h_{22} + \lambda_{1233} h_{12}h_{33} + \lambda_{1323} h_{13}h_{23} \\
&\hspace{20pt} + \lambda_{1332} h_{13} h_{32} + \lambda_{1213} h_{12} h_{13} + \lambda_{1232} h_{12} h_{32} + \mathrm{h.c.}  \Big\}.
\end{aligned}
\end{equation}
The first-order derivatives are:
\begin{subequations}
\begin{align}
\begin{split} \dfrac{\partial V}{\partial \eta_1} \Bigg|_v ={}& \mu_{11}^2 \hat v_1 + (\mu_{12}^2)^\mathrm{R} \hat v_2+ \frac{1}{2} \Bigg[\left( 2 \lambda_{1213}^\mathrm{R} \hat v_1 + \lambda_{1232}^\mathrm{R} \hat v_2 \right) \hat v_2 \hat v_3 + 2 \lambda_{1111} \hat v_1^3\\
& \hspace{40pt}   + \Big( 3 \lambda_{1112}^\mathrm{R} \hat v_1^2 + \left[ \lambda_{1122} + 2 \lambda_{1212}^\mathrm{R} + \lambda_{1221} \right]\hat v_1 \hat v_2  + \lambda_{1222}^\mathrm{R} \hat v_2^2 \Big) \hat v_2\\
& \hspace{40pt} + \Big( \left[ \lambda_{1133} + 2 \lambda_{1313}^\mathrm{R} + \lambda_{1331} \right] \hat v_1 + \left[ \lambda_{1233}^\mathrm{R} + \lambda_{1323}^\mathrm{R} + \lambda_{1332}^\mathrm{R}\right]\hat v_2 \Big)\hat v_3^2 \Bigg],
\end{split}\\
\begin{split} \dfrac{\partial V}{\partial \eta_2} \Bigg|_v ={}& \mu_{22}^2 \hat v_2 + (\mu_{12}^2)^\mathrm{R} \hat v_1+ \frac{1}{2} \Bigg[\left( 2 \lambda_{1232}^\mathrm{R} \hat v_2 + \lambda_{1213}^\mathrm{R} \hat v_1 \right) \hat v_1 \hat v_3 + 2\lambda_{2222} \hat v_2^3 +  \lambda_{1112}^\mathrm{R} \hat v_1^3\\
& \hspace{40pt} + \Big( 3 \lambda_{1222}^\mathrm{R} \hat v_1 \hat v_2 + \left[ \lambda_{1122} + 2 \lambda_{1212}^\mathrm{R} + \lambda_{1221} \right]\hat v_1^2   \Big) \hat v_2\\
& \hspace{40pt} + \Big( \left[ \lambda_{2233} + 2 \lambda_{2323}^\mathrm{R} + \lambda_{2332} \right] \hat v_2 + \left[ \lambda_{1233}^\mathrm{R} + \lambda_{1323}^\mathrm{R} + \lambda_{1332}^\mathrm{R}\right]\hat v_1 \Big)\hat v_3^2 \Bigg],
\end{split}\\
\begin{split} \dfrac{\partial V}{\partial \eta_3} \Bigg|_v ={}& \mu_{33}^2 \hat v_3 + \frac{1}{2} \Bigg[
\left( \lambda_{1213}^\mathrm{R} \hat v_1 + \lambda_{1232}^\mathrm{R} \hat v_2 \right) \hat v_1 \hat v_2 + 2\lambda_{3333} \hat v_3^3\\
& \hspace{30pt} + 2 \left( \lambda_{1233}^\mathrm{R} + \lambda_{1323}^\mathrm{R} + \lambda_{1332}^\mathrm{R}\right) \hat v_1 \hat v_2 \hat v_3\\
& \hspace{30pt} + \Big( \left[ \lambda_{1133} + 2 \lambda_{1313}^\mathrm{R} + \lambda_{1331} \right] \hat v_1^2  +  \left[ \lambda_{2233} + 2 \lambda_{2323}^\mathrm{R} + \lambda_{2332} \right] \hat v_2^2\Big)\hat v_3 \Bigg],
\end{split}\\
\begin{split} \dfrac{\partial V}{\partial \chi_1} \Bigg|_v ={}& (\mu_{12}^2)^\mathrm{I} \hat v_2+ \frac{1}{2} \Bigg[\left( 2 \lambda_{1213}^\mathrm{I} \hat v_1 + \lambda_{1232}^\mathrm{I} \hat v_2 \right) \hat v_2 \hat v_3 \\
& \hspace{70pt}+ \Big( \lambda_{1112}^\mathrm{I} \hat v_1^2 + \left[ 2\lambda_{1212}^\mathrm{I} \hat v_1 + \lambda_{1222}^\mathrm{I} \hat v_2 \right] \hat v_2  \Big) \hat v_2\\
& \hspace{70pt}+ \Big(  2 \lambda_{1313}^\mathrm{I} \hat v_1 + \left[ \lambda_{1233}^\mathrm{I} + \lambda_{1323}^\mathrm{I} + \lambda_{1332}^\mathrm{I} \right]\hat v_2 \Big) \hat v_3^2 \Bigg],
\end{split}\\
\begin{split} \dfrac{\partial V}{\partial \chi_2} \Bigg|_v ={}& - (\mu_{12}^2)^\mathrm{I} \hat v_1 - \frac{1}{2} \Bigg[\left( \lambda_{1213}^\mathrm{I} \hat v_1 + 2 \lambda_{1232}^\mathrm{I} \hat v_2 \right) \hat v_1 \hat v_3 \\
& \hspace{80pt}+ \Big( \lambda_{1112}^\mathrm{I} \hat v_1^2 + \left[ 2\lambda_{1212}^\mathrm{I} \hat v_1 + \lambda_{1222}^\mathrm{I} \hat v_2 \right] \hat v_2  \Big) \hat v_1\\
& \hspace{80pt}- \Big(  2 \lambda_{2323}^\mathrm{I} \hat v_2 + \left[ \lambda_{1323}^\mathrm{I} - \lambda_{1332}^\mathrm{I} - \lambda_{1233}^\mathrm{I} \right]\hat v_1 \Big) \hat v_3^2 \Bigg],
\end{split}\\
\begin{split} \dfrac{\partial V}{\partial \chi_3} \Bigg|_v ={}& -\frac{1}{2}\left( \lambda_{1213}^\mathrm{I} \hat v_1 - \lambda_{1232}^\mathrm{I} \hat v_2  \right) \hat v_1 \hat v_2 - \left[ \lambda_{1313}^\mathrm{I} \hat v_1^2 + \left( \lambda_{1323}^\mathrm{I} \hat v_1 + \lambda_{2323}^\mathrm{I} \hat v_2 \right) \hat v_2 \right] \hat v_3.
\end{split}
\end{align}
\end{subequations}

\section{Three symmetries in the bilinear space formalism}\label{App:Bilinear_3symm}

The quartic couplings of the scalar potential can be represented in the bilinear space formalism~\cite{Nishi:2006tg,Maniatis:2006fs,Nishi:2007nh,Maniatis:2007vn,Ivanov:2010ww,Maniatis:2014oza} by:
\begin{equation}
V = \Lambda_0 r_0^2 + L_i r_0 r_i + \Lambda_{ij} r_i r_j,
\end{equation}
where $r_i$ are the gauge-invariant bilinear combinations in the Gell-Mann basis. We follow the notation of Ref.~\cite{deMedeirosVarzielas:2019rrp} since it presents some essential tools for identification of the underlying symmetries in a basis-independent way. We shall present the quartic couplings for three symmetries in the form:
\begin{equation}
\Lambda = \begin{pmatrix}
\Lambda_0 & L_i \\
L_i^\mathrm{T} & \Lambda_{ij}
\end{pmatrix}.
\end{equation}

As identified in Sections~\ref{Sec:Ident_O2_U1}, \ref{Sec:Ident_U1_U1_S3} and \ref{Sec:Ident_U1_Z2_S2}, there are three cases which have not been discussed in the literature. These are given below in the bilinear formalism.

\bigskip\textbf{$\bullet$ Case of \boldmath$O(2) \times U(1)$ [Sections~\ref{Sec:Ident_O2_U1} and \ref{Sec:Pot_O2_U1}]}


\begin{equation}
\Lambda=
\begin{pmatrix}
 \lambda_a & 0 & 0 & 0 & \multirow{4}{*}{\vdots}  & \lambda_b \\
 0 & \lambda_{1221} & 0 & 0 &  & 0 \\
 0 & 0 & \lambda_{1221} & 0 &  & 0 \\
 0 & 0 & 0 & 2 \lambda_{1111}-\lambda_{1122} &  & 0\\
   \multicolumn{4}{c}{\dots}& \mathcal{I}_4 \otimes \lambda_{1331} & \dots \\[10pt]
 \lambda_b & 0 & 0 & 0 & \vdots & \lambda_c
\end{pmatrix},
\end{equation} 

where
\begin{subequations}\label{Eq:lambda_abc}
\begin{align}
\lambda_a ={}& \frac{1}{3} \left( 2 \lambda_{1111} + \lambda_{1122} + 2 \lambda_{1133} + \lambda_{3333} \right),\\
\lambda_b ={}& \frac{1}{3} \left( 2 \lambda_{1111} + \lambda_{1122} - \lambda_{1133} - 2\lambda_{3333} \right),\\
\lambda_c ={}& \frac{1}{3} \left( 2 \lambda_{1111} + \lambda_{1122} - 4 \lambda_{1133} + 4\lambda_{3333} \right).
\end{align}
\end{subequations}

The degeneracy pattern of eigenvalues of $\Lambda_{ij}$ is $1+1+2+4$.

\bigskip\textbf{ $\bullet$ Case of \boldmath$\left[ U(1) \times U(1) \right] \rtimes S_3$ [Sections~\ref{Sec:Ident_U1_U1_S3} and \ref{Sec:Pot_U1_U1_S3}]}
\begin{equation}
\Lambda = \mathrm{diag} \left( \lambda_{1111} + \lambda_{1122},\, \mathcal{I}_2 \times \lambda_{1221},\, 2 \lambda_{1111} - \lambda_{1122},\, \mathcal{I}_4 \times \lambda_{1221},\, 2 \lambda_{1111} - \lambda_{1122}  \right).
\end{equation}

The degeneracy pattern of eigenvalues of $\Lambda_{ij}$ is $2+6$.

\bigskip\textbf{ $\bullet$ Case of \boldmath$U(1) \times D_4$ [Sections~\ref{Sec:Ident_U1_Z2_S2} and \ref{Sec:Pot_U1_Z2_S2}]}

\begin{equation}
\Lambda=
\scriptstyle\begin{pmatrix}
 \lambda_a & 0 & 0 & 0 & 0 & 0 & 0 & 0 & \lambda_b \\
 0 & \lambda_{1221}+2\lambda_{1212} & 0 & 0 & 0 & 0 & 0 & 0 & 0 \\
 0 & 0 & \lambda_{1221}-2\lambda_{1212} & 0 & 0 & 0 & 0 & 0 & 0 \\
 0 & 0 & 0 & 2 \lambda_{1111}-\lambda_{1122} & 0 & 0 & 0 & 0 & 0 \\
 0 & 0 & 0 & 0 & \lambda_{1331} & 0 & 0 & 0 & 0 \\
 0 & 0 & 0 & 0 & 0 & \lambda_{1331} & 0 & 0 & 0 \\
 0 & 0 & 0 & 0 & 0 & 0 & \lambda_{1331} & 0 & 0 \\
 0 & 0 & 0 & 0 & 0 & 0 & 0 & \lambda_{1331} & 0 \\
 \lambda_b & 0 & 0 & 0 & 0 & 0 & 0 & 0 & \lambda_c
\end{pmatrix},
\end{equation}
where the $\lambda_i$ are those of eqs.~\eqref{Eq:lambda_abc}.

The degeneracy pattern of eigenvalues of $\Lambda_{ij}$ is $1+1+1+1+4$.

\bibliographystyle{JHEP}
\bibliography{ref}

\end{document}